\documentclass[fleqn,usenatbib]{mnras}

\usepackage{newtxtext,newtxmath}

\usepackage[T1]{fontenc}
\usepackage{ae,aecompl}

\usepackage{subfigure}
\usepackage{graphicx}

\usepackage{graphicx}	
\usepackage{amsmath}	
\usepackage{amssymb}	
\usepackage{bm}       
\usepackage{etoolbox}
\makeatletter 
  \patchcmd{\NAT@citex}
    {\@citea\NAT@hyper@{%
      \NAT@nmfmt{\NAT@nm}%
      \hyper@natlinkbreak{\NAT@aysep\NAT@spacechar}{\@citeb\@extra@b@citeb}%
      \NAT@date}}
    {\@citea\NAT@nmfmt{\NAT@nm}%
    \NAT@aysep\NAT@spacechar\NAT@hyper@{\NAT@date}}{}{}

  \patchcmd{\NAT@citex}
    {\@citea\NAT@hyper@{%
      \NAT@nmfmt{\NAT@nm}%
      \hyper@natlinkbreak{\NAT@spacechar\NAT@@open\if*#1*\else#1\NAT@spacechar\fi}%
        {\@citeb\@extra@b@citeb}%
      \NAT@date}}
    {\@citea\NAT@nmfmt{\NAT@nm}%
    \NAT@spacechar\NAT@@open\if*#1*\else#1\NAT@spacechar\fi\NAT@hyper@{\NAT@date}}
    {}{}
\makeatother

\newcommand{\beq}{\begin{equation}}
\newcommand{\eeq}{\end{equation}}
\newcommand{\bea}{\begin{eqnarray}}
\newcommand{\eea}{\end{eqnarray}}

\newcommand{\avg}[1]{\left\langle #1 \right\rangle}

\newcommand{\iso}[2]{\mbox{$^{#2}$#1}}

\title[Dust formation in SN 1987A]{Molecular nucleation theory of dust formation in core-collapse supernovae applied to SN 1987A}

\author[A. Sluder et al.]
{Alan~Sluder$^1$, Milo\v  s~Milosavljevi\'c$^1$, 
and Michael~H.~Montgomery$^{1,2}$ \\
$^1$Department of Astronomy, University of Texas at Austin, Austin, TX 78712, USA \\
$^2$McDonald Observatory, University of Texas at Austin, TX 78712, USA 
 }

\renewcommand{\u}[1]{\,\textrm{#1}}
\newcommand{\f}[2]{\frac{#1}{#2}}
\renewcommand{\(}[1]{\left(#1 \right)}
\renewcommand{\[}[1]{\left[ #1\right]}
\renewcommand{\=}{&=&}
\newcommand{\m}{\mathcal}
\newcommand{\h}{\\\hline}
\newcommand{\kboltzmann}{{k_{\rm B}}}
\newcommand{\pw}[1]{\left\{
\begin{array}{ll} #1
\end{array}
\right.}

\begin{document}

\label{firstpage}

\maketitle

\topmargin-1cm

\begin{abstract}

We model dust formation in the core collapse supernova explosion SN 1987A by treating the gas-phase formation of dust grain nuclei as a chemical process.  To compute the synthesis of fourteen species of grains we integrate a non-equilibrium network of nucleating and related chemical reactions and follow the growth of the nuclei into grains via accretion and coagulation. The effects of the radioactive \iso{Co}{56}, \iso{Co}{57}, \iso{Ti}{44}, and \iso{Na}{22} on the thermodynamics and chemistry of the ejecta are taken into account. The grain temperature, which we allow to differ from the gas temperature, affects the surface-tension-corrected evaporation rate. We also account for He$^+$, Ne$^+$, Ar$^+$, and O weathering. We combine our dust synthesis model with a crude prescription for anisotropic \iso{Ni}{56} dredge-up into the core ejecta, the so-called ``nickel bubbles'', to compute the total dust mass and molecular-species-specific grain size distribution. The total mass varies between $0.41\,M_\odot$ and $0.73\,M_\odot$, depending on the bubble shell density contrast. In the decreasing order of abundance, the grain species produced are: magnesia, silicon, forsterite, iron sulfide, carbon, silicon dioxide, alumina, and iron. The combined grain size distribution is a power law $dN/da\propto a^{-4.39}$. Early ejecta compaction by expanding radioactive \iso{Ni}{56} bubbles strongly enhances dust synthesis. This underscores the need for improved understanding of hydrodynamic transport and mixing over the entire pre-homologous expansion.

\end{abstract}


\section{Introduction}
\label{sec:intro}

Dust grains are important throughout astrophysics. They absorb ultraviolet (UV) and visible light and radiate in the infrared (IR). This produces the extinction and reddening \citep{1990ARA&A..28...37M} that must be taken into account when inferring the properties of astronomical sources such as the star formation rates of galaxies \citep{1998ARA&A..36..189K,Calzetti:00,Dunne:11}. Dust grains are a component of the interstellar medium (ISM) that is essential to the star formation process \citep{Draine:03,2007ARA&A..45..565M,Draine:11,2012ARA&A..50..531K}. 
They shield the interiors of dense molecular clouds from molecule-dissociating radiation. They act as cooling agents in star-forming gas clouds and as catalysts for formation of the molecules, such as H$_2$, that do not form efficiently in the gas phase \citep{2002ApJ...575L..29C}.  Grains are also essential in planet formation \citep{2011ARA&A..49...67W}. Since grains are composed of refractory elements such as carbon, oxygen, and silicon, these elements are depleted from the gas phase. In the densest, coldest molecular gas, volatile compounds such as H$_2$O and CO$_2$ form icy mantles on refractory grain cores. Radiation pressure on grains can drive winds from cool, evolved stars and potentially also drive outflows in active galactic nuclei and starbursts.

Given the ubiquity and importance of dust grains in the cosmos, it is vital that we understand how they are produced, modified, and destroyed. In principle, dust grains can form in any environment where an initially hot, dense gas expands and cools, as in explosions and outflows, or where a gas cloud is being compressed isothermally to high densities, as in pre-stellar cores. Decades of research, however, point to the stellar winds from the cool atmospheres of asymptotic giant branch and supergiant stars \citep[e.g.,][]{Ferrarotti:06} and the expanding ejecta of supernovae \citep[e.g.,][]{Clayton:97,Derdzinksi:17} as the main contributors of dust. Interstellar dust grains may also form in novae \cite[e.g.,][]{Mitchell:84,Rawlings:89}, in outflows from active galactic nuclei \citep[e.g.,][]{Elvis:02}, in the material ejected in stellar mergers and common envelope systems \citep[e.g.,][]{Lu:13}, in the colliding winds of Wolf-Rayet stars \citep[e.g.,][]{Crowther:03}, and in extreme mass loss events in luminous blue variables \citep[e.g.,][]{Kochanek:11}. It is unknown exactly what fraction of dust mass comes from each of these classes of events.  The origin of this uncertainty seems to be an incomplete theoretical understanding of the astrophysics of dust formation.

Dust formation can be directly observed in nearby core-collapse supernovae through the dust's imprint on supernova spectra \citep{Sugerman:06,Fox:09,Fox:10,Kotak:09,Sakon:09,Inserra:11,Meikle:11,Szalai:11,Maeda:13}. As dust grains condense in supernova ejecta, the spectrum of the supernova changes in three characteristic ways. The optical luminosity of the ejecta decreases due to absorption by grains. The IR luminosity increases as the grains reradiate the absorbed energy in the IR while the total luminosity of the ejecta decreases following the progression of radioactive decay. The peaks of optical emission lines are blueshifted as optical photons from the far side of the ejecta are more likely to be absorbed \citep[e.g.,][]{Smith:12}.   

 Due to its proximity, SN 1987A in the Large Magellanic Cloud has been the best studied case of dust formation in supernovae \citep[e.g.,][]{Gehrz:87,Gehrz:90,Dwek:88,Kozasa:89,Kozasa:91,Moseley:89,McCray:93,Colgan:94,Ercolano:07,VanDyk:13,Indebetouw:14,Matsuura:15,Wesson:15}. \citet{Wesson:15} used three-dimensional radiation transfer calculations to simulate the evolution of the spectral energy distribution (SED) of SN 1987A while varying the dust mass, grain chemical composition, grain size distribution, and location of dust in the ejecta. They are able to reproduce the observed SEDs if: (1) dust mass increases from $0.001\,M_\odot$ at 615 days to $0.8\,M_\odot$ at 9200 days after the explosion, (2) while the dust mass always increases, most of the dust forms well after 1000 days, (3) the dust is mostly carbon with some silicates (perhaps 85\% carbon and 15\% silicates), (4) the grain radius distribution has a logarithmic slope of $-3.5$ but the typical grain radius increases from $\approx 0.04\,\mu\text{m}$ at 615 days after the explosion to $3\,\mu\text{m}$ at 9200 days, (5) dust forms in clumps that occupy $\sim 10\%$ of the volume of the ejecta and have clump radii $\sim1/30$ of the ejecta radius (so there are $\approx 2700$ clumps), and (6) the clumps expand sub-homologously. \citet{2016MNRAS.456.1269B} have confirmed these relatively large dust masses by modeling the observed emission line profiles with Monte Carlo radiation transfer calculations. In contrast, \citet{Dwek:15} infer a sharply different, early and rapid dust mass evolution. By 615 days, the dust mass has already reached $0.45\,M_\odot$, with $0.4\,M_\odot$ in silicates and $0.05\,M_\odot$ in amorphous carbon, and that over the following two decades, the dust mass does not increase appreciably.  In this work we attempt to shed light on this apparent disagreement.
 
There are a few other supernovae that have shown evidence of dust formation, mostly through blueshifted emission lines \citep{2012ApJ...751...25M}. However, due to the large distances, it is typically difficult to observe the signatures of dust formation.  Alternatively, one can search for evidence of dust in supernova remnants in the Milky Way and its satellite galaxies \citep[e.g.,][]{Sandstrom:09,Rho:09}. Probably the most studied such object is Cassiopeia A, the remnant of a supernova at a distance of $\sim 3.5\,\text{kpc}$ that was observed to explode about 300 years ago \citep[see, e.g.,][]{Dunne:09}. Detecting alumina, carbon, enstatite, forsterite, magnesium protosilicates, silicon dioxide, silicon, iron, iron oxide, and iron sulfide with the Spitzer Space Telescope spectroscopy, \citet{Rho:08} showed that $0.02-0.054\,M_\odot$ of dust has formed in its ejecta.
\citet{DeLooze:17} used spatially resolved Herschel and Spitzer observations of Cas A to infer a cold dust mass of $0.1\, M_\odot < M_{\rm dust} < 0.6\,M_\odot$ in the unshocked ejecta.
\citet{Bevan:16b} used the \textsc{damocles} Monte Carlo radiation transfer code and observations of the blueshifted emission lines in the spectrum of SN 1980K, SN 1993J, and Cas A to infer a dust mass of  $0.12\,M_\odot < M_{\rm dust} < 0.3\,M_\odot$ in SN 1980K, $0.08\,M_\odot < M_{\rm dust} < 0.1\, M_\odot$ in SN 1993J, and $M_{\rm dust} \approx 1.1 M_\odot$ in Cas A.
 A significant dust mass has also been detected in the Crab nebula \citep{Gomez:12,Owen:15}. Its IR spectrum can be fitted with a dust size distribution that is a power law with slope between $-3.5$ and $-4$ \citep{Temim:13}.

Supernovae provide unique physical conditions for the production of dust grains. While the average metal mass fraction in galaxies is of the order of 1\%, supernova ejecta can be 100\% metal. The ejecta is exposed to the $\gamma$ rays, X-rays, and nonthermal electrons and positrons produced in the radioactive decay of \iso{Co}{56}, \iso{Co}{57}, \iso{Ti}{44}, and \iso{Na}{22}. These nonthermal particles ionize atoms and dissociate molecules and thus modify the chemistry of the ejecta.  For example, destruction of molecules can liberate metals to become incorporated in grains, whereas ionization of noble gas atoms provides agents for grain weathering.

The grains produced in the ejecta must ultimately survive destruction in the reverse shock before becoming a part of the ISM \citep[e.g.,][]{Biscaro:16,Micelotta:16}.   How much of the dust made in a supernova makes it to the ISM depends strongly on the grain size distribution, with larger grains in denser clumps more likely to survive the reverse shock \citep[e.g.,][]{Bianchi:07,Nozawa:07,Nozawa:10,Bocchio:16}. After newly formed grains enter the ISM, they are modified by shock waves created by supernovae, by coagulation, by cosmic ray sputtering, and by accretion of gas phase metals (and volatiles such as H$_2$O and CO). Grains can also be destroyed if they become incorporated in stars.

The dust grains' effects depend on the chemical composition and size. These properties should not be spatially uniform in the ISM because grains form in some environments and are modified in others. For example, extinction curve variation shows that that grains in dense molecular cloud cores have different properties than those in the diffuse ISM \citep{2009ApJ...690..496C}.  In an attempt to model the grain properties, theoretical computations of dust grain formation have been attempted at various levels of physical realism, each one adding a formidable layer of complexity.  Specifically, in the 30 years since SN1987A, three significantly different approaches to simulating dust formation in supernovae have emerged.

 The simplest approach is the classical nucleation theory (CNT) that treats grain formation as a barrier-crossing problem in which the free energy of a small cluster of atoms first increases as atoms are added to the cluster.  When a critical cluster size is reached, the free energy then begins to decrease as further atoms are added. The CNT provides the rate per unit volume, called the nucleation current, at which critical-size clusters come into existence, as well as the rate at which the nucleated clusters grow into grains by accreting gas-phase atoms. To estimate the nucleation current, the CNT assumes that a steady state has been attained between monomer attachment and detachment.  The CNT ignores the actual chemical reactions participating in the formation of the cluster. It assumes that clusters have thermodynamic properties of the bulk material from which they are made and are subject to surface tension.  It ignores chemical reactions that can destroy grains and ignores grain growth by coagulation.  Thanks to its simplicity, the CNT has been widely used, for example by \citet{Kozasa:89,Kozasa:91} in the modeling of dust grain formation in SN 1987A. \citet{Todini:01} used the CNT to model dust formation in core collapse supernovae from Population III star progenitors and \citet{Schneider:04} for dust formation in pair-instability supernovae (also from Population III stellar progenitors). \citet{Bianchi:07} used it to calculate the amount of dust produced in a SN 1987A-like explosion. Recently, \citet{Marassi:15} used it on a grid of progenitor and explosion models to compute the properties of grains formed in Population III supernovae.

The second method of modeling dust formation in supernovae is kinetic nucleation theory (KNT). In the KNT, the number densities $c_n$ of clusters of $n\geq 2$ atoms (called $n$-mers) are explicitly tracked. Grains are allowed to grow by addition of atoms (condensation) and erode by removal of atoms (evaporation). The condensation rate is computed from kinetic theory and the evaporation rate by applying the principle of detailed balance. The KNT is more realistic than the CNT in that it does not assume a steady state between condensation and evaporation. However it still ignores the actual chemical reactions participating in the formation of the initial seed nucleus of a dust grain.  In modeling the evaporation rate, it assumes that the grains has thermodynamic properties of a solid bulk material with surface tension correction.  It ignores chemical reactions contributing to grain destruction and also ignores grain growth by coagulation. The elements of this technique can be found in \citet{Nozawa:03} and \citet{Nozawa:13}. \citet{Nozawa:08a} used the KNT to model dust formation in SN 2006jc and \citet{Lazzati:16} for the formation of carbon grains in core-collapse supernova ejecta.

The third approach to modeling dust formation, one that we will adopt, could be called `molecular nucleation theory' (MNT). It explicitly tracks the abundance of each molecular species (such as CO and SiO) with a non-equilibrium chemical reaction network.  Specifically, it follows the chemical binding of clusters of monomers such as C$_4$ or Mg$_4$Si$_2$O$_8$ into larger $n$-mers that are still treated as molecular entities.  The molecules that have reached a certain size can then act as grain condensation and coagulation nuclei.  MNT was introduced in \citet{Cherchneff:08} that investigated molecular synthesis in a pair-instability-type Population III supernova.  \citet{Cherchneff:09} included the effects of radioactivity and \citet{Cherchneff:10} further computed the formation of condensation nuclei for various types of dust grains in both pair instability and core collapse supernovae.  \citet{Sarangi:13} extended this framework to cluster nucleation in Type II-P supernovae.  These applications of MNT did not treat dust grain growth, but only the formation of the molecular and cluster precursors that is the first stage of dust grain formation.  \citet{Sarangi:15} extended MNT to the grains themselves and began estimating the grain size distribution and total dust mass yield for various grain types. 

The cited studies of dust formation in supernovae assumed that supernova ejecta were either fully mixed (single zone models) or spherically symmetric (one-dimensional models). In one-dimensional models the ejecta are divided into concentric shells, each characterized by an initial elemental composition and prescribed thermal evolution. The shells at smaller radial mass coordinates contain heavier elements and expand from higher initial densities and temperatures. Recent realistic three-dimensional simulations of supernovae, however, suggest that the ejecta are not spherically symmetric \citep{Hammer:10,Wongwathanarat:10}. Heavy elements such as \iso{Ni}{56} can be ejected ahead of lighter elements such as \iso{C}{12} \citep{Wongwathanarat:13,Wongwathanarat:15,Wongwathanarat:16}.  This anisotropy is a consequence of the amplification of non-spherical perturbations by the Rayleigh-Taylor instability (also called Ritchmyer-Meshkov instability in impulsively accelerated fluid).
Sources of initial perturbations are turbulence and
convection during the unstable, dynamical inner-shell (e.g., silicon)
burning in the progenitor \citep{Arnett:11,Ono:13,Smith:14,Couch:15,Mueller:16} as well as neutrino-induced convection behind the stalled shock wave and the standing accretion shock instability \citep[e.g.,][and references therein]{Hanke:13,Abdikamalov:15,Lentz:15}.  Perturbations are amplified into nonlinear fingers in compositional interfaces where the mean molecular mass and the density drop sharply outward \citep[e.g.,][]{Mao:15,Wongwathanarat:15}.  The interfaces are unstable because in the aftermath of the shock crossing, they are where the acceleration vector (relative a local freely falling frame) aligns with a strong density gradient, both pointing inward.

A consequence of the strongly-aspherical explosion geometry is that blobs of the \iso{Ni}{56} synthesized during $\alpha$-rich freezeout of complete explosive silicon burning are ejected into, and become embedded within, lighter-element material. The \iso{Ni}{56} decays into \iso{Co}{56} which then decays into \iso{Fe}{56}.  These radioactive decays deposit thermal energy in the gas.   The heating raises the pressure in the blobs above the pressure in the surrounding ejecta. The overpressured  \iso{Ni}{56}-enriched blobs are termed ``bubbles'' \citep[e.g.,][]{Fryxell:91,Herant:92}.   The bubbles expand super-homologously with respect to the rest of the ejecta. The expanding bubbles can sweep up thin, high density shells. Once the bubbles and their shells become optically thin to the $\gamma$ rays emitted during radioactive decay, their interior pressure drops and they resume homologous expansion \citep{2005ApJ...626..183W}. The shell surrounding a bubble may itself become Rayleigh-Taylor unstable and fragment \citep{1994ApJ...425..264B}.  Once molecules and dust grains form in the shell, the shell may cool so rapidly that its pressure drops below that of the ambient ejecta.  In this case, the shell can enter contraction (in homologously expanding coordinates).  

This basic ``bubbly" structure is in fact observed in young supernova remnants such as the Cas A \citep{2015Sci...347..526M} and also the remnant B0049-73.6 in the Small Magellanic Cloud \citep{2005ApJ...622L.117H}.\footnote{More indirectly, based on the sizes of alumina grains in the pre-solar nebula, \citet{Nozawa:15} concluded that the alumina should have formed in dense clumps within core collapse supernova ejecta.}  Recently, \citet{Abellan:17} used observations with the Atacama Large Millimeter/submillimeter Array (ALMA) to create three dimensional maps of CO and SiO in the inner ejecta of SN 1987A. These maps definitively show that the distribution of molecules is not uniform but clumpy in the inner ejecta.  \citet{Matsuura:17} used ALMA observations at high frequency resolution to detect CO, SiO, SO, and HCO$^+$ in the ejecta of SN1987A. The distorted profiles of the emission lines from these molecules also imply that they are not uniformly distributed in the ejecta.

\begin{table*}
\begin{tabular}{ lcc }
\hline
Quantity & Symbol & Unit \\
\hline
bulk binding energy of atom to grain divided by Boltzmann's constant & $A$ & $\text{K}$ \\
Hamaker constant & $A_{\rm H}$ & $\text{erg}$ \\
Arrhenius rate coefficient &$A_j$ & $\text{cm}^{\{0,3,6\} }\,\text{s}^{-1}\,\text{molecule}^{-1}$ \\
radius of monomer & $a_1$ & $\text{cm}$ \\
radius of $n$-mer & $a_n$ &$\text{cm}$ \\
grain radius & $a$ & $\text{cm}$ \\
maximum grain radius & $a_{\rm max}$ &$\text{cm}$ \\
minimum grain radius & $a_{\rm min}$ & $\text{cm}$\\
$\ell$th grain radius grid point & $a_\ell$ & $\text{cm}$\\
number of times that species $i$ occurs as a product in reaction $j$ & $\alpha_{ij}$ & \\
parameter in vapor pressure approximation formula & $B$ & \\
number of times that species $i$ occurs as a reactant in reaction $j$ & $\beta_{ij}$ & \\
Coulomb correction factor & $\m C_{\ell_1\ell_2}$ & \\
number density of molecules of species $i$ & $c_i$ & $\text{cm}^{-3}$ \\
number density of electrons & $c_e$& $\text{cm}^{-3}$ \\
number density of ions & $c_{\rm ion}$ & $\text{cm}^{-3}$ \\
number density of monomers & $c_1$ & $\text{cm}^{-3}$ \\
number density of $n$-mers & $c_n$ & $\text{cm}^{-3}$ \\
number density of grains at radius grind point $\ell$ & $c_\ell$ & $\text{cm}^{-3}$\\
number density of the key species & $c_{\rm key}$ &$\text{cm}^{-3}$ \\
total gas number density & $c_{\rm tot}$ &$\text{cm}^{-3}$ \\
surface binding energy of an atom to a grain & $\m E_{\rm bind}$ & $\text{erg}$ \\
activation energy & $E_{\rm A,\it j}$ & $\text{erg}$ \\
energy emitted in X-rays, electrons, and positrons in one decay of isotope $n$ & $E_n^{\rm X}$ & $\text{keV}$  \\
energy emitted in $\gamma$-rays in one decay of isotope $n$ & $E_n^{\gamma}$ & $\text{MeV}$ \\
number of vibrational degrees of freedom in a grain & $F$ & \\
 fraction of deposited radioactive energy that goes into ionizing and dissociating atoms and molecules  & $f_{\rm ion}$ & \\
 fraction of deposited radioactive energy that goes into UV photons    & $f_{\rm UV}$ & \\
calibration ratio of gas temperature used in simulation to temperature from \textsc{cloudy} & $f_{\rm temp}$ & \\
rate at which electrons collide with a grain & $\Gamma_{e^-}$ & $\text{s}^{-1}\,\text{grain}^{-1}$ \\
rate at which electrons are ejected from a grain following absorption of a photon & $\Gamma_{\gamma}$ &  $\text{s}^{-1}\,\text{grain}^{-1}$ \\
rate at which ions collide with a grain &$\Gamma_{\rm ion}$ &  $\text{s}^{-1}\,\text{grain}^{-1}$ \\
average number of quanta per vibrational degree of freedom & $\gamma$ & \\

coagulation kernel & $K_{\ell_1\ell_2}$ &   $\text{cm}^3\,\text{s}^{-1}$  \\
accretion rate coefficient & $k_{\rm acc}$ & $\text{cm}^3\,\text{s}^{-1}$ \\
evaporation rate in monomers ejected per unit time per grain & $k_{\rm evap}$ & $\text{s}^{-1}$ \\
rate coefficient for reaction $j$ & $k_j$ & $\text{cm}^{\{0,3,6\}}\,\text{s}^{-1}$ \\
rate coefficient for Compton destruction of species $i$ via process $j$ & $k_{i,j}^{\rm C}$ &  $\text{s}^{-1}$ \\
rate coefficient for noble gas ion weathering & $k_{\rm He}$ & $\text{cm}^3\,\text{s}^{-1}$ \\
rate coefficient for oxygen weathering & $k_{\rm O}$ &   $\text{cm}^3\,\text{s}^{-1}$  \\
rate at which energy from all decaying isotopes is deposited in the ejecta & $L$ & $\text{erg}\,\text{s}^{-1}$ \\
rate at which a grain absorbs energy from the UV radiation field & $L_{\rm abs}$ &  $\text{erg}\,\text{s}^{-1}$ \\
rate at which a grain emits energy in the form of thermal radiation & $L_{\rm emit}$ &$\text{erg}\,\text{s}^{-1}$ \\
rate at which energy is transferred from the gas to a grain & $L_{\rm gas}$ & $\text{erg}\,\text{s}^{-1}$ \\
rate at which energy is deposited in the ejecta that destroys atoms and molecules & $L_{\rm ion}$ & \\
rate at which energy from decaying isotope $n$ is deposited in the ejecta &$L_n$ & $\text{erg}\,\text{s}^{-1}$  \\
rate at which energy is converted into UV radiation in the ejecta &$L_{\rm UV}$& $\text{erg}\,\text{s}^{-1}$\\
number of radius bins & $\ell_{\rm max}$ & \\
decay rate of isotope $n$ & $\lambda_n$ & $\text{yr}^{-1}$ \\  
mass coordinate in ejecta & $M$ & $M_\odot$ \\
total dust mass produced in simulation & $M_{\rm dust,tot}$ & $M_\odot$\\
total dust mass in SN 1987A as a function of time & $M_{\rm dust,tot,1987A}$ & $M_\odot$\\
enclosed mass coordinate in the \textsc{mesa} model &$M_{\text{ej},\textsc{mesa}}$&$M_\odot$ \\
mass of ejected helium and metal core in SN 1987A &$M_{\rm ej,tot,1987A}$&$M_\odot$ \\
mass of ejected helium and metal core in the \textsc{mesa} model &$M_{\text{ej,tot},\textsc{mesa}}$&$M_\odot$ \\
mean molecular weight & $\avg{m}$ & $\text{g}$ \\
mass of monomer & $m_1$ & $\text{g}$ \\
mass of $n$-mer & $m_n$ &g \\
mass of ions & $m_{\rm ion}$ & $\text{g}$ \\
molecular mass of the key species & $m_{\rm key}$ & $\text{g}$ \\
mass of grain at grid point $\ell$ & $m_\ell$ &g \\
reduced mass of colliding grains & $\mu_{\ell_1\ell_2}$ & g\\
reduced mass & $\mu$ & g\\
  \hline
\end{tabular}
\caption{Index of notation used in the paper.}
\label{tab:notation1}
\end{table*}

\begin{table*}
\begin{tabular}{ lcc }
\hline
Quantity & Symbol & Unit \\
\hline
total number of dependent variables in the system of ODEs that we are solving & $\m N$ & \\
number of grain species & $\m N_{\rm G}$ & \\
number of radioactive isotopes & $\m N_{\rm iso}$ & \\
number of reactions & $\m N_{\rm R}$ & \\
number of species (of molecules) in simulation & $\m N_{\rm S}$ & \\

number of atoms in a grain & $N$ & \\
number of atoms of isotope $n$ in ejecta & $N_n$ & \\
total number of particles in the ejecta & $N_{\rm tot}$ & \\

number of monomers & $n$ & \\
number density of UV photons per unit photon energy &$n_\gamma$& $\text{cm}^{-3}\,\text{eV}^{-1}$ \\
number of monomers in a grain at grid point $\ell$ & $n_\ell$ & \\
number of monomers that you have to add to $\m Z_{\ell-1}$ to obtain $\m Z_\ell$ & $\delta n_{\ell-1}$ & \\
number of monomers in largest cluster & $n_{\rm max}$ & \\

power law exponent in Arrhenius form of rate coefficient & $\nu_j$& \\
the $k$th stoichiometric coefficient in the formula for accretion/evaporation & $\nu_k$ & \\
stoichiometric coefficient of the key species & $\nu_{\rm key}$ & \\

angular frequency of vibrational degrees of freedom in a grain & $\omega_0$ & $\text{rad}\,\text{s}^{-1}$ \\
standard pressure & $p_{\rm s}$ & $\text{dyne}\,\text{cm}^{-2}$ \\
absorption coefficient & $Q_{\rm abs}$ &  \\
radius of the outer edge of core ejecta & $R$ & $\text{cm}$ \\
mass density of ejecta & $\rho$ & $\text{g}\,\text{cm}^{-3}$  \\
ejecta mass density at reference time $t_0$ & $\rho_0$ &$\text{g}\,\text{cm}^{-3}$ \\
mass density of grain & $\varrho$ & $\text{g}\,\text{cm}^{-3}$ \\
evaporation suppression factor & $S_N$ & \\
sticking coefficient & $s_n$ & \\
collision cross section & $\sigma$& $\text{cm}^2$ \\
surface tension & $\sigma_\text{ST}$ & $\text{erg}\,\text{cm}^{-2}$ \\
photon absorption cross section & $\sigma_{\rm abs}$ &$\text{cm}^2$ \\
temperature & $T$ & K \\
activation energy divided by Boltzmann's constant &$T_{\rm A,\it j}$& K \\
Debye temperature & $T_{\rm D}$ & K \\
grain temperature & $T_{\rm dust}$ & K \\
electron temperature & $T_e$ & K \\
gas temperature & $T_{\rm gas}$ & K  \\
temperature of ions & $T_{\rm ion}$ & K \\

time since explosion & $t$ & $\text{s}$ \\
reference time & $t_0$ & $=100\,\text{d}$ \\
 optical depth of ejecta & $\tau$  &  \\
  optical depth of ejecta to $\gamma$ photons emitted by isotope $n$  &  $\tau_n$&  \\
thermal energy in a grain & $U$ & $\text{erg}$ \\
total energy in UV radiation field &$U_{\rm UV}$& $\text{erg}$ \\
energy density in UV radiation field & $u_{\rm UV}$ & $\text{erg}\,\text{cm}^{-3}$ \\
potential energy due to van der Waals forces & $V_{\ell_1\ell_2}$ & $\text{erg}$ \\
amount of energy that an electron loses when it destroys a molecule of species $i$ via process $j$ & $W_i^j$ & $\text{eV}$ \\
van der Waals correction factor & $\m W_{\ell_1\ell_2}$ & \\
monomer of a single element grain & $\m X$ & \\
$n$-mer & $\m X_n$ & \\
condensation nucleus & $\m X_{n_{\rm max}}$ & \\
mass fraction of isotope $i$ & $X_i$ & \\
photoelectric yield & $Y$ & \\
grain at radial grid point $\ell$ &$\m Z_\ell$& \\
grain that results from coagulation of grains in bin $i$ and $j$ & $\m Z_{\ell_1\ell_2}$ & \\
equilibrium grain charge & $Z_{\rm dust}$ &$e$ \\
grain charge & $Z$& $e$ \\
charge of grain at radial grid point $\ell$ in units of $e$ & $Z_\ell $ &$e$ \\
charge of ions & $z_{\rm ion}$ &$e$ \\

  \hline
\end{tabular}
\caption{Index of notation used in the paper (continued).}
\label{tab:notation2}
\end{table*}

Here we present a model of dust formation in supernovae constructed within the framework of MNT. We precompute initial data for local temperature and mass density evolution of the ejecta and for the local nucleosynthetic yields and then follow the formation of molecules with a fully nonequilibrium chemical reaction network. The model includes reactions such as: three-body molecular association, thermal fragmentation, neutral-neutral and ion-molecule reactions, radiative association, charge exchange, recombination with electrons in the gas-phase, and destruction by energetic electrons produced by the radioactive decay of \iso{Co}{56}, \iso{Co}{57}, \iso{Na}{44}, and \iso{Ti}{44}. The formation of large molecules that act as condensation nuclei for grains is followed as part of this network. The grains grow by accreting gas-phase molecules and coagulating. The coagulation rate accounts for the effects of the van der Waals force and also the Coulomb force due to grain electric charge. Grains lose mass through evaporation, reaction with noble gas ions, and oxidation. The grain temperature, which is needed for the evaporation rate, is allowed to differ from the gas temperature and depend on the grain radius and post-explosion epoch. The evaporation rate computation includes the effects of surface tension and the lack of gas-grain thermal coupling (the latter implying that the thermodynamic fluctuation leading to evaporation must come from within the grain itself).  We consider several representative ejecta fluid elements, each of which has its own chemical composition and thermodynamic evolution, and track molecule and grain formation in each element. Our simulation calculates the abundance of each species (atoms, molecules, ions, free electrons, and dust condensation nuclei) as a function of time from the explosion. The calculation provides us with the properties of dust formed in each representative fluid element.

The paper is organized as follows. In Section \ref{sec:supernova_model} we present our modeling of the radioactive heat and ionization sources in the ejecta and of the structure and thermal evolution of the ejecta.  In Section \ref{sec:chemistry} we describe our chemical framework and in Section \ref{sec:grain_physics} we describe our modeling of grain growth and destruction processes.  In Section \ref{sec:time_integration} we describe our time integration scheme.  In Section \ref{sec:results} we present the results and in Section \ref{sec:discussion} we discuss the implications of our results and delineate desirable next steps.  Finally, in Section \ref{sec:conclusions} we summarize our main conclusions.  In Tables \ref{tab:notation1}--\ref{tab:notation2} we provide overview of the important mathematical notation used in the paper.

\section{Supernova model}
\label{sec:supernova_model}

\subsection{Radioactive decay}    
\label{sec:radioactive_decay}

Explosive nucleosynthesis produces large quantities of radioactive nuclei and their decay has profound consequences for the thermal and chemical evolution of the ejecta.  Let $N_n(0)$ be the total number of atoms of radioactive isotope $n$ immediately after the explosion. Then at time $t$ after the explosion, the remaining number of atoms of the radioactive isotope is $N_n(t) = N_n(0) e^{-\lambda_n t}$,
where $\lambda_n$ is the decay rate. The number of decays per unit time is $|dN_n/dt| = \lambda_n N_n(0) e^{-\lambda_n t}$. There are four radioactive decay chains that affect the ejecta during the period that dust grains are forming, between $\sim100$ and $\sim10^4$ days after the explosion: $\iso{Ni}{56} \to \iso{Co}{56} \to \iso{Fe}{56}$, $\iso{Ni}{57} \to \iso{Co}{57} \to \iso{Fe}{57}$, $\iso{Ti}{44} \to \iso{Sc}{44} \to \iso{Ca}{44}$, and $\iso{Na}{22} \to \iso{Ne}{22}$ \citep[e.g.,][]{Woosley:89}.
The half-life of \iso{Ni}{56} is 6 days, much shorter the the time scale on which dust forms, so we can assume that it has decayed into \iso{Co}{56}. Similarly, the half-life of \iso{Ni}{57} is only 36 hours and we can take that it has decayed into \iso{Co}{57}. The half-life of \iso{Sc}{44}, the immediate product of \iso{Ti}{44} decay, is only 4 hours; since this is much shorter than the 60 year half-life of \iso{Ti}{44}, we can assume that  \iso{Ti}{44} decays directly into \iso{Ca}{44}. Thus, the effective radioactive decays included in the simulation are:
\bea
\iso{Co}{56} \to \iso{Fe}{56}, \ \ 
\iso{Co}{57} \to \iso{Fe}{57}, \ \ 
\iso{Ti}{44} \to \iso{Ca}{44}, \ \ 
\iso{Na}{22} \to \iso{Ne}{22} .
\eea

Each radioactive decay releases an energy $Q_n$ that is distributed among $\gamma$-ray and X-ray photons, electrons, positrons, and neutrinos. The chain of processes begins when a parent nucleus decays into an excited state of the daughter nucleus by electron capture or positron emission. Then, the excited daughter nucleus decays to its ground state by emitting $\gamma$ photons or by transferring energy to bound electrons that are ejected (internal conversion). If bound electrons are removed by electron capture or internal conversion, then higher energy bound electrons can lose energy and fill the vacancy.  The electronic transitions occur via X-ray emission or Auger ionization.   The $\gamma$ photons repeatedly Compton scatter on bound and free electrons, each time losing some energy and producing a high energy ``Compton'' electron. Eventually the $\gamma$ photon either escapes the ejecta or is photoelectrically absorbed.  

The high energy electrons produced by Compton scattering, photoelectric absorption, internal conversion, Auger ionization, and secondary ionization lose energy by ionizing atoms and molecules, dissociating molecules, electronically exciting atoms and molecules, and undergoing Coulomb collisions with charged particles in the gas, the latter process converting the electron's kinetic energy into heat. The electrons produced by electron-impact ionization of atoms and molecules are called secondary electrons and themselves must lose energy via the above processes.  The positrons emitted during positron emission decays lose energy in the same way as electrons. After they lose all of their kinetic energy they bind with electrons into positronium, which decays into two 511 keV photons. These $\gamma$ photons lose energy in the same way as those produced directly in nuclear decays.  The X-rays produced when a bound electron transitions to a lower energy level due to a vacancy opened up by electron capture on a proton or ejection in an internal nuclear conversion are photoelectrically absorbed.  The neutrinos leave the ejecta without depositing any of their energy.

\subsection{Heating and ionization}

\begin{table}
\begin{center}
\begin{tabular}{ llll }
\hline
Isotope        & $\lambda_n$ (yr$^{-1}$) & $E_n^\text{X}$ (keV) & $E_n^\gamma$ (MeV) \h
\iso{Co}{56} & $3.3$     & 125                   & 3.6 \\
\iso{Co}{57} & $0.93$     & 22.6                  & 0.122  \\
\iso{Ti}{44}  & $0.011$    & 644                  & 2.27 \\
\iso{Na}{22} & $0.27$     &195                   & 2.2 \\
  \hline
\end{tabular}
\end{center}
\caption{Radioactive decay parameters: the decay rate $\lambda_n$, the energy $E_n^\text{X}$ emitted per decay in the form of X-rays, electrons, and positrons, and the energy $E_n^\gamma$ emitted per decay in the form of $\gamma$ photons including the photons produced by electron-positron annihilation \citep{Be2004TableofRadionuclides}.}
\label{tab:halflife}
\end{table}

In homologous expansion in which the radius of the ejecta increases linearly in time $R\propto t$, density decreases as $\rho\propto t^{-3}$, therefore the optical depth decreases as $\tau \propto R\rho\propto t^{-2}$, and we can set $\tau_n=\tau_{n,0} (t/t_0)^{-2}$ where $t_0$ is a reference time.  With this,
the rate at which energy is deposited in the ejecta via radioactive decay of isotope $n$ at time $t$ can be approximated as
\bea
L_n(t) &=& \lambda_{n}N_{n}(0) e^{-\lambda_{n} t}\nonumber\\  &\times&
[E_{n}^{\rm X} + (1 - e^{-\tau_{n,0} (t/t_0)^{-2}}) E_{n}^\gamma ],
\eea
where $E_{n}^{\rm X}$ is the energy emitted per decay in electrons, positrons, and X-ray photons, $E_{n}^\gamma$ is the energy emitted per decay in $\gamma$ photons, and $\tau_{n,0}$ is the optical depth from the center of the ejecta at $t_0$. The values of $\lambda_n$,  $E_n^\text{X}$, and  $E_n^\gamma$  are given in Table \ref{tab:halflife}. 

We take the initial quantities of the radioactive isotopes to be $N_{56}(0)= 1.62\times 10^{54}\ (0.076\,M_\odot)$, $N_{57}(0)=8.77 \times 10^{52}\  (4.2\times10^{-3}\,M_\odot)$, $N_{22}(0)=5.5 \times 10^{48}\ (10^{-7}\,M_\odot)$, and $N_{44}(0)=5.69 \times 10^{51}\ (2.1\times10^{-4}\,M_\odot)$.  These values are from the computation of explosive nucleosynthesis in SN 1987A by \citet{1990ApJ...349..222T}.  In particular, our adopted $^{44}$Ti mass of $\approx 2\times10^{-4}\,M_\odot$ is just somewhat larger than the mass $1.5\pm 0.3\times10^{-4}\,M_\odot$ recently inferred directly from spectroscopy with NuSTAR \citep{Boggs:15}, the latter consistent with \citet{2011A&A...530A..45J}, and below the $\approx 3\times10^{-4}\,M_\odot$ inferred from spectoscopy with INTEGRAL \citep{Grebenev:12}.  We refer the reader to \citet{McCray:16} for further discussion of the $^{44}$Ti mass. For the optical depth coefficients $\tau_{n,0}$ we adopt the estimates from \citet{1993ApJ...419..824L} for SN 1987A: $\tau_{56,0}=13.2$, $\tau_{57,0}=31.7$, and $\tau_{44,0}=\tau_{22,0}=16$.

The total radioactive decay energy deposited per unit time is
\beq
\label{eq:total_luminosity}
L(t) = \sum_{n=1}^{\m N_{\rm iso}}L_n(t) ,
\eeq 
where $\m N_{\rm iso}$ is the number of radioactive isotopes in the ejecta (here, $\m N_{\rm iso}=4$ for $n=\iso{Co}{56}, \iso{Co}{57}, \iso{Ti}{44}, \iso{Na}{22}$). 
We assume that some fraction $f_{\rm ion}$ of the deposited energy goes into ionizing atoms and ionizing as well as dissociating molecules; the rest goes into exciting atoms and molecules and heating the gas.  On the basis of the model of \citet{1995ApJ...454..472L} we crudely estimate $f_{\rm ion}\approx 0.35$ and use this value in all of our calculations.

\subsection{UV radiation}

A consequence of radioactive energy deposition is a buildup of UV radiation that permeates the ejecta. The UV photons are produced when atoms and molecules excited by Compton electrons de-excite by spontaneous emission, when atoms that have been ionized by Compton electrons radiatively recombine with thermal electrons, and when molecules that have been dissociated by Compton electrons reform by radiative association. Although we do not include the effect in our present calculations, this UV radiation is important for dust synthesis because it heats the dust grains, it influences the electric charge of dust grains via photoelectric absorption, and dissociates molecules. 

To model the UV radiation, let $f_{\rm UV}$ denote the fraction of deposited radioactive energy converted into UV radiation; we set $f_{\rm UV}=0.35$, within the range of values found in \citet{1992ApJ...390..602K}. The UV luminosity is then $L_{\rm UV}(t) = f_{\rm UV} L(t)$ where $L(t)$ is given in Equation (\ref{eq:total_luminosity}). If we assume that a UV photon spends a time $\Delta t = R / c$ in the ejecta before escaping, where $R$ is the radius of the ejecta and $c$ is the speed of light, then the energy in the UV radiation is $U_{\rm UV}(t) = L_{\rm UV}(t) \Delta t$ and the energy density is:
\beq
\label{eq:uUV}
u_{\rm UV} = \f{3f_{\rm UV} L}{4\pi c R^2} .
\eeq   
 
Let $n_\gamma(E)dE$ be the number density of UV photons with energy between $E$ and $E+dE$. Ideally, this should be computed with a Monte Carlo simulation that explicitly follows the degradation of energy deposited from radioactive decay and the subsequent radiative transfer \citep[e.g.,][]{Swartz:95,Kasen:06,2011A&A...530A..45J}.  Here, instead, we crudely approximate $n_\gamma(E)$ such that the total energy density in the UV radiation equals $u_{\rm UV}$.   We choose the photon number density per unit energy to be Gaussian:
\beq
n_\gamma(E) = A_{\gamma,0} e^{-(E-E_0)^2/2\sigma_0^2} ,
\eeq 
where $A$ is a normalization constant, $E_0$ is the mean photon energy, and $\sigma_0$ is the spread. The total energy density is:
\bea
u_{\rm UV} \approx\int_{-\infty}^\infty  n_\gamma(E) E dE = \sqrt{2\pi} A_{\gamma,0} \sigma_0 E_0 ,
\eea
where for convenience we have extended the lower integration limit to $-\infty$. This matches the energy density in Equation~(\ref{eq:uUV}) with:
\beq
A_{\gamma,0} = \f{3f_{\rm UV} L}{4\sqrt{2} \pi^{3/2} c R^2 \sigma_0 E_0} .
\eeq
In this work we use $E_0 = 4.431 \u{eV}$ and $\sigma_0 = 1 \u{eV}$, where these values are motivated by the analysis in \citet{2011A&A...530A..45J}.

\subsection{Progenitor model, ejecta composition, and kinematics}
\label{sec:ejecta_model}

To calculate the properties of dust that forms in SN 1987A we need the elemental composition as well as the mass density and gas temperature as a function of time at each point in the ejecta. The elemental composition of the ejecta was determined by simulating the evolution and explosion of a star with initial mass of $M_\text{ZAMS}=20\,M_\odot$ and initial absolute metallicity equal to that of the Large Magellanic Cloud $Z_\text{LMC}=0.007$ using the stellar evolution code \textsc{mesa} \citep{2015ApJS..220...15P}.   Stellar mass loss rate was parametrized to reduce the stellar mass to a pre-explosion value of $14.5\,M_\odot$, a target mass chosen to approximate the pre-explosion mass of $\sim 14\,M_\odot$ inferred from the early observations of SN 1987A \citep[see][and references therein]{McCray:16}.

\begin{figure*}
\begin{center}
\includegraphics[trim=0cm 4cm 0cm 5cm,clip=true,width=0.49\textwidth]{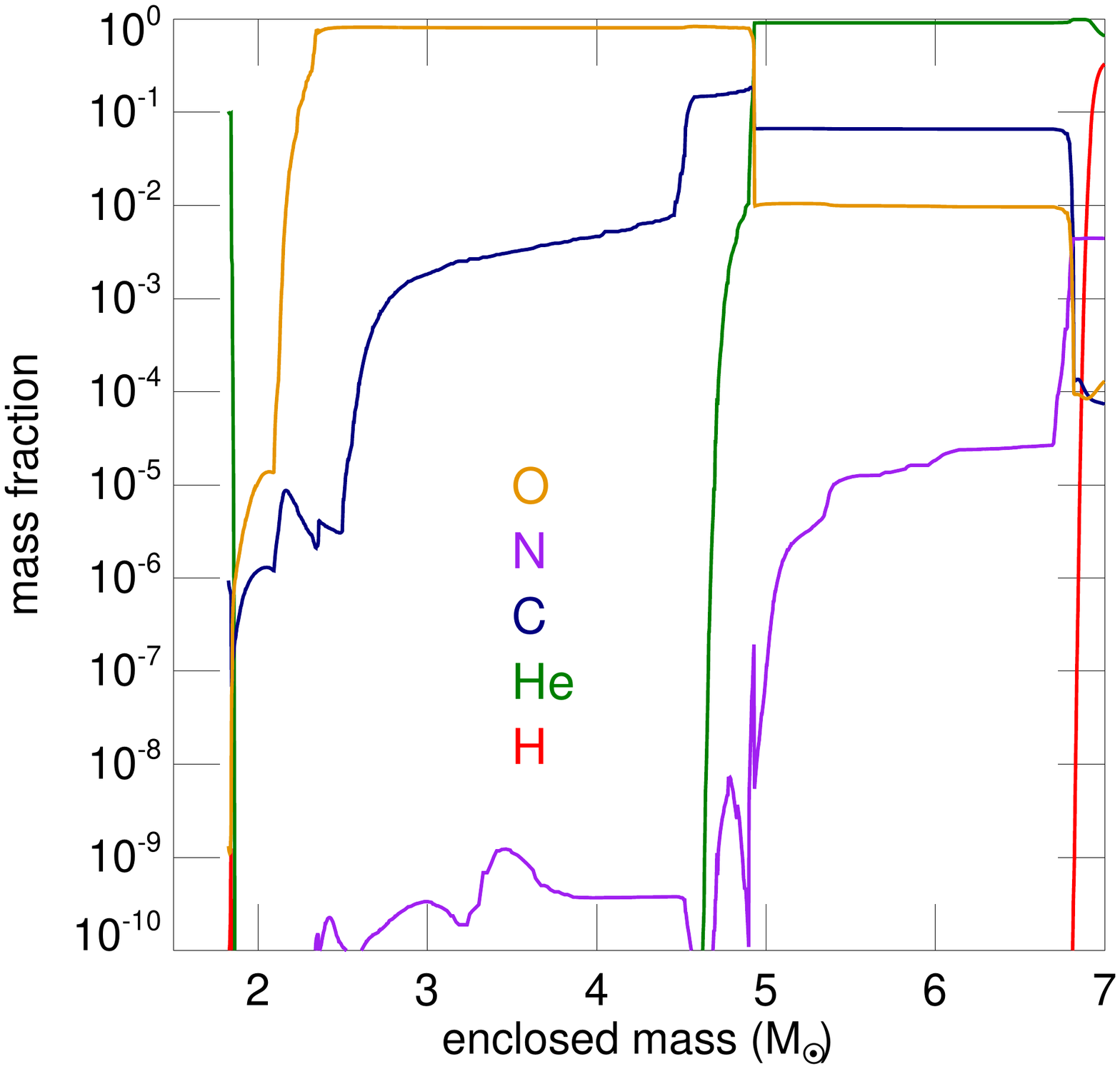}
\includegraphics[trim=0cm 4cm 0cm 5cm,clip=true,width=0.49\textwidth]{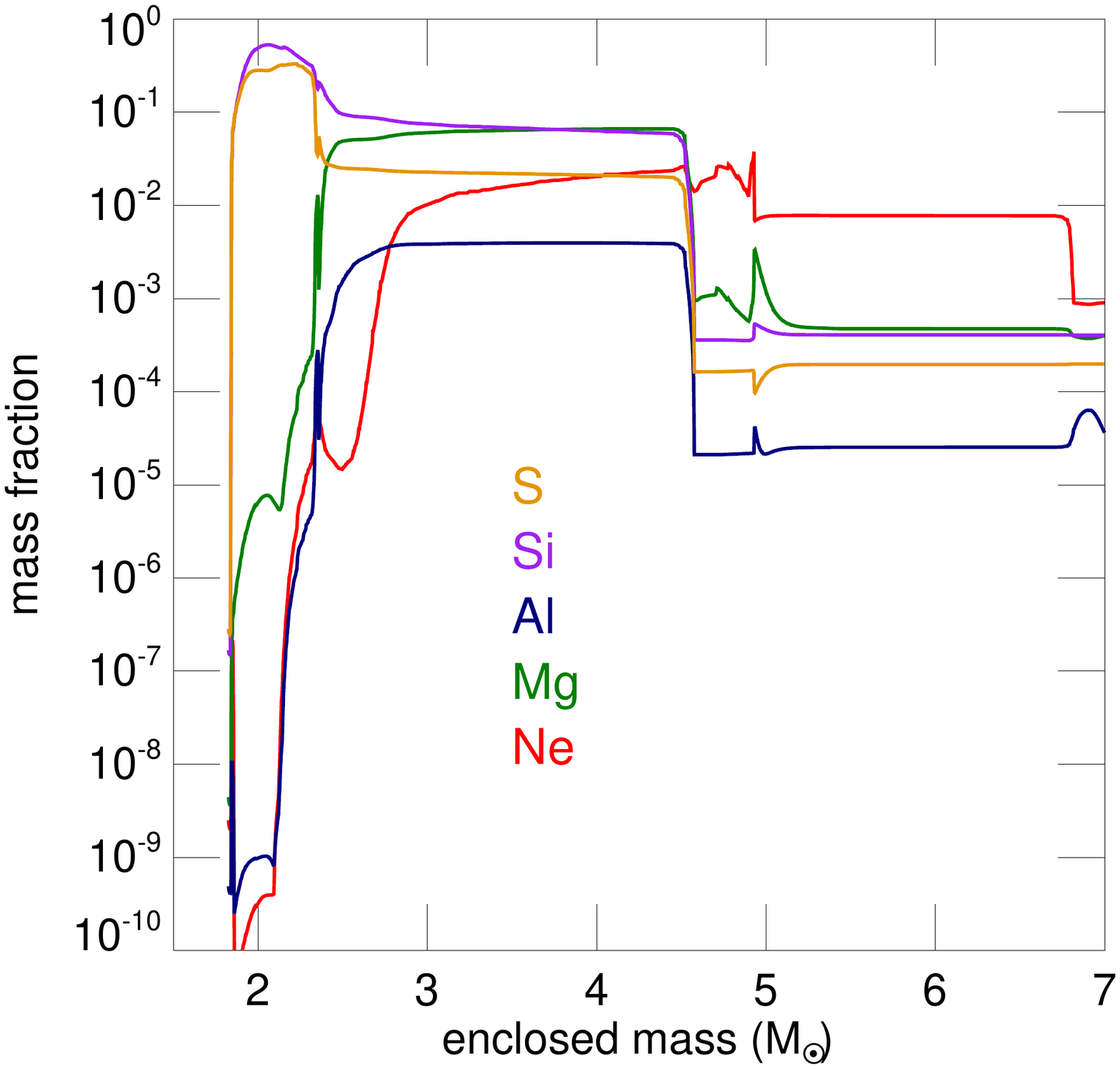}
\includegraphics[trim=0cm 4cm 0cm 5cm,clip=true,width=0.49\textwidth]{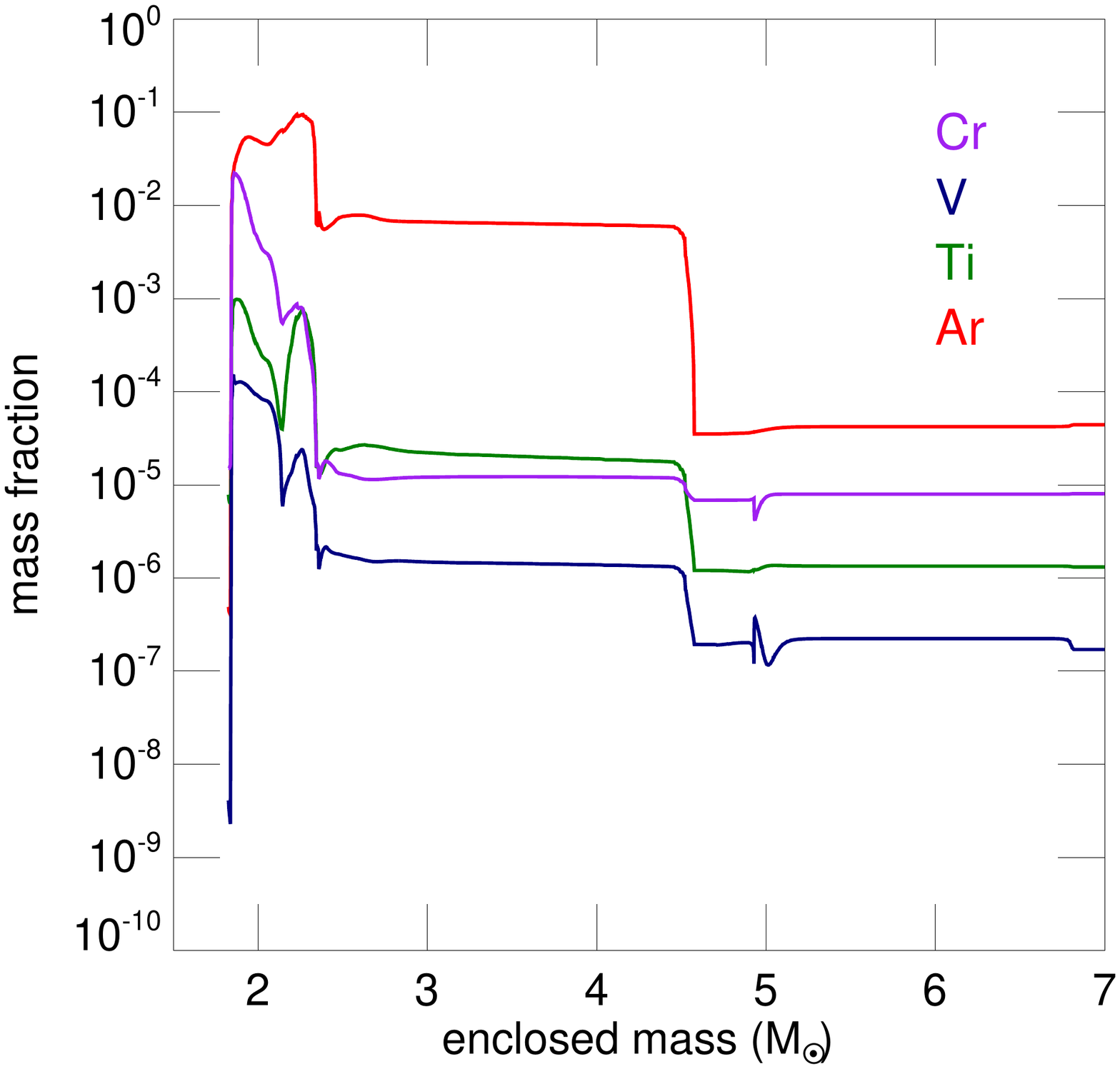}
\includegraphics[trim=0cm 4cm 0cm 5cm,clip=true,width=0.49\textwidth]{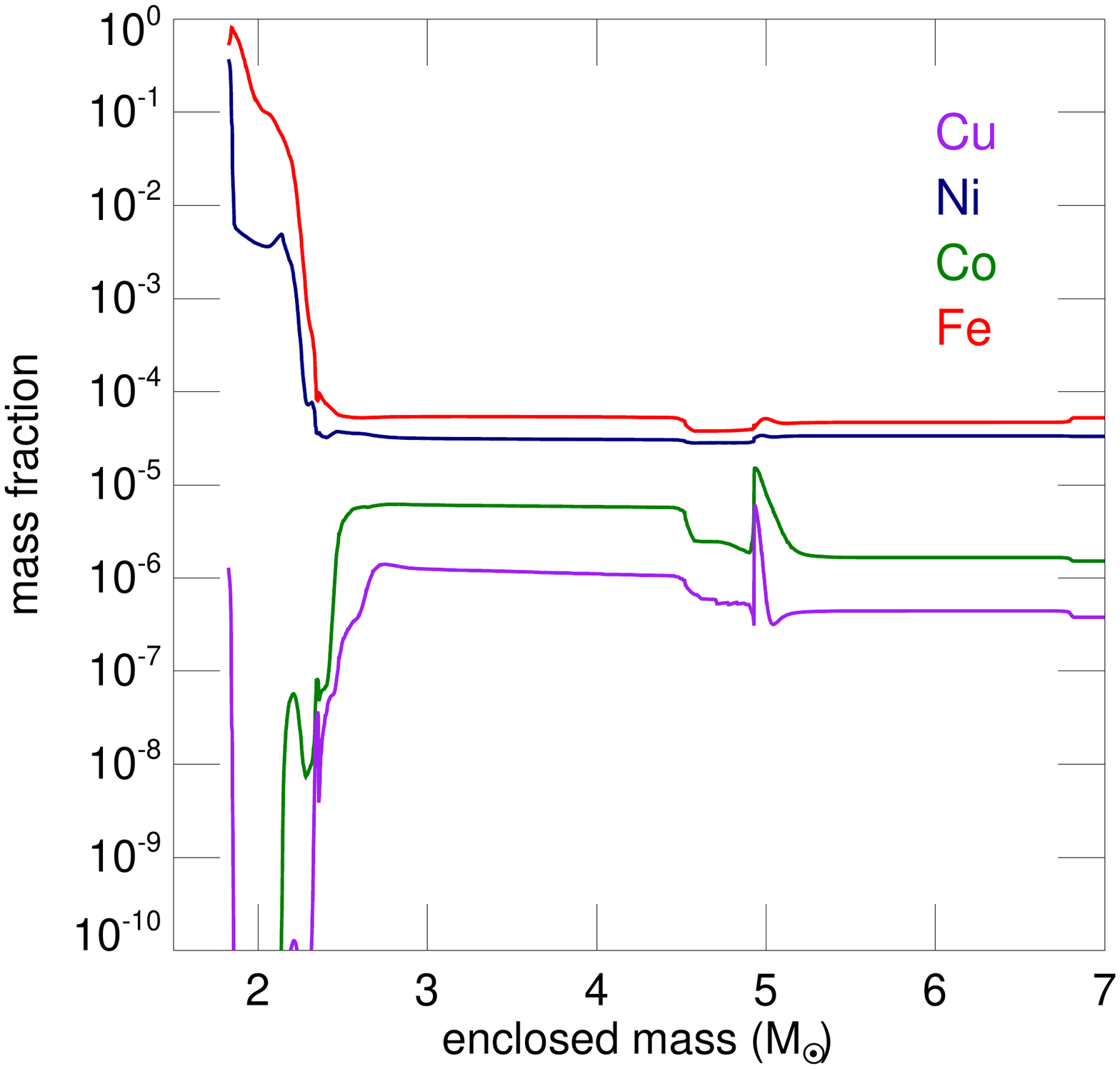}
\end{center}
\caption{The curves show elemental mass fractions as a function of the enclosed mass in our supernova ejecta model computed with the \textsc{mesa} code for the $20\,M_\odot$ progenitor model of \citet{2015ApJS..220...15P}. Radioactive isotopes with half-lives $<10\,\text{yr}$ were replaced into their stable daughter decay products (see Appendix, Table~\ref{tab:isoTOelement}).}
\label{fig:xVSm}
\end{figure*}

The core collapse was simulated by excising the central $1.825\,M_\odot$ and placing a reflecting hydrodynamic boundary at that mass coordinate.  The explosion was driven by depositing $2.3\times 10^{51} \u{ergs}$ of thermal energy over a mass coordinate range $\Delta M=0.05\,M_\odot$ adjacent to the excised region over the course of $1$ second. The deposition launched an outward-propagating shock wave.  Explosive nucleosynthesis in the shock-heated ejecta was followed through freezeout for $60$ seconds when the kinetic energy had dropped to $1.27\times 10^{51} \u{ergs}$. The \textsc{mesa} calculation gives isotope-specific yields with isotopic half-lives varying over a wide range. For simplicity, we converted the isotopes with half-lives shorter than $10\,\text{yr}$ into their stable daughter isotopes. The resulting isotopes all had half-lives longer than $30\,\text{yr}$. In the dust synthesis calculation we do not distinguish between isotopes; the isotope (unstable and stable) to element conversion scheme is given in Table~\ref{tab:isoTOelement} in the Appendix. In Figure~\ref{fig:xVSm} we show elemental mass fraction $X_i$ as a function of the enclosed mass $M_\text{ej,1987A}$.

The model implies the following stratification of the progenitor star: the neutron star ($M < 1.825\,M_\odot$), the explosively synthesized iron group ($1.825\,M_\odot < M < 1.892\,M_\odot$), the lighter element ejecta ($1.892\,M_\odot < M < 6.8\,M_\odot$), and the hydrogen envelope ($6.8\,M_\odot < M < 14.5\,M_\odot$). The mass coordinate extent of $^{56}$Ni was set found to be $M_{\rm Ni} = 0.068\,M_\odot$, a value consistent with the range allowed by observations of SN 1987A. The boundary between the helium core and the hydrogen envelope was set where the hydrogen mass fraction dropped to a negligible value $10^{-10}$.  The mass of helium and lighter element ``core'' ejecta was $M_{\rm core}\approx 5\,M_\odot$.  

\begin{figure}
\begin{center}
\includegraphics[trim=0cm 0cm 0cm 0cm,clip=true,width=0.45\textwidth]{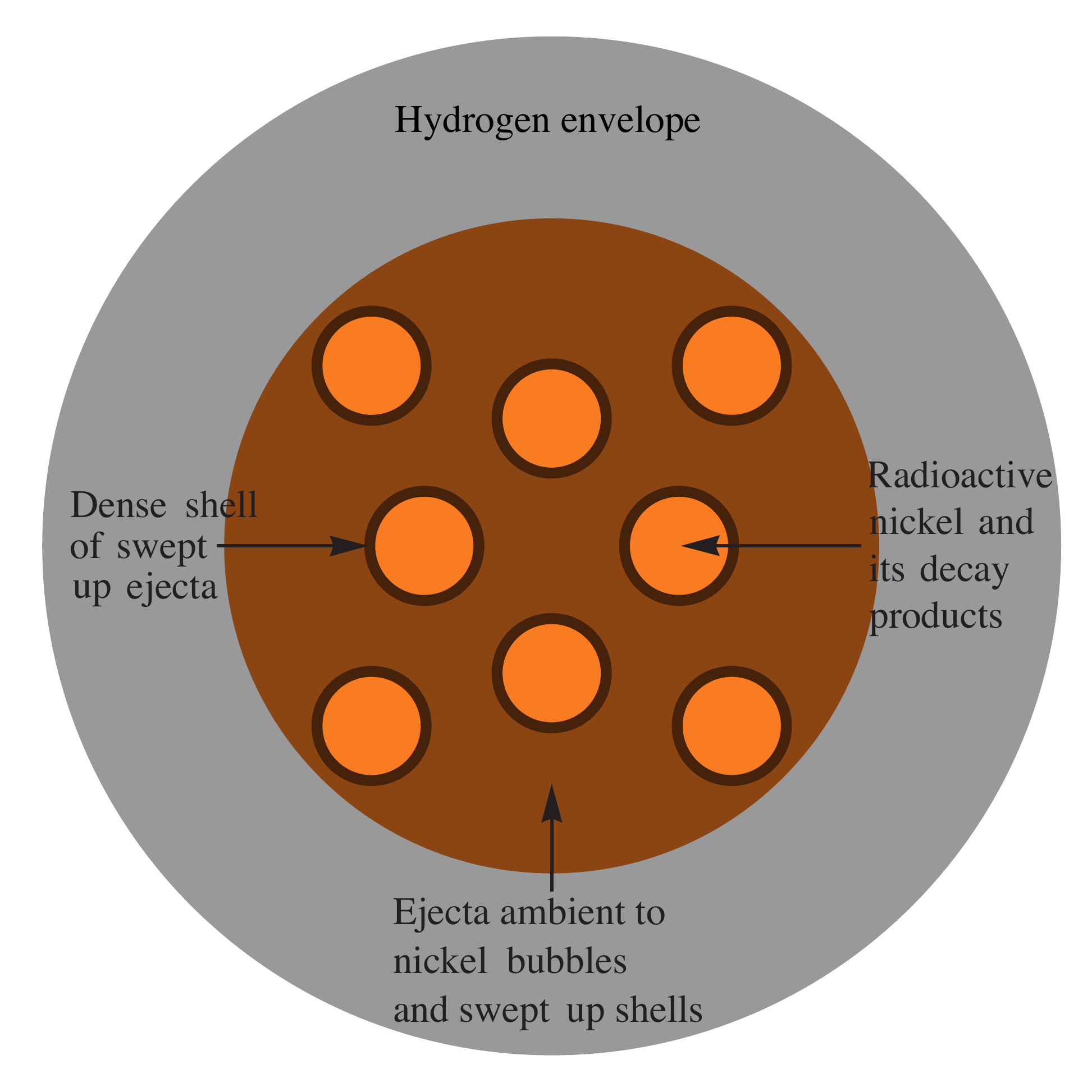}
\end{center}
\caption{The figure provides a schematic illustration, not to scale, of how we model the density structure of the ejecta.  We define three regions, each of uniform density, corresponding to the low density interiors of nickel bubbles, the high density shells of the material swept up by the expanding bubbles, and an intermediate density region unaffected by the bubbles.}
\label{fig:ejectaDiagram}
\end{figure}

While the \textsc{mesa} calculation preserves the initial elemental stratification, we take the $^{56}$Ni to be ``dredged-up'' into discrete clumps that end up randomly distributed in the core ejecta; we call these clumps ``bubbles'' (see Section \ref{sec:intro}).  The radioactive energy released when the \iso{Ni}{56} in the bubbles decayed into \iso{Co}{56} (and to a lesser extent when the \iso{Co}{56} decayed into \iso{Fe}{56}) over-pressured the bubbles against the surrounding core ejecta and for a period of time, the nickel bubbles expanded super-homologously. The super-homologous expansion stopped when the bubbles became optically thin to the $\gamma$-rays emitted in the radioactive decays. By the start our dust-synthesis simulations, at 100 days after the explosion, the bubbles have returned to homologous expansion but occupy an elevated fraction of the volume of the helium core. We assume that the bubble expansion has swept up thin shells of the surrounding core ejecta.  In Figure \ref{fig:ejectaDiagram} we show a schematic diagram of the geometry of our model.

At the end of super-homologous bubble expansion, the ejecta had the following structure: low density Ni, Co, and Fe bubbles with a total mass equal to the Ni mass $M_{\rm bub}=M_{\rm Ni}$
 occupying a fraction $f_{\rm Ni}$ of the volume of the helium core, multiple high density shells of swept up core ejecta with total mass:
\bea
M_{\rm shell} = \frac{\omega f_{\rm bub} - M_{\rm Ni}/M_{\rm core}}{\omega - M_{\rm bub}/ M_{\rm core}} (M_{\rm core} - M_{\rm bub}) ,
\eea
where the density in the bubbles immediately after the explosion (\emph{before} super-homologous expansion)
was assumed to be $\omega$ times the mean density,
and intermediate density ambient ejecta outside of the bubbles and shells with mass $M_{\rm amb} = M_{\rm core} - M_{\rm bub} - M_{\rm shell}$.

With these assumptions, the mass density in each of the three regions evolves under homologous expansion as:
\beq
\label{eq:density_evolution}
\rho_{\rm bub,shell,amb}(t) = \rho_{\rm bub,shell,amb}(t_0) \({\f{t}{t_0}}^{-3} .
\eeq
Let $\eta$ denote the ratio of shell thickness to bubble radius.
Then the density normalization factors are:
\bea
\label{eq:density_normalization}
\rho_{\rm bub}(t_0) \= \f{3M_{\rm Ni}}{4\pi f_{\rm Ni} v_{\rm core}^3 t_0^3} ,\nonumber \\
\rho_{\rm amb}(t_0) \=  \f{3({M_{\rm core} - M_{\rm Ni}})}{4\pi  v_{\rm core}^3 t_0^3} \({
1 - \f{M_{\rm Ni}}{\omega M_{\rm core}}
}^{-1} ,\nonumber \\
\rho_{\rm shell}(t_0) \= \frac{\rho_{\rm amb} (t_0)}{3 \eta + 3 \eta^2 + \eta^3}
\({1 - \f{M_{\rm Ni}}{\omega f_{\rm Ni} M_{\rm core}} } ,
\eea
where $v_{\rm core}$ is the expansion velocity at the edge of the helium core.\footnote{The number of nickel bubbles does not affect the mass density in shells and is inconsequential in our model.  For the interested reader, the number of bubbles could be $\approx 75$ for a filling factor of $0.5$ based on Figure 4 of \citet{1993ApJ...419..824L}.} 

For the above parameters we take $v_{\rm core} = 2100 \u{km}\u{s}^{-1}$ \citep{1989ApJ...340..414F}, $f_{\rm Ni} = 0.5$ \citep{1993ApJ...419..824L}, $\omega = 3$ \citep{1994ApJ...425..264B}, $M_{\rm core} = 5 \,M_\odot$, and $M_{\rm Ni} = 0.068\,M_\odot$. While the physically correct value of shell thickness could be $\eta\sim0.01$ \citep{1994ApJ...425..264B,2005ApJ...626..183W}, because the simulation of such thin, and therefore dense, shells is computationally expensive, for practical reasons we assume thicker shells $\eta\approx 0.1$ in our fiducial simulation, and separately explore the scaling of the results in the limit of thin shells. Using these parameters gives $M_{\rm shell} = 2.45\,M_\odot$ and $M_{\rm amb} = 2.48\,M_\odot$. Thus outside of the nickel bubbles a fraction $f_{\rm shell} = 0.498$ of the mass is in shells and $f_{\rm amb} = 0.502$ is in the ambient ejecta.   Evaluating Equation (\ref{eq:density_normalization}) we obtain, for the bubble and ambient densities, $\rho_{\rm bub}(t_0)=1.1\times10^{-14}\,\text{g}\,\text{cm}^{-3}$ and $\rho_{\rm amb}(t_0)=3.9\times10^{-13}\,\text{g}\,\text{cm}^{-3}$.  The shell densities depend on the shell thickness parameter.  For the shell densities we obtain $\rho_{{\rm shell},\eta=0.1}(t_0)=1.2\times10^{-12}\,\text{g}\,\text{cm}^{-3}$ and $\rho_{{\rm shell},\eta=0.01}(t_0)=1.3\times10^{-11}\,\text{g}\,\text{cm}^{-3}$.

We perform dust synthesis calculations on a grid of ejecta mass coordinates.  For the bubble ejecta within $1.825\,M_\odot\leq M<1.892\,M_\odot$ we lay a grid with spacing $\Delta M=0.01\,M_\odot$ and run our calculations separately at each coordinate. The elemental composition in each run is taken from the \textsc{mesa} calculation whereas the density and temperature are chosen as appropriate for bubble interiors.  The density is as given in Equations (\ref{eq:density_evolution}) and (\ref{eq:density_normalization}) and the temperature we discuss in the following section. The results of these runs are used to determine the properties of dust formed in the bubbles. For the shell and ambient ejecta within $1.892\,M_\odot\leq M<6.8\,M_\odot$ we choose a set of mass coordinates separated by $\Delta M=0.1\,M_\odot$ and run our code twice at each coordinate. The two runs at the same mass coordinate have the same composition but different densities and temperatures. One run has a temperature and density appropriate for the shells and the other for the ambient ejecta.

\subsection{Thermal evolution}
\label{sec:thermal_evolution}

The ejecta thermal evolution is governed by the radioactive energy input.\footnote{In SN 1987A, the luminosity by $\sim2000$ days is dominated by the ejecta's interaction with the circumstellar medium \citep[e.g.,][]{McCray:16}.  We do not model this effect.}  After \iso{Ni}{56} and \iso{Ni}{57} have decayed, the heating is due, in increasing order of the half-life, to \iso{Co}{56}, \iso{Co}{57}, \iso{Na}{22}, and finally \iso{Ti}{44}.   We used the radiation transfer code \textsc{cloudy} \citep{Ferland:13} to create a model for the gas temperature evolution.  The \textsc{cloudy} calculation is not designed to accurately capture the geometry of radiative energy input and transfer within the ejecta.  Therefore, it cannot be used to predict the normalization of the temperature, but only its variation in time.  We normalize the temperature evolution by recalibrating a \textsc{cloudy} integration to astronomical measurements of the temperature.  This approach allows us to extrapolate the temperature evolution past the first $1000\,\text{d}$ when measurements of the temperature are not available.  We believe that the temperature evolution obtained through this heuristic procedure is more realistic than the power-law models invoked in published computations of dust synthesis in supernovae.  In fact, we find that the ejecta temperature does not decrease in a power-law fashion.

\begin{table}
\begin{center}
\begin{tabular}{ lll }
\hline
Source & $E$ (keV) & $L$ (erg s$^{-1}$)  \h
$e^\pm$, X-ray & 1 &    $3.39\times 10^{40}\times e^{-t/111.3\:\rm d}$    \\
& & $+\, 9.4\times 10^{37}\times e^{-t/391.2\:\rm d}$ \\
& & $+\,1.45\times 10^{34}\times e^{-t/78\:\rm yr}$ \\
& & $+\,2.39\times 10^{36}\times e^{-t/3.75\:\rm yr}$ \\
\iso{Co}{56} $\gamma$-ray &1243 & $9.73\times 10^{41} \times e^{-t/111.3\:\rm d}$ \\
\iso{Co}{57} $\gamma$-ray &115.2 & $5.06\times 10^{38} \times e^{-t/391.2\:\rm d}$ \\
\iso{Ti}{44} $\gamma$-ray &73.24 & $5.14\times 10^{35} \times e^{-t/78\:\rm yr}$ \\
\iso{Sc}{44} $\gamma$-ray &738.3 & $7.91\times 10^{36} \times e^{-t/78\:\rm yr}$ \\
\iso{Na}{22} $\gamma$-ray &782.9 &$1.64\times 10^{35} \times e^{-t/3.75\:\rm yr}$ \\
  \hline
\end{tabular}
\end{center}
\caption{Radioactive emission energies and total luminosities used in the \textsc{cloudy} calculations.}
\label{tab:cloudy}
\end{table}

\begin{figure*}
\begin{center}
\includegraphics[trim=0cm 4cm 0cm 5cm,clip=true,width=0.33\textwidth]{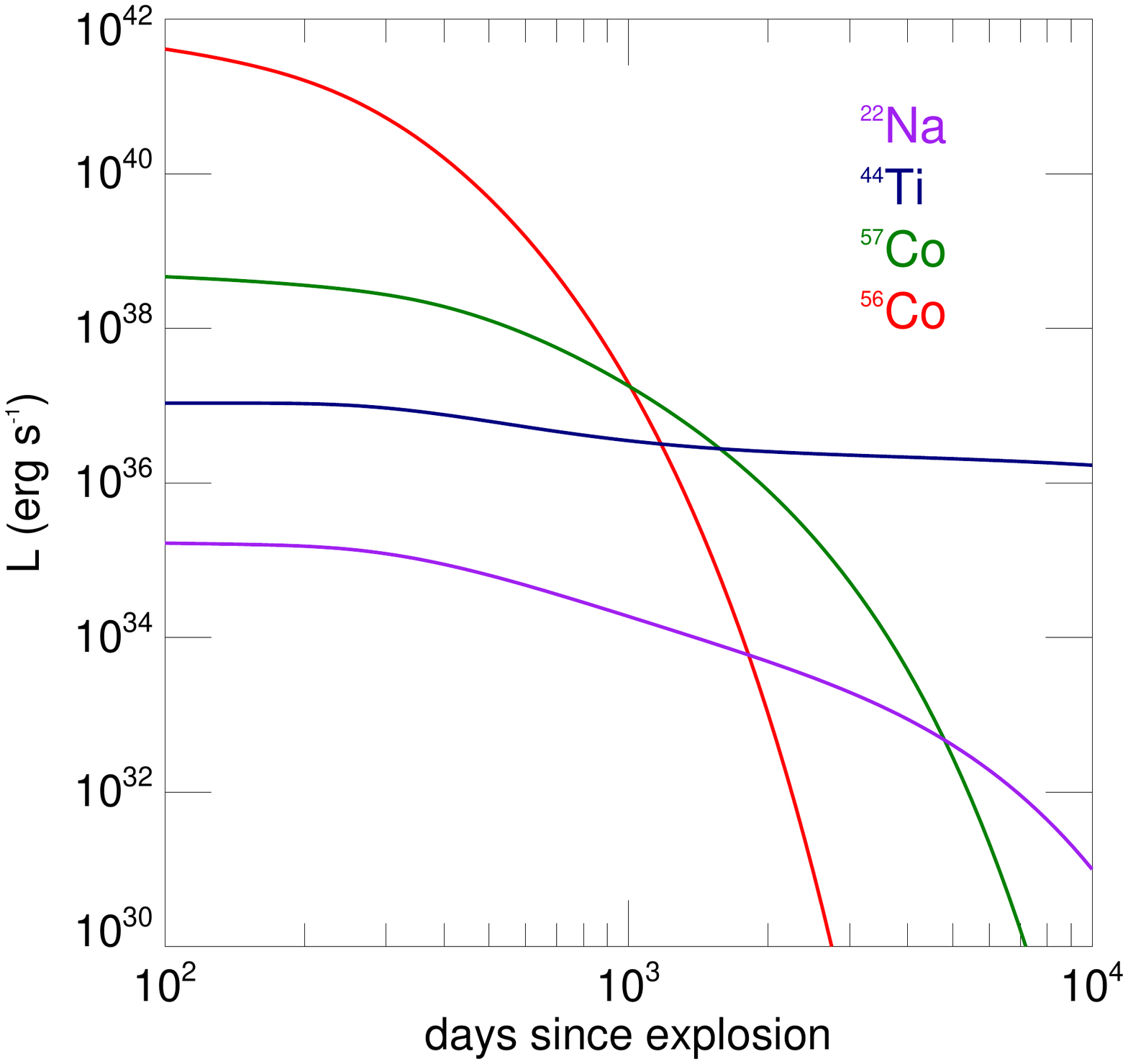}
\includegraphics[trim=0cm 4cm 0cm 5cm,clip=true,width=0.33\textwidth]{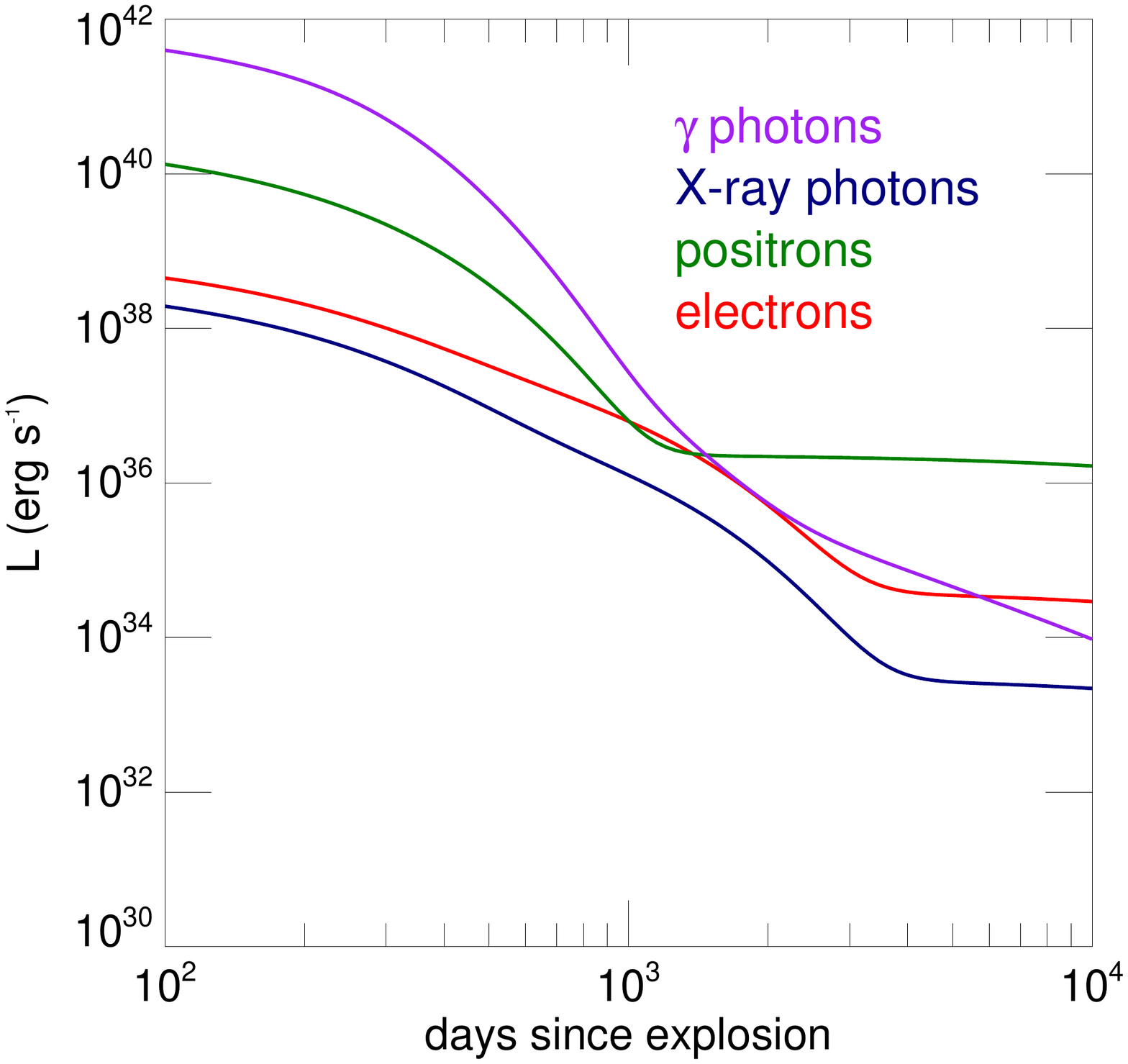}
\includegraphics[trim=0cm 4cm 0cm 5cm,clip=true,width=0.33\textwidth]{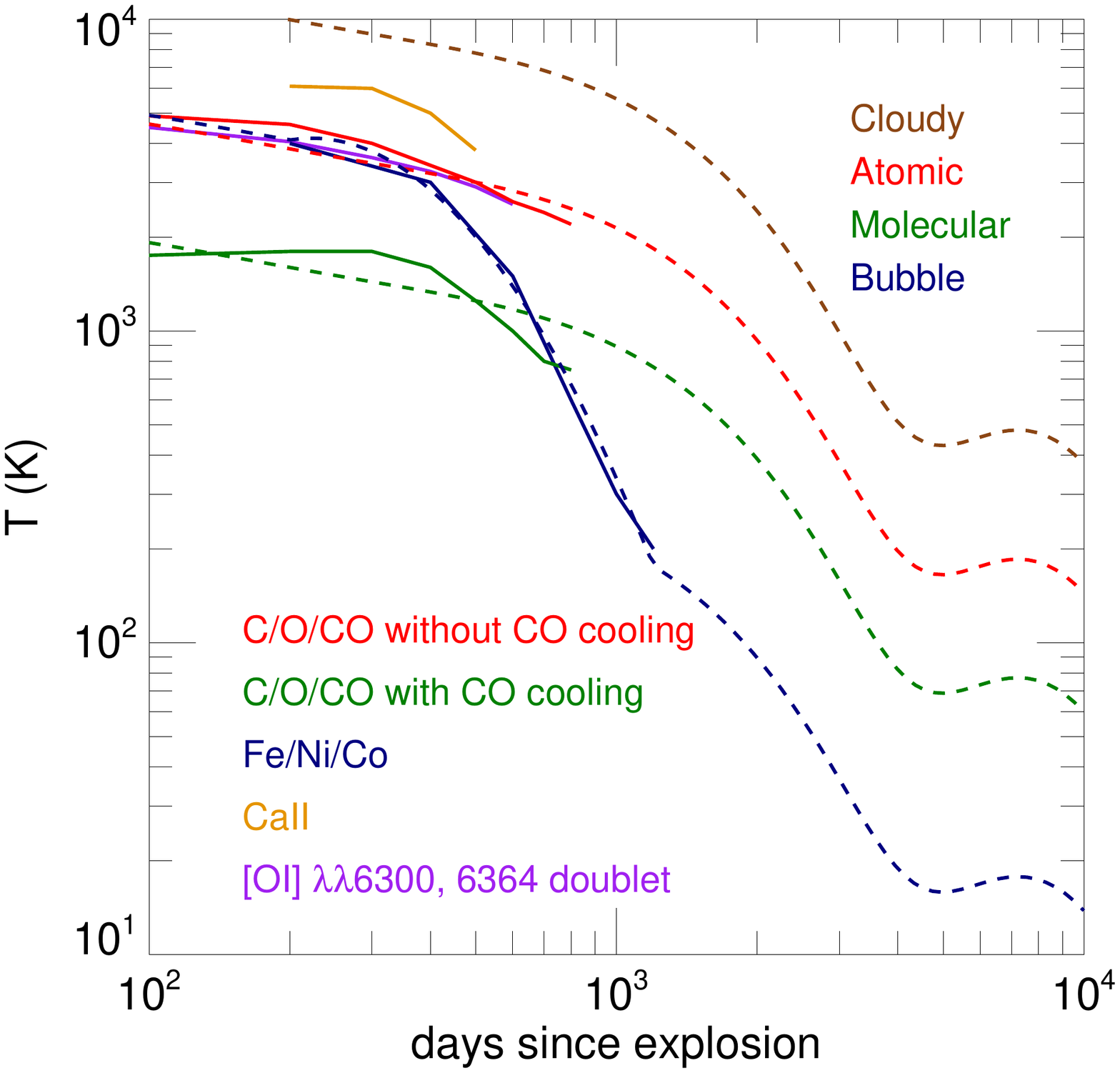}
\end{center}
\caption{The curves show the rate at which energy is deposited in the ejecta by the radioactive decay of four isotopes (left panel) and via four radioactive emission channels (middle panel), and temperature evolution in SN 1987A (right panel).  For a comparison of the energy deposition rates and compiled optical and far IR light curves of SN 1987A please see Figure 2 of \citet{McCray:16}. The energy deposition channels in the middle panel are classified by the initial energy injecting particle and account for all of the secondary processes excited by the injection event. The right panel shows: observationally inferred temperatures in SN 1987A (solid lines) and our rescaled \textsc{cloudy} temperature evolutions (dashed lines). The orange (assuming atomic C and O) and purple (assuming molecular CO) solid lines are from \citet{1995ApJ...454..472L}, the blue from \citet{1993ApJ...419..824L}, the green from \citet{1993ApJ...405..730L}, and the red from \citet{1992ApJ...387..309L}. Please see text for detail.}
\label{fig:radioactive}
\end{figure*}

Specifically, we first use \textsc{cloudy} to compute the temperature of an optically-thin singe zone at distance $R(t) = 2.16\times 10^{15}\u{cm}(t/100\u{days})$ from a point source and receding from the source with radial velocity $v = 2,500\,\text{km}\,\text{s}^{-1}$.  The zone has atomic number density $8.8\times 10^9 \u{cm}^{-3} (t/100\u{days})^{-3}$ and fiducial adopted atomic concentrations in the proportion $\text{C}:\text{O}:\text{Mg}:\text{Al}:\text{Si}:\text{S}:\text{Fe}=0.25:0.25:0.1:0.1:0.1:0.1:0.1$.\footnote{We exclude helium because it does not contribute to cooling.} The single zone is irradiated by a source with a luminosity equal to the rate at which energy is released during radioactive decay (excepting the portion of the radioactive energy released in neutrinos). We take the source to emit photons at 6 discrete energies, 5 of which are produced directly in the decays and the 6th is an artificial source of 1 keV photons crudely representing the energy deposited as X-rays, electrons, and positrons. The photon energies and luminosities as a function of time for each discrete energy are given in Table~\ref{tab:cloudy} and are plotted in Figure \ref{fig:radioactive}.  During the first 900 days our computed luminosity agrees with the empirical bolometric luminosity of SN 1987A \citep{Suntzeff:90,Bouchet:91}.
 
We ran the \textsc{cloudy} calculation on a temporal grid with 100 day spacing that spans the post-explosion period from $100$ to $10^4$ days. This provides an instantaneous temperature model $T_\textsc{cloudy}(t)$ at every epoch. The raw thermal evolution generated this way cannot be taken at face value because it ignores radiation transfer effects, spatial variation in chemical composition, and spatial variation in density. We \emph{recalibrate} by uniform rescaling the  \textsc{cloudy} model to empirical estimates of the temperature in SN 1987A. The recalibration is empirical and heuristic; it is justified by the close match between the recalibrated and measured temperature during the first $\sim1000$ days.  In particular, the recalibration can be construed as accounting for all the optical depth effects and the incompleteness of the inventory of molecular coolants included in \textsc{cloudy}.  The atomic temperature track is in reasonable agreement with the calculations shown in Figures 7 and 10 of \citet{Fransson:89} that cover the thermal evolution from 200 to 950 days.

In the right panel of Figure \ref{fig:radioactive} we show key measurements of the ejecta temperature.   \citet{1995ApJ...454..472L} estimated the temperature assuming that heating is from radioactive decay of \iso{Co}{56}, whereas cooling is from adiabatic expansion, free-free emission, recombination, C and O lines, and vibrational CO lines.  In their model, the ejecta consists of equal parts of C and O.  In this mixture, CO is destroyed by Compton electrons from radioactive decay and created by radiative association of C and O (they provide a separate estimate, also shown in the figure, with CO formation disabled).  \citet{1993ApJ...419..824L} estimated the temperature in the nickel-rich regions by taking into account radioactive heating and Fe, Co, and Ni line emission cooling. 
\citet{1993ApJ...405..730L} analyzed Ca\,II emission lines in the hydrogen envelope ejecta.  \citet{1992ApJ...387..309L} analyzed the flux and profile of two forbidden O\,I emission lines. They found that to match observations the oxygen must be absent at velocities $\lesssim 300\,\text{km}\,\text{s}^{-1}$ whereas $0.66\,M_\odot$ of oxygen occupies $10\%$ of the volume at velocities $300\,\text{km}\,\text{s}^{-1} < v < 1400\,\text{km}\,\text{s}^{-1}$ and $0.64\,M_\odot$ occupies $5\%$ of the volume at $1400\,\text{km}\,\text{s}^{-1} < v < 2100\,\text{km}\,\text{s}^{-1}$. The fact that the temperatures estimated from the forbidden O\,I lines overlap with those derived from \citet{1995ApJ...454..472L} with CO cooling disabled suggests that much of the oxygen may be where molecules are not able to form and cool the gas.

We use the estimates of \citet{1995ApJ...454..472L} to normalize our time evolution of temperature in the shells and the ambient gas.  For dense shells, where molecules should be able to form, we use the estimates with CO formation enabled. In Figure \ref{fig:radioactive}, right panel, we show the unnormalized $T_\textsc{cloudy}$ and the normalized shell temperature $T_{\rm shell}(t)=0.16\,T_\textsc{cloudy}(t)$. For the ambient ejecta, to match the atomic gas temperature estimate, we normalize as $T_{\rm amb}(t)=0.39\,T_\textsc{cloudy}(t)$. We use the observed Fe-Co-Ni temperatures from \citet{1993ApJ...419..824L} to construct a hybrid temperature model for the bubbles.  In the interval $200 \u{d}<t<1200\u{d}$, where temperature data are available, we use a parametric fit to the observations.   Outside of this period, we continuously extrapolate with an appropriately normalized \textsc{cloudy} temperature evolution:
\beq
T_{\rm bub}(t) = \pw{
\epsilon_1\,T_\textsc{cloudy}(t), & t < 200\u{d} ,\\
10^{\epsilon_2 + \epsilon_3 x+\epsilon_4 x^2} \u{K}, & 200 \u{d}<t<1200\u{d} ,\\
\epsilon_5\,T_\textsc{cloudy}(t), & t > 1200\u{d} ,
}
\eeq  
where $x = \log_{10}(t/\u{day})$ and the coefficients are: $\epsilon_1=0.411$, $\epsilon_2=-10.5$, $\epsilon_3=12.1$, $\epsilon_4=-2.57$, and $\epsilon_5=0.369$.  For the sake of reproducibility, here we provide a fitting function for $T_\textsc{cloudy}$, valid for post-explosion times $100\u{d} < t < 10,000 \u{d}$: $T_\textsc{cloudy}(t) = \sum_{i=1}^6 \xi_i e^{-t/t_i}$
where $t_i = 100\,\text{d} + 500\,(i-1)\,\text{d}$, $\xi_1 = 7012 \u{K}$, $\xi_2=-4588 \u{K}$, $\xi_3 = -27530 \u{K}$, $\xi_4=155300 \u{K}$, $\xi_5=-195200\u{K}$, and $\xi_6 = 81770 \u{K}$.

We assume that this temperature model applies throughout each of the three zones.  This is a very crude and ultimately incorrect assumption, though one without which our first attempt at a comprehensive dust synthesis calculation would have proven unmanageable.  In reality, the energy emitted in the form of $e^{\pm}$ and X-rays is deposited essentially locally as the mean free path of an electron or an X-ray is much smaller, by a factor of at least $10^5$ (for $e^\pm$) and at least $10^2$ (for X-rays), than the radius of the ejecta.  The mean free path for $\gamma$-ray absorption is much longer in comparison, e.g., for 1 MeV $\gamma$-rays it starts shorter than the radius of the ejecta but eventually becomes almost $10^3$ times longer. The ejecta become optically thin to $\gamma$-rays at $\sim360$--$560\,\text{days}$ depending on the photon energy.  Ideally, the radiation transfer effects should be modeled realistically.

In the course of revising this manuscript in response to the referee's comments we performed a test of the thermal evolution model presented in this section.  In the test we carried out a more accurate but simplified direct computation of the gas temperature.  We assumed that the ejecta consisted entirely of O and CO.  To compute the O cooling rate we take electron collision strengths for the lowest 5 energy levels of neutral oxygen from \citet{Draine:11}.  For the CO cooling rate we used the tabulated rotational and vibrational CO cooling rates from \citet{Neufeld:93} with corrections from \citet{Glover:10}.  In this simplified calculation we assumed that the molecular and ionization fractions were both 1\% and the density equaled the average density of the ejecta.  We obtained the temperature by equating the cooling rate to the radioactive heating rate using the Sobolev approximation for the optical depth.  We found that compared to the simplified calculation not relying on \textsc{cloudy}, the  \textsc{cloudy}-based model overestimates the temperature between 1000 and 3000 days and underestimates the temperature thereafter.  The simplified model also exhibits a more substantial late increase of temperature which brings into focus the complicated interplay of cooling and radioactive heating.

\section{Chemistry}
\label{sec:chemistry}

Immediately after a supernova explosion, the ejecta is ionized gas. As the ejecta expand and cool, the ions recombine into atoms and molecules form via gas-phase chemical reactions. Some molecules grow large enough to become what might be called condensation nuclei, which can grow into small grains via accretion. These grains can then grow into larger grains by accretion and coagulation or diminish by evaporation and chemical weathering.  To model the initial steps of the gas-to-dust transformation in supernova ejecta, the abundances of molecular species must be explicitly followed. The processes that modify molecular abundances, such as gas-phase chemical reactions, reactions with and accretion onto grains, and destruction by Compton electrons, are incorporated into the abundance evolution calculations. We proceed to describe how this is done in our simulation.

\subsection{Reaction network}

\begin{table}
\begin{tabular}{ ll }
\hline
Category & Species \\ \hline
Atoms &  He, C, O,  Ne,  Mg, Al, Si,  S, Ar,  Fe
\\ 
Molecules & CO, C$_2$, O$_2$, SO, SiO, SiC, Fe$_2$, FeO, \\
 & Mg$_2$, MgO, Si$_2$, AlO, FeS, MgS
\\ 
& C$_3$, SiO$_2$, Fe$_3$, Mg$_3$, Si$_3$
 \\ 
& C$_4$, Fe$_4$, Mg$_4$, Si$_4$, Si$_2$O$_2$, Al$_2$O$_2$, \\
& Fe$_2$O$_2$, Fe$_2$S$_2$, Mg$_2$O$_2$, Mg$_2$S$_2$, Si$_2$C$_2$
\\ 
 & Si$_2$O$_3$, Al$_2$O$_3$, Fe$_2$O$_3$
\\ 
& MgSi$_2$O$_3$, Si$_3$O$_3$, Fe$_3$O$_3$, Fe$_3$S$_3$,\\
&  Mg$_3$O$_3$, Mg$_3$S$_3$, Si$_2$O$_4$
\\ 
 & MgSi$_2$O$_4$, Si$_3$O$_4$, Fe$_3$O$_4$
\\ 
 & Mg$_2$Si$_2$O$_4$, Si$_4$O$_4$, Fe$_4$O$_4$, Fe$_4$S$_4$, \\ 
& Mg$_4$O$_4$, Mg$_4$S$_4$
\\ 
& Mg$_2$Si$_2$O$_5$, Si$_4$O$_5$
\\ 
& Mg$_3$Si$_2$O$_5$, Mg$_2$Si$_2$O$_6$, Si$_5$O$_5$, Al$_4$O$_6$
\\ 
 & Mg$_3$Si$_2$O$_6$
\\ 
 & Mg$_3$Si$_2$O$_7$
\\ 
 & Mg$_4$Si$_2$O$_7$
\\ 
 & Mg$_4$Si$_2$O$_8$, Fe$_6$O$_8$
\\ 
Ions & $e^-$ 
\\ 
&  He$^+$, C$^+$, O$^+$,  Ne$^+$,  Mg$^+$, Al$^+$,  Si$^+$, S$^+$, Ar$^+$, Fe$^+$
\\ 
& CO$^+$, C$_2^+$, O$_2^+$, SO$^+$, SiO$^+$, SiC$^+$, Fe$_2^+$, FeO$^+$, \\
 & Mg$_2^+$, MgO$^+$, Si$_2^+$, AlO$^+$, FeS$^+$, MgS$^+$
\\ 

\hline
\end{tabular}
\caption{Atomic, molecular, and ionic species included in the simulation classified by the number of atoms per molecule.}
\label{tab:species_table}
\end{table}

Let $\m N_{\rm S}$ be the total number of chemical species that we do not treat as dust grains (grains will be added to the picture in Section \ref{sec:grain_physics}). The chemical species include atoms, molecules, ions, free electrons, and atomic and molecular clusters, the latter being grain condensation nuclei.  The species are listed in Table~\ref{tab:species_table}.  We explicitly follow the number density $c_i(t)$ of species $i$, where $i=1,\,...,\,\m N_{\rm S}$,  as a function of post-explosion time $t$.  The number density of each species changes due to gas-phase chemical reactions, chemical reactions with, or catalyzed by dust grains, accretion onto and evaporation from grains, collisions with Compton electrons, and expansion of the ejecta.  The number densities $c_i(t)$ obey a system of coupled ordinary differential equations.

We include a total of $\m N_{\rm R}=341$ gas-phase chemical reactions.  We allow up to 3 reactants and up to 3 products in each reaction.  Reaction $j$, where $j=1,\,...,\,\m N_{\rm R}$, can be written in the form:
\beq
R_1(j) + R_2(j) + R_3(j) \to P_1(j) + P_2(j) + P_3(j) ,
\eeq  
where $R_k(j)$ is the $k$th reactant in reaction $j$ and $P_k(j)$ is the $k$th product in reaction $j$. If there are fewer than three reactants or products, we set the extra ones to zero. 

Each reaction $j$ has a gas-temperature-dependent rate coefficient $k_j(T)$. The rate coefficients are written in the Arrhenius form:
\beq
k_j(T) = A_j \({\f{T}{300\u{K}}}^{\nu_j} e^{-E_{\rm a,\it j}/\kboltzmann T} ,
\eeq
where $A_j$, $\nu_j$, and $E_{\rm a,\it j}$ are reaction-specific constants. The exponent $\nu_j$ is dimensionless. The activation energy $E_{\rm a,\it j}$ has the units of energy but is usually expressed as a temperature $T_{\rm A,\it j} = E_{\rm a,\it j}/\kboltzmann$. The units of the coefficient $A_j$ depend on the number of reactants: s$^{-1}$ for one reactant, cm$^3$ s$^{-1}$ for two, and cm$^6$ s$^{-1}$ for three. Reaction $j$ has a rate per unit volume $k_j(T) c_{R_1(j)} c_{R_2(j)} c_{R_3(j)}$, where $c_{R_3(j)} = 1$ if there are two reactants and $c_{R_2(j)} = c_{R_3(j)} = 1$ if there is only one reactant. 

The time derivative of the number density of species $i$ due to gas-phase chemical reactions is:
\beq
\label{eq:dndt_chem}
\left(\f{dc_i}{dt}\right)_\text{chem} = \sum_{j=1}^{\m N_{\rm R}} (\alpha_{ij}-\beta_{ij} ) k_j(T) c_{R_1(j)} c_{R_2(j)} c_{R_3(j)} 
\eeq
where $\alpha_{ij}$ and $\beta_{ij}$ are the number of times that species $i$ occurs as, respectively, a product and a reactant in reaction $j$.

\subsection{The rate coefficients}

\label{sec:estimatingRates}

The chemical reactions that we include in our simulations are given in Tables~\ref{tab:bigReactionTableFirst} through~\ref{tab:bigReactionTableLast} (hereafter referred to as the ``reaction tables"). The tables give, for each reaction $j$, the Arrhenius rate coefficient parameters $A_j$, $\nu_j$, and $T_{0,j}$. The numerical values of these parameters were taken from the literature when possible. Unfortunately, not all reactions relevant to dust formation have measured or calculated rates, and the rates of those that do are only valid in certain temperature and pressure range.  In some cases, as we outline here, we have had to perform informed extrapolations of the measured or calculated rates.


The coefficient for a \emph{two-body} reaction $\mbox{A} + \mbox{B} \to \mbox{C} + \mbox{D}$ is given by a thermal average of the reaction cross section multiplied by the relative velocity $v_\text{AB}$
\beq
\label{eq:k2body} 
k_{\rm AB} = \pi(r_{\rm A} + r_{\rm B})^2 \sqrt{\frac{8  \kboltzmann T}{\pi\mu_{\rm AB}}}\({1 + \f{E_{\rm a,AB}}{\kboltzmann T}} e^{-E_{\rm a,AB} / \kboltzmann T} ,
\eeq
where $r_i$ is the molecular radius of species $i$, $\mu_{AB}$ is the reduced mass, and $E_{\rm a,AB}$ is the activation energy (which may be zero). 
This expression for the rate coefficient is not in Arrhenius form. If the activation energy is zero or much less than $\kboltzmann T$, then all terms involving $E_{\rm a,AB}$ vanish and the rate coefficient is in Arrhenius form. On the other hand, if $E_{\rm a,AB}\gg\kboltzmann T$ then $1 + E_{\rm a,AB} / \kboltzmann T$ can be replaced with $E_{\rm a,AB} / \kboltzmann T$ and again the rate coefficient is in Arrhenius form. 


A \emph{three-body} reaction of the form $\mbox{A} + \mbox{B} + \mbox{M} \to \mbox{AB} + \mbox{M}$, where M is any gas particle, takes place in two steps. First an A particle collides with a B particle forming an unstable transition state AB$^*$. Then a gas particle M collides with the transition state and removes enough of the energy of the transition state to leave it in the form of a stable AB molecule.  
The rate coefficient $k^*_{\rm AB}$ for the formation of the transition state is obtained by setting the activation energy to zero in Equation~(\ref{eq:k2body}).  

Once the transition state forms, it has a lifetime $\tau^*_{\rm AB}$ during which it can fragment back into A and B. The mean time between collisions with a third body for a given transition state molecule is:   
\beq
\tau_{\rm col,AB} = \f{1}{c_{\rm tot} (\avg{r} + r_{\rm AB})^2}  \({
\f{8\pi\kboltzmann T}{\avg{m}} + \f{8\pi\kboltzmann T}{m_{\rm AB}} }^{-1/2} ,
\eeq
where $c_{\rm tot}$ is the total number density of all gas species, $\avg{r}$ is the average radius of a gas particle, $\avg{m}$ is the mean molecular weight, $r_{\rm AB}$ is the radius of the transition state, and $m_{\rm AB} = m_{\rm A} + m_{\rm B}$ is the mass of the transition state.

If $\tau_{\rm col,AB} > \tau_{\rm AB}^*$, the number density of transition state molecules is $c_{\rm AB}^* = c_{\rm A}c_{\rm B} k_{\rm AB}^* \tau_{\rm AB}^*$ and so the rate per unit volume of the overall three-body reaction is $c_{\rm AB}^* c_{\rm tot} k' = c_{\rm A}c_{\rm B} k_{\rm AB}^* \tau_{\rm AB}^* c_{\rm tot} k_{\rm AB}'$, where:
\beq
k_{\rm AB}' = (r_{\rm AB} + \avg{r})^2  \({
\f{8\pi\kboltzmann T}{\avg{m}} + \f{8\pi\kboltzmann T}{m_{\rm AB}} }^{1/2} .
\eeq
Thus the overall volumetric three-body reaction rate can be written as $k_{3,{\rm AB}} c_{\rm A}c_{\rm B} c_{\rm tot}$ with $k_{3,{\rm AB}} = k_{\rm AB}^* \tau_{\rm AB}^* k_{\rm AB}'$.  If, on the other hand, $\tau_{\rm col,AB} < \tau_{\rm AB}^*$, then we can assume that a third body M will collide with the transition state before it fragments and thus we can replace the reaction $\mbox{A} + \mbox{B} + \mbox{M} \to \mbox{AB} + \mbox{M}$ with $\mbox{A} + \mbox{B}  \to \mbox{AB} $. The rate coefficient for this reaction is again as given in Equation~(\ref{eq:k2body}).  

For simplicity we assume that $\tau_{\rm AB}^* = 2\pi /\omega_0 = 1.52\times 10^{-13}\u{s}$, where $\omega_0 = 4.12\times10^{13}\u{s}^{-1}$ is the angular frequency of bonds in an Einstein solid made of carbon. Effectively we are approximating the lifetime of the transition state as the vibrational period of the bond holding the transition state molecule together. We are also assuming that this vibrational period is similar to that of a carbon Einstein solid. Then, with $r_{\rm A} = r_{\rm B}=6.7\times 10^{-9}\u{cm}$ (the radius of a carbon atom), $m_{\rm A} = m_{\rm B} = 12.0\u{amu}$ (the mass of a carbon atom), $\avg{m}=12.0\u{amu}$, and $\avg{r} = 4.69\times 10^{-9}\u{cm}$ gives a three-body reaction coefficient with Arrhenius parameters $A=4.253\times 10^{-34}\u{cm}^6 \u{s}^{-1}$, $\nu =+1$, and $T_0=0$ (where the latter indicates that we are assuming there is no activation energy). We use these parameters for all three-body reactions; they agree with experimental values in the common temperature range of validity but give rates that behave well at low temperatures.

The reactions involving the formation of enstatite and forsterite dimers were taken from \citet{2012MNRAS.420.3344G}.  The reference describes how gas phase reactions can build up silicates such as enstatite and forsterite dimers starting with SiO as the seed. Each reaction in their paper involves either addition of a Si or Mg atom or oxidation by H$_2$O.  Since in our model, hydrogen is absent in the dust-forming-ejecta of SN 1987A, we follow the approach of \citet{Sarangi:13} and substitute O$_2$ and SO for the H$_2$O in  \citet{2012MNRAS.420.3344G}.  Tables providing all the reactions rates are available in electronic form at MNRAS online.

\subsection{Destruction by Compton electrons}
\label{sec:destruction_Compton}

In Section \ref{sec:supernova_model} we discussed how radioactivity produces a population of high energy electrons (and positrons), the so-called Compton electrons, that can go on to ionize atoms and molecules and dissociate molecules. We include the following ``destruction by Compton electron" reactions. For each neutral atomic species X we include a reaction of the form: 
\beq
\mbox{X} + e^-_* \to \mbox{X}^+ + e^- + e^-_* ,
\eeq
where $e^-_*$ denotes the Compton electron. For each neutral diatomic molecule AB we include the following four reactions:
\bea
\mbox{AB} + e^-_* &\to& \mbox{AB}^+ + e^- + e^-_* ,\nonumber\\ 
\mbox{AB} + e^-_* &\to& \mbox{A} + \mbox{B} + e^-_* ,\nonumber\\
 \mbox{AB} + e^-_* &\to& \mbox{A}^+ + \mbox{B} + e^- + e^-_* ,\nonumber\\
 \mbox{AB} + e^-_* &\to& \mbox{A} + \mbox{B}^+ + e^- + e^-_* .
\eea
For triatomic and larger molecules we ignore reactions with Compton electrons. 

\begin{table}
\begin{center}
\begin{tabular}{ llllllll }
\hline
Type & Reaction & $W_i^j$ (eV)\\
\hline
 Ionization (Atoms) &$\mbox{X}  \to \mbox{X}^+ + e^-$ & 47 \\
 Ionization (Molecules) & $\mbox{AB}  \to \mbox{AB}^+ + e^-$ & 34 \\
Dissociation & $\mbox{AB}  \to \mbox{A} + \mbox{B}$ & 125 \\
Dissociative Ionization & $\mbox{O}_2  \to \mbox{O}^+ + \mbox{O} + e^-$ & 768 \\
  & $\mbox{AO}  \to \mbox{A}^+ + \mbox{O} + e^-$ & 247 \\
    & $\mbox{AO}  \to \mbox{A} + \mbox{O}^+ + e^-$ & 768 \\
&$\mbox{AB}  \to \mbox{A}^+ + \mbox{B} + e^-$ & 247 \\
 \hline
\end{tabular}
\end{center}
\caption{Energy lost by Compton electrons per reaction.  In dissociative ionization, $\text{A},\text{B}\neq\text{O}$. The values of $W_i^j$ are from \citet{Cherchneff:09}.}
\label{tab:Compton}
\end{table}

To find the rate coefficient for reactions with Compton electrons we model the total energy per unit time that goes into ionization and dissociation by multiplying $L(t)$ in Equation (\ref{eq:total_luminosity}) with a dimensionless factor $L_{\rm ion}(t) = f_{\rm ion}\, L(t) $.  We assume that this energy is equally distributed among all gas particles. If $N_{\rm tot}$ is the total number of particles in the ejecta, then the rate at which ionizing energy is deposited directly onto an atom or molecule is $L_{\rm ion}/N_{\rm tot}$. For each species $i$ there are $\m N_i^{\rm C}$ possible reactions that can be induced by Compton electrons (ionization, dissociation, etc.). In each reaction $j$ the Compton electron loses an energy $W_i^j$ so the rate coefficient for reaction $j$ is
\beq
k_{i,j}^{\rm C} = \f{L_{\rm ion}}{N_{\rm tot}\,W_i^{j}} .
\eeq   
The rate per unit volume of reaction $j$ is $k_{i,j}^{\rm C} c_i$.  See Table~\ref{tab:Compton} for the values of $W_i^j$.

For convenience we give the time derivatives of the number densities of the species affected by Compton electron destruction reactions. For each neutral atomic species X we have 
\bea
\label{eq:Compton_1}
\({ \f{dc_{\rm X}}{dt}}_{\rm Compt} =
-\({ \f{dc_{{\rm X^+},e^-}}{dt}}_{\rm Compt} =-k_0c_{\rm X}  ,
\eea
where $k_0 = L_{\rm ion}/(N_{\rm tot} \times 47.0 \u{eV})$.
For each neutral diatomic molecular species AB we have:
\bea
\label{eq:Compton_2}
 \({ \f{dc_{\rm AB}}{dt}}_{\rm Compt} \= -  (k_1 + k_2 + k_3 + k_4)c_{\rm AB}  ,\nonumber\\
\({ \f{dc_{\rm AB^+}}{dt}}_{\rm Compt} \= k_1 c_{\rm AB} , \nonumber\\
\({ \f{dc_{\rm A^+}}{dt}}_{\rm Compt} \= k_3 c_{\rm AB} ,\nonumber \\
\({ \f{dc_{\rm B^+}}{dt}}_{\rm Compt} \=  k_4 c_{\rm AB} ,\nonumber\\
\({ \f{dc_{\rm A}}{dt}}_{\rm Compt} \=  (k_2 + k_4)c_{\rm AB}, \nonumber\\
\({ \f{dc_{\rm B}}{dt}}_{\rm Compt} \=   (k_2 + k_3)c_{\rm AB},\nonumber \\
\({ \f{dc_{e^-}}{dt}}_{\rm Compt} \= (k_1 + k_3 + k_4) c_{\rm AB}  ,
\eea
where $k_i = L_{\rm ion}/N_{\rm tot} W_i$.
Here, $W_{1...4}=(34,\,125,\,247,\,247)\,\text{eV}$ in reactions not producing ionized oxygen, and the same except for the replacement $247\,\text{eV}\rightarrow 768\,\text{eV}$ in reactions producing O$^+$.

\subsection{The nucleation of C clusters}

As an example we mention the main chemical reactions participating in the formation of carbon clusters. The main route to forming carbon grain condensation nuclei, which we take to be the clusters with 4 carbon atoms (C$_4$), involves the following monomer inclusion reactions $\text{C}+\text{C}_n \rightarrow\text{C}_{n+1}+\gamma$ for $n=(1,2,3)$ as well as $\text{C}_2 + \text{C}_2 \rightarrow \text{C}_4 + \gamma$.  Oxygen atoms disrupt this chain at several points. First, $
\text{C} + \text{O} \rightarrow \text{CO} +\gamma$ sequesters some carbon atoms so that they cannot be incorporated in clusters. This is somewhat offset by the Compton electron induced dissociation of carbon monoxide $\text{CO} + e_*^-\rightarrow \text{C} + \text{O} + e_*^-$.  Oxygen atoms also destroy carbon clusters via $\text{C}_n + \text{O}\rightarrow \text{C}_{n-1} + \text{CO}$.  Carbon clusters can also be destroyed by reactions with noble gas ions and Compton electrons
\begin{eqnarray}
& & \text{C}_n + \text{He}^+ \rightarrow \text{C}_{n-1} + \text{He} + \text{C},  \ \ \ (n=2,3,4) , \nonumber\\
& & \text{C}_n + e_*^- \rightarrow \text{C}_{n-1} + \text{C} + e_*^-, \ \ \ (n=2) .
\end{eqnarray}
Once C$_4$ forms dust grains are produced by $\text{C}_4+\text{C}_n\rightarrow \text{dust grains}$, where $n=(1,2,3,4)$. 

At this point a caveat is in order.  Carbon clusters, including those with $n\gg 4$, inhabit a complex configuration space where they form chains, rings, and fullerenes. The cluster geometry has a strong effect on the cluster stability as can be seen in the detailed calculations of, e.g., \citet{Mauney:15}.  The calculations presented in the present work could be improved (at the cost of substantial additional numerical development) by resolving all the $n\lesssim 100$ clusters by their atomic number and distinct geometric configuration with configuration-specific cohesion energies.  In particular some specific chains are less stable and dust nucleation then proceeds via other pathways.

\section{Grains}
\label{sec:grain_physics}

A dust grain is a collection of atoms and molecules that behaves as a solid body. We classify grains based on the molecular formula of the fundamental molecular constituent of the grain. When the molecular formula equals the fundamental molecular formula of a grain species, we call the molecule a monomer of the species. An object consisting of $n$ monomers of some grain species is an $n$-mer of the species. A dust grain is just an $n$-mer with a sufficiently large number of monomers.

\subsection{Classification}

\begin{figure*}
\begin{center}
\includegraphics[trim=0cm 0cm 0cm 0cm,clip=true,width=0.8\textwidth]{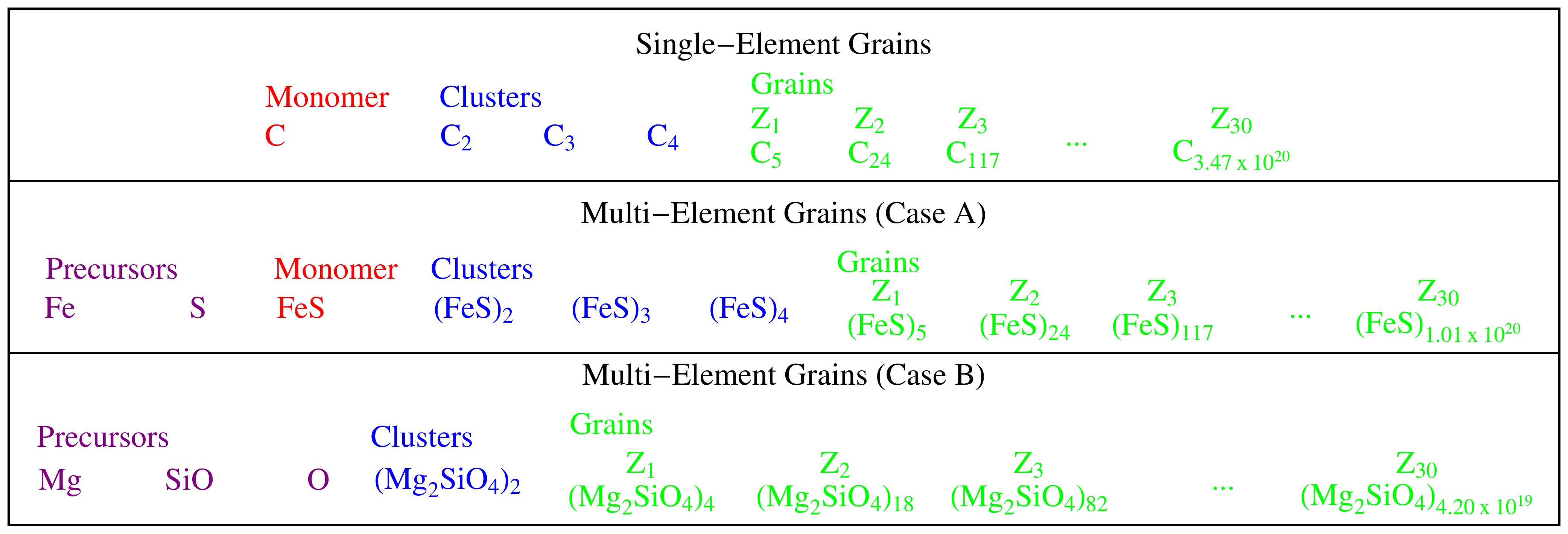}
\end{center}
\caption{The diagram shows the division of dust grains into categories based on their chemical synthesis pathway.  The single element grains nucleate directly from the atomic phase. The Case A multi-element grains nucleate from monomers existing in the gas phase, whereas the Case B grains nucleate from dimers that have formed in the gas phase.}
\label{fig:clusterClassification}
\end{figure*}

Grains of pure carbon, silicon, magnesium, and iron contain only one element and are referred to as single-element grains. Let $\m X$ denote a monomer of a single-element grain (which is just a single atom such as C, Si, Mg, or Fe) and $\m X_n$ denote an $n$-mer. In our simulation we treat $\m X$, $\m X_2$, $\m X_3$, and $\m X_4$ as molecular species in our non-equilibrium chemistry network. Clusters with $n>4$ are treated as grains.  We track the grain size distribution on a discrete grid of grain radii. The single-element grain synthesis chain consists of monomers, clusters, and grains as illustrated in Figure~\ref{fig:clusterClassification}.  

The monomers of multi-element grains are composed of more than one element. They can be as simple as iron sulfide (FeS) or as complex as forsterite (Mg$_2$SiO$_4$). It is convenient to divide multi-element grains into two sub-classes, Cases A and B, based on whether their monomer exists in the gas phase.  Case A multi-element grains are those that have a monomer that \emph{can} form in the gas phase. Here, such grains include: FeS, FeO, SiC, Al$_2$O$_3$, SiO$_2$, MgO, Fe$_3$O$_4$, and MgS. The first step in the formation of a Case A multi-element grain is the formation of a monomer $\m X$ from chemical reactions involving gas-phase precursors. Later $n$-mers $\m X_n$ are built from the monomers by accretion and coagulation and can eventually grow into grains.  The Case A multi-element grain synthesis chain consists of precursors, monomers, clusters, and grains, as shown in the middle section of Figure~\ref{fig:clusterClassification}.

Case B multi-element grains consist of monomers that \emph{cannot} exist in the gas phase. We consider only two such species, enstatite (MgSiO$_3$) and forsterite (Mg$_2$SiO$_4$). The first step in the synthesis of Case B multi-element grains is direct gas-phase formation of a dimer $\m X_2$.  The dimer serves as the nucleus for $n$-merization by accretion and coagulation. The Case B multi-element grain synthesis chain also consists of precursors, clusters, and grains, as shown in the lower section of Figure~\ref{fig:clusterClassification}.

The grains formed in supernovae can be heterogeneous mixtures aggregating different species, e.g., mixtures of carbon and enstatite. To curtail computational complexity, we did not consider heterogeneous grains.  The effect of heterogeneity would be to deplete refractory elements from the gas phase more efficiently than predicted here.

\subsection{Grain size discretization}

We model dust grains as balls of densely packed monomers in the solid phase. The treatment of molecular clusters as spherical and densely packed is of course highly artificial---the clusters can in fact have linear and other aspherical geometries---but it is a necessary oversimplification that makes our comprehensive calculation tractable. Let $m_1$, $a_1$, $x_1=\frac{4}{3}\pi a_1^3$, and $\varrho = m_1 /x_1$ denote the grain-species-specific mass, radius,  volume, and density of one monomer in the solid phase.  With these, $m_n=nm_1$, $a_n=n^{1/3}a_1$, $x_n=nx_1$, and $\varrho$ are the mass, radius, volume, and density of an $n$-mer.

For each grain species we track the number density of precursor $n$-mers, which we call `clusters', for all consecutive $n$ up to some maximum value $n_{\rm max}$.   Larger $n$-mers with $n > n_{\rm max}$ we refer to as `grains'.  Since we cannot separately track the number densities of grains for all consecutive, large $n$, for each grain species we discretize the grain density as a function of the $n$-mer number on a logarithmic grid of grain radii labeled by index $\ell$ (recall that the radius is in one-to-one relation with the $n$-mer number and the grain volume).  Let $\ell_{\rm max}$ be the number of the grid points.  We set the smallest radial grid point $a_{\rm min}$ to the radius of the $(n_{\rm max}+1)$mer particle for all species except for enstatite and forsterite, for which we use the radius of the $(n_{\rm max}+2)$mer. For the maximum radius and number of grid points we use $a_{\rm max}=100\u{$\mu$m}$ and $\ell_{\rm max}=50$ for all grain species.  Specifically the $\ell$th grid point is at $a_\ell=a_{\rm min} (a_{\rm max}/a_{\rm min})^{(\ell-1)/(\ell_{\rm max}-1)}$.  To distinguish between clusters and grains, a cluster with $n$ monomers is denoted with $\m X_n$, has radius $a_n$, mass $m_n$, and number density $c_n$. A grain associated with radial grid point $\ell$ is denoted with $\m Z_\ell$, has radius $a_\ell$, mass $m_\ell$, and number density $c_\ell$.

%

\subsection{Coagulation}
\label{sec:coagulation}

In our simulation clusters form via chemical binding via the reactions included in our chemical reaction network. The largest $n$-mer cluster $\m X_{n_{\rm max}}$ acting as precursor of a given grain species is referred to as the condensation nucleus. The condensation nucleus can grow into a grain by accretion of gas phase precursors or by coagulation with other clusters. It is also possible that two clusters, either or both of which can be smaller than the condensation nucleus, can merge to form a grain. 
We classify coagulation events based on the coagulating species and the coagulation product.
The events where two clusters collide and produce a cluster are handled by the chemical reaction network. The events in which two clusters collide and result in a grain, however, are not considered chemical reactions but are genuine coagulation events. The other possibilities, including cluster-grain and grain-grain coagulation events, always lead to grains.

Coagulation is the process:
\beq
\m X_{n_1} + \m X_{n_2} \to  \m X_{n_1+n_2} 
\eeq
in which grains $\m X_{n_1}$ and $\m X_{n_2}$ with volumes $x_{n_1}$ and $x_{n_2}$ collide and adhere to form a new grain $\m X_{n_1+n_2}$ with volume $x_{n_1+n_2} = x_{n_1} + x_{n_2}$.
The contribution of coagulations to the rate of change of the concentration $c_n$ of grains $\m X_n$ is:
\begin{equation}
\label{eq:coag_discrete}
\left(\frac{dc_n}{dt}\right)_{\rm coag} =  \frac{1}{2}\sum_{i=1}^{n-1} K_{i,n-i} c_i c_{n-i} - \sum_{i=1}^\infty K_{i,n} c_i c_n ,
\end{equation}
where $K_{i,n}$ is the temperature-dependent coagulation kernel.  

%

We coarse-grain coagulation on our grid of grain radii (or volumes) as follows. Consider the process:
\beq
\label{eq:coag0}\m Z_{\ell_1} + \m Z_{\ell_2} \to  \m Z_{\ell_1\ell_2} 
\eeq
in which grains $\m Z_{\ell_1}$ and $\m Z_{\ell_2}$ with radii $a_{\ell_1}$ and $a_{\ell_2}$ combine to form a grain $\m Z_{\ell_1\ell_2}$ with radius $a_{\ell_1\ell_2} = (a_{\ell_1}^3 + a_{\ell_2}^3)^{1/3}$.
Find the radial grid point interval $[a_k,a_{k+1})$ containing $a_{\ell_1\ell_2}$ by computing the index $\ell(\ell_1,\ell_2) =  \lfloor (\ell_{\rm max}-1) \ln(a_{\ell_1\ell_2}/a_{\rm min})/\ln(a_{\rm max}/a_{\rm min})\rfloor + 1$ and then distribute the new grain density increase between the flanking grid points:
\beq\label{eq:partition1}
\m Z_{\ell_1} + \m Z_{\ell_2} \to\eta_{\ell_1\ell_2} \m  Z_{\ell(\ell_1,\ell_2)} + (1-\eta_{\ell_1\ell_2}) \m  Z_{\ell(\ell_1,\ell_2)+1} ,
\eeq 
where the mass-conserving weighting is
$\eta_{\ell_1\ell_2} = (a_{\ell(\ell_1,\ell_2)+1}^3 - a_{\ell_1\ell_2}^3)/(a_{\ell(\ell_1,\ell_2)+1}^3 - a_{\ell(\ell_1,\ell_2)}^3)$. 
If the result of coagulation produces a grain with radius exceeding $a_{\rm max}$, we apportion the coagulation product to the largest radius grid point in mass conserving fashion $\m Z_{\ell_1} + \m Z_{\ell_2} \to \hat\eta_{\ell_1\ell_2} \m Z_{\ell_{\rm max}}$,
where $\hat \eta_{\ell_1\ell_2}= (a_{\ell_1\ell_2}/a_{\rm max})^3$.

The collision rate per unit volume between grains at radial grid points $\ell_1$ and $\ell_2$ with number densities $c_{\ell_1}$ and $c_{\ell_2}$ is $c_{\ell_1} c_{\ell_2} K_{\ell_1\ell_2}$ where $K_{\ell_1\ell_2}$ is the coagulation kernel:
\beq
\label{eq:coagkij} K_{\ell_1\ell_2} = \m W_{\ell_1\ell_2} \m C_{\ell_1\ell_2}(a_{\ell_1} + a_{\ell_2})^2 \sqrt{\frac{8\pi \kboltzmann T}{\mu_{\ell_1\ell_2}}} .
\eeq 
Here, $\mu_{\ell_1\ell_2}=m_{\ell_1}m_{\ell_2}/(m_{\ell_1}+m_{\ell_2})$ is the reduced mass, $\m W_{\ell_1\ell_2}$ is the factor by which the van der Waals force enhances the adhesion cross section, and $\m C_{\ell_1\ell_2}$ is the factor by which the Coulomb force between electrically charged grains modifies the collision cross section. The corresponding formulas for cluster-cluster and cluster-grain collisions can be found by replacing grain number densities, radii, and masses with the corresponding cluster values as appropriate.

The rate of change of the density $c_\ell$ of grains $\m Z_\ell$ is:
\begin{eqnarray}
\label{eq:coag_grid}
\left(\frac{dc_\ell}{dt}\right)_{\rm coag} &=& \sum_{\ell_1=1}^\ell \sum_{\ell_2=1}^{\ell_1} F_{\ell_1\ell_2;\ell} K_{\ell_1\ell_2}c_{\ell_1} c_{\ell_2} \nonumber\\& &- c_\ell \sum_{\ell_1=1}^{\ell_{\rm max}} K_{\ell_1\ell}c_{\ell_1} ,
\end{eqnarray}
where $F_{\ell_1\ell_2;\ell}$ is the fraction of the mass of $\m Z_{\ell_1}$ and $\m Z_{\ell_2}$ deposited in $\m Z_\ell$:
\beq
F_{\ell_1\ell_2;\ell} = \begin{cases}
\eta_{\ell_1\ell_2} , & \text{if }  \ell=\ell(\ell_1,\ell_2) < \ell_{\rm max} ,\\
1-\eta_{\ell_1\ell_2} , &  \text{if } \ell=\ell(\ell_1,\ell_2)+1 \le  \ell_{\rm max},\\
\hat\eta_{\ell_1\ell_2},  &  \text{if }\ell =\ell_{\rm max}\leq \ell(\ell_1,\ell_2)  ,\\
0 , &  \mbox{otherwise} .
\end{cases}
\eeq

\begin{figure*}
\begin{center}
\includegraphics[trim=3cm 3cm 2cm 6.5cm,clip=true,width=0.33\textwidth]{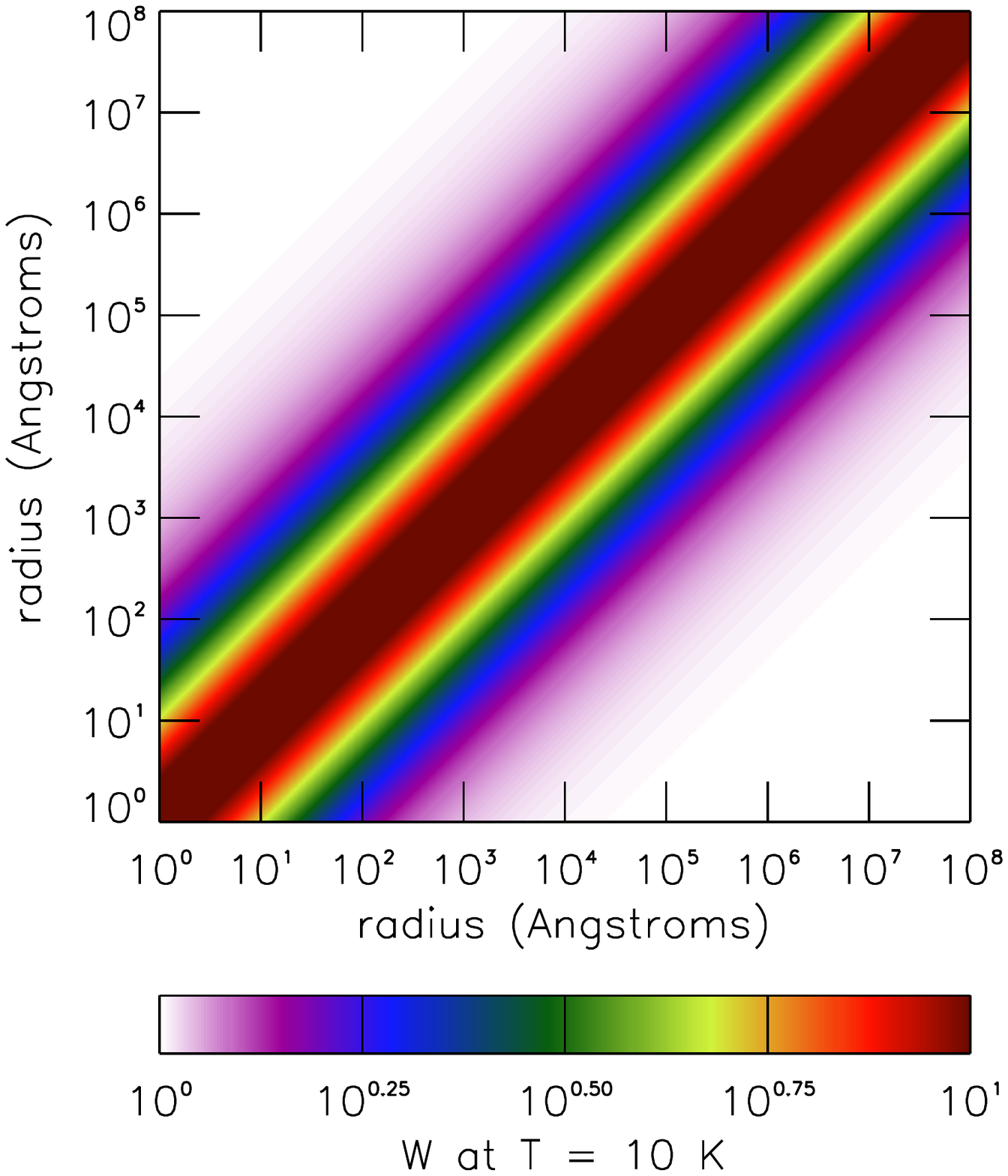}
\includegraphics[trim=3cm 3cm 2cm 6.5cm,clip=true,width=0.33\textwidth]{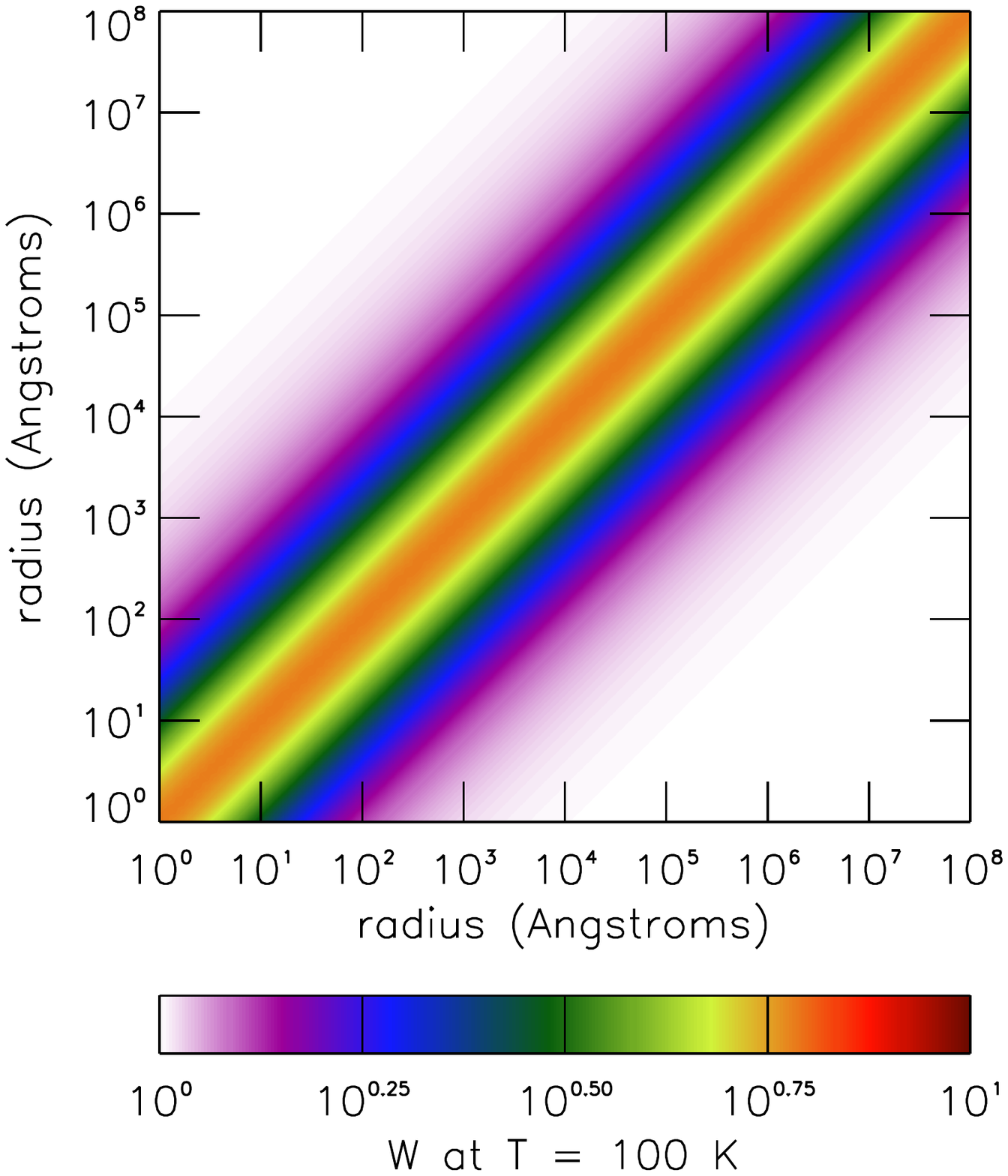}
\includegraphics[trim=3cm 3cm 2cm 6.5cm,clip=true,width=0.33\textwidth]{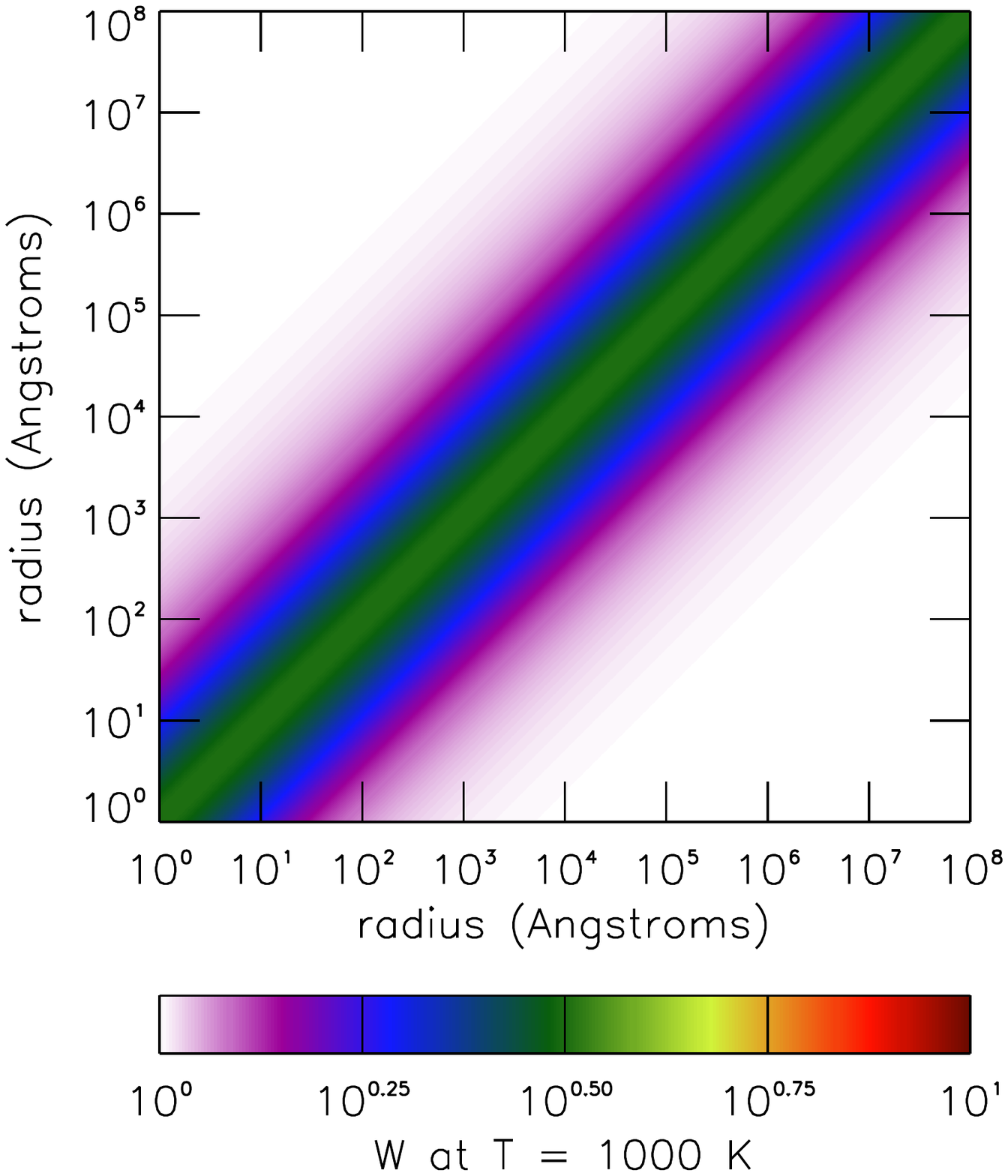}
\end{center}
\caption{The contour plot shows the Van der Waals correction factor $\m W_{\ell_1\ell_2}$ for carbon grains at four different temperatures (see Equation~\ref{eq:Wvanderwaals}). The horizontal and vertical axes give the radii of the colliding grains. Each panel corresponds to a particular gas temperature which is given below the legend.}
\label{fig:vanDerWaals}
\end{figure*}

\begin{table*}
\begin{center}
\begin{tabular}{ llllllll }
\hline
Species & Formula &  $A$ ($10^4$ K) $^1$ & $B$ $^1$ &  $\sigma_\text{ST}$ (erg cm$^{-2}$) $^1$ &  $r_1$ (\AA) $^1$ & $\varrho$ (g cm$^{-3}$) $^2$ & Condensation Nucleus \\ \hline
Iron & Fe & 4.8418 & 16.5566 & 1800 & 1.411 & 7.88069&Fe$_4$ \\
Silicon & Si & 5.36975 & 17.4349 & 800 & 1.684 & 2.3314&Si$_4$ \\
Carbon & C & 8.64726 & 19.0422 & 1400 & 1.281 & 2.26507& C$_4$\\
Magnesium $^3$ & Mg & 7.0085 & 18.2386 & 1100 & 1.76917 & 1.74&Mg$_4$ \\
Forsterite & Mg$_2$SiO$_4$ & 37.24 & 104.872 & 436 & 2.589 & 3.21394 &Mg$_4$Si$_2$O$_8$\\
Iron Sulfide & FeS & 9.31326 & 30.7771 & 380 & 1.932 & 4.83256& Fe$_4$S$_4$\\
Silicon Carbide & SiC & 14.8934 & 37.3825 & 1800 & 1.702 & 3.22393&Si$_2$C$_2$ \\
Alumina & Al$_2$O$_3$ & 18.4788 & 45.3543 & 690 & 1.718 & 7.97125 &Al$_4$O$_6$\\
Enstatite & MgSiO$_3$ & 25.0129 & 72.0015 & 400 & 2.319 & 7.97125 & Mg$_2$Si$_2$O$_6$ \\
Silicon Dioxide & SiO$_2$ & 12.6028 & 38.1507 & 605 & 2.08 & 2.64686 &Si$_2$O$_4$\\
Magnesia & MgO & 11.9237 & 33.1593 & 1100 & 1.646 & 3.58281&Mg$_4$O$_4$ \\
Magnetite & Fe$_3$O$_4$ & 13.2889 & 39.1687 & 400 & 1.805 & 15.6078 &Fe$_6$O$_8$\\
Iron Oxide & FeO & 11.129 & 31.985 & 580 & 1.682 & 5.98516& Fe$_4$O$_4$\\
Magnesium Sulfide $^4$ & MgS & 9.9783 & 31.9071 & 720.69 & 1.89065 & 3.30655&Mg$_4$S$_4$ \\
  \hline
\multicolumn{7}{l}{$^1$ The parameters $A$, $B$, $r_1$, and $\sigma$ are from \citet{Nozawa:03} for all grain species except Mg and MgS. }  \\
\multicolumn{7}{l}{$^2$  The mass density was taken to be the mass of a monomer divided by $4\pi a_1^3/3$.} \\
\multicolumn{7}{l}{$^3$  For Mg we simply averaged the parameters for C and Si.} \\
\multicolumn{7}{l}{$^4$  The parameters for MgS were scaled from those for MgO using the FeS to FeO parameter ratios.  } \\
\end{tabular}
\end{center}
\caption{Grain properties showing the species name, molecular formula, the vapor pressure coefficients $A$ and $B$, the monomer radius $r_1$, the surface tension $\sigma$, the mass density $\rho$, and the condensation nucleus.  }
\label{tab:grainProperties}
\end{table*}

\begin{table*}
\begin{center}
\begin{tabular}{ llllll }
\hline
Species & Smallest Grain & $a_{\rm min}$ (\AA) & $A_{\rm H}$ ($10^{-12}\u{erg}$) $^{1,2}$ & $T_\text{D}$ (K) $^{3,4,5}$ & Evaporation/Accretion $^6$\\ \hline
Iron &Fe$_5$ &2.41278 & 30 & 470 & $\mbox{Fe}_n + \mbox{Fe} \rightleftarrows \mbox{Fe}_{n+1}$\\
Silicon &Si$_5$ &2.8796 & 21 & 692 &  $\mbox{Si}_n + \mbox{Si} \rightleftarrows \mbox{Si}_{n+1}$\\
Carbon &C$_5$ &2.19048 & 4.7 & 420 &  $\mbox{C}_n + \mbox{C} \rightleftarrows \mbox{C}_{n+1}$\\
Magnesium &Mg$_5$ &3.02524 & 3.0 & 330 &  $\mbox{Mg}_n + \mbox{Mg} \rightleftarrows \mbox{Mg}_{n+1}$ \\
Forsterite &Mg$_8$Si$_4$O$_{16}$ &4.10978 & 0.65 & 470 &  $(\mbox{Mg$_2$SiO$_4$})_n + 2\mbox{Mg} + \mbox{SiO} + 3\mbox{O}\rightleftarrows (\mbox{Mg$_2$SiO$_4$})_{n+1}$  \\
Iron Sulfide &Fe$_5$S$_5$ &3.30367 &2.606 & 470 &  $(\mbox{FeS})_n + \mbox{Fe} + \mbox{S} \rightleftarrows (\mbox{FeS})_{n+1}$\\
Silicon Carbide &Si$_3$C$_3$ & 2.45471& 4.4 & 470 & $(\mbox{SiC})_n + \mbox{Si} + \mbox{C} \rightleftarrows (\mbox{SiC})_{n+1}$ \\
Alumina &Al$_6$O$_9$ &2.47778 & 1.50 & 470 & $(\mbox{Al$_2$O$_3$})_n + 2\mbox{Al} + 3\mbox{O} \rightleftarrows (\mbox{Al$_2$O$_3$})_{n+1}$\\
Enstatite &Mg$_4$Si$_4$O$_{12}$ &3.68118 &2.606 & 470 &  $(\mbox{MgSiO$_3$})_n + \mbox{Mg} + \mbox{SiO} + 2\mbox{O}\rightleftarrows (\mbox{MgSiO$_3$})_{n+1}$  \\
Silicon Dioxide &Si$_3$O$_6$ &2.99988 &2.606 &470 &  $(\mbox{SiO$_2$})_n + \mbox{SiO} + \mbox{O}\rightleftarrows (\mbox{SiO$_2$})_{n+1}$\\
Magnesia &Mg$_5$O$_5$ &2.81462 & 2.606& 470 &  $(\mbox{MgO})_n + \mbox{Mg} + \mbox{O} \rightleftarrows (\mbox{MgO})_{n+1}$\\
Magnetite &Fe$_9$O$_{12}$ &2.60326 &2.606 & 470 &  $(\mbox{Fe$_3$O$_4$})_n + 3\mbox{Fe} + 4\mbox{O} \rightleftarrows (\mbox{Fe$_3$O$_4$})_{n+1}$\\
Iron Oxide &Fe$_5$O$_5$ &2.87618 & 2.606& 470 &  $(\mbox{FeO})_n + \mbox{Fe} + \mbox{O} \rightleftarrows (\mbox{FeO})_{n+1}$\\
Magnesium Sulfide &Mg$_5$S$_5$ &3.23297 &2.606 & 470 &  $(\mbox{MgS})_n + \mbox{Mg} + \mbox{S} \rightleftarrows (\mbox{MgS})_{n+1}$ \\
  \hline
  \multicolumn{6}{l}{$^1$ The Hamaker constant is from \citet{Sarangi:15} for forsterite, alumina, carbon, magnesium, silicon carbide, silicon, and iron. }  \\
    \multicolumn{6}{l}{$^2$ For the grain species not listed in $^1$, we use an average value of $A_{\rm H} = 2.606 \times 10^{-12}\u{erg}$.  }\\
    \multicolumn{6}{l}{$^3$ For the Debye temperature we take the value for carbon and forsterite from \citet{1989ApJ...345..230G}.} \\
        \multicolumn{6}{l}{$^4$ The value of $T_\text{D}$ for magnesium and silicon is from values originally in \citet{Stewart:83} that have since been updated various sources (not cited). }  \\
         \multicolumn{6}{l}{$^5$ For the species not identified in $^4$ and $^5$ we use $T_\text{D} = 470\u{K}$. }  \\
     \multicolumn{6}{l}{$^6$ The evaporation and accretion reactions were taken from \citet{Nozawa:03}.} \\
\end{tabular}
\end{center}
\caption{Additional grain properties showing the smallest grain formula and radius $a_\text{min}$, the Hamaker constant $A_{\rm H}$, the Debye temperature $T_\text{D}$, and the evaporation-accretion reaction.}\label{tab:grainProperties2}
\end{table*}

\subsubsection{Van der Waals correction}

The van der Waals enhancement factor is \citep{Jacobson (2005)}:
\begin{eqnarray}
\label{eq:Wvanderwaals}
\m W_{\ell_1\ell_2} &= & -\f{1}{2\kboltzmann T(a_{\ell_1} + a_{\ell_2})^2}  \int_{a_{\ell_1}+a_{\ell_2}}^\infty   dr\, r^2 \frac{d}{dr}\({r\f{dV_{\ell_1\ell_2}}{dr}}\nonumber\\
& & \times e^{-(\frac{1}{2} rdV_{\ell_1\ell_2}/dr + V_{\ell_1\ell_2})/\kboltzmann T} ,
\end{eqnarray}
where $T$ is the gas temperature, $r$ is the distance between grain centers, and $V_{\ell_1\ell_2}$ is the potential energy associated with the van der Waals force. The potential energy is:
\begin{eqnarray}
V_{\ell_1\ell_2}(r) &=& -\f{A_{\rm H}}{6} \left[
 \f{2a_{\ell_1} a_{\ell_2}}{r^2 - (a_{\ell_1} + a_{\ell_2})^2}  + \f{2a_{\ell_1} a_{\ell_2}}{r^2 - (a_{\ell_1} - a_{\ell_2})^2} \right. \nonumber\\
 & & \left. +\ln\({ \f{r^2 - (a_{\ell_1} + a_{\ell_2})^2}{r^2 - (a_{\ell_1} - a_{\ell_2})^2}  }\right] ,
\end{eqnarray}
where $A_{\rm H}$ is the grain-species-specific Hamaker constant. The values of $A_{\rm H}$ can be found in Table~\ref{tab:grainProperties2}.
For the specific case of carbon grains, in Figure~\ref{fig:vanDerWaals} we plot the van der Waals enhancement factor as a function of the colliding grain radii. The enhancement factor is maximum when the colliding grains have similar radii and decreases with increasing temperature.

\subsubsection{Coulomb correction}

To find the Coulomb correction factor in Equation~(\ref{eq:coagkij}) we consider an infinitely massive ``target'' sphere of radius $a$ and charge $Ze$, where $e$ is the proton charge. The Coulomb-force-corrected collision cross section for a point ``projectile" of mass $m$, charge $ze$, and velocity $v$ to collide with the target is:
\beq
\label{eq:sigma_coulomb} \sigma (v) = \max\left[\pi a^2 \({ 1 - \f{2Zze^2}{v^2 m a} }, 0\right] .
\eeq
The rate at which projectiles with number density $c_0$ and temperature $T$ collide with the target is obtained by integrating the cross section $\sigma(v)$ over the Maxwell-Boltzmann distribution and equals: 
\bea
\Gamma = \m C_{\ell_1\ell_2} c_0 a^2 \sqrt{\f{8\pi \kboltzmann T}{m}} ,
\eea
where we expressed the rate in terms of the Coulomb correction factor that equals:
\begin{equation}
\m C_{\ell_1\ell_2}  = 
\begin{cases}
e^{- Zz e^2/a\kboltzmann T} & \mbox{ if } Z z > 0 , \\
 1 - \f{Zze^2}{a\kboltzmann T}  & \mbox{ if } Zz < 0 .
\end{cases}
\end{equation} 
This expression can be generalized to the collision of two grains with charges $Z_{\ell_1}e$ and $Z_{\ell_2}e$ by replacing $a$ with $a_{\ell_1} + a_{\ell_2}$ and $Zz$ with $Z_{\ell_1}Z_{\ell_2}$, namely,
\begin{equation}
\m C_{\ell_1\ell_2} = 
\begin{cases}
e^{- Z_{\ell_1} Z_{\ell_2} e^2/(a_{\ell_1} + a_{\ell_2})\kboltzmann T} &\mbox{ if } Z_{\ell_1} Z_{\ell_2} > 0 ,\\
 1 - \f{Z_{\ell_1}Z_{\ell_2}e^2}{(a_{\ell_1} + a_{\ell_2})\kboltzmann T}  & \mbox{ if } Z_{\ell_1} Z_{\ell_2} < 0 .
\end{cases}
\end{equation} 

\subsection{Grain charging} 

 To compute the Coulomb correction factor we estimate the average net electric charge on grains at each radial grid point. Grains become charged due to photoelectric absorption of UV photons and thermal electron and ion capture. The rate at which electrons with number density $c_e$ and temperature $T_e$ collide with a grain with radius $a$ and charge $Ze$ is: 
\begin{equation}
\Gamma_{e^-} = c_e a^2 \sqrt{\f{8\pi \kboltzmann T_e}{m_e}} 
\begin{cases}
e^{+Z e^2/a\kboltzmann T_e}  &  \mbox{ if } Z < 0 ,\\
1 + \f{ Z e^2}{a\kboltzmann T_e} & \mbox{ if } Z > 0 .
\end{cases}
\end{equation}
We assume that the electrons stick to the grain and ignore secondary electron emission.  The time derivative of grain charge due to collisions with thermal, free electrons is $({dZ}/{dt})_{e^-} = -\Gamma_{e^-}$.

The rate at which ions with number density $c_{\rm ion}$, charge $z_{\rm ion}e$, molecular mass $m_{\rm ion}$, and temperature $T_{\rm ion}$ collide with a grain with radius $a$ and charge $Ze$ is: 
\begin{eqnarray}
\Gamma_{\rm ion} &=&c_{\rm ion} a^2 \sqrt{\f{8\pi \kboltzmann T_{\rm ion}}{m_{\rm ion}}} \nonumber\\ 
& &\times
\begin{cases}
 1 - \f{Z z_{\rm ion} e^2}{a\kboltzmann T_{\rm ion}}  & \mbox{ if } Zz_{\rm ion} < 0 , \\
 e^{- Z z_{\rm ion} e^2/a\kboltzmann T_{\rm ion}} &\mbox{ if } Zz_{\rm ion} > 0 .
\end{cases}
\end{eqnarray}
We assume that an ion that hits the grain sticks to it so that the time derivative of grain charge due to collisions with ions is $({dZ}/{dt})_{\rm ion} = + z_{\rm ion} \Gamma_{\rm ion}$.

 \begin{figure*}
\begin{center}
\includegraphics[trim=3cm 3cm 2cm 6.5cm,clip=true,width=0.33\textwidth]{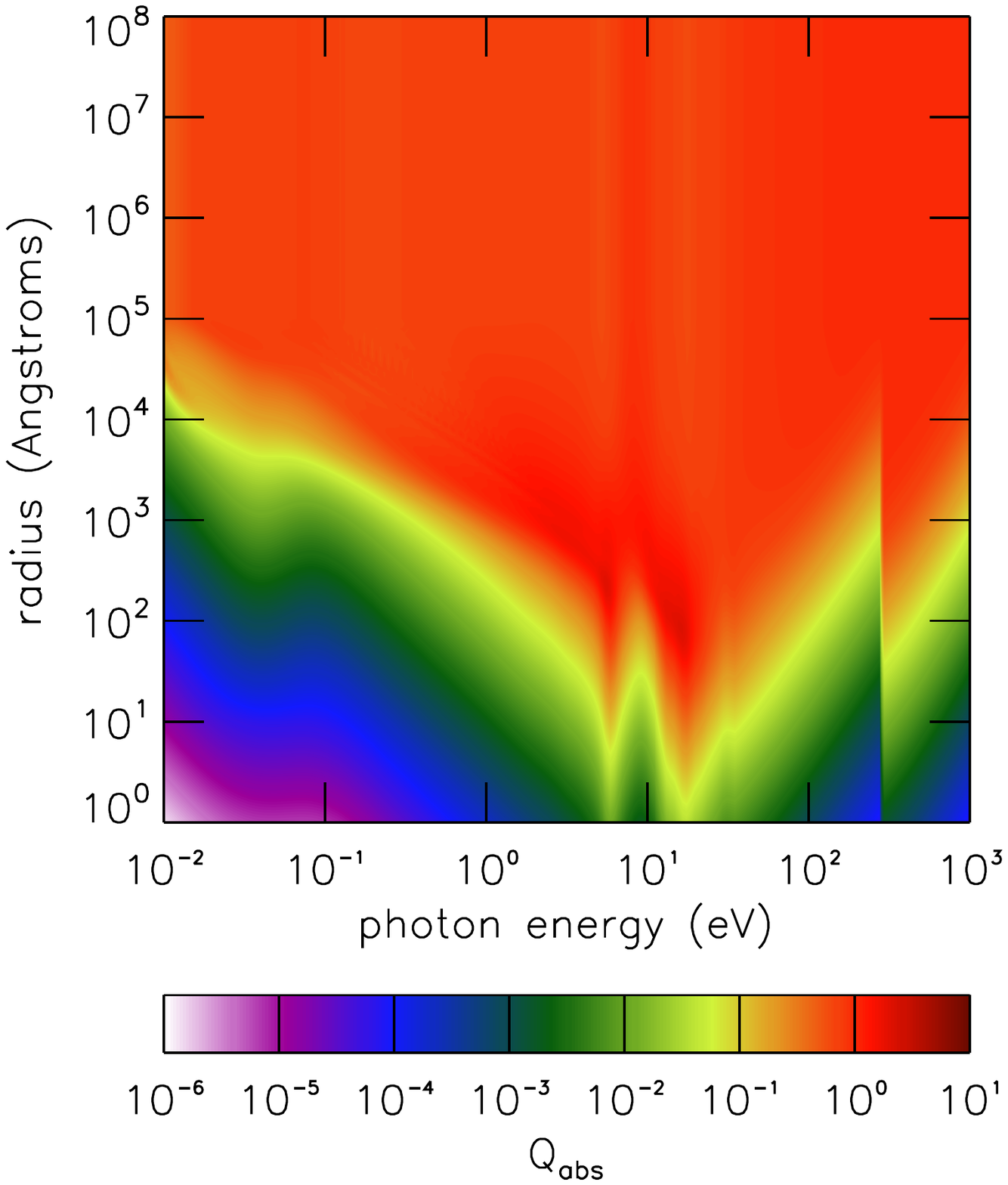}
\includegraphics[trim=3cm 3cm 2cm 6.5cm,clip=true,width=0.33\textwidth]{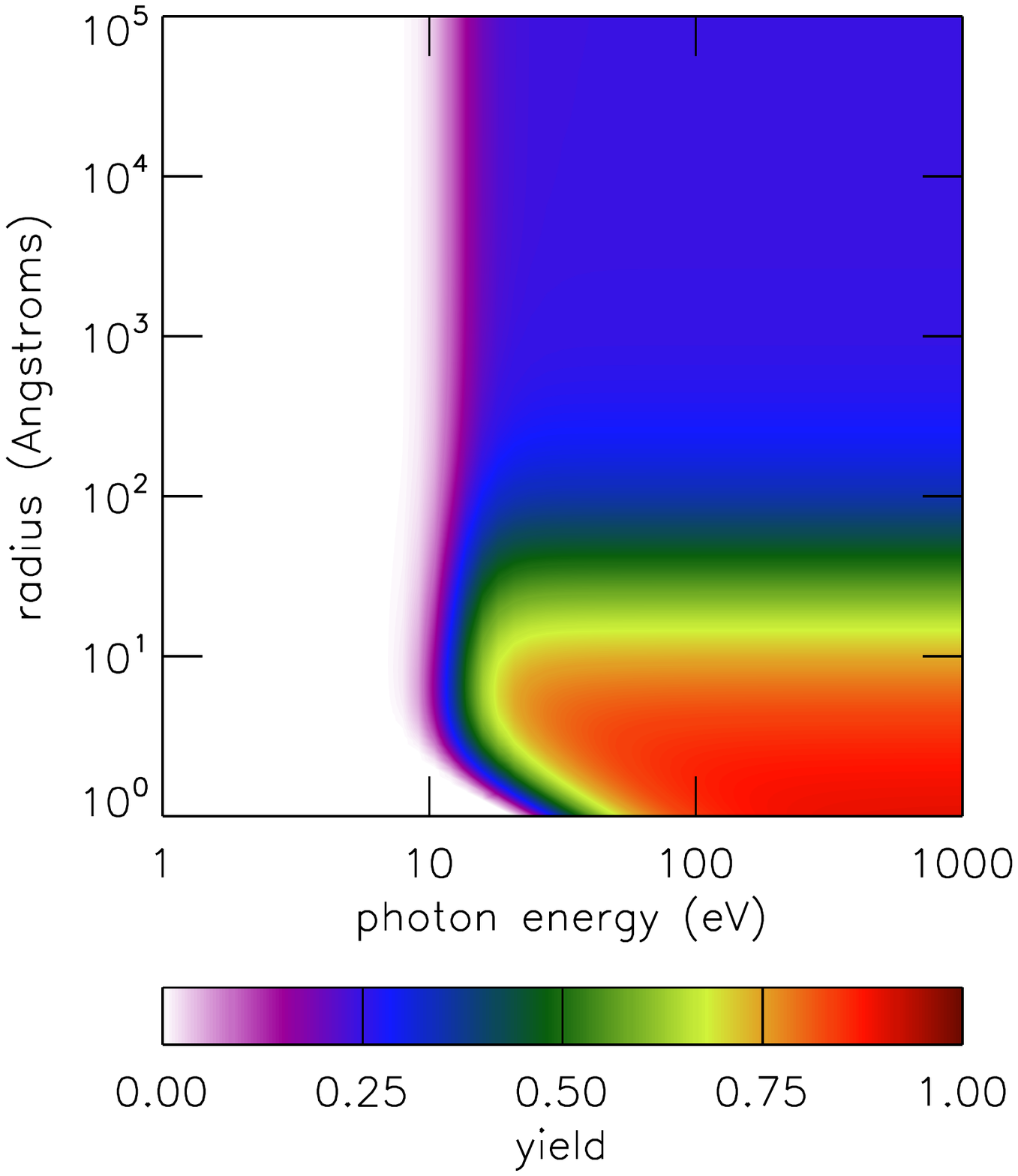}
\includegraphics[trim=3cm 3cm 2cm 6.5cm,clip=true,width=0.33\textwidth]{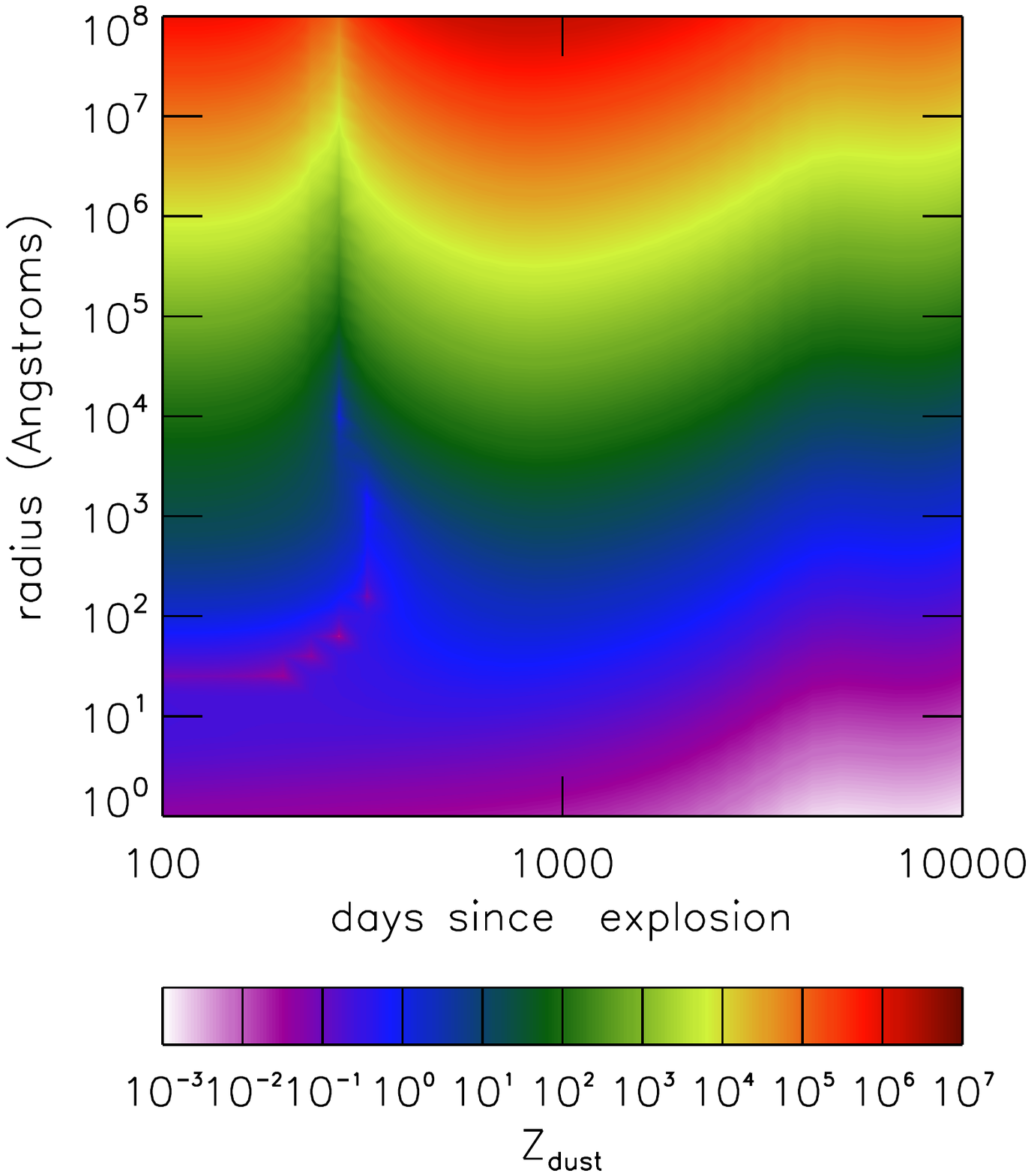}
\end{center}
\caption{Grain properties as a function of grain radius and photon energy or time since the explosion, including: the absorption coefficient $Q_{\rm abs}$ (left panel), the photoelectric yield $Y$ for neutral ($Z=0$) carbon grains (middle panel), equilibrium grain charge $Z_{\rm dust}$ (right panel).}
\label{fig:grain_properties}
\end{figure*}

We follow \citet{2006ApJ...645.1188W} and \citet{1987ApJ...320..803D} to find an expression for the rate at which electrons are ejected from a grain by photoelectric absorption. The general formula for the photoelectric ejection rate is:
\beq
\Gamma_{\gamma} = \int_0^\infty n_\gamma(E) Q_{\rm abs}(E,a)\pi a^2 c Y(E,a,Z) dE,
\eeq
where $Y(E,a,Z)$ is the photoelectric yield, namely, the average number of electrons ejected from a grain of radius $a$ and charge $Ze$ when it absorbs a photon with energy $E$, and $Q_{\rm abs}(E,a)$ is the absorption coefficient defined such that the cross section for a grain to absorb a photon of energy $E$ is $\sigma_{\rm abs}(E,a)=Q_{\rm abs}(E,a)\pi a^2$. For the absorption coefficient of all grain species we used an online table\footnote{https://www.astro.princeton.edu/$\sim$draine/dust/dust.diel.html} based on \citet{1984ApJ...285...89D} and \citet{1993ApJ...402..441L}. This table provides grain radius and photon energy dependent absorption coefficients for graphite spheres with radii $1\u{nm}\leq a \leq 10\u{$\mu$m}$. For smaller grains we assume that $Q_{\rm abs}(E,a) \propto a$ and for larger grains that $Q_{\rm abs}(E,a)$ is independent of radius.  We plot the absorption coefficient as a function of the grain radius and photon energy in the left panel of Figure~\ref{fig:grain_properties}.

To find the yield for given $E$, $a$, and $Z$, we write the valence band ionization potential
\beq
I = W + \({Z + \f{1}{2}}\f{e^2}{a} + (Z+2)\f{e^2}{a} \f{0.3\u{\AA}}{a} ,
\eeq
where $W$ is the work function that we take to equal $W=4.4\u{eV}$ \citep{2001ApJS..134..263W}. The minimum energy an electron must have to escape a negatively charged grain is:
\bea
E_{\rm min}= \f{e^2 | Z+1 | }{1 + | Z+1 |^{-1/2}} \[{1 - \frac{0.3}{|Z+1|^{0.26}} \({\f{a}{10\u{\AA}}}^{-0.45}}
\eea
if $Z < -1$ and $E_{\rm min}=0$ otherwise.
The threshold energy for photoelectric emission is $E_{\rm PET} = I +  E_{\rm min}$ if $Z < 1$ and $E_{\rm PET}=I$ otherwise. We  define auxiliary quantities: $\Theta = E - E_{\rm PET}$  if  $Z < 0$ and $\Theta=E - E_{\rm PET} + (Z+1)e^2/a$ othrewise, $\alpha = a/l_{\rm a} + a/l_{\rm e}$, $\beta = a/l_{\rm a}$,
where $l_{\rm a}$ is the photon attenuation length that we take to be $l_{\rm a} = 27.8\u{\AA}$ and $l_{\rm e}$ is the electron escape length which we take to be $l_{\rm e} = 10.0\u{\AA}$. We define further auxiliary quantities:
\bea
y_0 \= 0.009 \times \f{(\Theta/W)^5}{1+0.037 (\Theta/W)^5} ,\\
y_1 \= \({\f{\beta}{\alpha}}^2 \f{\alpha^2 -2\alpha + 2 -2e^{-\alpha}}{\beta^2 -2\beta + 2 -2 e^{-\beta}} ,\\
y_2 \= 
\begin{cases}
\f{E_{\rm high}^2 (E_{\rm high} - 3 E_{\rm low})}{(E_{\rm high} -  E_{\rm low})^3} , & \mbox{ if } Z \ge 0 ,\\
 1 , & \mbox{ if } Z < 0 ,
\end{cases}
\eea
as well as $E_{\rm high} = E_{\rm min} + E - E_{\rm PET}$ if $Z < 0$ and
$E_{\rm high}= E - E_{\rm PET}$ otherwise, $E_{\rm low} = E_{\rm min}$ if $ Z < 0$ and $E_{\rm low}=
-(Z+1)e^2/a$ otherwise.
The yield is non-zero when $E>E_{\rm PET}$ and equals $Y = y_2 \times\mbox{min}\{ \,y_0 y_1,\,1\, \}$.
The photoelectric yield of a neutral carbon grain as a function of grain radius and photon energy is shown in Figure~\ref{fig:grain_properties}, middle panel.

The total time derivative of the electric charge of a grain is:
\beq
\f{dZ}{dt} = -\Gamma_{e^-} + z_{\rm ion} \Gamma_{\rm ion}+\Gamma_{\gamma} .
\eeq
The equilibrium charge is found by setting $dZ/dt=0$ and solving for $Z$.  We assume that all grain species have the same grain-radius-dependent charge computed according to the just described procedure substituting the values of $Q_{\rm abs}$, $Y$, and $W$ specific to carbon grains. We set $T_e = T_{\rm ion} = T_{\rm gas}$, $z_{\rm ion}=+1$, $m_{\rm ion}=20.3 \u{amu}$, and $c_{\rm ion} = c_e = 2.94\times 10^8\u{cm}^{-3}(t/100\u{days})^{-3}$ in all calculations (including those pertaining to the bubble and ambient density regions). We plot the equilibrium grain charge $Z_\text{dust}$ in the right panel of Figure~\ref{fig:grain_properties}.

\subsection{Grain temperature}

The temperature of a dust grain can be different from the temperature of the gas and is an important parameter because the evaporation rate, as we discuss below, depends on the grain temperature exponentially. We explicitly compute the grain temperature as a function of grain radius.  For the purpose of this calculation only, we assume the grain is pure carbon regardless of its actual composition.   Dust grains heat by absorbing photons from the UV radiation field and cool by emitting IR photons. They also exchange energy with the surrounding gas via gas-grain collisions. We ignore the heating and cooling due to evaporation, accretion, coagulation, and chemical reactions.

The rate at which the internal energy $U$ of a grain with radius $a$ increases through absorption of UV photons is:
\beq
L_{\rm abs} (a)= \pi a^2 c \int_0^\infty Q_{\rm abs} (E,a) E n_\gamma(E) dE ,
\eeq
where $c$ is the speed of light.
The rate at which a grain with temperature $T_{\rm dust}$ cools by emitting IR photons is:
\beq
L_{\rm emit} (a,T_\text{dust}) = \f{4\pi^6 a^2 \kboltzmann T_{\rm dust}}{15 h^3 c^2 \zeta(3)} \int_0^\infty \f{E^2 Q_{\rm abs}(E,a)}{e^{E/\kboltzmann T_{\rm dust}}-1} dE ,
\eeq
where $h$ is Planck's constant and $\zeta(x)$ is the Riemann-Zeta function.  Collisions with gas particles effect energy transfer (heating or cooling) at the rate \citep{1983ApJ...265..223B}: 
\begin{eqnarray}
L_{\rm gas} (a,T_\text{gas},T_\text{dust}) &=& \pi a^2 n_{\rm tot} \sqrt{\f{8\kboltzmann T_{\rm gas}}{\pi \avg{m}}} \nonumber\\
& &\times
(0.1 + 0.35 \,e^{-\sqrt{(T_{\rm dust} + T_{\rm gas})/500\:\rm K}})\nonumber\\
& & \times 2\kboltzmann (T_{\rm gas} - T_{\rm dust}) .
\end{eqnarray}
 The equilibrium grain temperature is found by setting
\beq
\label{eq:grain_temperature_equilibrium}
L_{\rm abs} (a) - L_{\rm emit} (a,T_\text{dust}) + L_{\rm gas}  (a,T_\text{gas},T_\text{dust}) = 0
\eeq
and solving for $T_{\rm dust}$.  

For the gas density we use $n_{\rm tot} = 2.23\times 10^{11}\u{cm}^{-3}(t/100\u{days})^{-3}$ and for average gas particle mass we use $\avg{m} = 3.07\times 10^{-23}\u{g}$.  These are valid in the bubble shells but we also use them for ambient and nickel bubble regions.  Fixing the gas density here to the density in the shells is justified by noting that in the period $100$--$1000$ days after the explosion when the dust forms, the collisional gas-dust coupling is negligible compared to the dust's thermal coupling to radiation.\footnote{At 500 days the number density of UV photons is $\approx 2.3\times10^7\,\text{cm}^{-3}$. If a 1 nm grain has an absorption coefficient of $Q_\text{abs} = 0.001$ and all photons have energy $4.43\,\text{eV}$ then that grain will absorb 22 photons per second which translates to a heating rate of $\approx1.6\times10^{-10}\,\text{erg}\,\text{s}^{-1}$. At 500 days the mean gas density is $\approx 1.2\times10^{8}\,\text{cm}^{-3}$ and the molecular gas temperature is $\approx900\,\text{K}$. Thus the mean speed of a gas molecules, assuming that each molecule has a mass of one oxygen atom, is $\approx 2\,\text{km}\,\text{s}^{-1}$. The rate at which gas molecules collide with the grain is $\approx 0.004\,\text{s}^{-1}$. If each gas molecule transfers an energy $\frac{1}{2} k_\text{B} T$ to the grain, the heating rate is $\approx 2.5\times10^{-16}\,\text{erg}\,\text{s}^{-1}$. Thus the radiative heating rate is ${\cal O}(10^6)$ times larger than the gas heating rate and thus, evidently, the gas heating is negligible.}  E.g., for small $1\,\text{nm}$ grains we have $L_\text{abs}\gtrsim 100\times L_\text{gas}$. Therefore the dust temperature is not sensitive to gas density. In Figure~\ref{fig:grainTEMP} we plot grain temperature as a function of time and grain radius.

\begin{figure*}
\begin{center}
\includegraphics[trim=3.5cm 3cm 1cm 6.5cm,clip=true,width=0.33\textwidth]{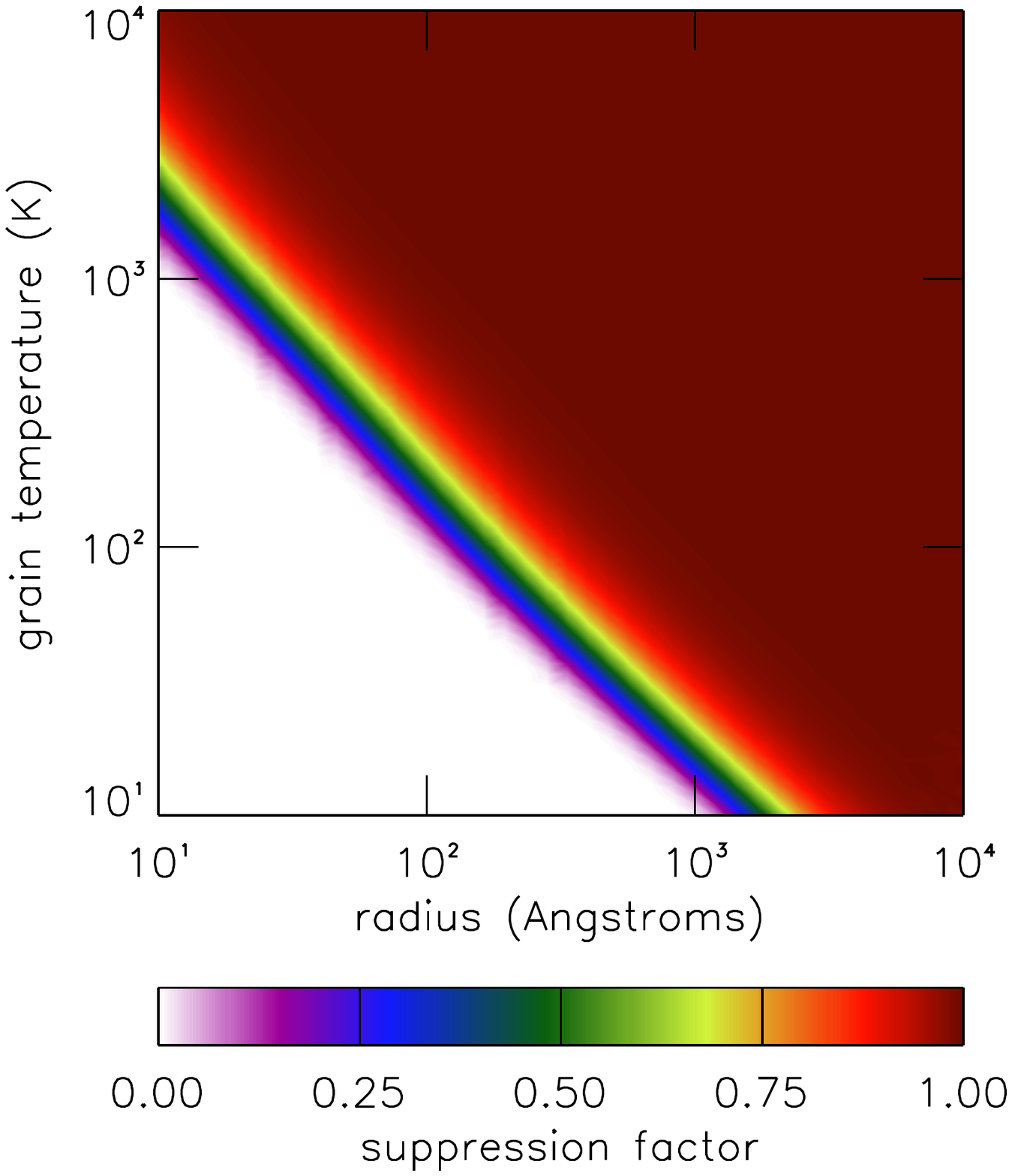}\hspace{1cm}
\includegraphics[trim=3.5cm 3cm 1cm 6.5cm,clip=true,width=0.33\textwidth]{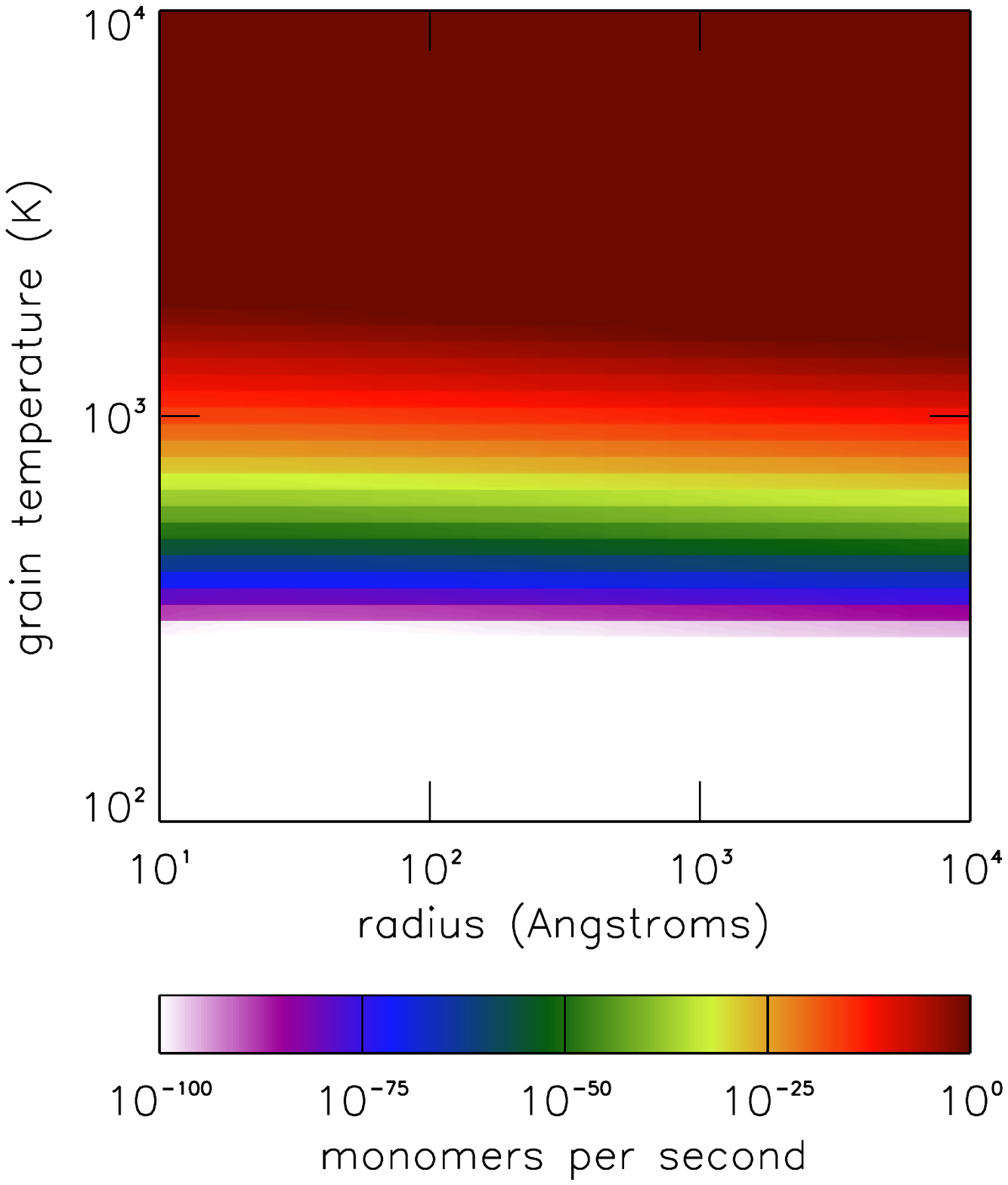}
\end{center}
\caption{Evaporation suppression factor (left panel) and evaporation rate (right panel) as a function of grain radius and temperature.}
\label{fig:grain_evaporation}
\end{figure*}

\begin{figure*}
\begin{center}
\includegraphics[trim=3.5cm 3cm 1cm 6.5cm,clip=true,width=0.33\textwidth]{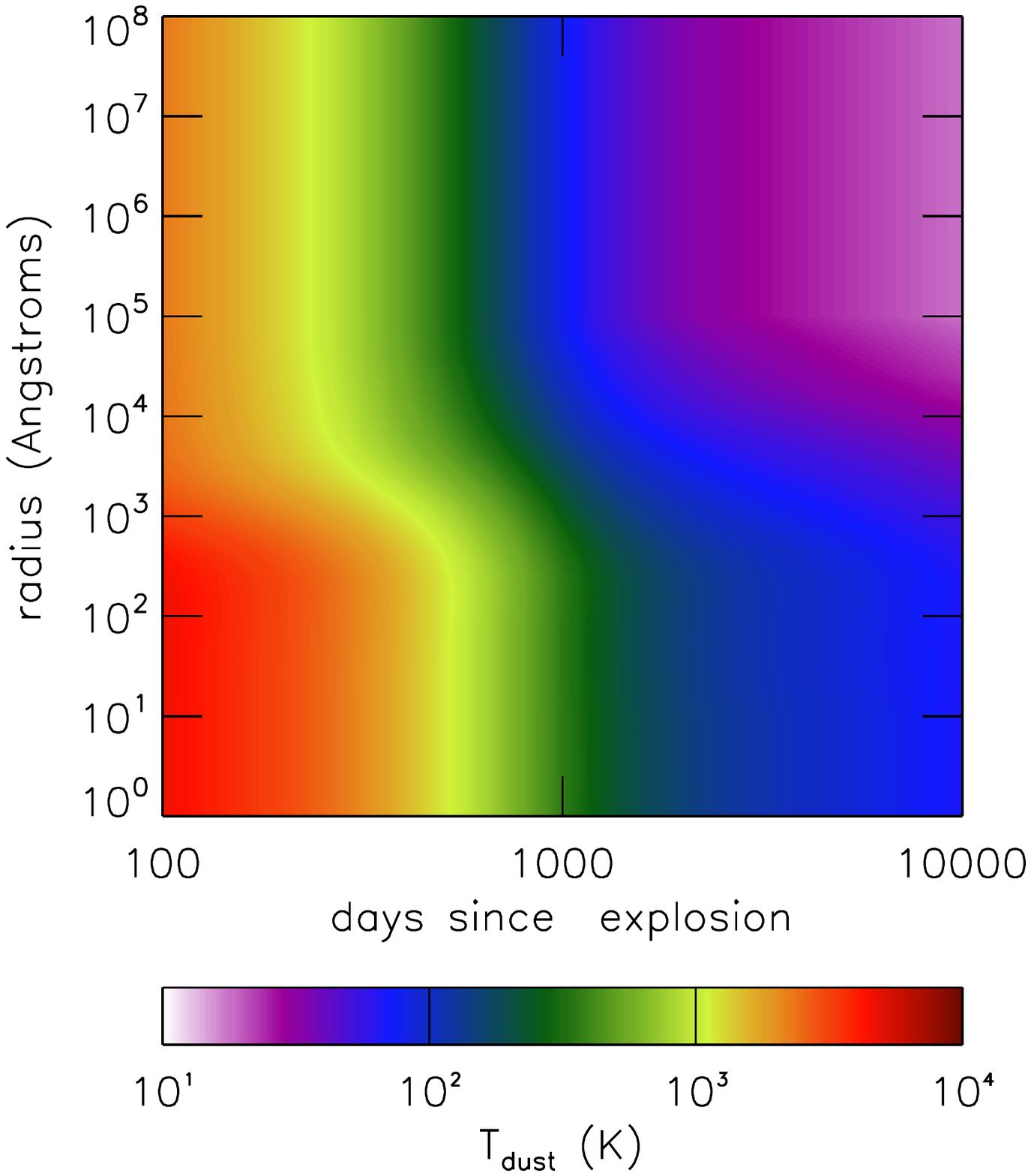}\hspace{1cm}
\includegraphics[trim=1cm 1cm 0cm 2cm,clip=true,width=0.34\textwidth]{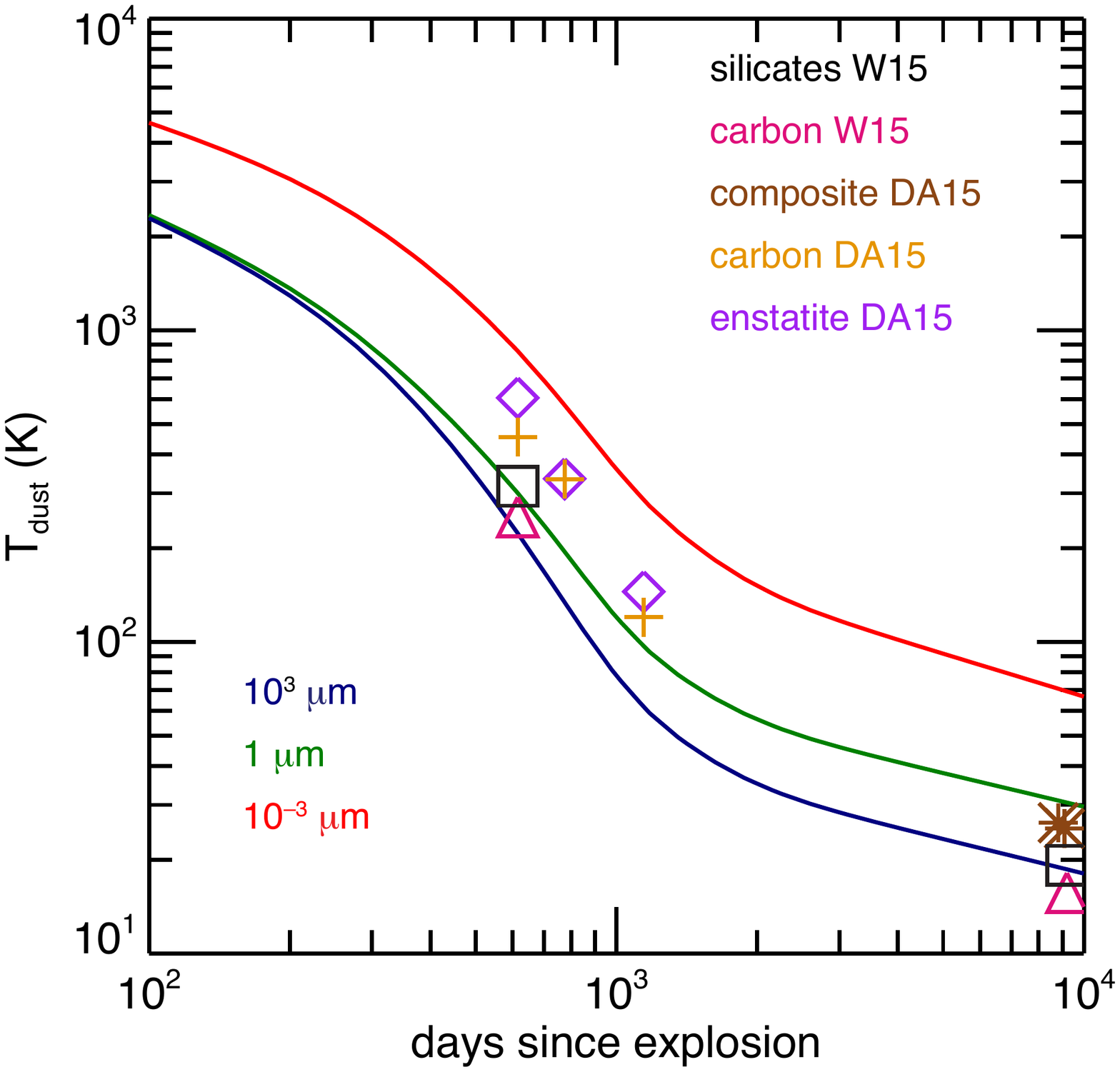}
\end{center}
\caption{Grain temperature as a function of time and grain radius (left panel) and measured values dust temperature (right panel) from \citet{Dwek:15} (DA15) and \citet{Wesson:15} (W15), where in the right panel, the solid lines are solutions of Equation (\ref{eq:grain_temperature_equilibrium}).}
\label{fig:grainTEMP}
\end{figure*}

\subsection{Evaporation}
\label{sec:evaporation}

Atoms at the surface of a dust grain can be ejected into the gas phase in a process called evaporation (or sublimation). To calculate the evaporation rate as a function of radius and grain temperature we follow \citet{1989ApJ...345..230G} and idealize a grain as consisting of $N$ atoms, each connected to the lattice with a spring. Each atom can vibrate in three independent directions so the grain has $3N$ degrees of freedom. Since 6 of these correspond to translation and rotation of the grain as a whole, the grain has $F=3N-6$ vibrational degrees of freedom. The internal or thermal energy $U$ of the grain is distributed over these vibrational degrees of freedom.

We assume that each grain species behaves as an Einstein solid with Debye temperature $T_\text{D}$. The vibrational degrees of freedom are treated as quantum harmonic oscillators with angular frequency $\omega_0 = (\pi/6)^{1/3}\kboltzmann T_\text{D}/\hbar$. The total number of vibrational quanta in a grain is $U/\hbar\omega_0$ and the average number of quanta per vibrational degree of freedom is $\gamma = U/\hbar\omega_0 F$. The values of $T_\text{D}$ used in our simulation are given in Table~\ref{tab:grainProperties2}.

The number of quanta in a vibrational degree of freedom fluctuates as energy shifts between atoms in a grain. Every once in a while, one vibrational degree of freedom has so much energy that the atom becomes unbound and is ejected from the grain. The surface binding energy of the atom to the grain of radius $a$ and surface tension $\sigma_\text{ST}$ is:
\begin{equation}
\m E_{\rm bind} = \kboltzmann A - 4\pi r_1^2 \sigma_\text{ST} [(n-1)^{2/3} - (n-2)^{2/3}] ,
\end{equation}
where $A$ is the bulk binding energy of an atom to the grain divided by the Boltzmann constant.
Then, for evaporation to occur, the number of quanta that must be concentrated in a single vibrational degree of freedom is $b = \m E_{\rm bind}/\hbar\omega_0$. 

For single-element grains evaporation is the process:
\beq
\m X_n \to \m X_{n-1} + \m X .
\eeq
This occurs at a rate per unit volume $k_{\rm evap} (T,n) c_n$, where,  in the limit in which the grains are in thermal equilibrium with the gas, the evaporation rate coefficient is:
\begin{eqnarray}
\label{eq:kevap} k_{\rm evap} (T,n) &=& p_{\rm s} s_n a_n^2 \sqrt{\f{8\pi}{\kboltzmann T m_1}} e^{-\m E_\text{bind}/\kboltzmann T+ B} .
\end{eqnarray}
Here $p_{\rm s} = 10^6 \u{dyne}\u{cm}^{-2}$ is the standard pressure, $s_n$ is the sticking coefficient (which we take to be unity), and $B$ is a parameter that depends on the grain composition (i.e., the species). This expression was derived by setting the evaporation rate equal to the condensation rate when the partial pressure equals the vapor pressure \citep{Nozawa:13}. The values of the parameters $A$, $B$, $\sigma_\text{ST}$, and $r_1$ that we use can be found in Table~\ref{tab:grainProperties}.

In reality, grains may not be in thermal equilibrium with the gas.  The gas density can be so low that thermal fluctuations within a grain occur much faster than the grain exchanges energy with the the ambient atoms and molecules.  Therefore, the energy needed to remove an atom from a grain may have to come from a thermal fluctuation within the grain itself.  We can account for this thermal isolation by multiplying the evaporation rate coefficient with a dimensionless suppression factor $S_N$. The suppression factor is the ratio of the actual probability that there will be at least $b$ quanta in one vibrational degree of freedom to what that probability would be if the grain were in thermal equilibrium with the gas. The probability that there is at least $b$ quanta in one vibrational degree of freedom is \citep{1989ApJ...345..230G}: 
\beq
\label{eq:probability_nobath}
\tilde p_F(\gamma,b) = \f{(\gamma F)!(\gamma F - b+F-1)!}{(\gamma F + F - 1)!(\gamma F - b)!} .
\eeq 
The same probability if the grain were coupled to a heat bath and thus $F\to\infty$ would be:
\beq
\label{eq:probability_bath}
\tilde p_\infty(\gamma, b) = \({\f{\gamma}{1+\gamma}}^b .
\eeq 
Dividing Equation (\ref{eq:probability_nobath}) with (\ref{eq:probability_bath}) gives the suppression factor \citep{1989ApJ...345..230G}:
\bea
S_N(T_{\rm dust}) &=& \f{\tilde p_F(\gamma, b)}{\tilde p_\infty(\gamma, b)} \nonumber\\
&=& \({\f{1+\gamma}{\gamma}}^b \f{(\gamma F)!(\gamma F - b+F-1)!}{(\gamma F+ F - 1)! (\gamma F - b)!} .
\eea
Note that if $\gamma F < b$, there is not enough internal energy in the grain for evaporation and the suppression factor (and thus the evaporation rate) should be set to zero.

We discretize the evaporation process on the grid of grain radii:
\beq
\label{eq:evapFromBin}\m Z_{\ell} \to \m Z_{\ell-1} + \delta n_{\ell-1}\m X ,
\eeq
where
 \beq
\delta n_{\ell-1} =  \({\f{a_\ell}{a_1}}^3 - \({\f{a_{\ell-1}}{a_1}}^3
\eeq 
is the number of monomers that would need to be added to an $(i-1)$mer to obtain an $i$mer.
The rate per unit volume of the reaction in Equation~(\ref{eq:evapFromBin}) is:
\begin{equation}
\m R^{\rm evap}_\ell = \frac{ k_{\rm evap}(T_{\rm dust},n_\ell) S_N(T_{\rm dust}) c_\ell}{\delta n_{\ell-1}} .
\end{equation}

We turn to  multi-element grains which evaporate by losing atoms as well as molecules via the reaction:
\beq
\label{eq:evapFromBin_multi} \m Z_{\ell} \to \m Z_{\ell-1} + \delta n_{\ell-1} (\nu_1 \m A_1 + ... + \nu_N \m A_N),
\eeq
where $\m A_i$ denote molecular species.
Table~\ref{tab:grainProperties2} gives the coefficients of evaporative reactions for all grain species. We assume that the evaporation rate, in monomers per unit time, can be approximated with Equation~(\ref{eq:kevap}), but now with `$n$' interpreted as the number of monomers in the grain. The parameters needed to calculate the evaporation rate of a multi-element grain can also be found in Table~\ref{tab:grainProperties}--\ref{tab:grainProperties2}. The suppression factor and evaporation rate for carbon grains are plotted in Figure~\ref{fig:grain_evaporation}.

The change of molecular abundances due to grain evaporation is given by:
\bea
\label{eq:evaporation_1}
\({\f{dc_{\m A_k}}{dt}}_{\rm evap} \= \nu_k  \sum_{\ell = 1}^{\ell_{\rm max}}\delta n_{\ell-1} \m R^{\rm evap}_\ell 
\eea
and the corresponding change of grain abundances is given by:
\bea
\label{eq:evaporation_2}
\({\f{dc_\ell}{dt}}_{\rm evap} \= \m R^{\rm evap}_{\ell + 1} - \m R^{\rm evap}_\ell .
\eea

\subsection{Accretion}
\label{sec:accretion}

Accretion is the inverse process of evaporation in which sub-monomeric fragments (or molecular monomer precursors; e.g., Mg, O, or SiO for a forsterite Mg$_2$SiO$_4$ grain) collide with and are absorbed by a grain.  
Suppose that accretion can be parametrized by stoichiometric coefficients $\nu_1$, ..., $\nu_N$:
\beq
\m X_n + \nu_1 \m A_1 + ... + \nu_N \m A_N \to \m X_{n+1} .
\eeq
Each molecular species $\m A_i$ collides with the grain at a rate $a^2 c_{\m A_i} \sqrt{8\pi \kboltzmann T/m_{\m A_i}}$. We define the \emph{key} species as the one with the lowest collision rate. The rate at which monomers are accreted to the grain is:
\beq
k_{\rm acc} = \f{a^2 c_{\rm key}}{\nu_{\rm key}}  \sqrt{\f{8\pi \kboltzmann T}{m_{\rm key}}} .
\eeq
where the subscript `key' designates the quantities associated with the key species.
Discretizing this on the grid of grain radii we obtain the reaction:  
\beq
\m Z_\ell + \delta n_\ell (\nu_1 \m A_1 + ... + \nu_N \m A_N) \to \m Z_{\ell+1} .
\eeq
This reaction occurs at a rate per unit volume:
\bea
\m R^{\rm acc}_\ell \=  \frac{k_{\rm acc} c_\ell }{\delta n_\ell} .
\eea
The change of molecular abundances due to accretion is given by:
\bea
\label{eq:accretion_1}
\({\f{dc_{\m A_k}}{dt}}_{\rm acc} \= - \nu_k  \sum_{\ell = 1}^{\ell_{\rm max}}\delta n_\ell \m R^{\rm acc}_\ell 
\eea
and the corresponding change of grain abundances is given by:
\bea
\label{eq:accretion_2}
\({\f{dc_\ell}{dt}}_{\rm acc} \= \m R^{\rm acc}_{\ell - 1} - \m R^{\rm acc}_\ell .
\eea

\subsection{Weathering}
\label{sec:weathering}

Chemical weathering refers to the process in which atoms (and ions) in the gas phase collide with a grain and chemically react with an atom in the grain resulting in the removal of the latter atom from the grain. Like evaporation, chemical weathering is a grain destruction process.
A prominent form of weathering is by noble gas ions.
When an He$^+$ ion collides with a grain it can steal an electron to recombine. The recombination energy goes into the grain and can result the ejection of an atom from the grain. In single element grains this corresponds to the reaction:
\beq
\m X_n + \mbox{He}^+ \to \m X_{n-1} + \m X^+ + \mbox{He} .
\eeq 
We take the rate per unit volume of this reaction to be $k_{\rm He} c_{\rm He^+} c_n$, where the coefficient is \citep{Lazzati:16}:  
\beq
k_{\rm He} = 1.6\times 10^{-9} \u{cm}^3 \u{s}^{-1} \times n^{2/3} .
\eeq
This value is for carbon grains but we adopt it for all single-element grains. 

For multi-element grains we assume that He$^+$ ions collide with grains and remove one atom at a time at the same rate as in single element grains. More precisely, let (X$_1$)$_{\eta_1}$(X$_2$)$_{\eta_2}$(X$_3$)$_{\eta_3}$ be the chemical formula for the monomer of some grain species, where X$_k$ is the $k$th element and $\eta_k$ the number of atoms of that element. Then helium weathering corresponds to the reaction: 
\bea
&& \m Z_\ell + \delta n_{\ell - 1} \eta_0  \mbox{He}^+\nonumber \\
&& \to \m Z_{\ell -1} + \delta n_{\ell - 1} \({\eta_1 \mbox{X}_1^+ +  \eta_2 \mbox{X}_2^+ + \eta_3 \mbox{X}_3^+ + \eta_0 \mbox{He} }
\eea 
with rate per unit volume:
\beq
\m R^{\rm weath,He}_\ell = \f{1}{\delta n_{\ell - 1}\eta_0}\m K_0 \({\f{a_\ell}{a_0}}^2 c_\ell c_{\rm He^+} ,
\eeq
where $\m K_0 = 1.6 \times 10^{-9} \u{cm}^3 \u{s}^{-1}$ and $\eta_0 = \eta_1 + \eta_2 + \eta_3$. The time derivatives of the number densities of the affected species are:
\bea
\label{eq:weathering_He}
\({\f{dc_{\rm He^+}}{dt}}_{\rm weath} \= - \sum_{\ell = 1}^{\ell_{\rm max}} \delta n_{\ell - 1} \eta_0\m R^{\rm weath,He}_\ell \nonumber \\
 \= -\({\f{dc_{\rm He}}{dt}}_{\rm weath} ,\nonumber \\
\({\f{dc_{\rm X_k}}{dt}}_{\rm weath} \= \eta_k \sum_{\ell = 1}^{\ell_{\rm max}} \delta n_{\ell - 1} \m R^{\rm weath,He}_\ell ,\nonumber\\ 
\({\f{dc_{\ell}}{dt}}_{\rm weath,He} \= \m R^{\rm weath,He}_{\ell + 1} - \m R^{\rm weath,He}_\ell .
\eea
In principle, Ne$^+$ and Ar$^+$ can also contribute to weathering. We assume that this occurs with the same rate coefficient as for helium.

When a neutral oxygen atom collides with a carbon grain it can remove a carbon atom and form a CO molecule that is ejected into the gas phase. This corresponds to the reaction:
\beq
\mbox{C}_n + \mbox{O} \to \mbox{C}_{n-1} + \mbox{CO} .
\eeq
We take the rate of this reaction from \citet{Lazzati:16}:
\beq
k_\text{O} = 10^{-11}  e^{-1130.0\:\rm K\it/T}  n^{2/3} \u{cm}^3 \u{s}^{-1} .
\eeq
We only include this process, which is called oxygen weathering, for carbon grains.

Oxygen weathering is discretized on the grid of grain radii as follows. The effective reaction on grain radial grid points is 
\bea
\m Z_\ell + \delta n_{\ell - 1}   \mbox{O} \to \m Z_{\ell -1} + \delta n_{\ell - 1} \mbox{CO}
\eea 
This occurs at a rate per unit volume of 
\beq
\m R_\ell^{\rm weath,O} = \f{k_\text{O} c_\ell c_{\rm O}}{\delta n_{\ell - 1}}
\eeq
and affects the number density time derivatives as follows:
\bea
\label{eq:weathering_O}
\({\f{dc_{\rm O}}{dt}}_{\rm weath} \= - \delta n_{\ell - 1} \m R_\ell^{\rm weath,O} ,\nonumber \\
\({\f{dc_{\rm CO}}{dt}}_{\rm weath} \= \delta n_{\ell - 1} \m R_\ell^{\rm weath,O} ,\nonumber\\
\({\f{dc_{\ell}}{dt}}_{\rm weath,O} \= R_{\ell+1}^{\rm weath,O} - \m R_\ell^{\rm weath,O} .
\eea

\begin{figure*}
\begin{center}
\includegraphics[trim=0cm 4cm 0cm 5cm,clip=true,width=0.33\textwidth]{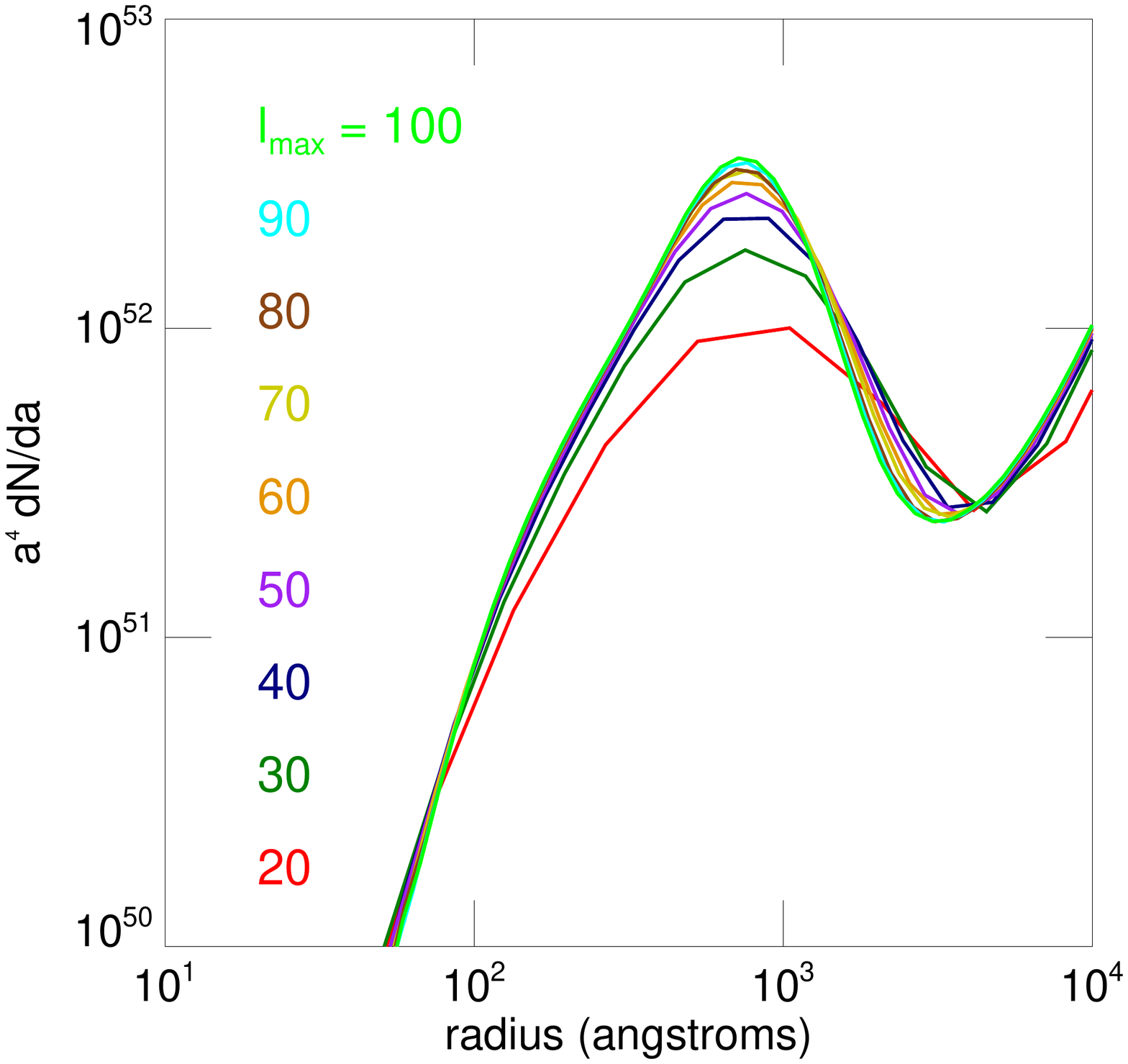}\hspace{1cm}
\includegraphics[trim=0cm 4cm 0cm 5cm,clip=true,width=0.33\textwidth]{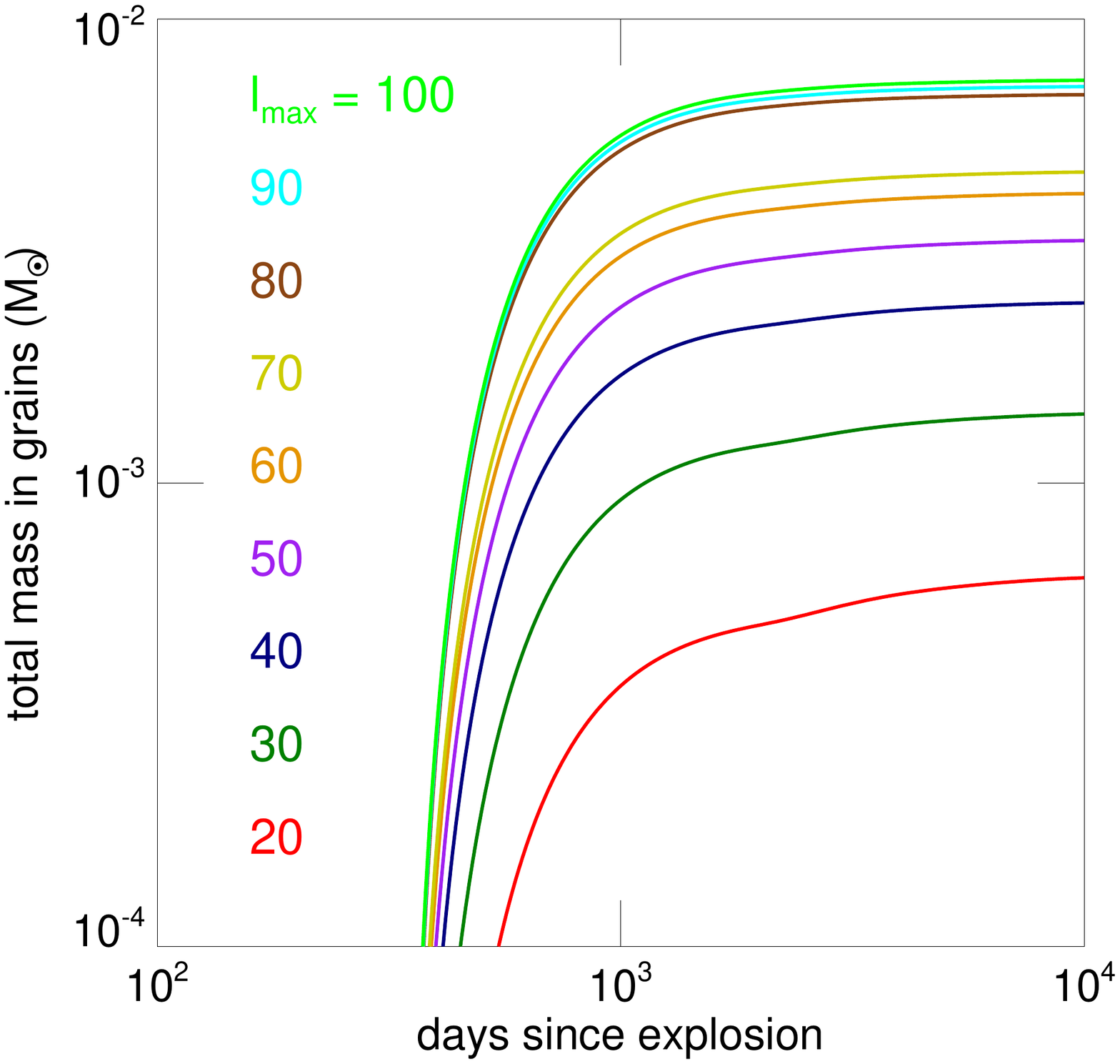}
\end{center}
\caption{
The curves show the final size distribution (left) and grain mass versus time (right) of carbon grains from a series of test simulations in which all parameters were held fixed except for the number of radial grid points $\ell_{\rm max}$.  The upturn in $a^4 dN/da$ at grain sizes $a\gtrsim 0.3\,\mu\text{m}$ is a numerical artifact resulting from the finite extent of the radial grid.}
\label{fig:ConvergenceTests}
\end{figure*}

\section{Time integration}
\label{sec:time_integration}

In our time integration calculation, an ejecta fluid element is described by a state vector of dimension $
\m N = \m N_\text{S} + \m N_\text{G} \times\ell_{\rm max}= 95+ 14\times 50= 795$.
We track $\m N_\text{R} = 341$ chemical reactions that modify the state vector and solve $\m N$ coupled ordinary differential equations (ODEs) of the form
\begin{eqnarray}
\frac{dc_i}{dt} &=& \left(\frac{dc_i}{dt}\right)_\text{chem} + \left(\frac{dc_i}{dt}\right)_\text{Compt}
+\left(\frac{dc_i}{dt}\right)_\text{coag} \nonumber\\ &+& \left(\frac{dc_i}{dt}\right)_\text{evap} + \left(\frac{dc_i}{dt}\right)_\text{acc}  + \left(\frac{dc_i}{dt}\right)_\text{weath} - 3\f{c_i}{t} .
\end{eqnarray}
Here, $(dc_i/dt)_\text{chem}$ is from Equation (\ref{eq:dndt_chem}), $(dc_i/dt)_\text{Compt}$ is from Equations (\ref{eq:Compton_1}) and (\ref{eq:Compton_2}), $(dc_i/dt)_\text{coag}$ is from Equation (\ref{eq:coag_grid}), $(dc_i/dt)_\text{evap}$ is from Equations (\ref{eq:evaporation_1}) and (\ref{eq:evaporation_2}), $(dc_i/dt)_\text{acc}$ is from Equations (\ref{eq:accretion_1}) and (\ref{eq:accretion_2}), $(dc_i/dt)_\text{weath}$ is from Equations (\ref{eq:weathering_He}) and (\ref{eq:weathering_O}), and the last term on the right is the overall decrease of number density due to homologous expansion.
The ODEs are integrated with a Bulirsch-Stoer-type, semi-implicit extrapolation mid-point method \citep{Bader:83}.

Our science suite of consists of 108 integrations, specifically 8, 50, and 50 integrations in the nickel bubbles, ambient ejecta, and dense shells, respectively. The integrations covered mass coordinate ranges $1.82 M_\odot < M < 1.89 M_\odot $ in the nickel bubbles and $1.89 M_\odot < M < 6.81 M_\odot $ for the ambient ejecta and dense shells. Each integration was started at 100 days after the explosion and carried out until a post-explosion time of $10^4\,\text{d}$.  Since at $10^4\,\text{d}$, the specific dust formation rate is $|d\ln M_{\rm dust,tot}/dt|^{-1}  \sim 4000 \u{years}$, no significant dust growth is expected at still later times.

The density in the shells depends on their thickness relative to the bubble radius. This thickness is not constrained by observations and one should vary the shell thickness as a parameter. We found very overdense shell integrations too computationally expensive, and thus, simulated shells only as dense as four times the ambient ejecta density.  Specifically, we ran the shell integrations at three different densities, 1, 2, and 4 times the density in the ambient ejecta.

\subsection{Convergence test}

We performed a convergence test by running a suite of integrations that are identical except for the number of radial mesh points $\ell_{\rm max}$. In the test, the density was set to the ambient ejecta density, the temperature was set to the atomic temperature evolution track (Figure \ref{fig:radioactive}, right panel), the mass fraction of C was set to 0.3, the mass fractions of He, O, Mg, Al, Si, S, and Fe were set to 0.1.

The left panel of Figure~\ref{fig:ConvergenceTests}  shows the carbon grain size distribution in the form $a^4dN/da$ at the end of the simulations, $10^4$ days after the explosion. The maximum fractional residual 
\begin{equation}
\max_\ell \frac{|(dN/da)_\ell^{(\ell_{\rm max})}-(dN/da)_\ell^{(\ell_{\rm max}=100)}|}{(dN/da)_\ell^{(\ell_{\rm max}=100)}}
\end{equation}
 in the interesting range of grain radii $10\u{\AA}\leq a<1\u{$\mu$m}$ decreases monotonically in $\ell_{\rm max}$ and drops to $\approx 0.5$ for the resolution we adopt, $\ell_{\rm max} = 50$; the average residual is much smaller. The location of the peak of $a^4dN/da$ is accurate independent of $\ell_{\rm max}$.

The right panel of Figure~\ref{fig:ConvergenceTests} shows the mass in carbon grains as a function of time. The maximum fractional residual in the final carbon grain mass drops to $\approx 0.5$ for $\ell_{\rm max} =50$. The fractional residual in the CO number density at the end of the simulation, not shown, drops below $0.01$ for the same number of grid points.

\section{Results}
\label{sec:results}

The model described in the preceding sections can be applied to arbitrary core collapse supernovae. The specific objective of the numerical integrations we present here is to assess the efficiency of dust production in SN 1987A.  From each integration we derive the dust mass synthesized as a function of grain composition and size.  Different integrations sample the various chemically and hydrodynamically distinct zones.  

Our presentation of the results is structured as follows. We first analyze how dust mass is partitioned as a function of grain size and composition and how refractory element mass is partitioned between molecules and grains of different sizes.  Then we examine the evolution of grain mass by species, ejecta zone, and mass coordinate within the ejecta.  We analyze the shape of the grain size (or, equivalently, grain mass) distribution at $10^4$ days. Next we turn the evolution of the abundance of important molecules and how their abundances are distributed throughout the ejecta.  Finally, we discuss how the grain properties depend on the mass density in the dense shells compressed by nickel bubbles.


\begin{table*}
\centering
\begin{tabular}{llllllll}
 \hline 
Species & Very Small & Small & Medium & Large & Very Large & Giant & Total \\\hline
Alumina & 0  & 3.5($-$7) & 1.2($-$2) & 7.5($-$4) & 1.8($-$10) & 4.0($-$7) & 1.4($-$2) \\
Carbon & 1.5($-$10) & 4.7($-$7) & 1.9($-$3) & 2.2($-$2) & 8.8($-$3) & 1.0($-$3) & 3.4($-$2) \\
Enstatite & 0  & 3.8($-$7) & 2.8($-$5) & 1.3($-$6) & 0  & 0  & 3.0($-$5) \\
Forsterite & 0  & 2.0($-$5) & 8.9($-$2) & 6.0($-$3) & 2.1($-$10) & 4.2($-$7) & 0.1 \\
Iron & 1.2($-$8) & 3.0($-$5) & 4.1($-$3) & 2.5($-$3) & 9.3($-$6) & 0  & 6.6($-$3) \\
Iron Oxide & 0  & 1.7($-$10) & 3.7($-$6) & 1.5($-$4) & 6.4($-$9) & 0  & 1.5($-$4) \\
Iron Sulfide & 0  & 7.3($-$6) & 1.3($-$2) & 2.7($-$2) & 2.0($-$3) & 0  & 4.1($-$2) \\
Magnesia & 8.0($-$10) & 1.1($-$3) & 0.13 & 4.6($-$2) & 9.7($-$10) & 0  & 0.17 \\
Magnetite & 0  & 0  & 3.5($-$7) & 2.7($-$4) & 3.3($-$7) & 0  & 2.6($-$4) \\
Silicon Carbide & 0  & 0  & 0  & 0  & 0  & 1.8($-$7) & 1.8($-$7) \\
Silicon Dioxide & 7.3($-$9) & 6.6($-$3) & 1.7($-$2) & 3.0($-$7) & 0  & 0  & 2.0($-$2) \\
Silicon & 9.8($-$9) & 4.4($-$4) & 6.4($-$2) & 5.7($-$3) & 8.9($-$2) & 1.2($-$3) & 0.16 \\
\hline 
Total & 3.0($-$8) & 8.8($-$3) & 0.31 & 0.11 & 9.9($-$2) & 2.2($-$3) & 0.59 \\
\hline 
\end{tabular}
\caption{Final dust mass classified by grain species and size for ($a < 1\u{nm}$), small ($1\u{nm} \leq a < 10 \u{nm}$), medium ($10\u{nm} 
\leq a < 0.1 \u{$\mu$m}$), large ($0.1\u{$\mu$m} \leq a < 1 \u{$\mu$m}$), very large ($1\u{$\mu$m} \leq a < 10 \u{$\mu$m}$), and giant ($ a \geq 10 \u{$\mu$m}$) grains.
\label{tab:DustmassHistogrambySize}}
\end{table*}

\subsection{Grains}

Figure~\ref{fig:grainTEMP} shows the dust temperature as a function of time and grain radius.  At a given radius the temperature drops rapidly with time.  For example, the temperature of $1\,\text{nm}$ grains is $\sim 4500\,\text{K}$ at $100$ days; it drops to $\sim 350\,\text{K}$ at $1000$ days and further to $\sim 70\,\text{K}$ at $10^4$ days.  The temperature of grains with radii above a micron starts at $\sim 2000\,\text{K}$ at $100$ days and already drops to $\sim 60\,\text{K}$ by $1000$ days.  At $10^4$ days the dust temperature is $\lesssim 20\,\text{K}$ which is consistent with the dust temperature inferred by \citet{Zanardo:14} from observations with Australia Telescope Compact Array (ATCA) and ALMA.  The impact of the dust temperature on the evaporation rate can seen in the right panel of Figure \ref{fig:grain_evaporation}.  The evaporation rate, which depends on the dust temperature, drops quickly to become completely negligible after $300$ days.  The rate of accretion onto grains, which depends on the higher gas temperature, does not drop as fast.  Therefore, after $300$ days, grain growth proceeds unimpeded by evaporation and is only modulated by noble gas weathering.

In Table~\ref{tab:DustmassHistogrambySize} we provide the dust mass synthesized a function of grain size for each of the grain species.  At the end of the simulation the total dust mass produced was $\approx 0.5\,M_\odot$, the bulk of which, $\approx 0.3\,M_\odot$, was in medium size grains ($10\,\text{nm}\leq a<100\,\text{nm}$), and another $\approx 0.1\,M_\odot$ each in large ($0.1\,\mu\text{m}\leq a<1\,\mu\text{m}$) and very large ($1\,\mu\text{m}\leq a<10\,\mu\text{m}$) grains. 


\begin{table*}
\centering
\begin{tabular}{llllllllllll}
 \hline 
Species & 1 & 2 & 3 & 4 & $5+$ & Small & Medium & Large & Very Large & Giant \\\hline
Carbon & 0.13 & 2.6($-$2) & 0  & 0  & 0  & 3.4($-$7) & 1.2($-$3) & 1.4($-$2) & 3.4($-$3) & 3.9($-$4) \\
Iron & 5.9($-$2) & 1.0($-$4) & 1.8($-$8) & 0   & 0  & 8.1($-$6) & 4.5($-$3) & 3.9($-$3) & 1.9($-$4) & 0 \\
Magnesium & 3.5($-$3) & 3.1($-$4) & 0 & 0  & 0  & 4.5($-$4) & 7.5($-$2) & 2.3($-$2) & 5.1($-$10) & 1.1($-$7) \\
Oxygen & 2.0 & 3.9($-$2) & 8.1($-$8) & 0  & 1.7($-$5) & 3.1($-$3) & 7.2($-$2) & 1.5($-$2) & 1.7($-$8) & 2.6($-$7) \\
Sulfur & 0.15 & 4.7($-$5) & 0  & 0   & 0  & 1.2($-$6) & 2.2($-$3) & 1.9($-$3) & 1.0($-$4) & 0  \\
Silicon & 0.17 & 2.8($-$4) & 1.5($-$7) & 0  & 2.7($-$5)  & 2.4($-$3) & 4.2($-$2) & 2.6($-$3) & 3.9($-$2) & 7.6($-$4) \\
Aluminum & 4.1($-$3) & 9.8($-$6) & 0  & 0  & 0  & 1.4($-$7) & 4.5($-$3) & 2.9($-$4) & 0  & 1.5($-$7) \\
\hline 
\end{tabular}
\caption{Ejecta mass in solar masses by element and number of atoms or grain size in the same grain size bins as in Table~\ref{tab:DustmassHistogrambySize}.
\label{tab:ElementmassHistogram}}
\end{table*}

Table~\ref{tab:ElementmassHistogram} provides the partitioning of the elemental mass into different chemical products. Overall, the simulated ejecta contains noble gases (He, Ne, and Ar) and refractory elements (C, O, Mg, Al, Si, S, and Fe). In the simulation the noble gases comprise $1.8\,M_\odot$ or $37\%$ of the core mass, whereas the refractory elements, which can be incorporated in grains, comprise $3.1\,M_\odot$ or $62\%$. Approximately $16\%$ of this refractory mass has ended up in grains at $10^4$ days. The rest of the refractory mass remained in atomic form ($2.5\,M_\odot$ or 82\%) with a small fraction locked in molecules, predominantly diatomic ($0.069\,M_\odot$ or 2\%). 

\begin{figure*}
\begin{center}
\includegraphics[trim=0cm 4cm 0cm 5cm,clip=true,width=0.49\textwidth]{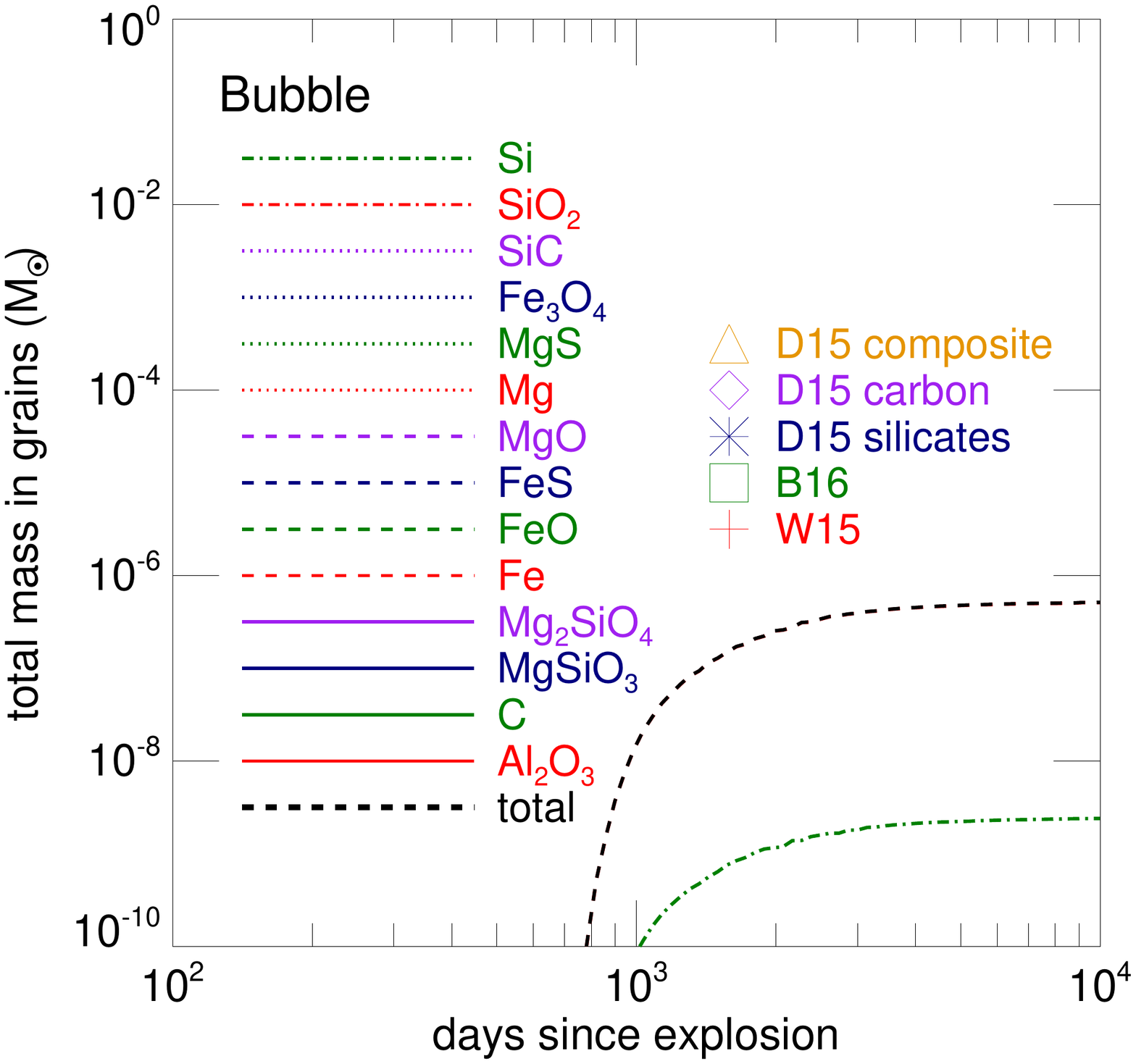}
\includegraphics[trim=0cm 4cm 0cm 5cm,clip=true,width=0.49\textwidth]{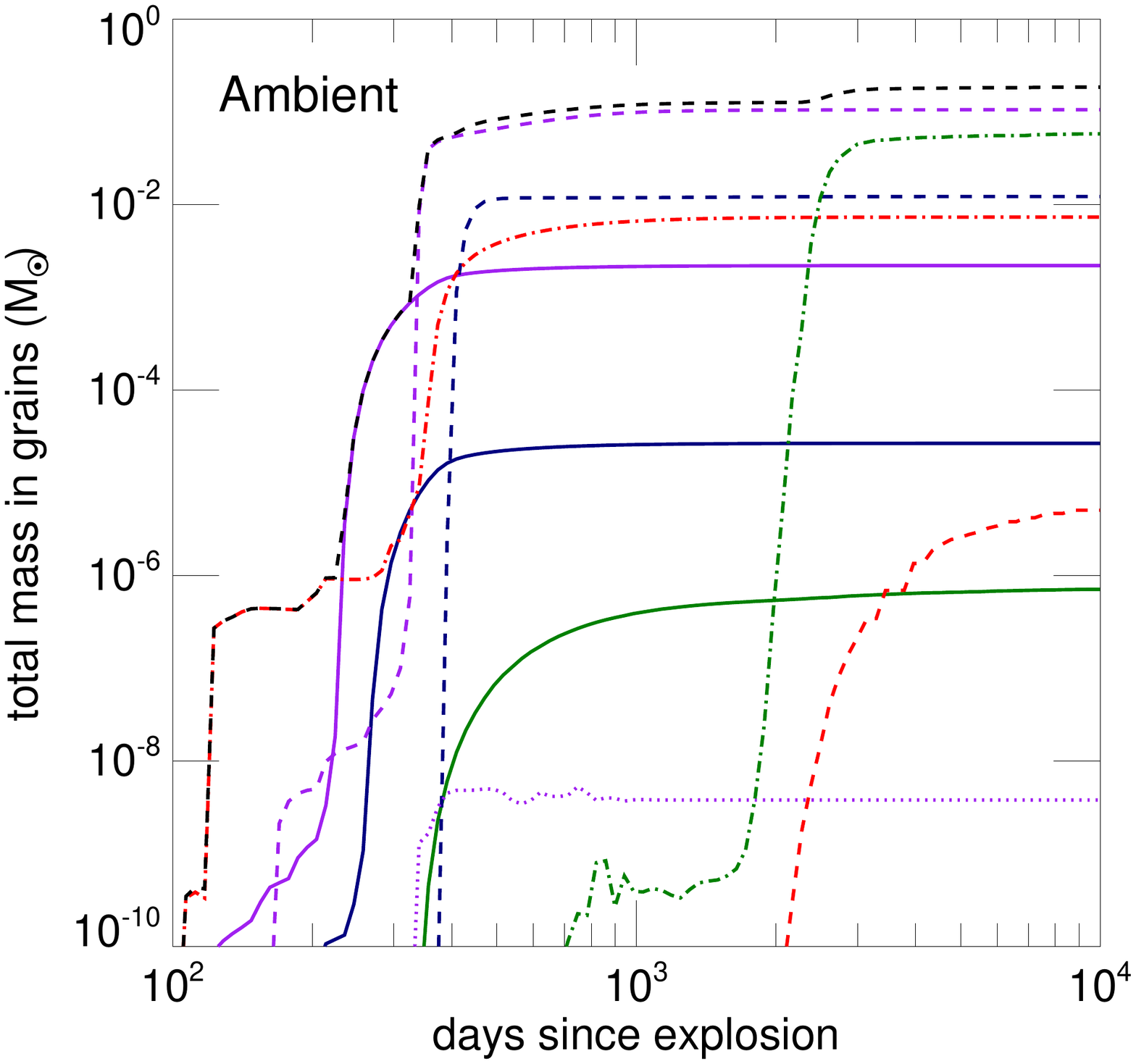}
\includegraphics[trim=0cm 4cm 0cm 5cm,clip=true,width=0.49\textwidth]{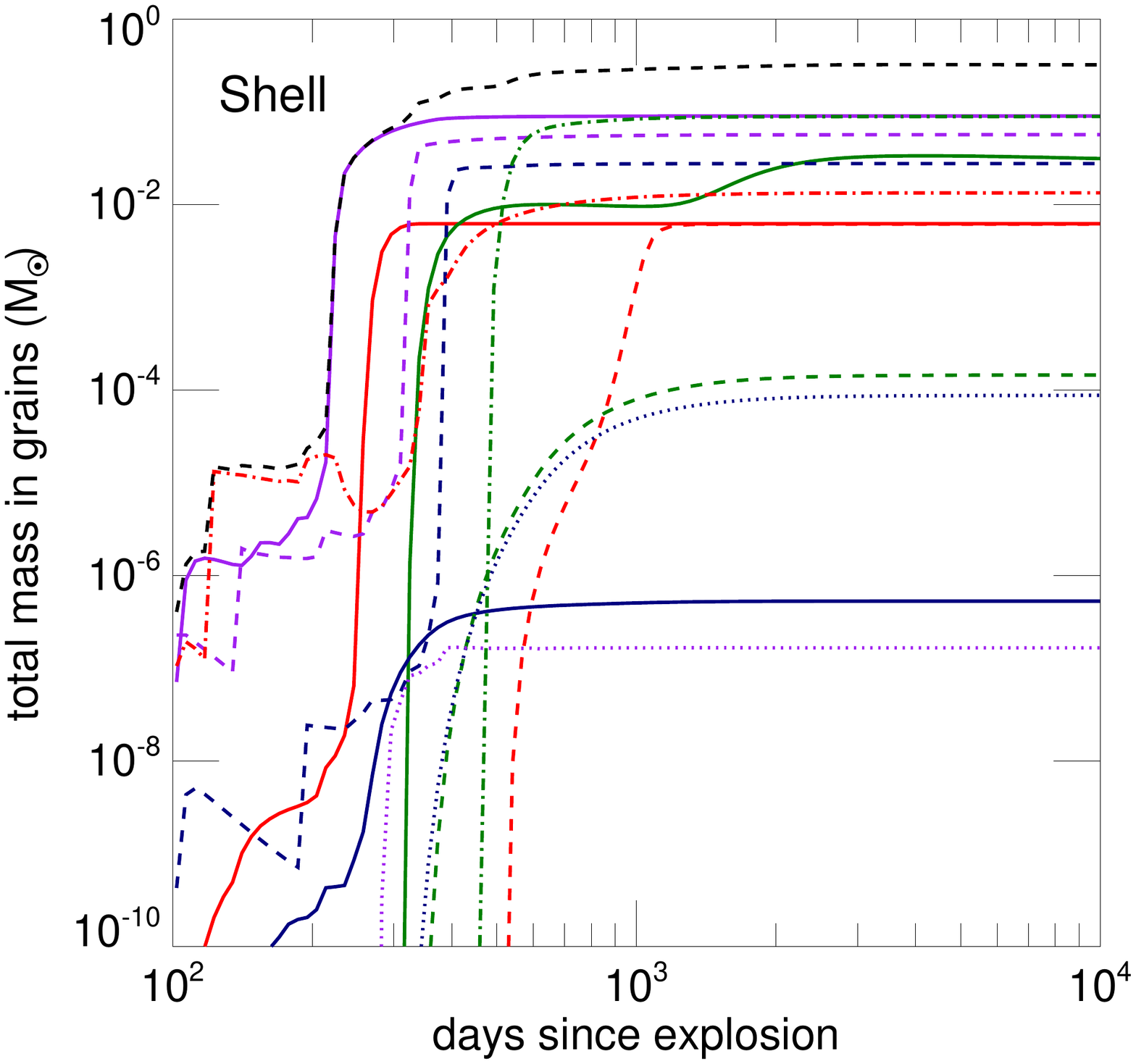}
\includegraphics[trim=0cm 4cm 0cm 5cm,clip=true,width=0.49\textwidth]{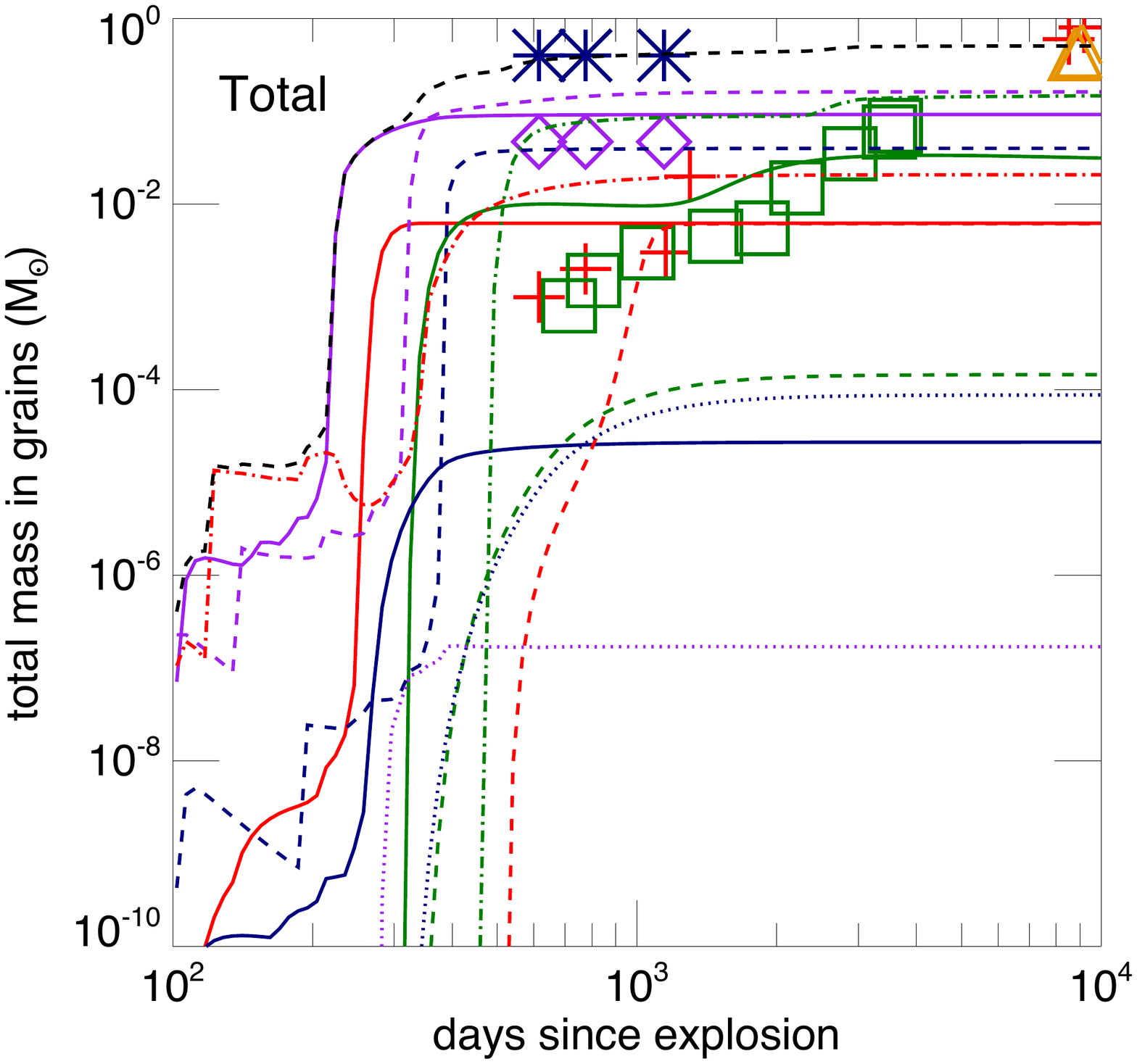}
\end{center}
\caption{The curves show the evolution of dust mass in the nickel bubbles (upper left), ambient ejecta (upper right), and thin shells at the bubble-ambient interface (lower left), and the sum of the three (lower right). Observations of dust mass in SN 1987A are shown as green squares \citep{2016MNRAS.456.1269B} and blue crosses \citep{Wesson:15}. The solid magenta line is a power law fit to the observations. The solid dashed line is the total dust mass summed over all grain species.}
\label{fig:GrainMassVsTime}
\end{figure*}

In Figure~\ref{fig:GrainMassVsTime} we plot the evolution of the dust mass in each of the three ejecta zones: the nickel bubbles, the ambient ejecta, and the dense shells at the bubble and ambient ejecta interface. The total dust mass in all zones and the observations from \citet{2016MNRAS.456.1269B}, \citet{Wesson:15}, and \citet{Dwek:15} are also shown on the lower right panel; we compare the computed and observationally inferred masses in Section \ref{sec:observations} below.   

Dust formation is rapid between 215 and 620 days after the explosion. The average dust growth rate is $\sim 0.3\,M_\odot \u{yr}^{-1}$.  After 620 days, dust formation continues at a much slower rate $\sim 0.006\, M_\odot \u{yr}^{-1}$. There are actually five distinct epochs when the dust growth rate $dM_{\rm dust}/dt$ spikes. Peak dust formation rate, mostly due to the formation in shells, occurs at 229 days for forsterite, 332 days for magnesia, 382 days for iron sulfide, 528 days for silicon, and 2551 days for carbon. The overall dust formation rate peaks at $dM_{\rm dust}/dt \approx1\, M_\odot \u{yr}^{-1}$ at 332 days.

Most of the dust forms in the dense shells ($0.32\, M_\odot$ or 63\% of the total mass). Less forms in the ambient ejecta ($0.18\,M_\odot$ or 37\%), and very little in the bubbles ($5\times 10^{-7}\,M_\odot$). This reflects the strong density dependence of dust yield: the nucleation rate is proportional to the square of the number density  $(dc_i/dt)_\text{nuc}\propto c_1^2$ and the accretion rate is proportional to the density $(dc_i/dt)_\text{acc}\propto c_1$. It is also because at high densities the ionization fraction, and thus the weathering rate, is lower. 

Not only the grain mass produced but also its chemical composition differs between the ambient and shell zones. Carbon and iron grains have a significantly higher mass in dense shells than in ambient ejecta.  Alumina, iron oxide, and magnetite grains do not form at all in the ambient ejecta but do form in significant amounts in the dense shells. All other grain species are produced in similar quantities in the ambient ejecta and shells.


\begin{table}
\centering
\begin{tabular}{lllll}
 \hline 
Species & Bubble & Ambient & Shell & Total  \\\hline
Alumina & 0  & 0  & 6.1($-$3) & 6.1($-$3) \\
Carbon & 0  & 7.6($-$7) & 3.0($-$2) & 3.0($-$2) \\
Enstatite & 0  & 2.7($-$5) & 5.7($-$7) & 2.6($-$5) \\
Forsterite & 0  & 2.0($-$3) & 9.0($-$2) & 9.0($-$2) \\
Iron & 5.0($-$7) & 5.0($-$6) & 5.8($-$3) & 6.4($-$3) \\
Iron Oxide & 0  & 0  & 1.4($-$4) & 1.4($-$4) \\
Iron Sulfide & 0  & 1.2($-$2) & 2.6($-$2) & 3.8($-$2) \\
Magnesia & 0  & 0.1 & 5.6($-$2) & 0.16 \\
Magnetite & 0  & 0  & 8.5($-$5) & 8.5($-$5) \\
Silicon Carbide & 0  & 3.7($-$9) & 1.7($-$7) & 1.7($-$7) \\
Silicon Dioxide & 0  & 7.8($-$3) & 1.4($-$2) & 2.2($-$2) \\
Silicon & 2.6($-$9) & 5.6($-$2) & 9.5($-$2) & 0.15 \\
\hline 
Total & 5.0($-$7) & 0.19 & 0.33 & 0.51 \\
\hline 
\end{tabular}
\caption{The same as Table \ref{tab:DustmassHistogrambySize}, but now providing the final dust mass classified by grain species as well as the ejecta zone.
\label{tab:DustmassHistogram}}
\end{table}

Table~\ref{tab:DustmassHistogram} provides the dust yield classified by grain species. The eight most abundant grain species are: magnesia ($0.16\,M_\odot$ or 32\% of the total dust mass), silicon ($0.15\,M_\odot$ or 29\%), forsterite ($0.092\,M_\odot$ or 18.0\%), iron sulfide ($0.04\,M_\odot$ or 7.78\%), carbon ($0.031\,M_\odot$ or 6.1\%), silicon dioxide ($0.021\,M_\odot$ or 4\%), alumina ($0.013\,M_\odot$ or 2.4\%), and iron ($0.0062\,M_\odot$ or 1.2\%).

A variety of competing effects influences the evolution of the mass in a grain species. At the earliest times grains are hot and evaporation completely inhibits grain growth. As the grain temperature drops, the evaporation rate decreases exponentially, and at some point it becomes negligible. Then, chemical weathering becomes the main grain destruction process. But even this is progressively less significant as the ionized fraction  gradually decreases (it never becomes completely negligible). Dust formation ceases when the accretion time exceeds the expansion time.  Coagulation does not affect the total dust mass although it modifies the size distribution.  However, the coagulation rate is always lower than the accretion rate because the abundance of material in clusters that can accrete onto a grain is always higher that in other grains.

The sudden rise of carbon dust mass at $\sim2000$ days is due to grain formation in the dense shells located in the mass coordinate range $4.5$--$7\,M_\odot$. There, the bulk of the mass is in helium and the abundance of He$^+$ is very high.  For example, at the mass coordinate $5.7\,M_\odot$ we find that the He$^+$ relative abundance is $10\%$ at 100 days and 0.06\% at 1000 days.  At 1830 days the He$^+$ abundance reaches a minimum of 0.04\% and thereafter increases steadily to 0.2\% at $10^4$ days. Thus, $\sim2000$ days is when the He$^+$ weathering is at a relative minimum and this is the optimal time to produce carbon dust.  The carbon dust mass increase between 300 and 400 days seen in Figure \ref{fig:GrainMassVsTime} occurs at mass coordinate $\sim 4.8\,M_\odot$ where the abundance of carbon exceeds that of helium and thus helium weathering is not an obstacle. Carbon is substantially more abundant than helium between $2\,M_\odot$ and $5\,M_\odot$, but inwards of $4.5\,M_\odot$, carbon dust formation is poor due to decreasing carbon mass fraction, the presence of the two other noble gas ions, Ne$^+$ and Ar$^+$, and an increasing oxygen abundance and thus stronger oxygen weathering.

\begin{figure*}
\begin{center}
\includegraphics[trim=0cm 4cm 0cm 5cm,clip=true,width=0.49\textwidth]{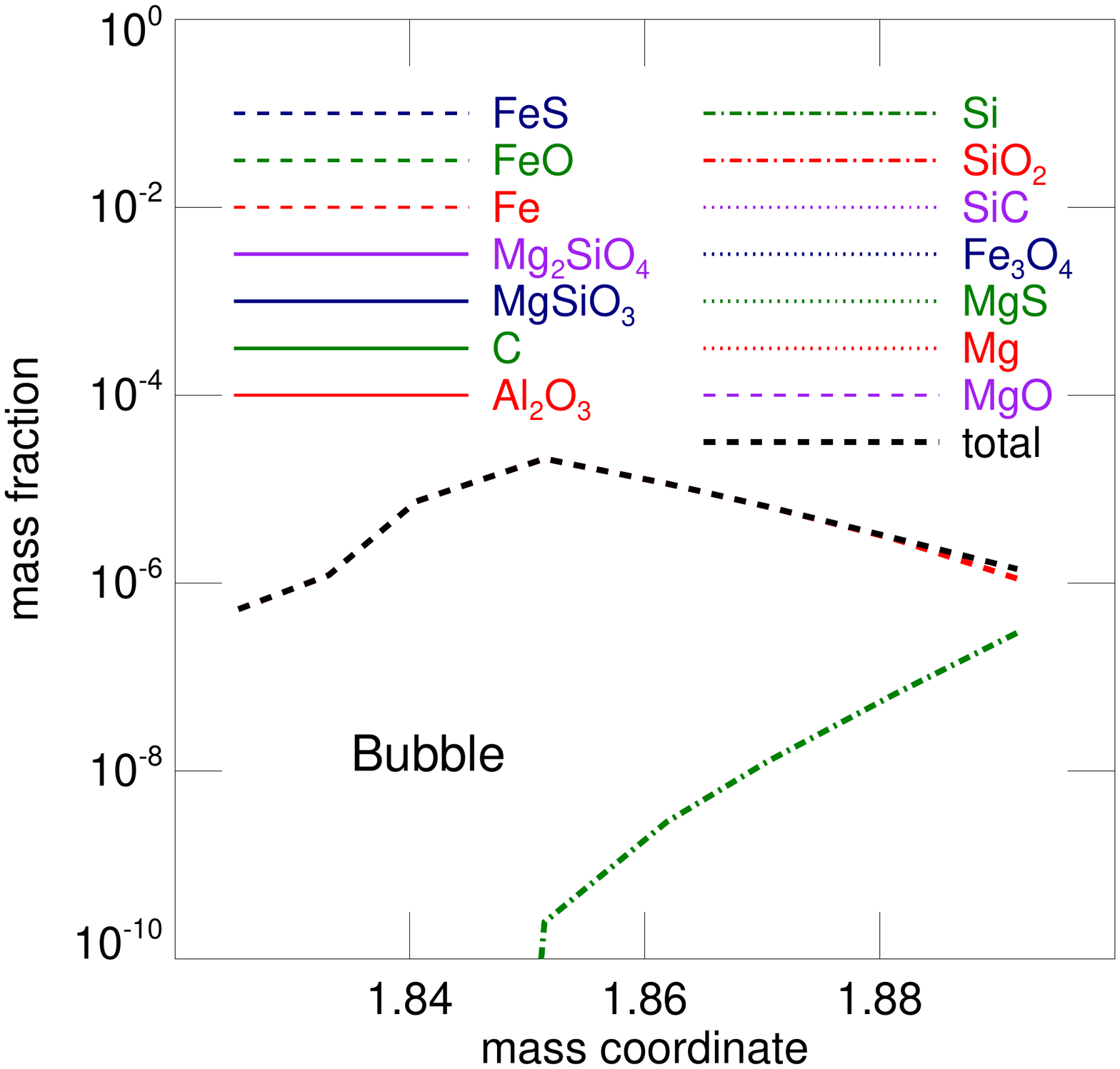}
\includegraphics[trim=0cm 4cm 0cm 5cm,clip=true,width=0.49\textwidth]{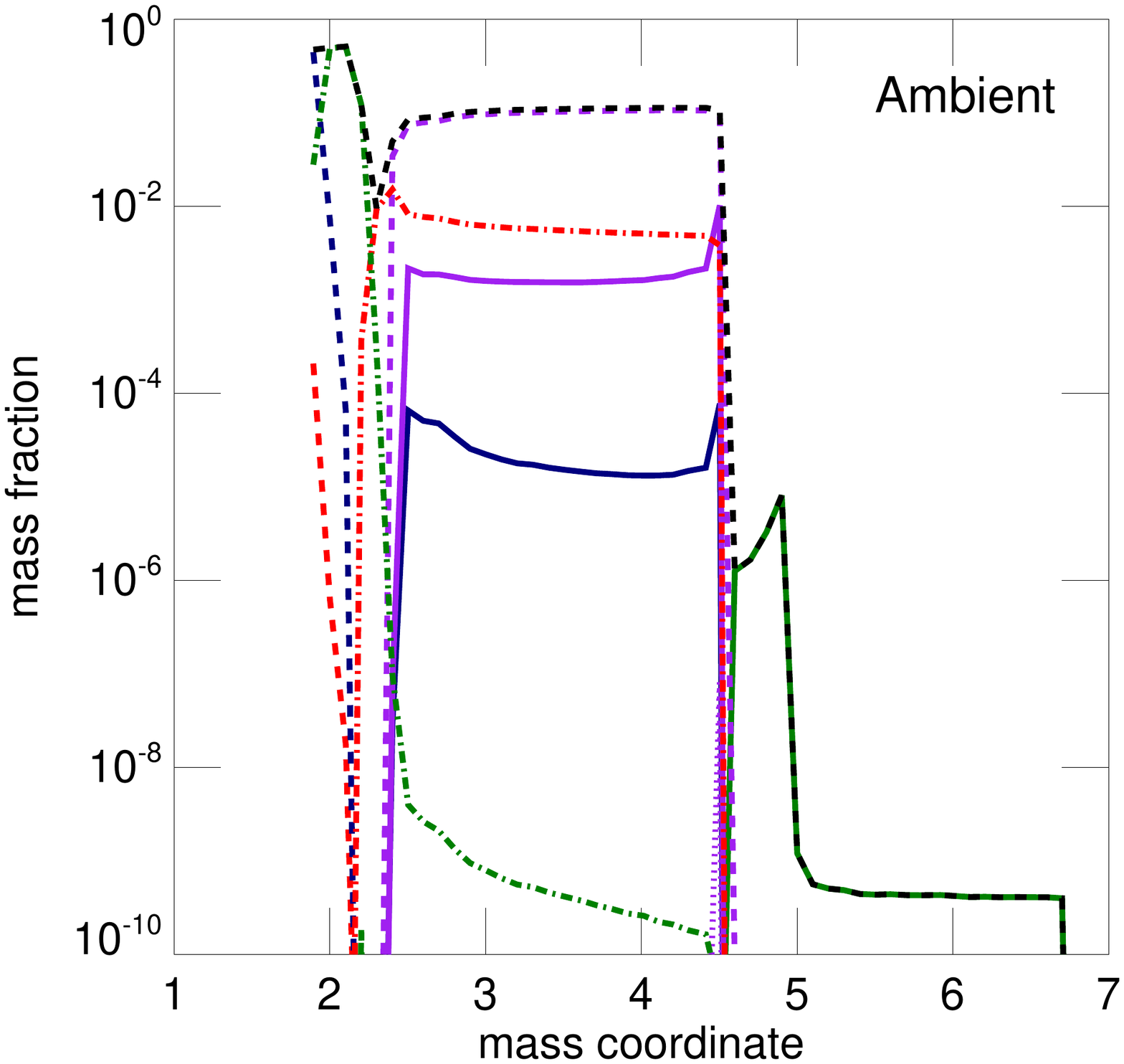}
\includegraphics[trim=0cm 4cm 0cm 5cm,clip=true,width=0.49\textwidth]{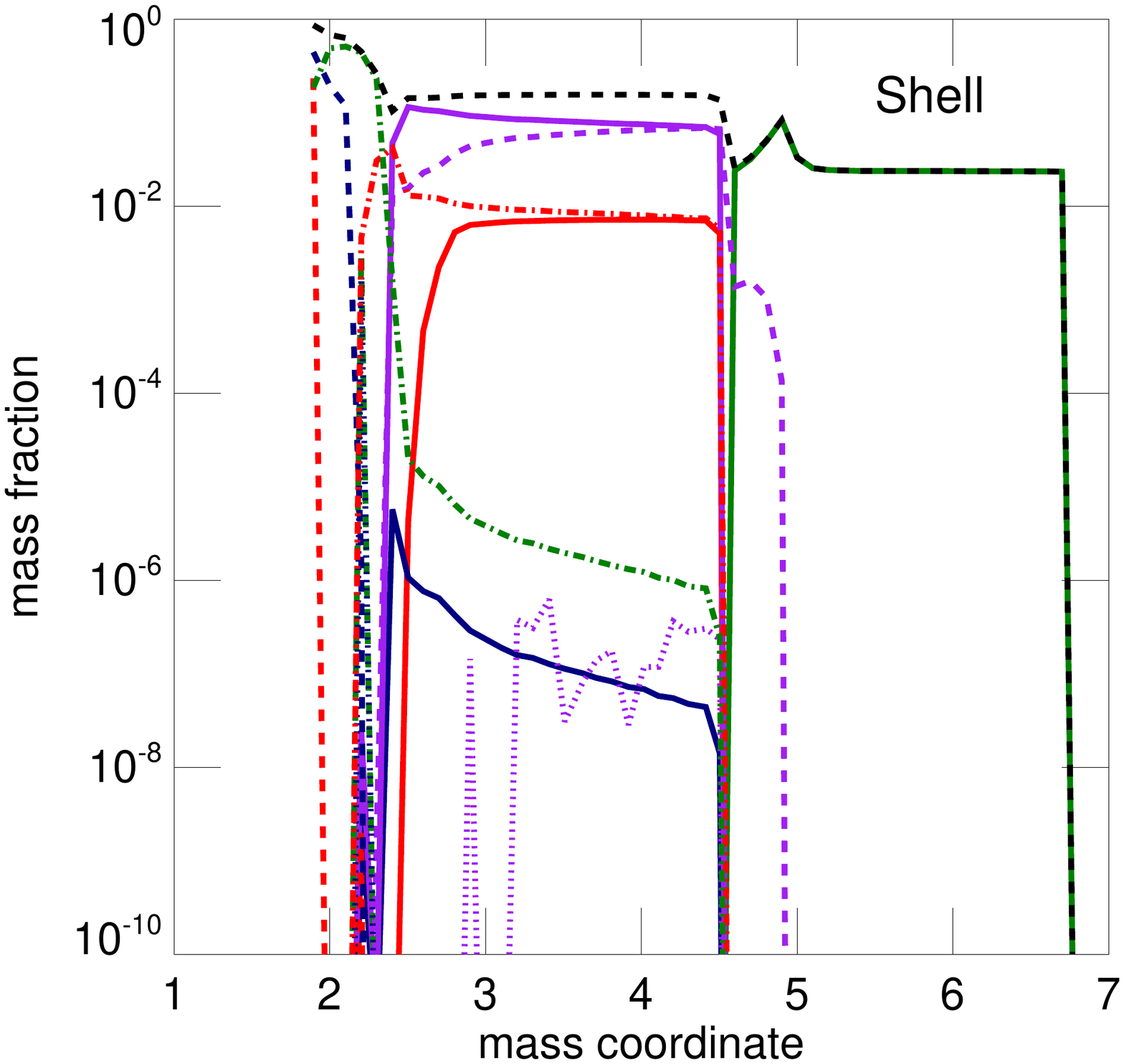}
\includegraphics[trim=0cm 4cm 0cm 5cm,clip=true,width=0.49\textwidth]{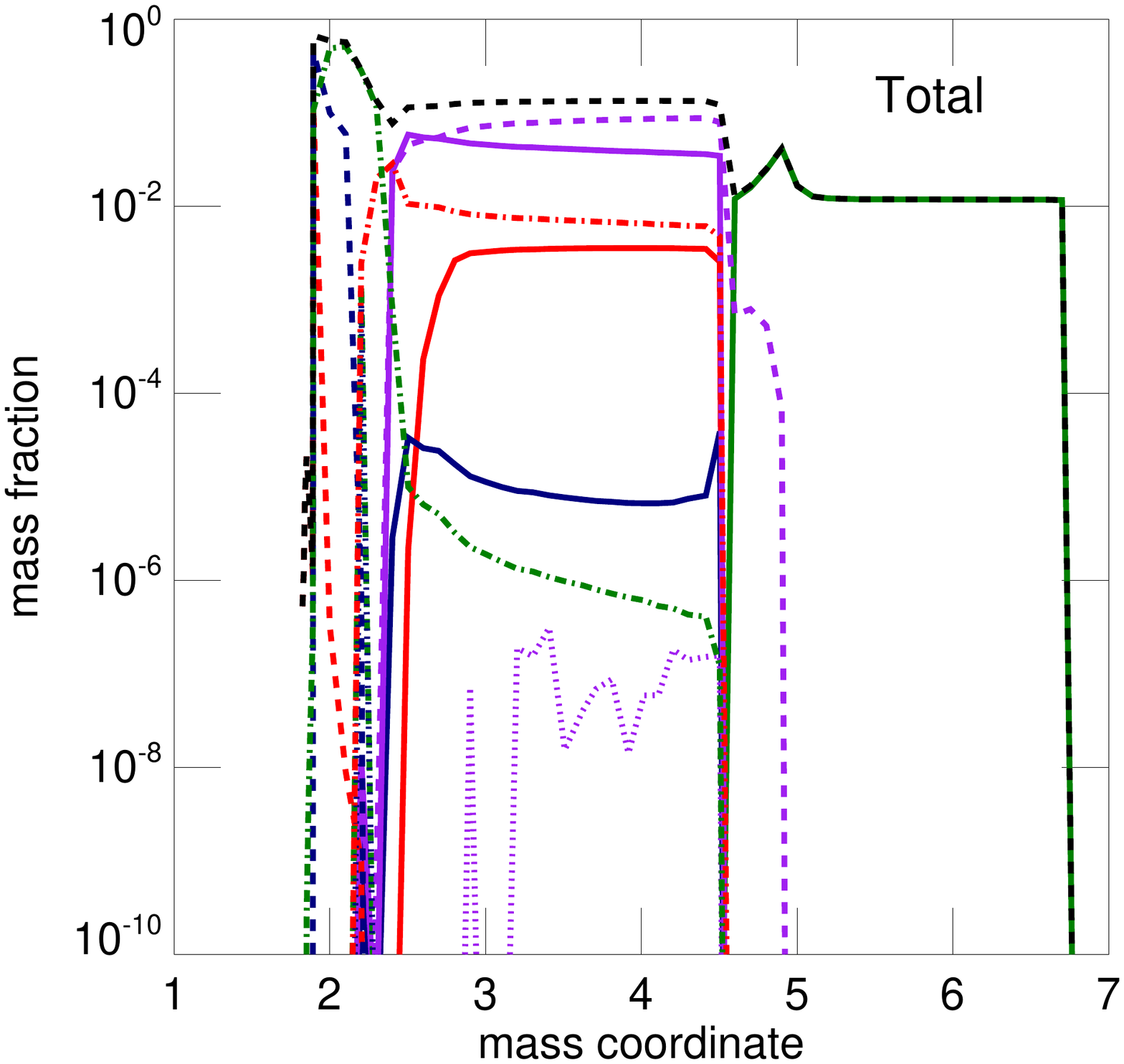}
\end{center}
\caption{Grain mass fraction as a function of the mass coordinate in bubbles (upper left), ambient ejecta (upper right), dense shells (lower left), and all zones combined (lower right).}
\label{fig:GrainMassVsRadius}
\end{figure*}

Figure~\ref{fig:GrainMassVsRadius} shows the mass fraction in each of the grain species as a function of the mass coordinate within the ejecta at the end of the simulation. The low-density bubbles form only a very small amount of silicon and iron grains with a total depletion fraction of $10^{-7}$--$10^{-5}$. In the ambient ejecta grains form in significant amounts for $M < 4.5\,M_\odot$. Specifically, silicon and iron sulfide dominate the grain mass fraction for $M < 2.25\,M_\odot$, and some iron grains are present there as well. For $2.25\,M_\odot < M < 4.5 \,M_\odot$, magnesia is the dominant grain species, followed by silicon dioxide, forsterite, and enstatite.
In the  carbon-rich region $4.5\,M_\odot < M < 5\,M_\odot$ some carbon grains manage to form, comprising only a small mass fraction ($\sim 10^{-6}$--$10^{-5}$).  
Grains are practically absent in the helium rich outer core $M > 5\,M_\odot$. 

The shell ejecta produce the highest diversity of grain species.  The innermost region ($M < 2.25\,M_\odot$) again forms silicon, iron sulfide, and iron grains. The next region ($2.25\,M_\odot < M < 4.5\,M_\odot$) forms, in decreasing order of yield: forsterite, magnesia, silicon dioxide, alumina, silicon, silicon carbide, and enstatite. The carbon-rich region $4.5\,M_\odot < M < 5\,M_\odot$ froms carbon and magnesia grains. The outermost helium core $M > 5\,M_\odot$ forms only carbon grains. In the entire ejecta carbon grains only form when $M > 4.5\,M_\odot$ and they mostly form in the dense shells.

\begin{figure*}
\begin{center}
\includegraphics[trim=0cm 4cm 0cm 5cm,clip=true,width=0.49\textwidth]{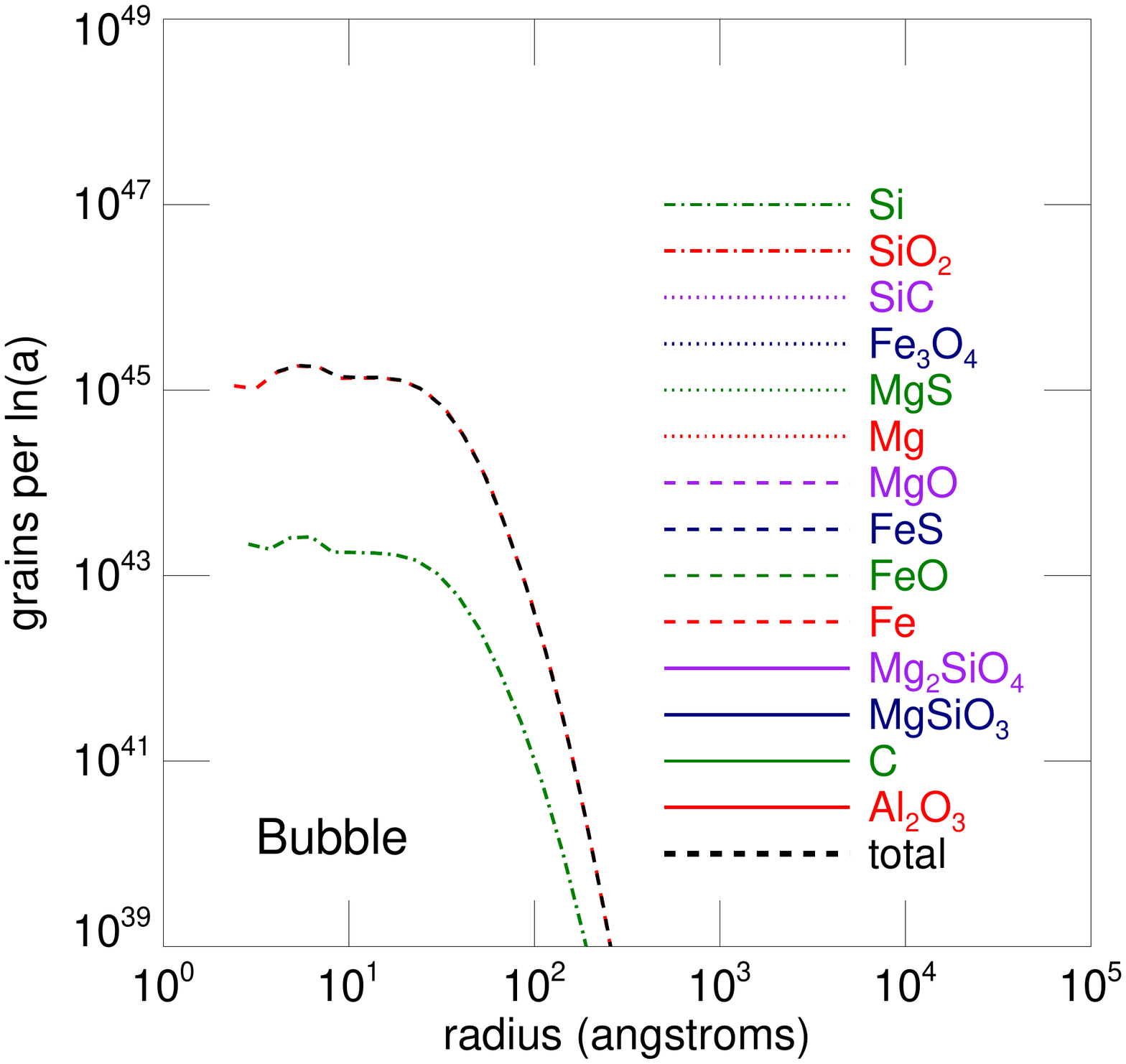}
\includegraphics[trim=0cm 4cm 0cm 5cm,clip=true,width=0.49\textwidth]{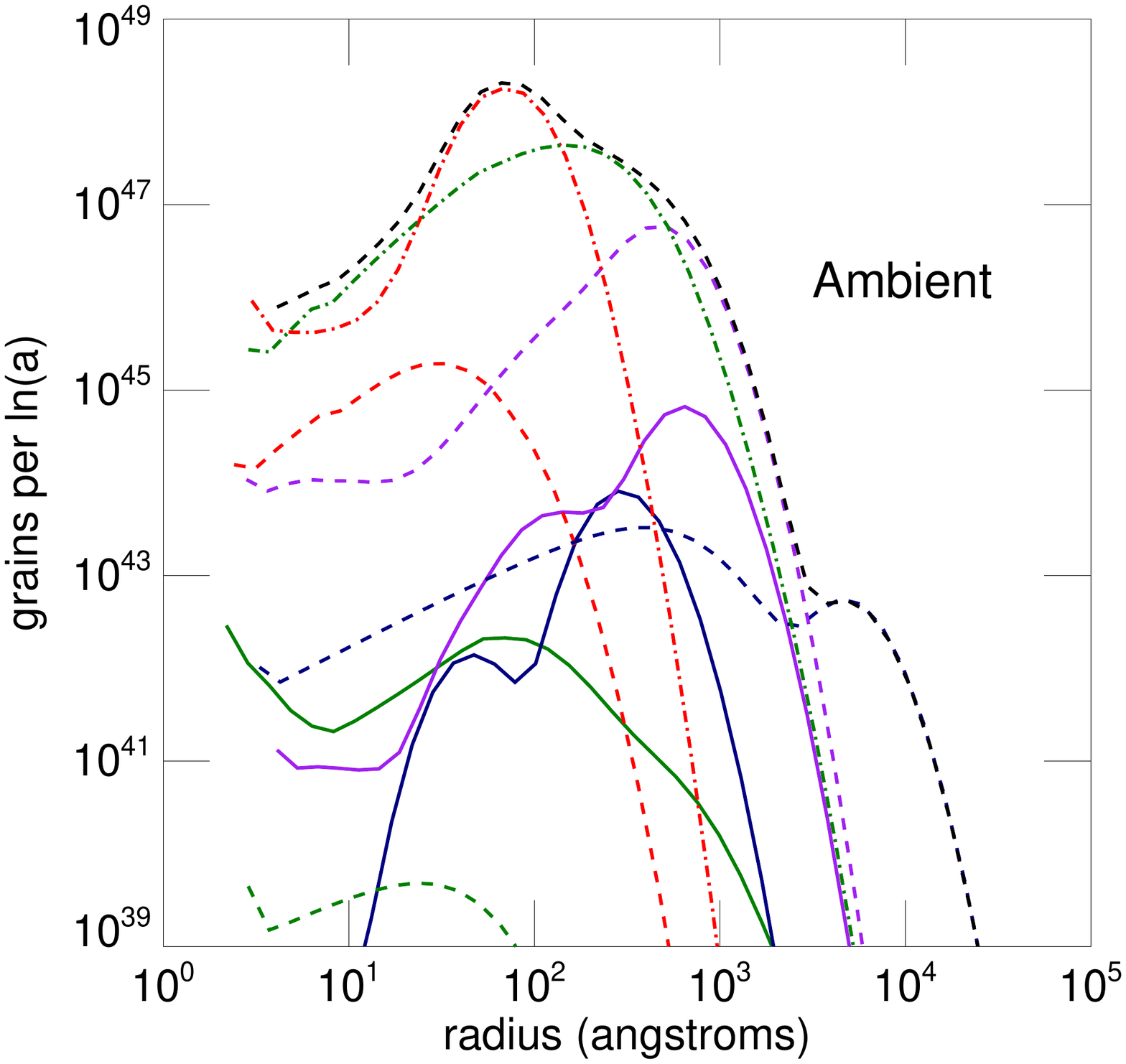}
\includegraphics[trim=0cm 4cm 0cm 5cm,clip=true,width=0.49\textwidth]{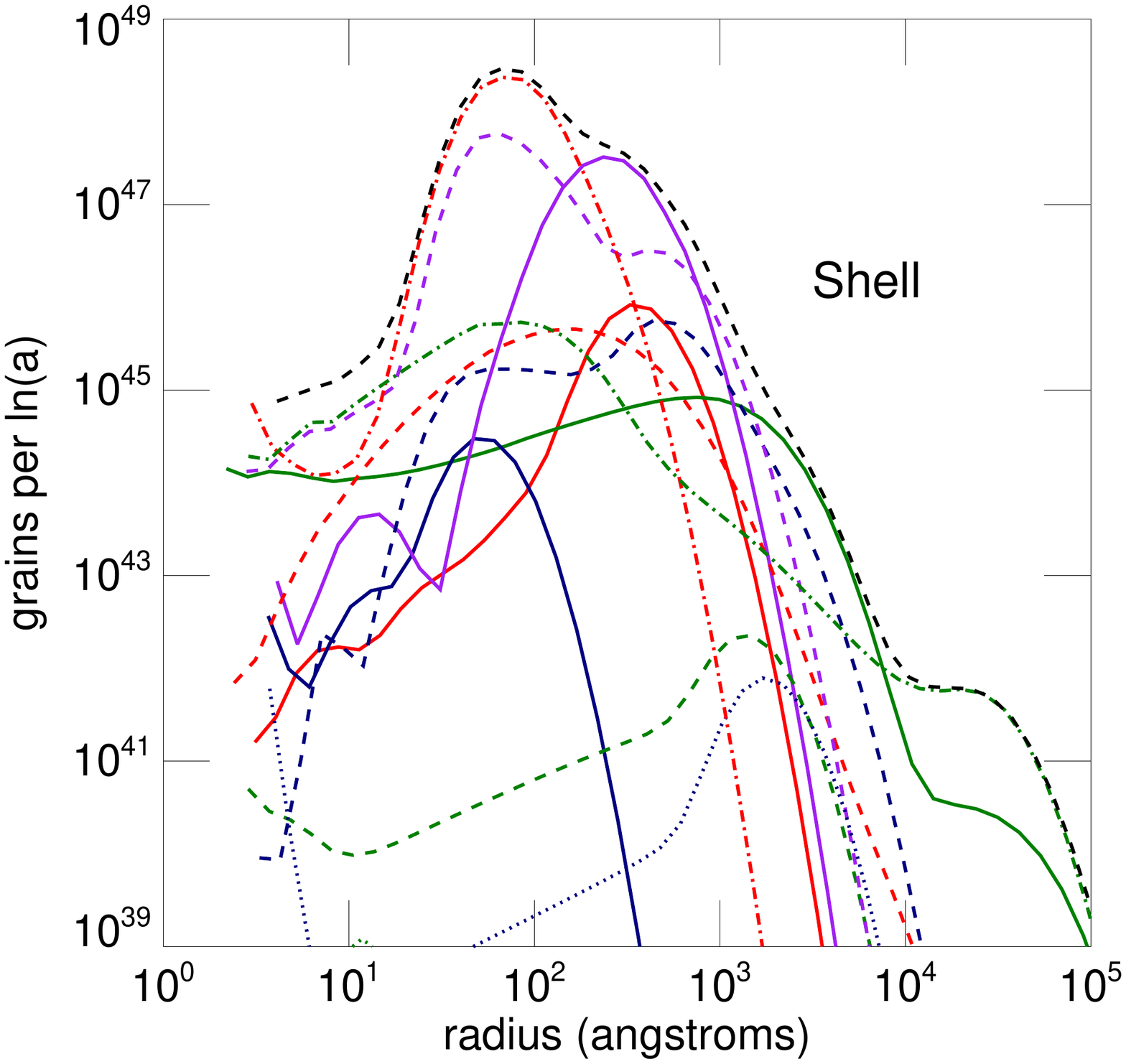}
\includegraphics[trim=0cm 4cm 0cm 5cm,clip=true,width=0.49\textwidth]{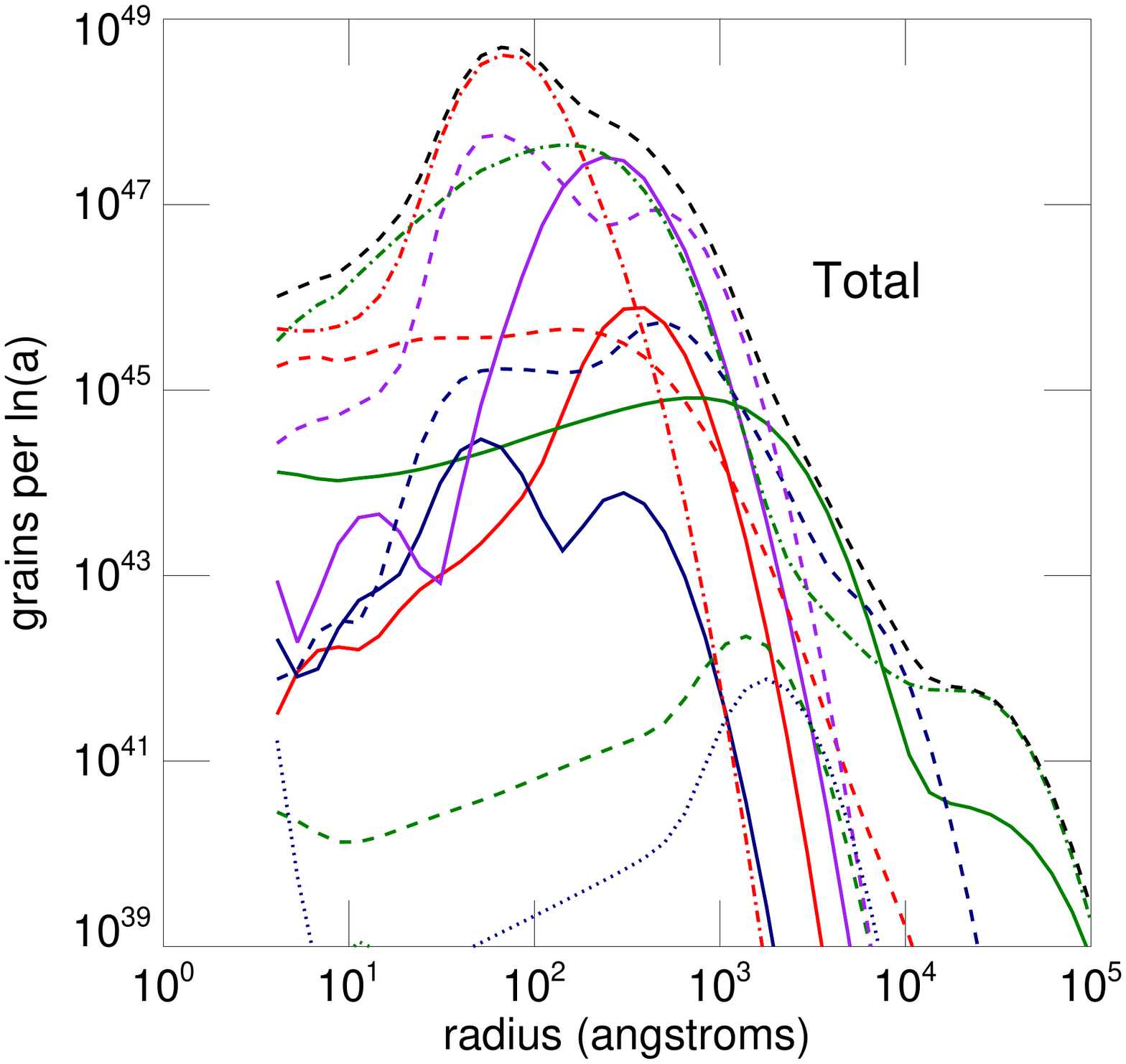}
\end{center}
\caption{
The curves show the grain size distribution $dN_{\text{dust},p}/d\ln a$ at the end of the simulation in the nickel bubbles (upper left), ambient ejecta (upper right), and thin shells at the bubble-ambient interface (middle left), and combined across the three zones (middle right).  }
\label{fig:LogGrainSizeDistribution}
\end{figure*}

\begin{figure*}
\begin{center}
\includegraphics[trim=0cm 4cm 0cm 5cm,clip=true,width=0.49\textwidth]{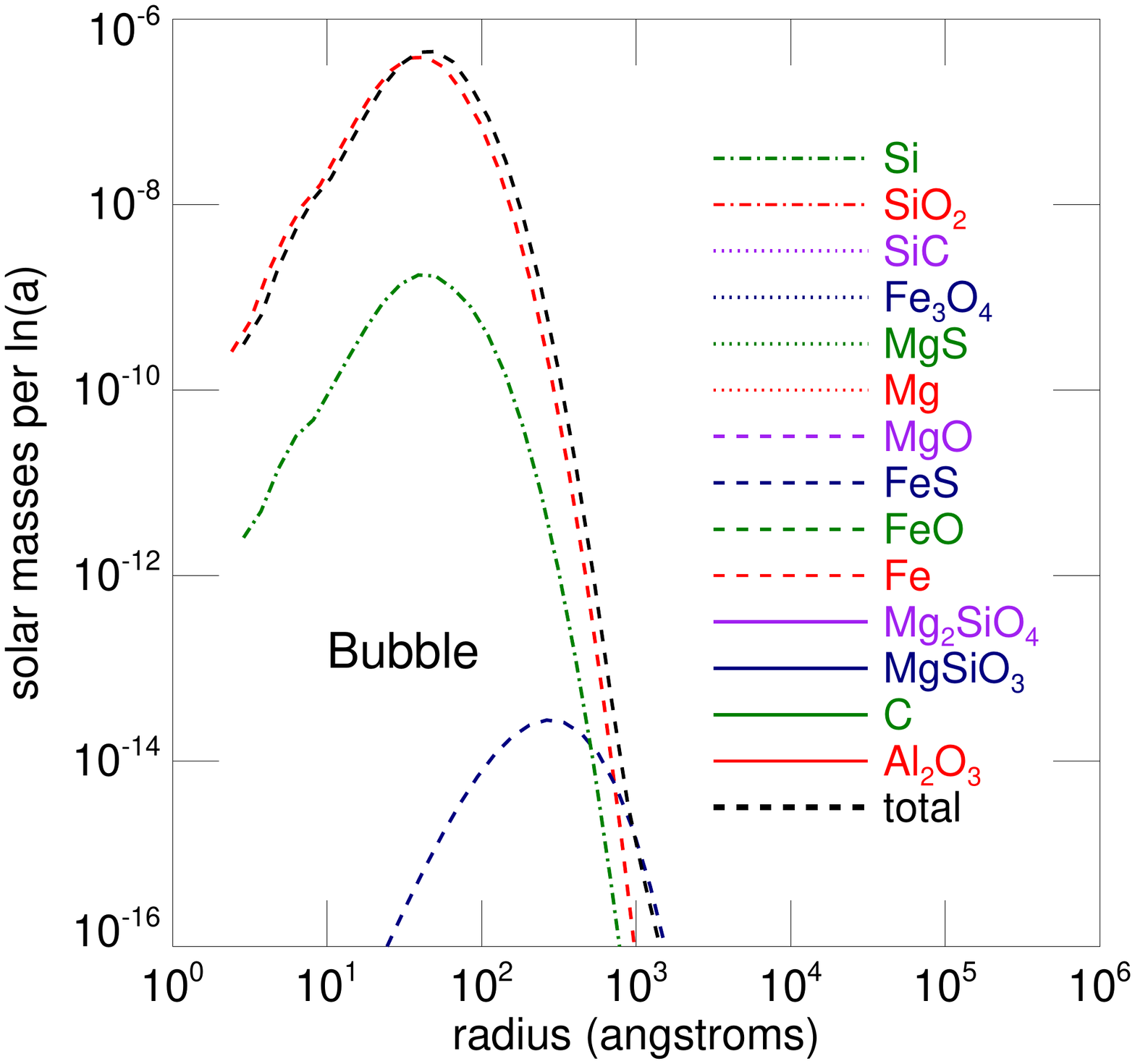}
\includegraphics[trim=0cm 4cm 0cm 5cm,clip=true,width=0.49\textwidth]{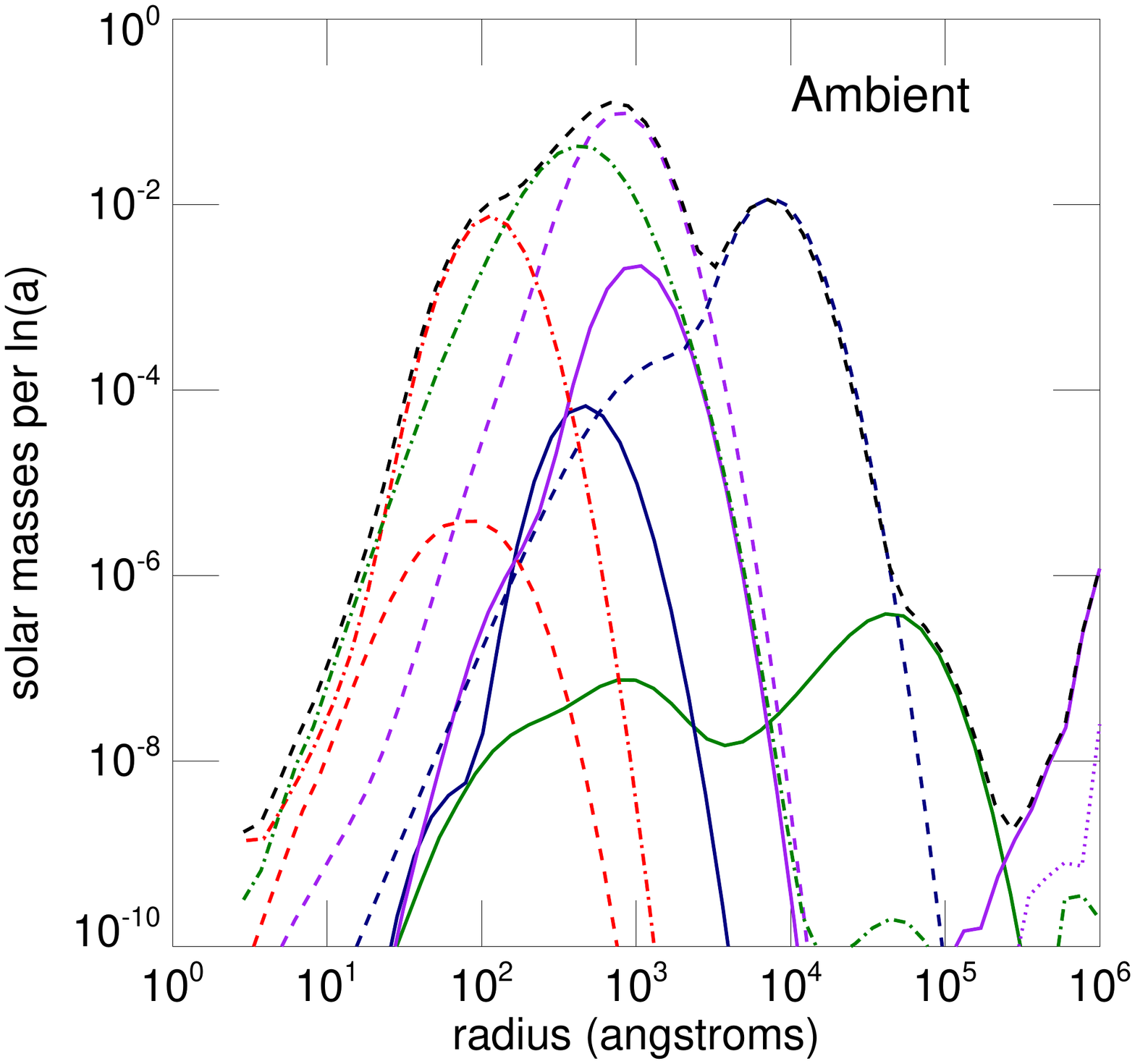}
\includegraphics[trim=0cm 4cm 0cm 5cm,clip=true,width=0.49\textwidth]{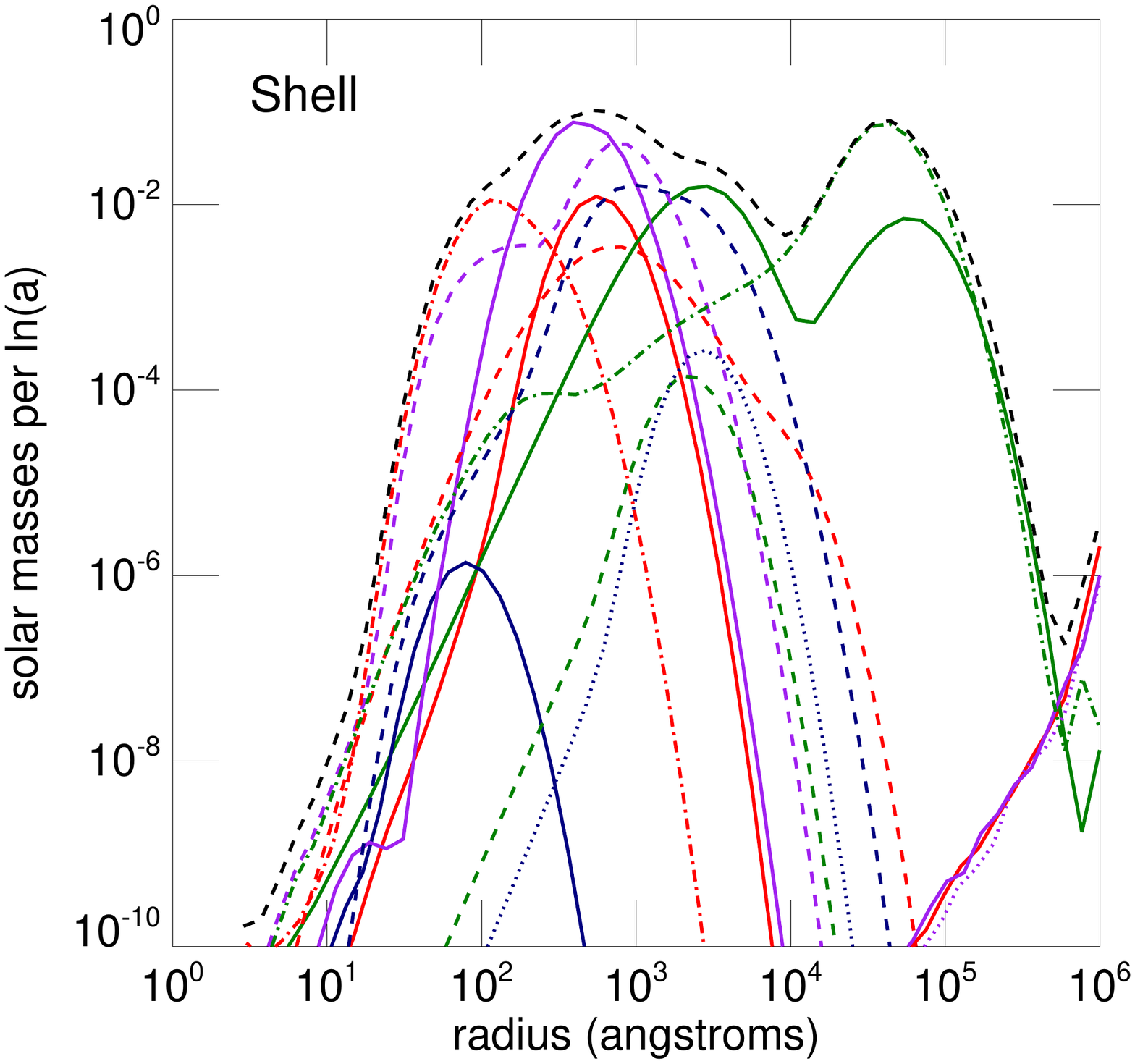}
\includegraphics[trim=0cm 4cm 0cm 5cm,clip=true,width=0.49\textwidth]{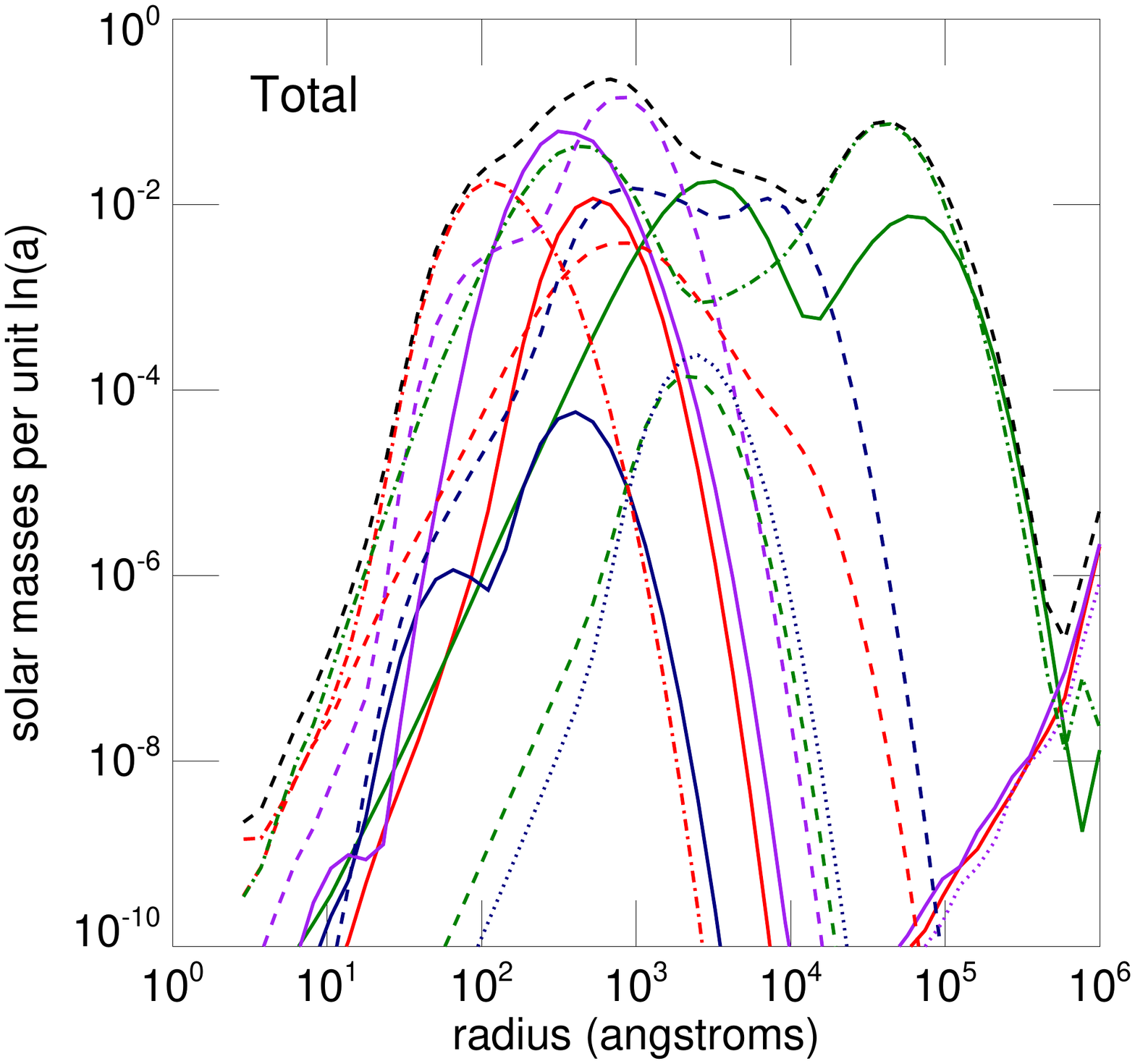}
\end{center}
\caption{
The same as Figure \ref{fig:LogGrainSizeDistribution}, but now showing the dust mass per logarithm of the radius $dM_{{\rm dust},p}/d\ln a$.}
\label{fig:LogGrainMassDistribution}
\end{figure*}

Figure~\ref{fig:LogGrainSizeDistribution} shows the logarithmic size distribution (grains per log grain radius) at the end of the simulation. The distribution peaks at $5.3\u{\AA}$ in the bubbles and $66\u{\AA}$ in the ambient ejecta, shells, and overall ejecta.  We also plot the mass in grains per unit logarithmic radius $dM_{{\rm dust},p} /d\ln a$ at the end of the simulation in Figure~\ref{fig:LogGrainMassDistribution}. This distribution peaks at $50.5\u{\AA}$, $682\u{\AA}$, $526\u{\AA}$, and $682\u{\AA}$ in the bubbles, ambient ejecta, shells, and total ejecta, respectively. The total grain size distribution beyond the peak is approximately a power law:
\bea
\f{dN}{d\ln a}\propto a^{-3.4} .
\eea 
This is steeper than the classical grain size distribution in the ISM, $dN/d\ln a \propto a^{-2.5}$ \citep{1977ApJ...217..425M}.  The sharp rise of $dM_{{\rm dust},p} /d\ln a$ with radius at the largest radii, $a\gtrsim 10\,\mu\text{m}$, is an artifact of a numerical instability of our time primitive discretization scheme.

\subsection{Molecules}

The molecules CO, SiO, SO, and O$_2$ play important roles in dust formation. CO locks up carbon atoms so they cannot be added to carbon and silicon carbide grains. SiO is required to form the condensation nuclei of enstatite, forsterite, and silicon dioxide, whereas SO and O$_2$ are oxidizing agents that play a role in the formation of enstatite, forsterite, iron oxide, alumina, iron sulfide, magnesia, magnesium sulfide, magnetite, and silicon carbide.

\begin{figure*}
\begin{center}
\includegraphics[trim=0cm 4cm 0cm 5cm,clip=true,width=0.33\textwidth]{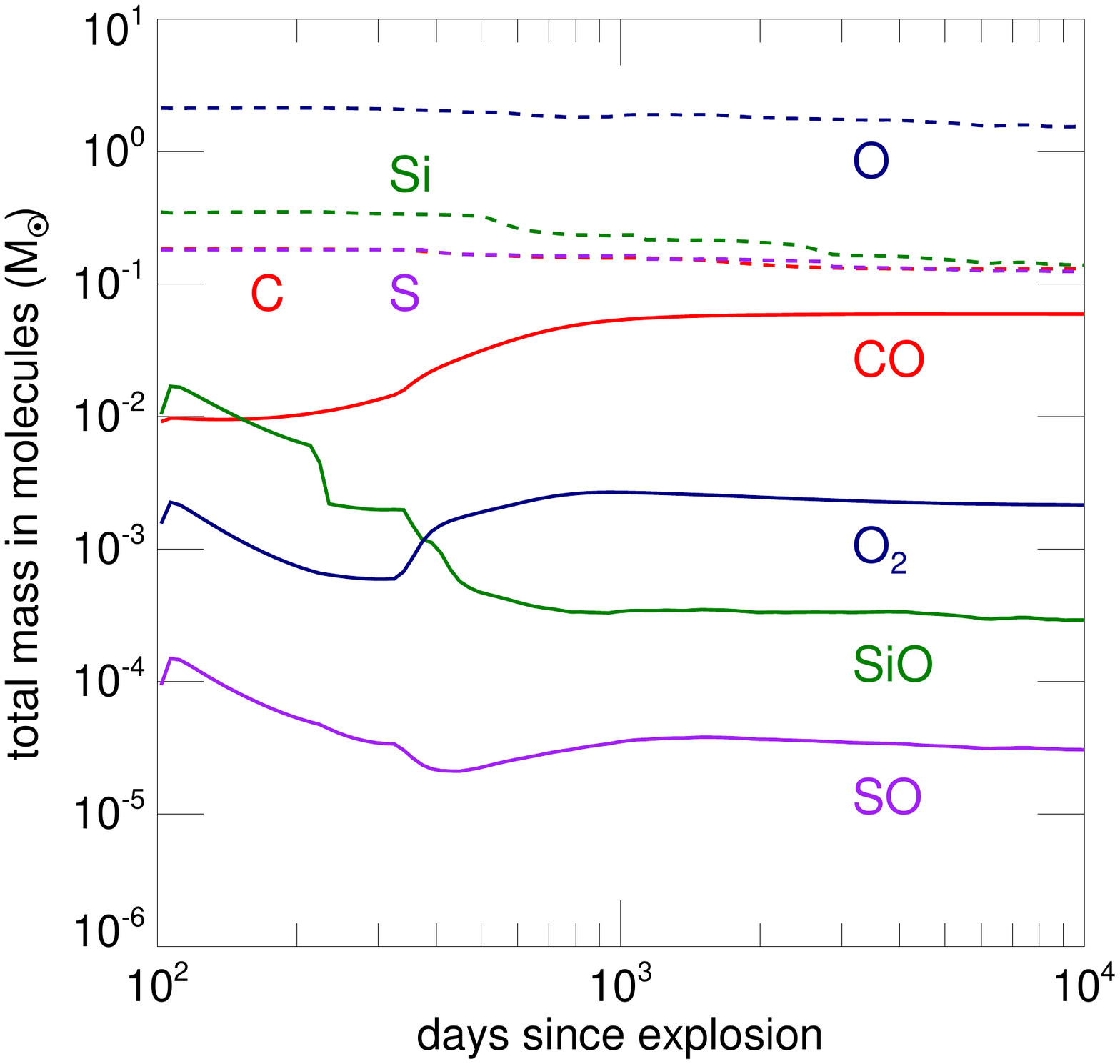}
\includegraphics[trim=0cm 4cm 0cm 5cm,clip=true,width=0.33\textwidth]{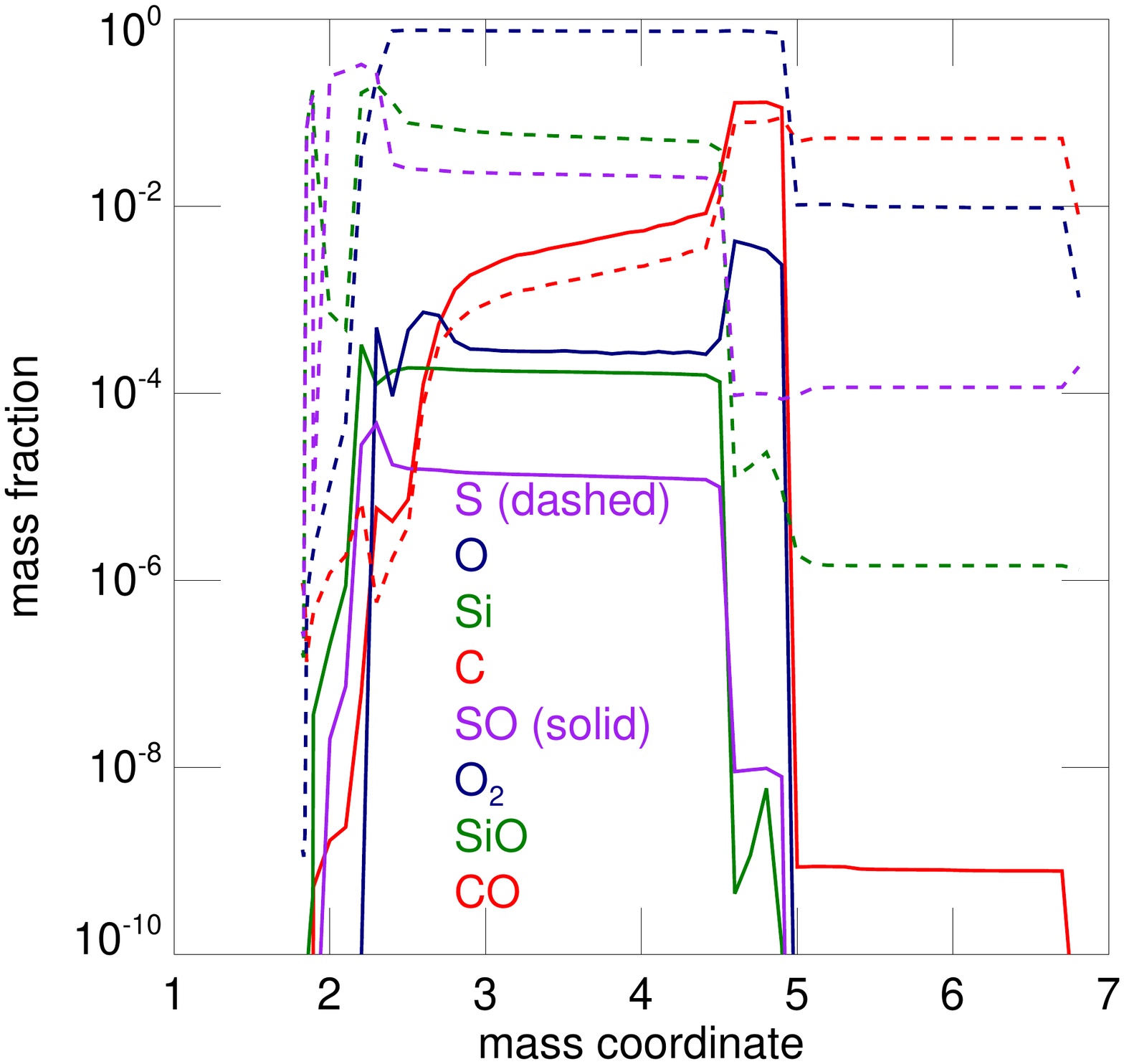}
\includegraphics[trim=0cm 4cm 0cm 5cm,clip=true,width=0.33\textwidth]{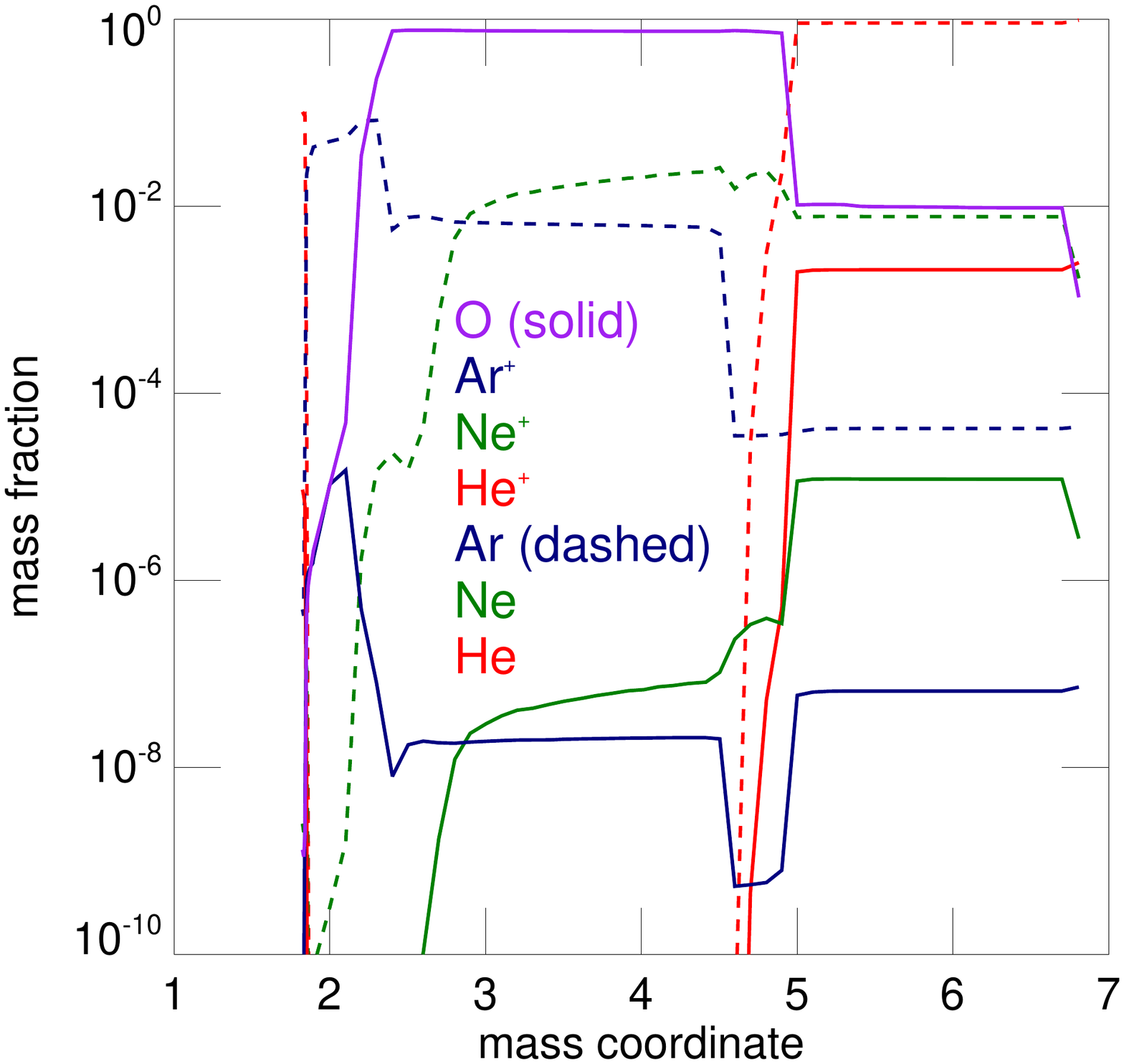}
\end{center}
\caption{Evolution of the mass in common molecular species (left), mass fraction versus mass coordinate in common molecular species (middle), and mass fraction versus mass coordinate in weathering agents (right), each for the entire ejecta.}
\label{fig:MoleculeProperties}
\end{figure*}

The left panel in Figure \ref{fig:MoleculeProperties} shows the mass of CO, SiO, SO, and O$_2$ as a function of time in each of the density zones, as well as the mass of atomic C, O, Si, and S. In contrast to the dust mass, the mass in molecules is relatively constant over the entire $100$--$10^4$ day period. At the end of the simulation, the total mass in CO, SiO, SO, and O$_2$ is ($0.059$, $2.9\times10^{-4}$, $3.1\times 10^{-5}$,  $2.2\times 10^{-3}) M_\odot$.\footnote{The CO mass is well above what we interpret as a lower limit of $5\times10^{-5}\,M_\odot$ inferred by \citet{Spyromilio:88} at 255 days.  At a similar age of the remnant \citet{1992ApJ...396..679L} inferred a higher value of $0.001\,M_\odot$, an order of magnitude smaller than our prediction.  At $\sim10^4$ days \citet{Kamenetzky:13} inferred a CO mass of $0.01\,M_\odot$, again an order of magnitude smaller than ours at the corresponding time. The SiO mass is well above the value $4\times10^{-6}\,M_\odot$ inferred by \citet{Roche:91} at 500 days.} Dissociation by Compton electrons and charge transfer reactions with noble gas ions, as well as depletion onto grains, keep the gas from turning fully molecular.

The middle panel in Figure \ref{fig:MoleculeProperties} shows the mass fraction of these species as a function of mass coordinate in each density zone. Virtually no molecules form in the bubbles due to the low density and high ionization fraction. Outside of the bubbles molecules mostly form at mass coordinates $2\,M_\odot<M<5\,M_\odot$, with more molecules forming in the shells than in the ambient ejecta. At $M>5\,M_\odot$ the high abundance of helium suppresses molecules in both the ambient ejecta and the shells.

In the right panel in Figure \ref{fig:MoleculeProperties} we show the mass fraction of the weathering agents He$^+$, Ne$^+$, Ar$^+$, and O as a function of mass coordinate. It can be seen that noble gas weathering is dominated by helium for $M > 5\,M_\odot$, neon for $3\,M_\odot < M < 5\,M_\odot$, and argon for $M < 3\,M_\odot$. Since the mass fraction of Ne$^+$ and Ar$^+$ for $M < 5\,M_\odot$ is $10^4$ times below that of He$^+$ for $M > 5\,M_\odot$, neon and argon weathering are relatively insignificant compared to helium weathering. The large diversity of grain species over the mass coordinates $2\,M_\odot < M < 5\,M_\odot$ is in part due to the lack of noble gas weathering agents there. However, oxygen is very abundant in this region and its weathering suppresses carbon grain formation.

\subsection{Dependance on shell density}

\begin{figure*}
\begin{center}
\includegraphics[trim=0cm 4cm 0cm 5cm,clip=true,width=0.33\textwidth]{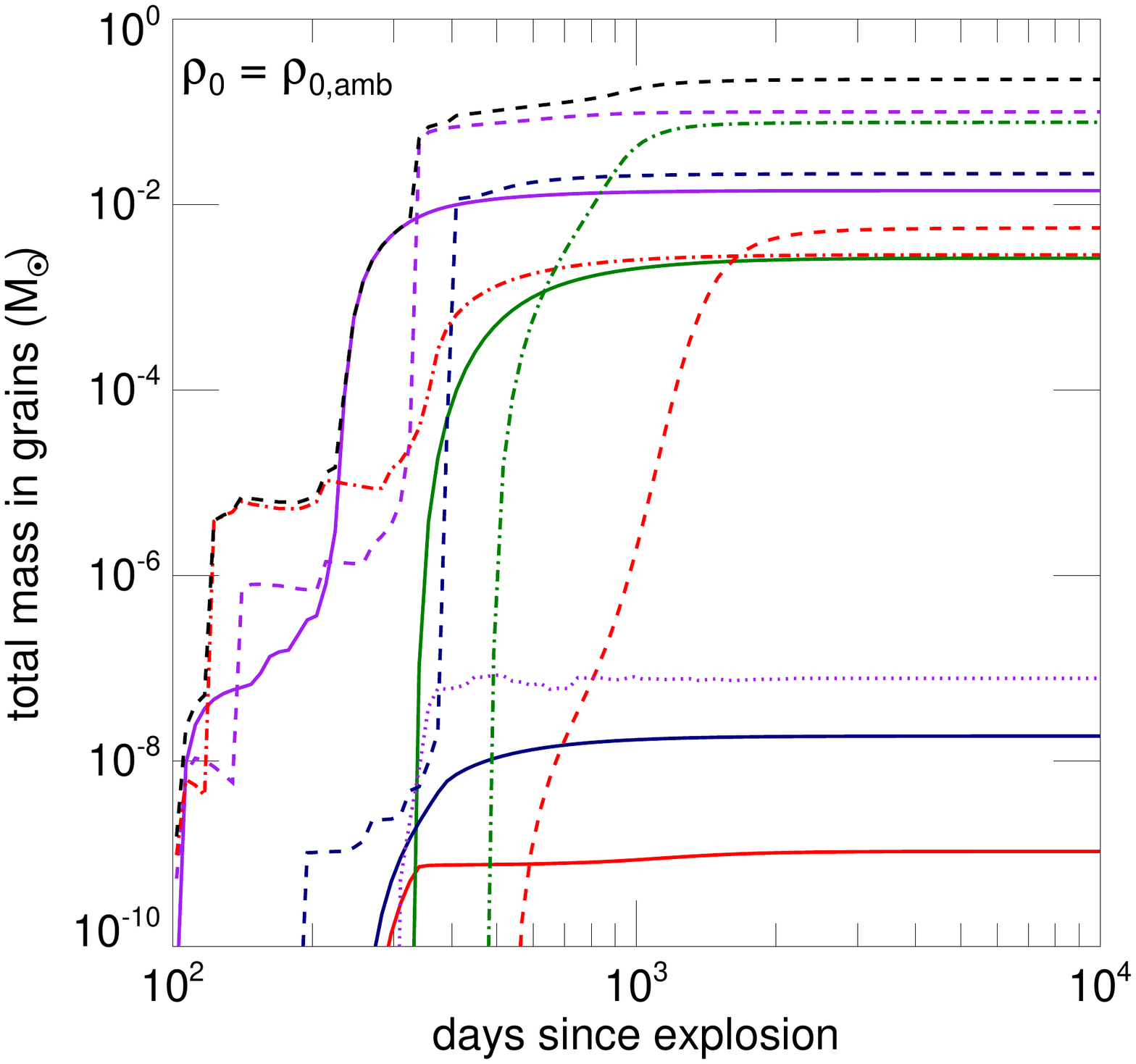}
\includegraphics[trim=0cm 4cm 0cm 5cm,clip=true,width=0.33\textwidth]{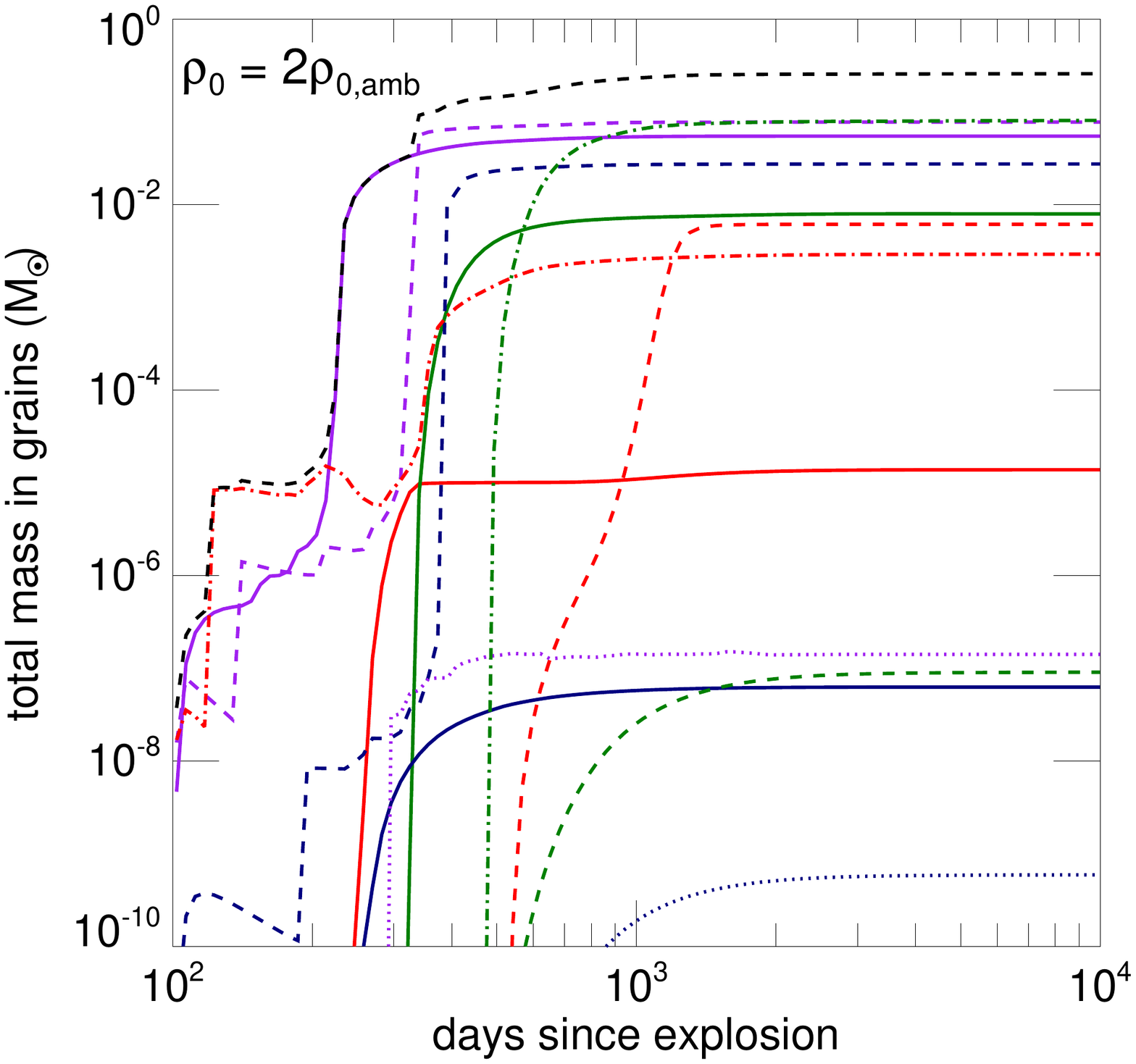}
\includegraphics[trim=0cm 4cm 0cm 5cm,clip=true,width=0.33\textwidth]{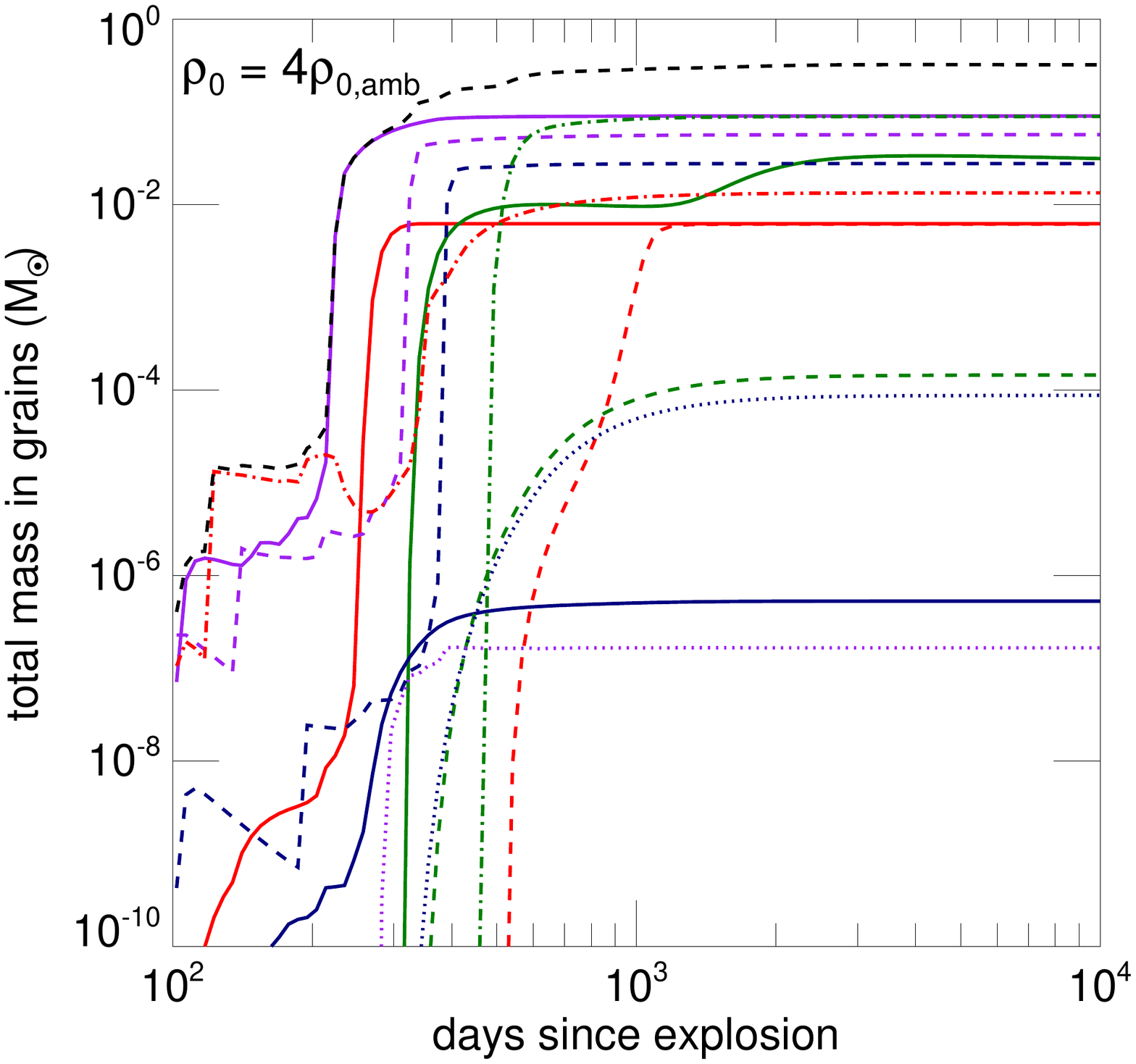}
\includegraphics[trim=0cm 4cm 0cm 5cm,clip=true,width=0.33\textwidth]{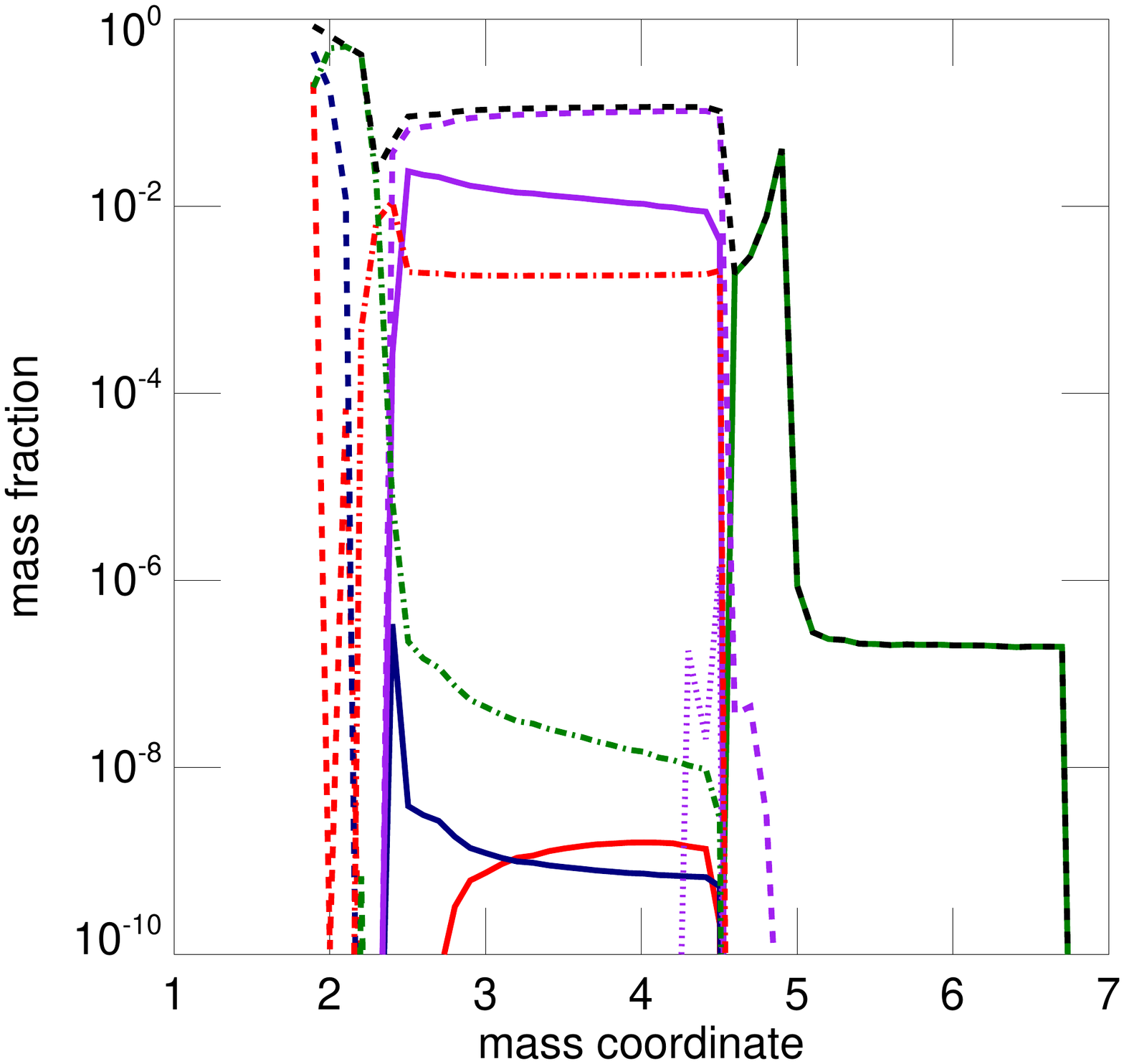}
\includegraphics[trim=0cm 4cm 0cm 5cm,clip=true,width=0.33\textwidth]{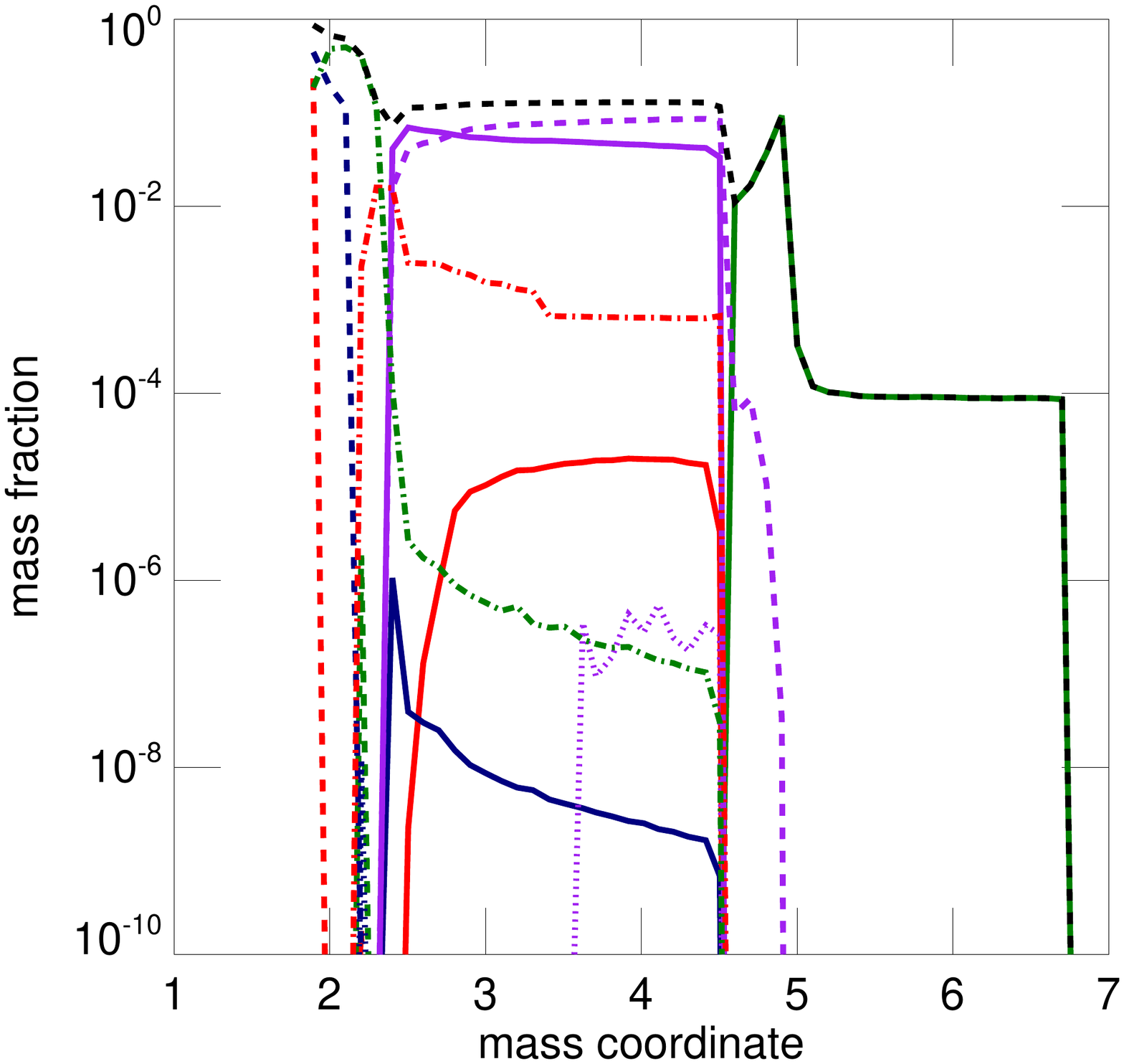}
\includegraphics[trim=0cm 4cm 0cm 5cm,clip=true,width=0.33\textwidth]{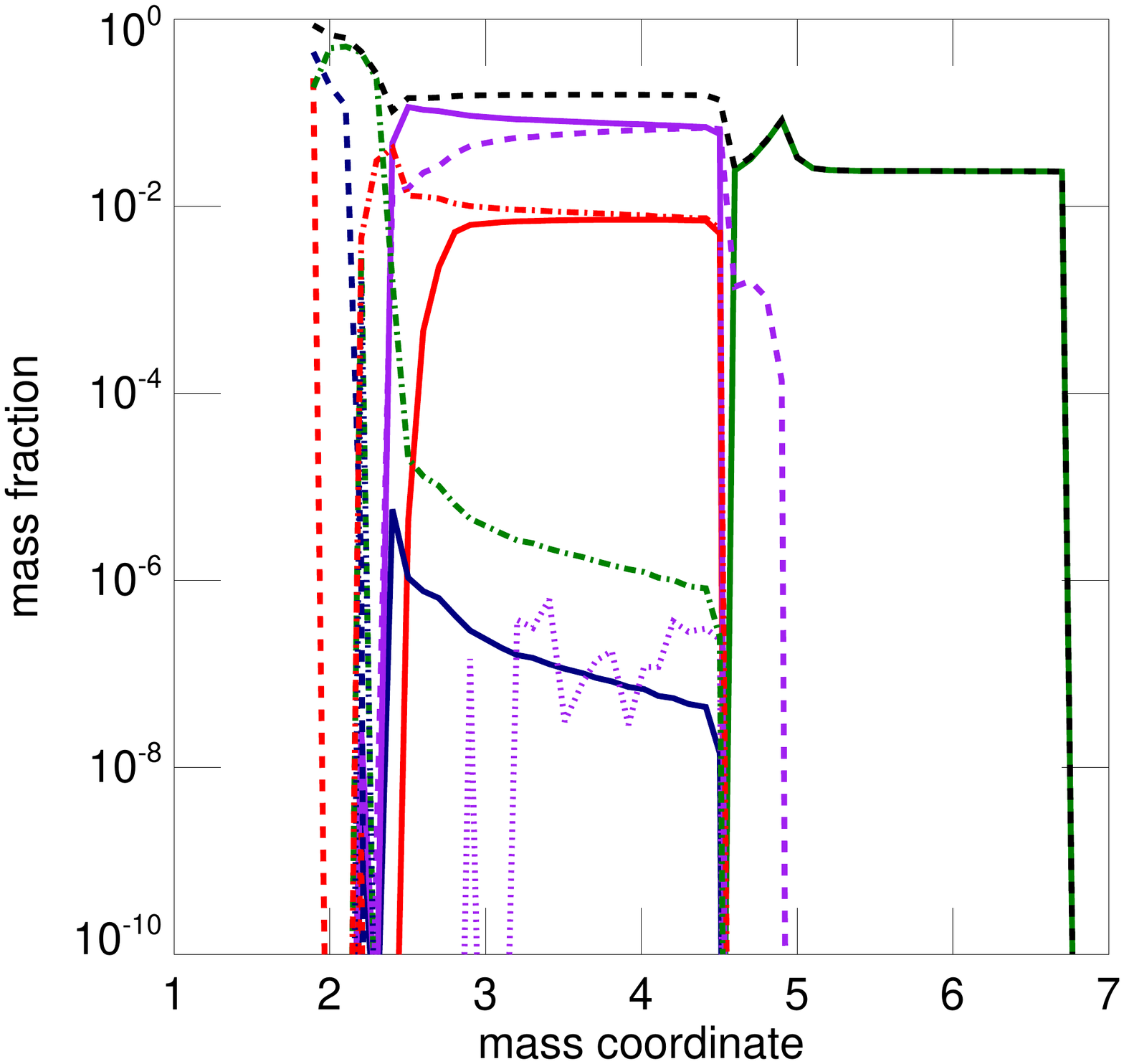}
\includegraphics[trim=0cm 4cm 0cm 5cm,clip=true,width=0.33\textwidth]{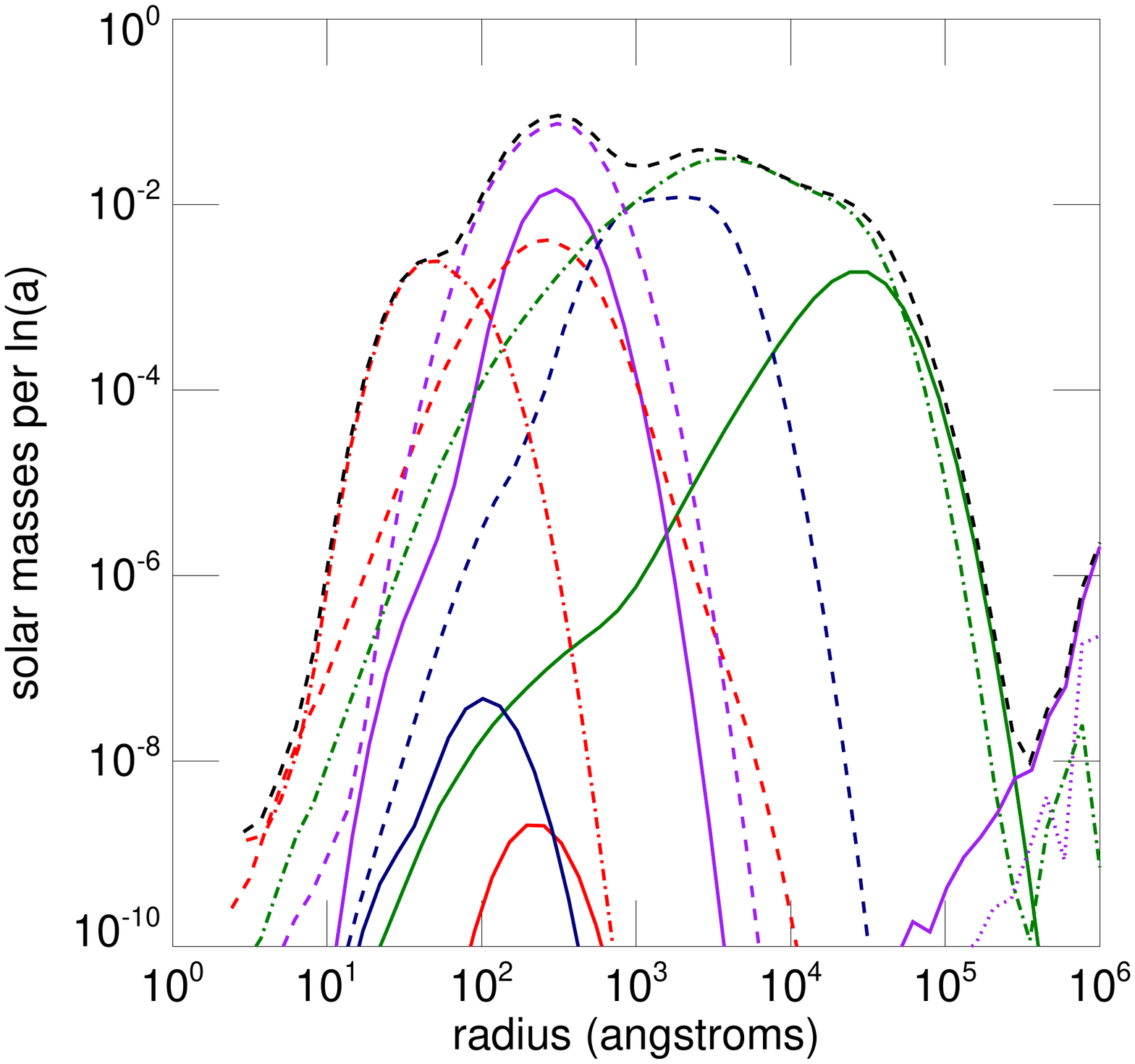}
\includegraphics[trim=0cm 4cm 0cm 5cm,clip=true,width=0.33\textwidth]{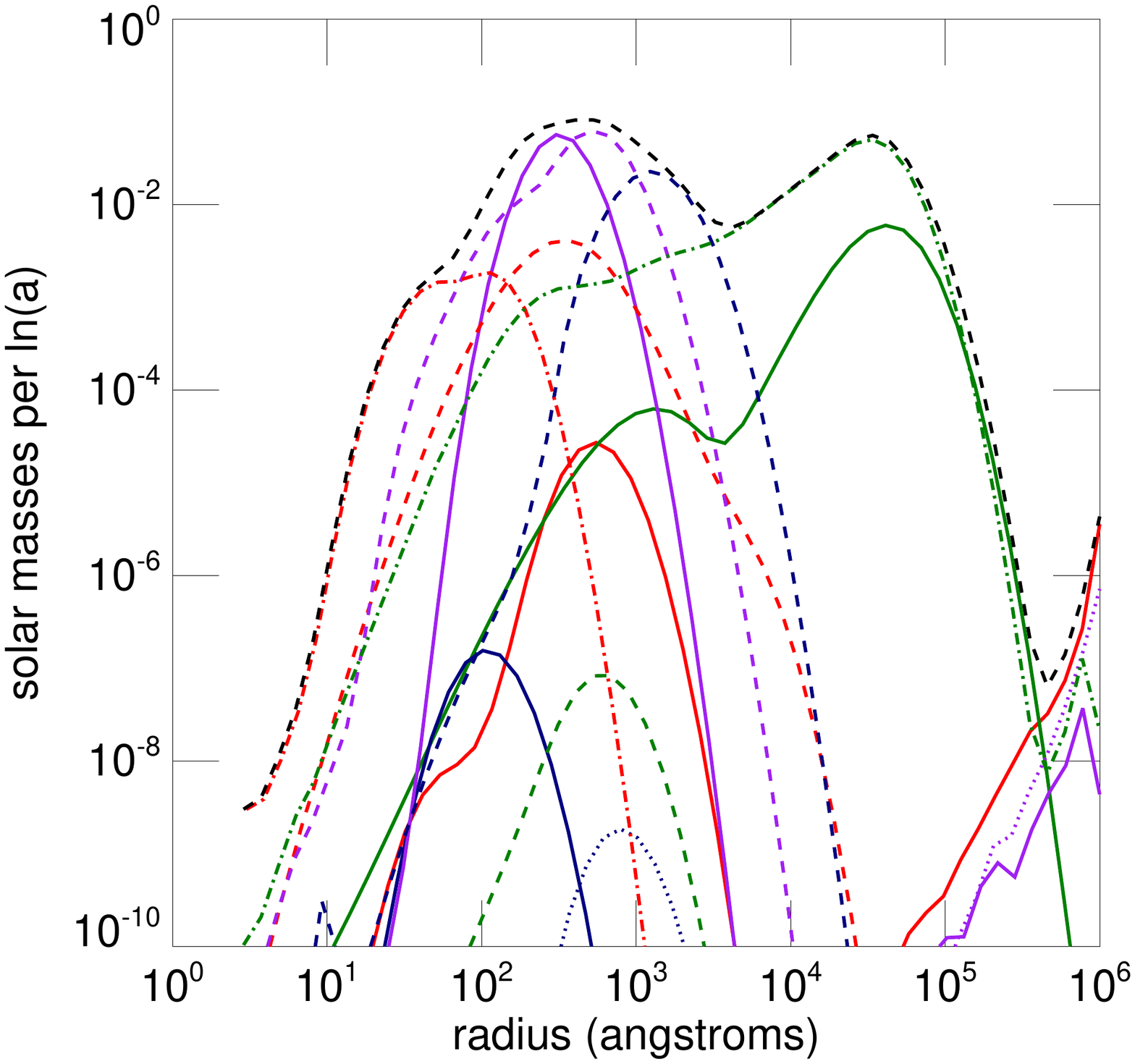}
\includegraphics[trim=0cm 4cm 0cm 5cm,clip=true,width=0.33\textwidth]{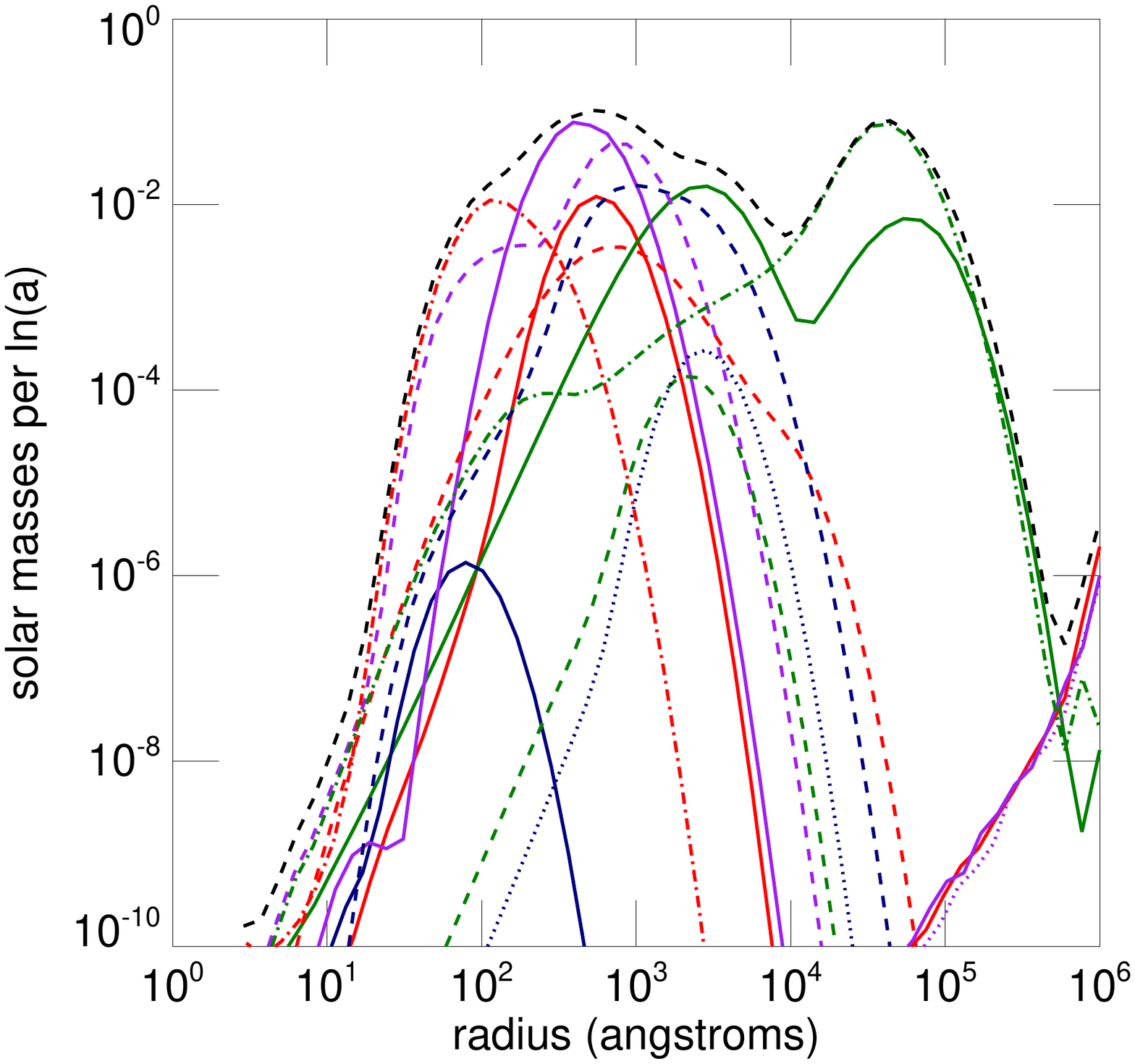}
\end{center}
\caption{
Grain properties in shells of different density. From left to right the shell mass density is 1, 2, and 4 times the ambient ejecta density. From top to bottom the grain properties shown are grain mass versus time, grain mass fraction versus mass coordinate, and (logarithmic) grain mass distribution. The grain species legend is the same as in Figure \ref{fig:LogGrainMassDistribution}.}
\label{fig:GrainMassVsTimeDifferentShells}
\end{figure*}

Recall that we varied the shell density between 1, 2, and 4 times the mass density in the ambient ejecta. The corresponding grain masses as a function of time, grain mass fractions as a function of the enclosed mass, and grain size distributions at the end of the simulation are shown in Figure~\ref{fig:GrainMassVsTimeDifferentShells}.  Grains form slightly earlier at higher shell densities and the total dust mass is larger. The dust mass created in the shells can be approximated with the power law:
\beq
M_{\rm dust,shell,1987A} = 0.22\,M_\odot \times \({\f{\rho_0}{4\times 10^{-13}\u{g}\u{cm}^{-3}}}^{0.26} ,
\label{eq:dust_shell_exponent}
\eeq
where $\rho_0$ is the shell density at 100 days. The normalization factor in Equation (\ref{eq:dust_shell_exponent}) is of course specific to our model of SN 1987A.
The grain species mass fractions at a given mass coordinate are not very sensitive to the mass density in the shell, except for those of alumina and carbon grains.  In the case of alumina, its abundance  in the region $2.5\,M_\odot < M < 4.5\,M_\odot$ sharply increases with density.  Similarly, the abundance of carbon grains increases in the helium shell $M > 5\, M_\odot$.

\section{Discussion}
\label{sec:discussion}

\subsection{Comparison with previous work in MNT}

Recall that there are three  frameworks for modeling dust formation in supernovae, in the increasing order of computational complexity: classical nucleation theory (CNT), kinetic nucleation theory (KNT), and molecular nucleation theory (MNT).   Our work is an extension of MNT, which was developed in \citet{Cherchneff:08}, \citet{Cherchneff:09,Cherchneff:10},  \citet{Sarangi:13,Sarangi:15}, and \citet{2014A&A...564A..25B,Biscaro:16}.

Our model is similar to that of \citet{Sarangi:15} (hereafter SC15). These authors used MNT to compute the grain mass versus time and grain size distribution for 8 grain species (forsterite, alumina, carbon, magnesium, silicon carbide, silicon, iron, and iron sulfide). They considered four explosion scenarios involving two progenitor masses, $15\,M_\odot$ and $19\,M_\odot$, and ran the simulations from 100 to 2000 days after the explosion. For the $15\,M_\odot$ progenitor they considered a case with normal \iso{Ni}{56} production and a case with a \iso{Ni}{56} mass of only $0.01\,M_\odot$. For the $19\,M_\odot$ progenitor they considered smooth as well as clumpy ejecta. In all cases they divided the ejecta into 6 or 7 annular shells, each of which was characterized by a density, temperature, elemental composition, and number and size of clumps. 

Before comparing our simulation results to SC15, we should note some model differences other than physical process prescriptions; we discuss those at the end of this subsection. The mass densities in their zones were approximately 1.6 times our shell density in their $15\,M_\odot$ progenitor models and 0.62 times our shell density in their $19\,M_\odot$ progenitor models. In the clumps in their $19\,M_\odot$ progenitor models the density varied from 8 to 162 times our shell density. Overall, their densities are higher than ours by as much as two orders of magnitude. The temperatures in SC15 are higher than ours initially, at $\sim100$ days, but as they follow power-laws, they drop below our temperatures after few hundred days (recall that we explicitly model, but the power-laws ignore, the late heating by $^{57}$Co and ultimately $^{44}$Ti radioactivity that slows the early cooling trend). The elemental compositions in SC15 are averaged over coarse annual shells, while ours are taken from discrete mass coordinates each having a different composition as provided by the \textsc{mesa} calculation; therefore, we do not assume that the explosion hydrodynamics can drive microscopic mixing.
  Finally, we include six additional grain species: enstatite, magnetite, silicon dioxide, magnesium sulfide, magnesia, and iron oxide.

For concreteness, we focus comparison on SC15's $19\,M_\odot$ model with clumpy ejecta that is the closest to our simulation other than for having much higher ejecta density. At 2000 days after the explosion, the dust mass in our model was $0.44\,M_\odot$ while it was $0.14\,M_\odot$ in theirs. Our higher mass can be attributed to grain growth by accretion; SC15 included only growth by coagulation. Our overall grain mass evolution for forsterite, alumina, iron sulfide, and silicon is similar to that in the SC15 model, although in each case these grain species form later in our simulations owing to lower densities. Carbon grains form much earlier in our simulations than in the SC15 model, while magnesium grains are completely absent in our simulation but are present in theirs. The lack of magnesium grains in our simulation is due to quick elemental magnesium depletion into magnesia grains, a species not included in SC15. 
The peak radii of the grain mass distributions agree to within a factor of a few for forsterite, carbon, alumina, iron, and iron sulfide. Our silicon mass distribution peaks at significantly larger radii than in the SC15 model. 

We now return to the physical process improvements over SC15. We include the effects of evaporation (sublimation) on grain growth and of finite grain size on the evaporation rate. We include the weathering of grains by noble gas ions and oxygen atoms (the latter in the case of carbon grains), whereas SC15 included only the corresponding processes for molecules. We compute the grain temperature to which the evaporation rate is sensitive separately from the gas temperature.  In fact, it seems that most previous investigations simply set the dust temperature equal to the gas temperature (SC15 did not need the grain temperature as they did not treat evaporation). We compute the grain charge and its effects on the coagulation rate. We treat grain growth by accretion and coagulation, whereas SC15 treated only the latter process. We extend our simulation to $10^4$ days and include the effects of radioactive decay of \iso{Co}{56,57}, \iso{Ti}{44}, and \iso{Na}{22} on the ejecta temperature evolution and chemistry and on grain temperature and charge; SC15 simulated to 2,000 days and only treated \iso{Co}{56} decay. 

\subsection{Observations} 
\label{sec:observations}

 As dust begins to form, it reprocesses the optical and UV into the IR. The thermal emission from dust is initially in the near-IR (for $T_{\rm dust} > 500 K$) but gradually shifts to the mid-IR (for $100\u{K} < T_{\rm dust} < 500\u{K}$) and ultimately the far-IR (for $T_{\rm dust} < 100 K$). Much of the thermal radiation emitted by grains cannot be observed from the earth's surface due to atmospheric absorption. At the time that SN 1987A was observed, there would not be infrared space telescopes for another 8 years. However, measurements of the infrared SED were made from the Kuiper Airborne Observatory (KAO) at 60, 250, 415, 615, and 775 days after the explosion \citep{1993ApJS...88..477W}. Note the 200 day gap in the observations between 415 and 615 days.  This is the period when most of the dust mass seems to have formed, as the percentage of the bolometric luminosity contributed by the IR jumps from 2\% at 415 days to 45\% at 615 days and then to 83\% at 775 days.  The much later mid-IR observations with the Spitzer Space Telescope (after 5,800 days), the far-IR observations with the Herschel Space Observatory (after 8,000 days), and sub-mm observations with ALMA (after 9,000 days) are now providing a much more complete picture of dust properties in the remnant.

Attempts have been made to use the observations from KAO, Spitzer, Herschel, and ALMA (and optical/UV observations from telescopes such as the Hubble Space Telescope) to determine the time evolution of grain properties in the ejecta of SN 1987A. This is done by fitting a dust reprocessing model to the SED. 
The ejecta is assumed to be divided into an inner heavy element core where grains form in high density clumps, a helium shell, and a hydrogen envelope. The grain mass, composition, size distribution, spatial distribution within clumps, and as well as the clump spatial distribution are varied until the resulting SED matches the observations. We discuss the three major attempts to do this in SN 1987A: \citet{Wesson:15}, \citet{2016MNRAS.456.1269B}, and \citet{Dwek:15}.

 \citet{Wesson:15} used three dimensional radiative transfer simulations with \textsc{mocassin} \citep{Ercolano:05} to calculate the SED of SN 1987A over the wavelength range $0.3\u{$\mu$m}< \lambda < 100\u{$\mu$m}$ at 615, 775, 1153, 1300, 8515, and 9200 days after the explosion.  They found that in order to match the observed SEDs, the grain mass should increase from $0.001\,M_\odot$ at 615 days to $0.8\,M_\odot$ at 9200 days and that the grain size distribution has power-law slope $dN/da\propto a^{-3.5}$ with grain sizes between $0.005\u{$\mu$m}$ and $0.25\u{$\mu$m}$ in the period between 615 and 1300 days and between $3.005\u{$\mu$m}$ and $3.25\u{$\mu$m}$ in the period between 8515 and 9200 days. The grain mass increases from $0.02\,M_\odot$ at 1300 days to $0.6\,M_\odot$ at 8515 days. This means that 72.5\% of the final dust mass formed between these two times and that 25\% of the dust mass formed between 8515 and 9090 days (so that 97.5\% of the dust mass formed after 1300 days). 
 
 \citet{2016MNRAS.456.1269B} performed three dimensional Monte Carlo radiative transfer to model the emission line blueshifting in the presence of dust. Their results fill in some of the gaps between days 1300 and 8515 in \citet{Wesson:15}. The results of \citet{Wesson:15} and \citet{2016MNRAS.456.1269B} can be combined to arrive at the conclusion---one that is controversial in view of the remaining of the three observational attempts and the theoretical predictions of SC15---that dust mass increases as a power law in time: $M_{\rm dust,tot,1987A}\propto t^{2.5}$.  These results, which assume spherically-symmetric ejecta, are questioned by \citet{Larsson:16}, who used spatially resolved spectra at optical and near IR wavelengths to determine the 3D distribution of H, He, O, Mg, Si, Ca, and Fe at $10^4$ days. The distribution of these elements was sufficiently anisotropic to explain on its own the spectral line distortion that \citet{2016MNRAS.456.1269B} attribute to dust.

 \citet{Dwek:15} used observations at 615, 775, 1144, 8815, and 9090 days and modeled the dust as amorphous carbon and enstatite.   At 615, 775, and 1144 days, the clumps are optically thick to the radiation emitted by the grains and the ejecta \emph{already} contains $0.4\, M_\odot$ of enstatite and $0.047M_\odot$ of carbon. By 8815 days the clumps become optically thin and the carbon and enstatite have coagulated to make composite grains that are essentially an enstatite matrix with carbon inclusions (18\% of the volume of the grains is occupied by the carbon inclusions). The total dust mass is $0.42\,M_\odot$ at 8815 days and $0.45\,M_\odot$ at 9090 days.\footnote{\citet{Matsuura:11} and \citet{Matsuura:15} estimated the dust mass at $\sim9000$ days with various assumptions about composition and size and inferred a mass of $\sim 0.5\,M_\odot$, in general agreement with the other observational estimates.}
 
While the estimates of \citet{2016MNRAS.456.1269B} and \citet{Wesson:15} agree, they differ drastically with those of \citet{Dwek:15}. The most important difference is that dust mass continuously increases for all times after 615 days in \citet{2016MNRAS.456.1269B} and \citet{Wesson:15} while almost all of the dust has already formed by 615 days in \citet{Dwek:15}. It is important to note that the discrepancy in inferred dust mass between the groups is highest at 615 days when the estimates differ by a factor of 447 and decreases with time, coming into near agreement at 9090 days. Another discrepancy is in the inferred composition: \citet{Wesson:15} find that 15\% of the dust mass is in silicates and 85\% is in carbon while \citet{Dwek:15} find that 89.5\% of the dust mass is in enstatite (a silicate) and 10.5\% in carbon.

The disagreement in the first 1000 days may be due to the very different approaches to radiative transfer by \citet{Wesson:15} and \citet{Dwek:15}. \citet{Wesson:15} used a fully 3D radiative transfer calculation to model the SED while \citet{Dwek:15} used a simple formula for the probability that a photon will escape from a dusty sphere. The two techniques give dramatically different values for what the flux at $100\u{$\mu$m}$ should be for a given amount of dust mass (the observed flux at 626 days at $100\u{$\mu$m}$ is $0.6\u{Jy}$). According to \citet{Wesson:15}, the dust mass at 615 days cannot exceed $0.01\,M_\odot$ since otherwise the flux at $100\u{$\mu$m}$ would exceed $0.6\u{Jy}$. On the other hand, according to \citet{Dwek:15} the observed flux at $100\u{$\mu$m}$ is exactly what it should be if there were $0.447\,M_\odot$ of dust at 615 days. At late times (9090 days) the ejecta is optically thin to dust thermal emission and so the dust mass is (relatively) straightforward to determine from far-IR Herschel and sub-mm ALMA observations. 

At the time of writing the discrepancy in dust mass inferred at early times by different groups is unresolved, and it seems that more realistic radiative transfer modeling is required to settle the matter of the early dust mass in SN 1987A.
Most authors assume that the grains are a mixture of carbon, iron, and enstatite (or other silicates), but our calculations suggest that at 615 days most of the dust mass is in magnesia, forsterite, silicon, and iron sulfide. Most groups also assume that the size distribution scales as $dN/da\propto a^{-3.5}$, but we find that each grain species has its own size distribution and that the total grain size is steeper $dN/da\propto a^{-4.39}$. Most groups assume that all grains are in dense clumps (what we would call dense shells), but we find that at 615 days the unclumped ambient ejecta contains 27\% of the dust mass. Radiative transfer should further take into account the very low density cavities of the fossil nickel bubbles and consider the possibility that dust rich clumps occupy much less volume than currently assumed.

We finally note that magnesia has not been detected in SN 1987A, nor has it been detected in the observations of other supernovae.  We speculate that the high abundance of MgO in our simulations is an artifact of not allowing MgO monomers and clusters to merge with other grain species.  As is evident in Figure \ref{fig:GrainMassVsRadius}, where MgO forms in significant amounts, forsterite is also plentiful.  It could be that in reality, the MgO monomers and clusters that form are aways absorbed into forsterite grains and do not exist as MgO monomers and grains.  If this hypothesis is true, it explains the absence of detection of MgO in existing observations and also diminishes the promise of detection of MgO in future observations.

\subsection{Future directions}

Many aspects of our model of dust formation in supernova could be improved and made more realistic.  We briefly mention a few.  The degradation of the energy released by decaying radioactive nuclei, starting with $\gamma$-rays, should be followed with a Monte Carlo simulation \citep[see, e.g.,][]{Hugerford:04,2011A&A...530A..45J}. Such a simulation would provide the local, time-dependent gas temperature and ionization fraction in the ejecta as a function of time and an accurate model for  the UV radiation field SED.  These simulations need to take into account the asymmetric distribution of radioactive material, such as from the three-dimensional, neutrino-driven core-collapse simulation of \citet{Wongwathanarat:16} modeling Cas A. In post-processing they computed the local $^{44}$Ti and $^{56}$Ni yield in the ejecta finding that the $^{44}$Ti distribution was very asymmetrical, with the bulk of isotope ejected into the hemisphere opposite to the neutron star kick.  

Grain temperature can fluctuate stochastically, which, owing to the exponential dependence of the evaporation rate on temperature, has been found to influence grain growth in AGB stellar outflows \citep{Kochanek:14}.  Compton-electron-induced destruction reactions should be included for all atoms, molecules, and dust grains.  Ionization and molecular dissociation by UV photons should also be included. The destructive reactions with He$^+$, Ne$^+$, and Ar$^+$ should be included for all neutral molecules. Grain-species-specific absorption coefficients and photoelectric yields should be used in the place of our carbon grain-based model.  

Perhaps more consequentially, we do not expect that the dust grains have spherical shapes \citep{Fallest:11}. They could have porous and even fractal-like structures. The shape complexity could imply some very different thermodynamic, chemical, and optical behavior \citep[e.g.,][]{Keith:11}.  While a fully realistic treatment of grain shapes will remain beyond reach, the salient effects of shape complexity should be assessed and incorporated in dust synthesis models.  We also note that many of the chemical reaction rates used in this and preceding investigations of astrophysical dust synthesis are unverified extrapolations of sparsely-cataloged laboratory measurements.  With the rapid development of {\it ab initio} electronic structure methods, it is now becoming possible to complement laboratory data with theoretical calculations.  For example, \citet{Mauney:15} recently reported density function theory (DFT) based calculations of the work of small carbon cluster formation and the nucleation rate in a saturated, hydrogen-poor carbon gas.  The new quantum electrodynamical time-dependent DFT \citep{Flick:15} should enable direct computation of collisional cross sections involving photon emission, such as for radiative association in the formation small grain-precursor atomic clusters.

The dust produced in our calculation must pass a reverse shock before joining the ISM.  Sputtering in the reverse shock destroys small grains and reduces the large ones.  Therefore, dust survival hinges on the grain size distribution \citep[e.g.,][]{Bianchi:09}.  This effect can be studied in Cassiopeia A where the reverse shock has already crossed a fraction of the dust-forming ejecta \citep{Nozawa:10}. The characteristic minimum size for a grain to survive the reverse shock may be $\sim 0.1\,\mu\text{m}$ \citep{Silvia:10,Silvia:12}.  Our calculation produces an abundance of such large grains.   Since the reverse shock diminishes in clumps, our hypothesized ejecta clumpiness, in the form of thin shells compressed by overpressured nickel bubbles, serves to protect the grains in clumps from the reverse shock \citep{Biscaro:16}.

\section{Conclusions}
\label{sec:conclusions}

We simulated the formation molecules and dust grains in the ejecta of SN 1987A using an improved molecular nucleation theory model. The model assumes that \iso{Ni}{56}-rich clumps are injected into the helium core where they expand and sweep up dense shells of ambient ejecta. We compute the abundance of molecules as a function of time using a nonequilibrium chemical reaction network including the effects of radioactive decay of \iso{Ni}{56}, \iso{Ni}{57}, \iso{Ti}{44}, and \iso{Na}{22}. Grain formation starts with the formation of condensation nuclei which are treated as molecular species in the chemical network.  
The nuclei grow into grains via accretion and coagulation. Grain charge and van der Waals interaction are explicitly calculated to correct the coagulation rate. Grain destruction by oxygen and noble gas weathering is included, as is evaporation. To get the evaporation rate, we explicitly compute the grain temperature as a function of radius and account for the finite grain size.    

The model produces a total dust mass of $0.51\,M_\odot$, which is 16\% of the mass of refractory elements. Grain formation is rapid between 200 and 600 days and slower thereafter. At 615 days, our computed total grain mass agrees with the observational estimate of \citet{Dwek:15} but not with those of \citet{Wesson:15} and \citet{2016MNRAS.456.1269B}. We find that the dust mass produced in the dense shells scales as the 0.26 power of the mass density in the shells. The final computed dust mass is close to, and bracketed by, the observationally inferred dust masses in SN 1987A. The mass distribution $dM_{\rm dust}/d\ln a$ peaks at the grain radius $a = 680\u{\AA}$. Beyond this peak, the size distribution scales as $dN/d\ln a \propto a^{-3.4}$, and the overall distribution is skewed toward larger grains. The most common grain species by mass are magnesia (32\%), silicon (29\%), forsterite (18\%), iron sulfide (7.8\%), and carbon (6.1\%). The dense shells produced more dust mass and a greater variety of grain species than the ambient ejecta whereas grain formation in the nickel bubbles was negligible.

Finally, we note the numerical complexity of dust synthesis calculations. To validate calculations such as ours, it is necessary to run different codes on the same or similar models and compare the results.  We are encouraged that dust composition and the timing of dust growth closely resemble those in \citet{Sarangi:15}.

\section*{Acknowledgments}

We acknowledge invaluable consultation with, comments from, and support by D.\ Lazzati,  J.\ Scalo, and J.\ C.\ Wheeler.  We also acknowledge continuous encouragements from V.\ Bromm, helpful comments from N.\ Evans and W.\ Lu, discussions with I.\ Cherchneff, C.\ Gall, R.\ McCray, and R.\ Schneider, and essential technical assistance by J.\ Ritter and B.\ Tsang.
The authors acknowledge the Texas Advanced Computing Center at The University of
Texas at Austin for providing HPC resources. This study was supported
by the NSF grant AST-1413501.
 
\footnotesize{

}

\section{Appendix}

In this Appendix we provide our isotope to element conversion table (Table \ref{tab:isoTOelement}) and tables of chemical reactions included in our model as well as some additional reactions (the remaining tables).  The additional reactions not included in our chemical network are those involving nitrogen, charged molecules with three or more atoms, and the molecules CS, CSO, C$_2$O, CO$_2$, SO$_2$, S$_2$, SiS, and their charged counterparts.  We provide these reactions for the convenience of the reader who wishes to use a more comprehensive reaction network.

The reaction source references indicated with acronyms are the University of Manchester Institute of Science and Technology (UMIST) Database for Astrochemistry \\ (U12; \citealt{2013A&A...550A..36M}; http://udfa.ajmarkwick.net/) and National Institute of Standards and Technology (NIST) Chemical Kinetics Database \\ (http://kinetics.nist.gov/kinetics/KineticsSearchForm.jsp).

 In the reaction tables the last column is the reference that describes how we obtained the rate coefficient. If a paper or database is cited in the column,  we either use the rate coefficient cited in the source or estimate it from information in the source. If a reaction has a reference with the statement ``$=$ X", then its rate coefficient could not be found in the literature and is duplicated from a similar reaction X. Reactions with reference ``O" use the rate coefficient for $\mbox{C}_n + \mbox{O} \to \mbox{C}_{n-1} + \mbox{CO}$ from \citet{Lazzati:16} while reactions with reference ``He$^+$" use the rate coefficient for $\mbox{C}_n + \mbox{He}^+ \to \mbox{C}_{n-1} + \mbox{C}^+ + \mbox{He}$, from the same reference. Reactions with reference ``Est" used the formulas in Section \ref{sec:estimatingRates}.
 All three-body reactions use the same rate coefficient derived in that section, even if they have rate coefficients given in the literature, because the published rates are only valid at high temperatures.

\begin{table}
\begin{tabular}{ ll }
\hline
Element & Isotopes \h
H  & \iso{H}{1}, \iso{H}{2} \\
He  & \iso{He}{3}, \iso{He}{4}, \iso{B}{8}  \\
Li  & \iso{Li}{6}, \iso{Li}{7}, \iso{Be}{7}  \\
Be  & \iso{Be}{9}, \iso{Be}{10}  \\
B  & \iso{B}{10}, \iso{B}{11}  \\
C  & \iso{C}{12}, \iso{C}{13}, \iso{N}{13}  \\
N  & \iso{N}{14}, \iso{N}{15}, \iso{O}{15}  \\
O  & \iso{O}{16}, \iso{O}{17}, \iso{O}{18}, \iso{N}{16}, \iso{F}{17}, \iso{F}{18}  \\
\hline
\end{tabular}
\caption{Isotope to element conversion.}
\label{tab:isoTOelement}
\end{table}

\clearpage

\begin{table*}
\begin{center}
\begin{tabular}{ llllll }
\hline
Code & Reaction & $A\ (\text{cgs})$ & $
\nu$ & $T_{\rm A}\ (\text{K})$ & Ref. \\ \hline
A1 & $\rm  C + O \to CO + \gamma$ & $4.69 \times 10^{-19} $ & $1.52$ & $-50.5$ & U12 \\
A2 & $\rm  C + C \to C_{2} + \gamma$ & $4.36 \times 10^{-18} $ & $0.35$ & $161.3$ & U12 \\
A4 & $\rm  C^+ + C_{3} \to C_{4}^+ + \gamma$ & $1 \times 10^{-13} $ & $-1$ & $0$ & U12 \\
A11 & $\rm  C^+ + C \to C_{2}^+ + \gamma$ & $4.01 \times 10^{-18} $ & $0.17$ & $101.5$ & U12 \\
A12 & $\rm  C^+ + N \to CN^+ + \gamma$ & $1.08 \times 10^{-18} $ & $0.07$ & $57.5$ & U12 \\
A13 & $\rm  C^+ + O \to CO^+ + \gamma$ & $3.14 \times 10^{-18} $ & $-0.15$ & $68$ & U12 \\
A14 & $\rm  C^+ + S \to CS^+ + \gamma$ & $3.07 \times 10^{-19} $ & $0.15$ & $0$ & U12 \\
\hline
\end{tabular}
\end{center}
\caption{Reaction Set A (Radiative Association Reactions). Note that AM05 refers to \citet{2005ApJ...624.1121A}.}
\label{tab:bigReactionTableFirst}
\end{table*}
    
\begin{table*}
\begin{center}
\begin{tabular}{ llllll }
\hline
Code & Reaction & $A\ (\text{cgs})$ & $
\nu$ & $T_{\rm A}\ (\text{K})$ & Ref. \\ \hline
B1 & $\rm  C + C + M \to C_{2} + M$ & $4.25 \times 10^{-34} $ & 1 & 0 & Derived in Section~\ref{sec:estimatingRates} \\
B2 & $\rm  C + N + M \to CN + M$ & $4.25 \times 10^{-34} $ & 1 & 0 & Derived in Section~\ref{sec:estimatingRates}  \\
B3 & $\rm  C + O + M \to CO + M$ & $4.25 \times 10^{-34} $ & 1 & 0 & Derived in Section~\ref{sec:estimatingRates}  \\
B4 & $\rm  C^+ + O + M \to CO^+ + M$ & $4.25 \times 10^{-34} $ & 1 & 0 & Derived in Section~\ref{sec:estimatingRates}  \\
B5 & $\rm  C + O^+ + M \to CO^+ + M$ & $4.25 \times 10^{-34} $ & 1 & 0& Derived in Section~\ref{sec:estimatingRates}  \\
B6 & $\rm  C + CO + M \to C_{2}O + M$ & $4.25 \times 10^{-34} $ & 1 & 0& Derived in Section~\ref{sec:estimatingRates}  \\
B7 & $\rm  O + O + M \to O_{2} + M$ & $4.25 \times 10^{-34} $ & 1 & 0 & Derived in Section~\ref{sec:estimatingRates}  \\
\hline
\end{tabular}
\end{center}
\caption{Reaction Set B (Three-Body Reactions).}
\end{table*}

\begin{table*}
\begin{center}
\begin{tabular}{ llllll }
\hline
Code & Reaction & $A\ (\text{cgs})$ & $
\nu$ & $T_{\rm A}\ (\text{K})$ & Ref. \\ \hline
C1 & $\rm  CO + M \to O + C + M$ & $2.86 \times 10^{-3} $ & $-3.52$ & $112700$ & U12 \\
C2 & $\rm  C_{2} + M \to C + C + M$ & $2.49 \times 10^{-8} $ & $0$ & $71561.6$ & NIST \\
C3 & $\rm  O_{2} + M \to O + O + M$ & $1.01 \times 10^{-8} $ & $-1$ & $59414.2$ & NIST \\
C4 & $\rm  SO + M \to S + O + M$ & $6.61 \times 10^{-10} $ & $0$ & $53881.7$ & NIST \\
C5 & $\rm  SiO + M \to Si + O + M$ & $2.86 \times 10^{-3} $ & $-3.52$ & $112700$ & $=$ C1 \\
C6 & $\rm  NO + M \to N + O + M$ & $4 \times 10^{-9} $ & $0$ & $74568.4$ & NIST \\
C7 & $\rm  CO_{2} + M \to CO + O + M$ & $4.17 \times 10^{-11} $ & $0$ & $43778.9$ & NIST \\
C8 & $\rm  CS + M \to C + S + M$ & $2.86 \times 10^{-3} $ & $-3.52$ & $112700$ & $=$ C1 \\
C9 & $\rm  CN + M \to C + N + M$ & $3.32 \times 10^{-10} $ & $0$ & $75049.5$ & NIST \\
C10 & $\rm  SiS + M \to Si + S + M$ & $2.86 \times 10^{-3} $ & $-3.52$ & $112700$ & $=$ C1 \\
C11 & $\rm  SiN + M \to Si + N + M$ & $3.32 \times 10^{-10} $ & $0$ & $75049.5$ & $=$ C9 \\
C12 & $\rm  S_{2} + M \to S + S + M$ & $7.95 \times 10^{-11} $ & $0$ & $38727.5$ & NIST \\
C13 & $\rm  N_{2} + M \to N + N + M$ & $2.52 \times 10^{-7} $ & $-1.6$ & $57008.8$ & NIST \\
\hline
\end{tabular}
\end{center}
\caption{Reaction Set C (Thermal Fragmentation Reactions).}
\end{table*}
     
\begin{table*}
\begin{center}
\begin{tabular}{ llllll }
\hline
Code & Reaction & $A\ (\text{cgs})$ & $
\nu$ & $T_{\rm A}\ (\text{K})$ & Ref. \\ \hline
D1 & $\rm  C + CO \to C_{2} + O$ & $2.94 \times 10^{-11} $ & $0.5$ & $58025$ & U12 \\
D2 & $\rm  O + C_{2} \to CO + C$ & $2 \times 10^{-10} $ & $-0.12$ & $0$ & U12 \\
D3 & $\rm  C_{2} + O_{2} \to CO + CO$ & $1.5 \times 10^{-11} $ & $0$ & $4300$ & U12 \\
D5 & $\rm  C_{2} + S \to CS + C$ & $1 \times 10^{-10} $ & $0$ & $0$ & U12 \\
D7 & $\rm  C + C_{2}O \to C_{2} + CO$ & $2 \times 10^{-10} $ & $0$ & $0$ & U12 \\
D10 & $\rm  C + C_{4} \to C_{2} + C_{3}$ & $9 \times 10^{-11} $ & $0$ & $0$ & U12 \\
D19 & $\rm  C + CN \to C_{2} + N$ & $4.98 \times 10^{-10} $ & $0$ & $18116$ & U12 \\
\hline
\end{tabular}
\end{center}
\caption{Reaction Set D (Neutral-Neutral Reactions).}
\end{table*}

\begin{table*}
\begin{tabular}{ llllll }
\hline
Code & Reaction & $A\ (\text{cgs})$ & $
\nu$ & $T_{\rm A}\ (\text{K})$ & Ref. \\ \hline
D61 & $\rm  N + CN \to N_{2} + C$ & $1 \times 10^{-10} $ & $0.18$ & $0$ & U12 \\
D62 & $\rm  N + CO_{2} \to NO + CO$ & $3.2 \times 10^{-13} $ & $0$ & $1710$ & U12 \\
D63 & $\rm  N + CS \to S + CN$ & $3.8 \times 10^{-11} $ & $0.5$ & $1160$ & U12 \\
D65 & $\rm  N + NO_{2} \to N_{2} + O + O$ & $2.41 \times 10^{-12} $ & $0$ & $0$ & U12 \\
D66 & $\rm  N + NO_{2} \to N_{2}O + O$ & $3 \times 10^{-12} $ & $0$ & $0$ & U12 \\
D67 & $\rm  N + NO_{2} \to NO + NO$ & $1 \times 10^{-12} $ & $0$ & $0$ & U12 \\
D68 & $\rm  N + NO_{2} \to O_{2} + N_{2}$ & $1 \times 10^{-12} $ & $0$ & $0$ & U12 \\
\end{tabular}
%
D85 & $\rm  O_{2} + CNO \to CO_{2} + NO$ & $1.32 \times 10^{-12} $ & $0$ & $0$ & U12 \\
D86 & $\rm  O_{2} + CNO \to NO_{2} + CO$ & $8.1 \times 10^{-11} $ & $0$ & $773$ & U12 \\
D87 & $\rm  O_{2} + S \to SO + O$ & $1.76 \times 10^{-12} $ & $0.81$ & $-30.8$ & U12 \\
D88 & $\rm  O_{2} + SO \to SO_{2} + O$ & $1.1 \times 10^{-14} $ & $1.89$ & $1538$ & U12 \\
D92 & $\rm  O + C_{2}O \to CO + CO$ & $8.59 \times 10^{-11} $ & $0$ & $0$ & U12 \\
D94 & $\rm  O + C_{3} \to CO + C_{2}$ & $5 \times 10^{-12} $ & $0$ & $900$ & U12 \\
D97 & $\rm  O + C_{4} \to C_{3} + CO$ & $1 \times 10^{-10} $ & $0$ & $0$ & U12 \\
D107 & $\rm  O + CN \to CO + N$ & $2.54 \times 10^{-11} $ & $0$ & $0$ & U12 \\
\hline
\end{tabular}
\caption{Reaction Set D (Neutral-Neutral Reactions, continued).}
\end{table*}

\begin{table*}
\begin{tabular}{ llllll }
\hline
Code & Reaction & $A\ (\text{cgs})$ & $
\nu$ & $T_{\rm A}\ (\text{K})$ & Ref. \\ \hline
D140 & $\rm  Si + NO \to SiO + N$ & $9 \times 10^{-11} $ & $-0.96$ & $28$ & U12 \\
D141 & $\rm  Si + O_{2} \to SiO + O$ & $1.72 \times 10^{-10} $ & $-0.53$ & $17$ & U12 \\
D142 & $\rm  O + CO \to O_{2} + C$ & $1 \times 10^{-16} $ & $0$ & $0$ & CD09 \\
D143 & $\rm  O + SiO \to O_{2} + Si$ & $1 \times 10^{-16} $ & $0$ & $0$ & CD09 \\
D144 & $\rm  C + SiO \to CO + Si$ & $1 \times 10^{-16} $ & $0$ & $0$ & CD09 \\
D145 & $\rm  Si + S_{2} \to SiS + S$ & $7 \times 10^{-11} $ & $0$ & $0$ & $=$ D30 \\
D146 & $\rm  C + CO_{2} \to CO + CO$ & $1 \times 10^{-15} $ & $0$ & $0$ & NIST \\
D147 & $\rm  N + CO \to CN + O$ & $3.84 \times 10^{-9} $ & $0$ & $35961.2$ & NIST \\
D148 & $\rm  N + CO \to NO + C$ & $3.84 \times 10^{-9} $ & $0$ & $35961.2$ & $=$ D147 \\
D149 & $\rm  S + CO \to SO + C$ & $1 \times 10^{-16} $ & $0$ & $0$ & CD09 \\
D150 & $\rm  S + CO \to CS + O$ & $1 \times 10^{-16} $ & $0$ & $0$ & CD09 \\
D151 & $\rm  S + CS \to S_{2} + C$ & $1.73 \times 10^{-11} $ & $0.5$ & $11500$ & $=$ D137 \\
D152 & $\rm  S + CN \to CS + N$ & $2.54 \times 10^{-11} $ & $0$ & $0$ & $=$ D107 \\
D153 & $\rm  S + SiS \to S_{2} + Si$ & $5.75 \times 10^{-11} $ & $0.1$ & $200$ & $=$ D133 \\
D154 & $\rm  N + SiO \to SiN + O$ & $3.84 \times 10^{-9} $ & $0$ & $35961.2$ & $=$ D148 \\
D155 & $\rm  N + SiO \to NO + Si$ & $1 \times 10^{-16} $ & $0$ & $0$ & CD09 \\
D156 & $\rm  N + SiS \to SiN + S$ & $3.8 \times 10^{-11} $ & $0.5$ & $1160$ & $=$ D63 \\
D157 & $\rm  O_{2} + N_{2} \to NO + NO$ & $1 \times 10^{-16} $ & $0$ & $0$ & CD09 \\
D158 & $\rm  CO + SiO \to CO_{2} + Si$ & $1 \times 10^{-16} $ & $0$ & $0$ &CD09 \\
D159 & $\rm  CO + NO \to CO_{2} + N$ & $1 \times 10^{-16} $ & $0$ & $0$ & CD09 \\
\hline
\end{tabular}
\caption{Reaction Set D (Neutral-Neutral Reactions, continued). Note that CD09 refers to \citet{Cherchneff:09}.}
\end{table*}
     
\begin{table*}
\begin{tabular}{ llllll }
\hline
Code & Reaction & $A\ (\text{cgs})$ & $
\nu$ & $T_{\rm A}\ (\text{K})$ & Ref. \\ \hline
E6 & $\rm  C^+ + CNO \to CO^+ + CN$ & $8.98 \times 10^{-9} $ & $-0.5$ & $0$ & U12 \\
E7 & $\rm  C^+ + CO_{2} \to CO^+ + CO$ & $1.1 \times 10^{-9} $ & $0$ & $0$ & U12 \\
E8 & $\rm  C^+ + N_{2}O \to NO^+ + CN$ & $9.1 \times 10^{-10} $ & $0$ & $0$ & U12 \\
E9 & $\rm  C^+ + NS \to CS^+ + N$ & $7.6 \times 10^{-10} $ & $-0.5$ & $0$ & U12 \\
E10 & $\rm  C^+ + O_{2} \to CO^+ + O$ & $3.42 \times 10^{-10} $ & $0$ & $0$ & U12 \\
E11 & $\rm  C^+ + O_{2} \to CO + O^+$ & $4.54 \times 10^{-10} $ & $0$ & $0$ & U12 \\
E14 & $\rm  C^+ + SO_{2} \to SO^+ + CO$ & $2.3 \times 10^{-9} $ & $-0.5$ & $0$ & U12 \\
\hline
\end{tabular}
\caption{Reaction Set E (Ion-Molecule Reactions).}
\end{table*}

\begin{table*}
\begin{tabular}{ llllll }
\hline
Code & Reaction & $A\ (\text{cgs})$ & $
\nu$ & $T_{\rm A}\ (\text{K})$ & Ref. \\ \hline
E38 & $\rm  N^+ + CO_{2} \to NO + CO^+$ & $2.5 \times 10^{-10} $ & $0$ & $0$ & U12 \\
E39 & $\rm  N^+ + CO \to NO^+ + C$ & $1.45 \times 10^{-10} $ & $0$ & $0$ & U12 \\
E42 & $\rm  N^+ + NO \to N_{2}^+ + O$ & $7.9 \times 10^{-11} $ & $0$ & $0$ & U12 \\
E43 & $\rm  N^+ + O_{2} \to NO^+ + O$ & $2.63 \times 10^{-10} $ & $0$ & $0$ & U12 \\
E44 & $\rm  N^+ + O_{2} \to NO + O^+$ & $3.66 \times 10^{-11} $ & $0$ & $0$ & U12 \\
E48 & $\rm  N + C_{2}^+ \to CN + C^+$ & $4 \times 10^{-11} $ & $0$ & $0$ & U12 \\
E49 & $\rm  N + CN^+ \to N_{2}^+ + C$ & $6.1 \times 10^{-10} $ & $0$ & $0$ & U12 \\
E50 & $\rm  N + O_{2}^+ \to NO^+ + O$ & $1.8 \times 10^{-10} $ & $0$ & $0$ & U12 \\
E51 & $\rm  N + SO^+ \to NS^+ + O$ & $5 \times 10^{-11} $ & $0$ & $0$ & U12 \\
E52 & $\rm  N + SiC^+ \to Si^+ + CN$ & $7.7 \times 10^{-10} $ & $0$ & $0$ & U12 \\
E53 & $\rm  N + SiO^+ \to NO^+ + Si$ & $9 \times 10^{-11} $ & $0$ & $0$ & U12 \\
E54 & $\rm  N + SiO^+ \to NO + Si^+$ & $2.1 \times 10^{-10} $ & $0$ & $0$ & U12 \\
E55 & $\rm  O^+ + C_{2} \to CO^+ + C$ & $4.8 \times 10^{-10} $ & $0$ & $0$ & U12 \\
E56 & $\rm  O^+ + CN \to NO^+ + C$ & $1 \times 10^{-9} $ & $-0.5$ & $0$ & U12 \\
E57 & $\rm  O^+ + CO_{2} \to O_{2}^+ + CO$ & $9.4 \times 10^{-10} $ & $0$ & $0$ & U12 \\
E58 & $\rm  O^+ + N_{2} \to NO^+ + N$ & $2.42 \times 10^{-12} $ & $-0.21$ & $-44$ & U12 \\
E59 & $\rm  O^+ + NO_{2} \to O_{2} + NO^+$ & $8.3 \times 10^{-10} $ & $0$ & $0$ & U12 \\
%
%
E61 & $\rm  O^+ + SO_{2} \to SO^+ + O_{2}$ & $3.6 \times 10^{-10} $ & $-0.5$ & $0$ & U12 \\
E62 & $\rm  O_{2}^+ + S \to SO^+ + O$ & $5.4 \times 10^{-10} $ & $0$ & $0$ & U12 \\
E63 & $\rm  O_{2} + C_{3}^+ \to C_{3}O^+ + O$ & $1.75 \times 10^{-10} $ & $0$ & $0$ & U12 \\
E64 & $\rm  O_{2} + CS^+ \to CSO^+ + O$ & $1.3 \times 10^{-10} $ & $0$ & $0$ & U12 \\
E65 & $\rm  O_{2} + S^+ \to SO^+ + O$ & $1.5 \times 10^{-11} $ & $0$ & $0$ & U12 \\
E66 & $\rm  O_{2} + SiS^+ \to SO^+ + SiO$ & $6.23 \times 10^{-11} $ & $0$ & $0$ & U12 \\
E67 & $\rm  O_{2} + SiS^+ \to SiO^+ + SO$ & $2.67 \times 10^{-11} $ & $0$ & $0$ & U12 \\
E68 & $\rm  O + C_{2}^+ \to CO^+ + C$ & $3.1 \times 10^{-10} $ & $0$ & $0$ & U12 \\
\hline
\end{tabular}
\caption{Reaction Set E (Ion-Molecule Reactions).  For all reactions involving He$^+$ we add an additional pair of reactions with He replaced by Ne and Ar. The rate coefficients for the reactions involving Ne and Ar are taken to be the same for He.}
\end{table*}

\begin{table*}
\begin{tabular}{ llllll }
\hline
Code & Reaction & $A\ (\text{cgs})$ & $
\nu$ & $T_{\rm A}\ (\text{K})$ & Ref. \\ \hline
E140 & $\rm  He^+ + O_{2} \to O^+ + O + He$ & $1.1 \times 10^{-9} $ & $0$ & $0$ & U12 \\
E142 & $\rm  He^+ + CNO \to CN + O^+ + He$ & $3 \times 10^{-9} $ & $0$ & $0$ & U12 \\
E147 & $\rm  He^+ + S_{2} \to S^+ + S + He$ & $2 \times 10^{-9} $ & $0$ & $0$ & U12 \\
E148 & $\rm  He^+ + SO_{2} \to S^+ + O_{2} + He$ & $9 \times 10^{-10} $ & $-0.5$ & $0$ & U12 \\
E149 & $\rm  He^+ + SO_{2} \to SO^+ + O + He$ & $2.97 \times 10^{-9} $ & $-0.5$ & $0$ & U12 \\
E150 & $\rm  He^+ + SO \to S^+ + O + He$ & $8.3 \times 10^{-10} $ & $-0.5$ & $0$ & U12 \\
E151 & $\rm  He^+ + SO \to S + O^+ + He$ & $8.3 \times 10^{-10} $ & $-0.5$ & $0$ & U12 \\
E155 & $\rm  He^+ + SiC \to Si^+ + C + He$ & $2 \times 10^{-9} $ & $-0.5$ & $0$ & U12 \\
E156 & $\rm  He^+ + SiC \to Si + C^+ + He$ & $2 \times 10^{-9} $ & $-0.5$ & $0$ & U12 \\
E157 & $\rm  He^+ + SiN \to Si^+ + N + He$ & $2 \times 10^{-9} $ & $-0.5$ & $0$ & U12 \\
E159 & $\rm  He^+ + SiO_{2} \to O_{2} + Si^+ + He$ & $2 \times 10^{-9} $ & $0$ & $0$ & U12 \\
E160 & $\rm  He^+ + SiO \to Si^+ + O + He$ & $8.6 \times 10^{-10} $ & $-0.5$ & $0$ & U12 \\
E161 & $\rm  He^+ + SiO \to Si + O^+ + He$ & $8.6 \times 10^{-10} $ & $-0.5$ & $0$ & U12 \\
E162 & $\rm  He^+ + SiS \to S^+ + Si + He$ & $3.8 \times 10^{-9} $ & $-0.5$ & $0$ & U12 \\
E163 & $\rm  He^+ + SiS \to S + Si^+ + He$ & $3.8 \times 10^{-9} $ & $-0.5$ & $0$ & U12 \\
\hline
\end{tabular}
\caption{Reaction Set E (Ion-Molecule Reactions, continued). For all reactions involving He$^+$ we add an additional pair of reactions with He replaced by Ne and Ar. The rate coefficients for the reactions involving Ne and Ar are taken to be the same for He.}
\end{table*}
   
\begin{table*}
\begin{tabular}{ llllll }
\hline
Code & Reaction & $A\ (\text{cgs})$ & $
\nu$ & $T_{\rm A}\ (\text{K})$ & Ref. \\ \hline
F1 & $\rm  C + CO^+ \to CO + C^+$ & $1.1 \times 10^{-10} $ & $0$ & $0$ & U12 \\
F2 & $\rm  C^+ + C_{2}O \to C_{2}O^+ + C$ & $1 \times 10^{-9} $ & $-0.5$ & $0$ & U12 \\
F7 & $\rm  C^+ + Fe \to Fe^+ + C$ & $2.6 \times 10^{-9} $ & $0$ & $0$ & U12 \\
F8 & $\rm  C^+ + Mg \to Mg^+ + C$ & $1.1 \times 10^{-9} $ & $0$ & $0$ & U12 \\
F11 & $\rm  C^+ + NO \to NO^+ + C$ & $7.05 \times 10^{-10} $ & $-0.03$ & $-16.7$ & U12 \\
F12 & $\rm  C^+ + NS \to NS^+ + C$ & $7.6 \times 10^{-10} $ & $-0.5$ & $0$ & U12 \\
F14 & $\rm  C^+ + SO \to SO^+ + C$ & $2.6 \times 10^{-10} $ & $-0.5$ & $0$ & U12 \\
\hline
\end{tabular}
\caption{Reaction Set F (Charge Exchange Reactions). Note Z04 refers to \citet{2004ApJ...615.1063Z}.  For all reactions involving He$^+$ we add an additional pair of reactions with He replaced by Ne and Ar. The rate coefficients for the reactions involving Ne and Ar are taken to be the same for He.}
\end{table*}

\clearpage

\begin{table*}
\begin{tabular}{ llllll }
\hline
Code & Reaction & $A\ (\text{cgs})$ & $
\nu$ & $T_{\rm A}\ (\text{K})$ & Ref. \\ \hline
F50 & $\rm  Mg + SO^+ \to SO + Mg^+$ & $1 \times 10^{-10} $ & $0$ & $0$ & U12 \\
F51 & $\rm  Mg + Si^+ \to Si + Mg^+$ & $2.9 \times 10^{-9} $ & $0$ & $0$ & U12 \\
F52 & $\rm  Mg + SiO^+ \to SiO + Mg^+$ & $1 \times 10^{-9} $ & $0$ & $0$ & U12 \\
F53 & $\rm  N^+ + C_{2} \to C_{2}^+ + N$ & $1 \times 10^{-9} $ & $0$ & $0$ & U12 \\
F54 & $\rm  N^+ + CN \to CN^+ + N$ & $1.1 \times 10^{-9} $ & $-0.5$ & $0$ & U12 \\
F55 & $\rm  N^+ + CO_{2} \to CO_{2}^+ + N$ & $7.5 \times 10^{-10} $ & $0$ & $0$ & U12 \\
F56 & $\rm  N^+ + CO \to CO^+ + N$ & $8.25 \times 10^{-10} $ & $0$ & $0$ & U12 \\
F57 & $\rm  N^+ + Fe \to Fe^+ + N$ & $1.5 \times 10^{-9} $ & $0$ & $0$ & U12 \\
F58 & $\rm  N^+ + Mg \to Mg^+ + N$ & $1.2 \times 10^{-9} $ & $0$ & $0$ & U12 \\
F59 & $\rm  N^+ + NO \to NO^+ + N$ & $4.51 \times 10^{-10} $ & $0$ & $0$ & U12 \\
F60 & $\rm  N^+ + O_{2} \to O_{2}^+ + N$ & $3.11 \times 10^{-10} $ & $0$ & $0$ & U12 \\
%
%
F62 & $\rm  N_{2}^+ + CO_{2} \to CO_{2}^+ + N_{2}$ & $7.7 \times 10^{-10} $ & $0$ & $0$ & U12 \\
F63 & $\rm  N_{2}^+ + Fe \to Fe^+ + N_{2}$ & $4.3 \times 10^{-10} $ & $0$ & $0$ & U12 \\
F64 & $\rm  N_{2}^+ + NO \to NO^+ + N_{2}$ & $4.4 \times 10^{-10} $ & $0$ & $0$ & U12 \\
F65 & $\rm  N_{2}^+ + O_{2} \to O_{2}^+ + N_{2}$ & $5 \times 10^{-11} $ & $0$ & $0$ & U12 \\
F67 & $\rm  N_{2}^+ + S \to S^+ + N_{2}$ & $1.1 \times 10^{-9} $ & $0$ & $0$ & U12 \\
F68 & $\rm  N + N_{2}^+ \to N_{2} + N^+$ & $1 \times 10^{-11} $ & $0$ & $0$ & U12 \\
F69 & $\rm  NO^+ + Fe \to Fe^+ + NO$ & $9.2 \times 10^{-10} $ & $0$ & $0$ & U12 \\
F70 & $\rm  NO + CO_{2}^+ \to CO_{2} + NO^+$ & $1.2 \times 10^{-10} $ & $0$ & $0$ & U12 \\
\hline
\end{tabular}
\caption{Reaction Set F (Charge Exchange Reactions, continued). }
\end{table*}

\begin{table*}
\begin{tabular}{ llllll }
\hline
Code & Reaction & $A\ (\text{cgs})$ & $
\nu$ & $T_{\rm A}\ (\text{K})$ & Ref. \\ \hline
G7 & $\rm  C_{2}^+ + e^- \to C + C$ & $3 \times 10^{-7} $ & $-0.5$ & $0$ & U12 \\
G12 & $\rm  C_{2}O^+ + e^- \to CO + C$ & $3 \times 10^{-7} $ & $-0.5$ & $0$ & U12 \\
G15 & $\rm  C_{3}^+ + e^- \to C_{2} + C$ & $3 \times 10^{-7} $ & $-0.5$ & $0$ & U12 \\
G17 & $\rm  C_{3}O^+ + e^- \to CO + C_{2}$ & $3 \times 10^{-7} $ & $-0.5$ & $0$ & U12 \\
G21 & $\rm  C_{4}^+ + e^- \to C_{2} + C_{2}$ & $1.2 \times 10^{-7} $ & $-0.5$ & $0$ & U12 \\
G22 & $\rm  C_{4}^+ + e^- \to C_{3} + C$ & $1.8 \times 10^{-7} $ & $-0.5$ & $0$ & U12 \\
G41 & $\rm  CN^+ + e^- \to N + C$ & $1.8 \times 10^{-7} $ & $-0.5$ & $0$ & U12 \\
\end{tabular}
%
G67 & $\rm  SiN^+ + e^- \to Si + N$ & $2 \times 10^{-7} $ & $-0.5$ & $0$ & U12 \\
G69 & $\rm  SiO^+ + e^- \to Si + O$ & $2 \times 10^{-7} $ & $-0.5$ & $0$ & U12 \\
G70 & $\rm  SiS^+ + e^- \to S + Si$ & $2 \times 10^{-7} $ & $-0.5$ & $0$ & U12 \\
\hline
\end{tabular}
\caption{Reaction Set G (Dissociative Recombination Reactions).}
\end{table*}
     
\begin{table*}
\begin{tabular}{ llllll }
\hline
Code & Reaction & $A\ (\text{cgs})$ & $
\nu$ & $T_{\rm A}\ (\text{K})$ & Ref. \\ \hline
H1 & $\rm  C^+ + e^- \to C + \gamma$ & $2.36 \times 10^{-12} $ & $-0.29$ & $-17.6$ & U12 \\
H2 & $\rm  Fe^+ + e^- \to Fe + \gamma$ & $2.55 \times 10^{-12} $ & $-0.69$ & $0$ & U12 \\
H3 & $\rm  Mg^+ + e^- \to Mg + \gamma$ & $2.78 \times 10^{-12} $ & $-0.68$ & $0$ & U12 \\
H4 & $\rm  N^+ + e^- \to N + \gamma$ & $3.5 \times 10^{-12} $ & $-0.53$ & $-3.2$ & U12 \\
H5 & $\rm  O^+ + e^- \to O + \gamma$ & $3.24 \times 10^{-12} $ & $-0.66$ & $0$ & U12 \\
H6 & $\rm  S^+ + e^- \to S + \gamma$ & $5.49 \times 10^{-12} $ & $-0.59$ & $0$ & U12 \\
H7 & $\rm  Si^+ + e^- \to Si + \gamma$ & $4.26 \times 10^{-12} $ & $-0.62$ & $0$ & U12 \\
H8 & $\rm  Al^+ + e^- \to Al + \gamma$ & $3.24 \times 10^{-12} $ & $-0.66$ & $0$ & $=$ H5 \\
H9 & $\rm  He^+ + e^- \to He + \gamma$ & $3.43 \times 10^{-12} $ & $-0.652$ & $0.735$ & HS98 \\
H10 & $\rm  Ne^+ + e^- \to Ne + \gamma$ & $3.43 \times 10^{-12} $ & $-0.652$ & $0.735$ & $=$ H9 \\
H11 & $\rm  Ar^+ + e^- \to Ar + \gamma$ & $3.43 \times 10^{-12} $ & $-0.652$ & $0.735$ & $=$ H9 \\
H12 & $\rm  Ti^+ + e^- \to Ti + \gamma$ & $2.55 \times 10^{-12} $ & $-0.69$ & $0$ & $=$ H2 \\
H13 & $\rm  V^+ + e^- \to V + \gamma$ & $2.55 \times 10^{-12} $ & $-0.69$ & $0$ & $=$ H2 \\
H14 & $\rm  Cr^+ + e^- \to Cr + \gamma$ & $2.55 \times 10^{-12} $ & $-0.69$ & $0$ & $=$ H2 \\
H15 & $\rm  Co^+ + e^- \to Co + \gamma$ & $2.55 \times 10^{-12} $ & $-0.69$ & $0$ & $=$ H2 \\
H16 & $\rm  Ni^+ + e^- \to Ni + \gamma$ & $2.55 \times 10^{-12} $ & $-0.69$ & $0$ & $=$ H2 \\
H17 & $\rm  Cu^+ + e^- \to Cu + \gamma$ & $2.55 \times 10^{-12} $ & $-0.69$ & $0$ & $=$ H2 \\
\hline
\end{tabular}
\caption{Reaction Set H (Radiative Recombination Reactions). Note that HS98 refers to \citet{1998MNRAS.297.1073H}.}
\end{table*}
     
\begin{table*}
\begin{tabular}{ llllll }
\hline
Code & Reaction & $A\ (\text{cgs})$ & $
\nu$ & $T_{\rm A}\ (\text{K})$ & Ref. \\ \hline
I1 & $\rm  Fe + Fe + M \to Fe_{2} + M$ & $2.76 \times 10^{-29} $ & $0$ & $0$ & G03 \\
I2 & $\rm  Fe_{2} + M \to Fe + Fe + M$ & $7.14 \times 10^{-5} $ & $0$ & $17800$ & G03 \\
I3 & $\rm  Fe + Fe_{2} + M \to Fe_{3} + M$ & $2.76 \times 10^{-29} $ & $0$ & $0$ & G03 \\
I4 & $\rm  Fe_{3} + M \to Fe_{2} + Fe + M$ & $1.66 \times 10^{-5} $ & $0$ & $19200$ & G03\\
I5 & $\rm  Fe + Fe_{3} + M \to Fe_{4} + M$ & $2.76 \times 10^{-29} $ & $0$ & $0$ & G03 \\
I6 & $\rm  Fe_{4} + M \to Fe_{3} + Fe + M$ & $8.3 \times 10^{-7} $ & $0$ & $21600$ & G03 \\
I7 & $\rm  Fe_{2} + Fe_{2} \to Fe_{3} + Fe$ & $8.3 \times 10^{-10} $ & $0$ & $0$ & G03 \\
\hline
\end{tabular}
\caption{Reaction Set I (Iron Nucleation Reactions).  For all reactions involving He$^+$ we add an additional pair of reactions with He replaced by Ne and Ar. The rate coefficients for the reactions involving Ne and Ar are taken to be the same for He. G03 refers to \citet{Giesen:03}. The oxygen and helium weathering, here denoted by ``O'' and ``He$^+$'', data are from \citep{Lazzati:16}.}
\end{table*}

\begin{table*}
\begin{tabular}{ llllll }
\hline
Code & Reaction & $A\ (\text{cgs})$ & $
\nu$ & $T_{\rm A}\ (\text{K})$ & Ref. \\ \hline
J1 & $\rm  Mg + Mg + M \to Mg_{2} + M$ & $2.76 \times 10^{-29} $ & $0$ & $0$ & $=$ I1 \\
J2 & $\rm  Mg_{2} + M \to Mg + Mg + M$ & $7.14 \times 10^{-5} $ & $0$ & $17800$ & $=$ I2 \\
J3 & $\rm  Mg + Mg_{2} + M \to Mg_{3} + M$ & $2.76 \times 10^{-29} $ & $0$ & $0$ & $=$ I3 \\
J4 & $\rm  Mg_{3} + M \to Mg_{2} + Mg + M$ & $1.66 \times 10^{-5} $ & $0$ & $19200$ & $=$ I4 \\
J5 & $\rm  Mg + Mg_{3} + M \to Mg_{4} + M$ & $2.76 \times 10^{-29} $ & $0$ & $0$ & $=$ I5 \\
J6 & $\rm  Mg_{4} + M \to Mg_{3} + Mg + M$ & $8.3 \times 10^{-7} $ & $0$ & $21600$ & $=$ I6 \\
J7 & $\rm  Mg_{2} + Mg_{2} \to Mg_{3} + Mg$ & $8.3 \times 10^{-10} $ & $0$ & $0$ & $=$ I7 \\
\hline
\end{tabular}
\caption{Reaction Set J (Magnesium Nucleation Reactions).  For all reactions involving He$^+$ we add an additional pair of reactions with He replaced by Ne and Ar. The rate coefficients for the reactions involving Ne and Ar are taken to be the same for He. The oxygen and helium weathering, here denoted by ``O'' and ``He$^+$'', data are from \citep{Lazzati:16}.}
\end{table*}

\begin{table*}
\begin{tabular}{ llllll }
\hline
Code & Reaction & $A\ (\text{cgs})$ & $
\nu$ & $T_{\rm A}\ (\text{K})$ & Ref. \\ \hline
K1 & $\rm  Si + Si + M \to Si_{2} + M$ & $2.76 \times 10^{-29} $ & $0$ & $0$ & $=$ I1 \\
K2 & $\rm  Si_{2} + M \to Si + Si + M$ & $7.14 \times 10^{-5} $ & $0$ & $17800$ & $=$ I2 \\
K3 & $\rm  Si + Si_{2} + M \to Si_{3} + M$ & $2.76 \times 10^{-29} $ & $0$ & $0$ & $=$ I3 \\
K4 & $\rm  Si_{3} + M \to Si_{2} + Si + M$ & $1.66 \times 10^{-5} $ & $0$ & $19200$ & $=$ I4 \\
K5 & $\rm  Si + Si_{3} + M \to Si_{4} + M$ & $2.76 \times 10^{-29} $ & $0$ & $0$ & $=$ I5 \\
K6 & $\rm  Si_{4} + M \to Si_{3} + Si + M$ & $8.3 \times 10^{-7} $ & $0$ & $21600$ & $=$ I6 \\
K7 & $\rm  Si_{2} + Si_{2} \to Si_{3} + Si$ & $8.3 \times 10^{-10} $ & $0$ & $0$ & $=$ I7 \\
\hline
\end{tabular}
\caption{Reaction Set K (Silicon Nucleation Reactions).  A97 refers to \citet{1997MNRAS.287..287A}. For all reactions involving He$^+$ we add an additional pair of reactions with He replaced by Ne and Ar. The rate coefficients for the reactions involving Ne and Ar are taken to be the same for He. The oxygen and helium weathering, here denoted by ``O'' and ``He$^+$'', data are from \citep{Lazzati:16}.}
\end{table*}

\begin{table*}
\begin{tabular}{ llllll }
\hline
Code & Reaction & $A\ (\text{cgs})$ & $
\nu$ & $T_{\rm A}\ (\text{K})$ & Ref. \\ \hline
L1 & $\rm  C_{3} + C + M \to C_{4} + M$ & $2.76 \times 10^{-29} $ & $0$ & $0$ & $=$ I5 \\
L2 & $\rm  C_{3} + M \to C_{2} + C + M$ & $1.66 \times 10^{-5} $ & $0$ & $19200$ & $=$ I4 \\
L3 & $\rm  C_{4} + M \to C_{3} + C + M$ & $8.3 \times 10^{-7} $ & $0$ & $21600$ & $=$ I6 \\
L4 & $\rm  C_{2} + C_{2} \to C_{3} + C$ & $8.3 \times 10^{-10} $ & $0$ & $0$ & $=$ I7 \\
L5 & $\rm  C_{3} + C_{2} \to C_{4} + C$ & $8.3 \times 10^{-10} $ & $0$ & $0$ & $=$ I8 \\
\hline
\end{tabular}
\caption{Reaction Set L (Carbon Nucleation Reactions).}
\end{table*}
     
\begin{table*}
\begin{tabular}{ llllll }
\hline
Code & Reaction & $A\ (\text{cgs})$ & $
\nu$ & $T_{\rm A}\ (\text{K})$ & Ref. \\ \hline
M1 & $\rm  Si_{2}O_{3} + Mg \to MgSi_{2}O_{3}$ & $1.59 \times 10^{-10} $ & $0.5$ & $0$ & Est. \\
M2 & $\rm  MgSi_{2}O_{4} + Mg \to Mg_{2}Si_{2}O_{4}$ & $2.01 \times 10^{-10} $ & $0.5$ & $0$ & Est. \\
M3 & $\rm  Mg_{2}Si_{2}O_{5} + Mg \to Mg_{3}Si_{2}O_{5}$ & $2.32 \times 10^{-10} $ & $0.5$ & $0$ & Est. \\
M4 & $\rm  Mg_{3}Si_{2}O_{7} + Mg \to Mg_{4}Si_{2}O_{7}$ & $2.6 \times 10^{-10} $ & $0.5$ & $0$ & Est. \\
M5 & $\rm  Mg_{2}Si_{2}O_{6} + Mg \to Mg_{3}Si_{2}O_{6}$ & $2.33 \times 10^{-10} $ & $0.5$ & $0$ & Est. \\
M6 & $\rm  Si_{2}O_{2} + O_{2} \to Si_{2}O_{3} + O$ & $8 \times 10^{-11} $ & $0.5$ & $0$ & Est. \\
M7 & $\rm  MgSi_{2}O_{3} + O_{2} \to MgSi_{2}O_{4} + O$ & $1.09 \times 10^{-10} $ & $0.5$ & $0$ & Est. \\
\hline
\end{tabular}
\caption{Reaction Set M (Silicate Nucleation Reactions). The reactions in this set are those listed in \citet{ZT:93}, \citet{2012MNRAS.420.3344G}, and \citet{Sarangi:13}.  The corresponding reaction rates were not taken from these cited works but were calculated as explained in Section~\ref{sec:estimatingRates}}.
\end{table*}

\begin{table*}
\begin{tabular}{ llllll }
\hline
Code & Reaction & $A\ (\text{cgs})$ & $
\nu$ & $T_{\rm A}\ (\text{K})$ & Ref. \\ \hline
M20 & $\rm  SiO + SiO + M \to Si_{2}O_{2} + M$ & $1.18 \times 10^{-26} $ & $-0.87$ & $0$ & Est. \\
M21 & $\rm  Si_{2}O_{2} + SiO \to Si_{3}O_{3}$ & $9.98 \times 10^{-11} $ & $0.5$ & $0$ & Est. \\
M22 & $\rm  Si_{3}O_{3} + SiO \to Si_{4}O_{4}$ & $1.1 \times 10^{-10} $ & $0.5$ & $0$ & Est. \\
M23 & $\rm  Si_{2}O_{2} + M \to SiO + SiO + M$ & $1.88 \times 10^{-8} $ & $-0.5$ & $24775$ & Est. \\
M24 & $\rm  Si_{3}O_{3} + M \to Si_{2}O_{2} + SiO + M$ & $2.38 \times 10^{-8} $ & $-0.5$ & $27096$ & Est. \\
M25 & $\rm  Si_{4}O_{4} + M \to Si_{3}O_{3} + SiO + M$ & $1.85 \times 10^{-8} $ & $-0.5$ & $18846$ & Est. \\
M26 & $\rm  Si_{2}O_{2} + Si_{2}O_{2} \to Si_{3}O_{3} + SiO$ & $1.01 \times 10^{-10} $ & $0.5$ & $0$ & Est. \\
\hline
\end{tabular}
\caption{Reaction Set M (Silicate Nucleation Reactions, continued).}
\end{table*}

     
\begin{table*}
\begin{tabular}{ llllll }
\hline
Code & Reaction & $A\ (\text{cgs})$ & $
\nu$ & $T_{\rm A}\ (\text{K})$ & Ref. \\ \hline
N1 & $\rm  AlO + AlO + M \to Al_{2}O_{2} + M$ & $1.18 \times 10^{-26} $ & $-0.87$ & $0$ & $=$ M20 \\
N2 & $\rm  Al_{2}O_{2} + O_{2} \to Al_{2}O_{3} + O$ & $8 \times 10^{-11} $ & $0.5$ & $0$ & $=$ M6 \\
N3 & $\rm  Al_{2}O_{2} + SO \to Al_{2}O_{3} + S$ & $9.05 \times 10^{-11} $ & $0.5$ & $0$ & $=$ M13 \\
N4 & $\rm  Al_{2}O_{3} + Al_{2}O_{3} \to Al_{4}O_{6}$ & $1.22 \times 10^{-10} $ & $0.5$ & $0$ & Est. \\
N5 & $\rm  Al_{2}O_{3} + M \to Al_{2}O_{2} + O + M$ & $4.3 \times 10^{-8} $ & $-0.5$ & $55354$ & $=$ M47 \\
N6 & $\rm  Al_{4}O_{6} + M \to Al_{2}O_{3} + Al_{2}O_{3} + M$ & $4.94 \times 10^{-8} $ & $-0.5$ & $55354$ & $\propto$ N5 \\
N7  & $\rm  Al_{2}O_{2} + M \to AlO + AlO + M$ &  $1.88\times 10^{-8}$  & $-0.5$ &  $24776$ & $\propto$ N5 \\
\hline
\end{tabular}
\caption{Reaction Set N (Alumina Nucleation Reactions).   The rates with ``Est.'' in the ``Ref.'' column were estimated as explained in Section~\ref{sec:estimatingRates}.}
\end{table*}
     
\begin{table*}
\begin{tabular}{ llllll }
\hline
Code & Reaction & $A\ (\text{cgs})$ & $$=$ M
\nu$ & $T_{\rm A}\ (\text{K})$ & Ref. \\ \hline
O1 & $\rm  Fe + O_{2} \to FeO + O$ & $1.72 \times 10^{-10} $ & $-0.53$ & $17$ & $=$ D141 \\
O2 & $\rm  Fe + CO_{2} \to FeO + CO$ & $2.72 \times 10^{-11} $ & $0$ & $282$ & $=$ D138 \\
O3 & $\rm  FeO + O \to Fe + O_{2}$ & $1 \times 10^{-16} $ & $0$ & $0$ & $=$ D143 \\
O4 & $\rm  FeO + FeO + M \to Fe_{2}O_{2} + M$ & $1.18 \times 10^{-26} $ & $-0.87$ & $0$ & $=$ M20 \\
O5 & $\rm  Fe_{2}O_{2} + M \to FeO + FeO + M$ & $1.88 \times 10^{-8} $ & $-0.5$ & $24776$ & $=$ M23 \\
O6 & $\rm  Fe_{2}O_{2} + FeO \to Fe_{3}O_{3}$ & $9.98 \times 10^{-11} $ & $0.5$ & $0$ & $=$ M21 \\
O7 & $\rm  Fe_{3}O_{3} + M \to Fe_{2}O_{2} + FeO + M$ & $2.38 \times 10^{-8} $ & $-0.5$ & $27096$ & $=$ M24 \\
O8 & $\rm  Fe_{3}O_{3} + FeO \to Fe_{4}O_{4}$ & $1.1 \times 10^{-10} $ & $0.5$ & $0$ & $=$ M22 \\
O9 & $\rm  Fe_{4}O_{4} + M \to Fe_{3}O_{3} + FeO + M$ & $1.85 \times 10^{-8} $ & $-0.5$ & $18846$ & $=$ M25 \\
O10 & $\rm  Fe_{4}O_{4} + M \to Fe_{2}O_{2} + Fe_{2}O_{2} + M$ & $5.43 \times 10^{-8} $ & $-0.5$ & $55354$ & $=$ M31 \\
\hline
\end{tabular}
\caption{Reaction Set O (Iron Oxide Nucleation Reactions).}
\end{table*}

\begin{table*}
\begin{tabular}{ llllll }
\hline
Code & Reaction & $A\ (\text{cgs})$ & $
\nu$ & $T_{\rm A}\ (\text{K})$ & Ref. \\ \hline
P1 & $\rm  Fe + SO \to FeS + O$ & $1.72 \times 10^{-10} $ & $-0.53$ & $17$ & $=$ D141 \\
P2 & $\rm  Fe + S_{2} \to FeS + S$ & $1.72 \times 10^{-10} $ & $-0.53$ & $17$ & $=$ D141 \\
P3 & $\rm  FeS + O \to Fe + SO_{2}$ & $9.97 \times 10^{-11} $ & $0.5$ & $0$ & Est.\\
P4 & $\rm  FeS + S \to Fe + S_{2}$ & $1.06 \times 10^{-10} $ & $0.5$ & $0$ & Est. \\
P5 & $\rm  FeS + FeS + M \to Fe_{2}S_{2} + M$ & $1.18 \times 10^{-26} $ & $-0.873$ & $0$ & $=$ M20 \\
P6 & $\rm  Fe_{2}S_{2} + M \to FeS + FeS + M$ & $1.88 \times 10^{-8} $ & $-0.5$ & $24776$ & $=$ M23 \\
P7 & $\rm  Fe_{2}S_{2} + FeS \to Fe_{3}S_{3}$ & $9.98 \times 10^{-11} $ & $0.5$ & $0$ & $=$ M21 \\
P8 & $\rm  Fe_{3}S_{3} + M \to Fe_{2}S_{2} + FeS + M$ & $2.38 \times 10^{-8} $ & $-0.5$ & $27096$ & $=$ M24 \\
P9 & $\rm  Fe_{3}S_{3} + FeS \to Fe_{4}S_{4}$ & $1.1 \times 10^{-10} $ & $0.5$ & $0$ & $=$ M22 \\
P10 & $\rm  Fe_{4}S_{4} + M \to Fe_{3}S_{3} + FeS + M$ & $1.85 \times 10^{-8} $ & $-0.5$ & $18846$ & $=$ M25 \\
P11 & $\rm  Fe_{4}S_{4} + M \to Fe_{2}S_{2} + Fe_{2}S_{2} + M$ & $5.43 \times 10^{-8} $ & $-0.5$ & $55354$ & $=$ M44 \\
\hline
\end{tabular}
\caption{Reaction Set P (Iron Sulfide Nucleation Reactions).   The rates with ``Est.'' in the ``Ref.'' column were estimated as explained in Section~\ref{sec:estimatingRates}.}
\end{table*}
     
\begin{table*}
\begin{tabular}{ llllll }
\hline
Code & Reaction & $A\ (\text{cgs})$ & $
\nu$ & $T_{\rm A}\ (\text{K})$ & Ref. \\ \hline
Q1 & $\rm  Mg + O_{2} \to MgO + O$ & $1.72 \times 10^{-10} $ & $-0.53$ & $17$ & $=$ D141 \\
Q2 & $\rm  Mg + CO_{2} \to MgO + CO$ & $2.72 \times 10^{-11} $ & $0$ & $282$ & $=$ D138 \\
Q3 & $\rm  MgO + O \to Mg + O_{2}$ & $1 \times 10^{-16} $ & $0$ & $0$ & $=$ D143 \\
Q4 & $\rm  MgO + MgO + M \to Mg_{2}O_{2} + M$ & $1.18 \times 10^{-26} $ & $-0.87$ & $0$ & $=$ M20 \\
Q5 & $\rm  Mg_{2}O_{2} + M \to MgO + MgO + M$ & $1.88 \times 10^{-8} $ & $-0.5$ & $24776$ & $=$ M23 \\
Q6 & $\rm  Mg_{2}O_{2} + MgO \to Mg_{3}O_{3}$ & $9.98 \times 10^{-11} $ & $0.5$ & $0$ & $=$ M21 \\
Q7 & $\rm  Mg_{3}O_{3} + M \to Mg_{2}O_{2} + MgO + M$ & $2.38 \times 10^{-8} $ & $-0.5$ & $27096$ & $=$ M24 \\
Q8 & $\rm  Mg_{3}O_{3} + MgO \to Mg_{4}O_{4}$ & $1.1 \times 10^{-10} $ & $0.5$ & $0$ & $=$ M22 \\
Q9 & $\rm  Mg_{4}O_{4} + M \to Mg_{3}O_{3} + MgO + M$ & $1.84 \times 10^{-8} $ & $-0.5$ & $18846$ & $=$ M25 \\
Q10 & $\rm  Mg_{4}O_{4} + M \to Mg_{2}O_{2} + Mg_{2}O_{2} + M$ & $5.43 \times 10^{-8} $ & $-0.5$ & $55354$ & $=$ M44 \\
\hline
\end{tabular}
\caption{Reaction Set Q (Magnesia Nucleation Reactions).}
\end{table*}
     
\begin{table*}
\begin{tabular}{ llllll }
\hline
Code & Reaction & $A\ (\text{cgs})$ & $
\nu$ & $T_{\rm A}\ (\text{K})$ & Ref. \\ \hline
R1 & $\rm  Mg + SO \to MgS + O$ & $1.72 \times 10^{-10} $ & $-0.53$ & $17$ & $=$ D141 \\
R2 & $\rm  Mg + S_{2} \to MgS + S$ & $1.72 \times 10^{-10} $ & $-0.53$ & $17$ & $=$ D141 \\
R3 & $\rm  MgS + O \to Mg + SO_{2}$ & $9.97 \times 10^{-11} $ & $0.5$ & $0$ & Est. \\
R4 & $\rm  MgS + S \to Mg + S_{2}$ & $1.06 \times 10^{-10} $ & $0.5$ & $0$ & Est. \\
R5 & $\rm  MgS + MgS + M \to Mg_{2}S_{2} + M$ & $1.18 \times 10^{-26} $ & $-0.87$ & $0$ & $=$ M20 \\
R6 & $\rm  Mg_{2}S_{2} + M \to MgS + MgS + M$ & $1.88 \times 10^{-8} $ & $-0.5$ & $24776$ & $=$ M23 \\
R7 & $\rm  Mg_{2}S_{2} + MgS \to Mg_{3}S_{3}$ & $9.98 \times 10^{-11} $ & $0.5$ & $0$ & $=$ M21 \\
R8 & $\rm  Mg_{3}S_{3} + M \to Mg_{2}S_{2} + MgS + M$ & $2.38 \times 10^{-8} $ & $-0.5$ & $27096$ & $=$ M24 \\
R9 & $\rm  Mg_{3}S_{3} + MgS \to Mg_{4}S_{4}$ & $1.1 \times 10^{-10} $ & $0.5$ & $0$ & $=$ M22 \\
R10 & $\rm  Mg_{4}S_{4} + M \to Mg_{3}S_{3} + MgS + M$ & $1.85 \times 10^{-8} $ & $-0.5$ & $18846$ & $=$ M25 \\
R11 & $\rm  Mg_{4}S_{4} + M \to Mg_{2}S_{2} + Mg_{2}S_{2} + M$ & $5.43 \times 10^{-8} $ & $-0.5$ & $55354$ & $=$ M44 \\
\hline
\end{tabular}
\caption{Reaction Set R (Magnesium Sulfide Nucleation Reactions).   The rates with ``Est.'' in the ``Ref.'' column were estimated as explained in Section~\ref{sec:estimatingRates}.}
\end{table*}
     
\begin{table*}
\begin{tabular}{ llllll }
\hline
Code & Reaction & $A\ (\text{cgs})$ & $
\nu$ & $T_{\rm A}\ (\text{K})$ & Ref. \\ \hline
S1 & $\rm  Fe_{2}O_{2} + O_{2} \to Fe_{2}O_{3} + O$ & $8 \times 10^{-11} $ & $0.5$ & $0$ & $=$ M6 \\
S2 & $\rm  Fe_{2}O_{2} + SO \to Fe_{2}O_{3} + S$ & $9.05 \times 10^{-11} $ & $0.5$ & $0$ & $=$ M6 \\
S3 & $\rm  Fe_{2}O_{3} + FeO \to Fe_{3}O_{4}$ & $9.97 \times 10^{-11} $ & $0.5$ & $0$ & $=$ M39 \\
S4 & $\rm  Fe_{3}O_{3} + O_{2} \to Fe_{3}O_{4} + O$ & $9.17 \times 10^{-11} $ & $0.5$ & $0$ & $=$ M35 \\
S5 & $\rm  Fe_{3}O_{3} + SO \to Fe_{3}O_{4} + S$ & $9.98 \times 10^{-11} $ & $0.5$ & $0$ & $=$ M36 \\
S6 & $\rm  Fe_{3}O_{4} + Fe_{3}O_{4} \to Fe_{6}O_{8}$ & $1.28 \times 10^{-10} $ & $0.5$ & $0$ & Est. \\
S7 & $\rm  Fe_{2}O_{3} + M \to Fe_{2}O_{2} + O + M$ & $4.3 \times 10^{-8} $ & $-0.5$ & $55354$ & $=$ M47 \\
S8 & $\rm  Fe_{3}O_{4} + M \to Fe_{3}O_{3} + O + M$ & $4.94 \times 10^{-8} $ & $-0.5$ & $55354$ & $=$ M48 \\
S9 & $\rm  Fe_{6}O_{8} + M \to Fe_{3}O_{4} + Fe_{3}O_{4} + M$ & $4.94 \times 10^{-8} $ & $-0.5$ & $55354$ & $=$ M48 \\
\hline
\end{tabular}
\caption{Reaction Set S (Magnetite Nucleation Reactions).    The rate with ``Est.'' in the ``Ref.'' column was estimated as explained in Section~\ref{sec:estimatingRates}.}
\end{table*}
     
\begin{table*}
\begin{tabular}{ llllll }
\hline
Code & Reaction & $A\ (\text{cgs})$ & $
\nu$ & $T_{\rm A}\ (\text{K})$ & Ref. \\ \hline
T1 & $\rm  Si + CO \to SiC + O$ & $2.94 \times 10^{-11} $ & $0.5$ & $58025$ & $=$ D1 \\
T2 & $\rm  SiC + O \to Si + CO$ & $2 \times 10^{-10} $ & $-0.12$ & $0$ & $=$ D2 \\
T3 & $\rm  SiO + C \to SiC + O$ & $2.94 \times 10^{-11} $ & $0.5$ & $58025$ & $=$ D1 \\
T4 & $\rm  SiC + O \to SiO + C$ & $2 \times 10^{-10} $ & $-0.12$ & $0$ & $=$ D2 \\
T5 & $\rm  C_{2} + Si \to SiC + C$ & $2 \times 10^{-10} $ & $-0.12$ & $0$ & $=$ D2 \\
T6 & $\rm  SiC + SiC + M \to Si_{2}C_{2} + M$ & $1.18 \times 10^{-26} $ & $-0.87$ & $0$ & $=$ M20 \\
T7 & $\rm  Si_{2}C_{2} + M \to SiC + SiC + M$ & $1.88 \times 10^{-8} $ & $-0.5$ & $24776$ & $=$ M23 \\
\hline
\end{tabular}
\caption{Reaction Set T (Silicon Carbide Nucleation Reactions).}
\end{table*}
     
\begin{table*}
\begin{tabular}{ llllll }
\hline
Code & Reaction & $A\ (\text{cgs})$ & $
\nu$ & $T_{\rm A}\ (\text{K})$ & Ref. \\ \hline
U1 & $\rm  O + SiO + M \to SiO_{2} + M$ & $1.2 \times 10^{-32} $ & $0$ & $2160$ & $=$ B12 \\
U2 & $\rm  SiO_{2} + M \to SiO + O + M$ & $4.17 \times 10^{-11} $ & $0$ & $43779$ & $=$ C7 \\
U3 & $\rm  SiO + O_{2} \to SiO_{2} + O$ & $5.99 \times 10^{-12} $ & $0$ & $24075$ & $=$ D43 \\
U4 & $\rm  O + SiO_{2} \to O_{2} + SiO$ & $2.46 \times 10^{-11} $ & $0$ & $26567$ & $=$ D110 \\
U5 & $\rm  Si + SiO_{2} \to SiO + SiO$ & $1 \times 10^{-15} $ & $0$ & $0$ & $=$ D146 \\
U6 & $\rm  SiO_{2} + SiO_{2} \to Si_{2}O_{4}$ & $8.43 \times 10^{-11} $ & $0.5$ & $0$ & Est. \\
U7 & $\rm  Si_{2}O_{4} + M \to SiO_{2} + SiO_{2} + M$ & $4.17 \times 10^{-11} $ & $0$ & $43779$ & $=$ C7 \\
\hline
\end{tabular}
\caption{Reaction Set U (Silicon Dioxide Nucleation Reactions).   The rate with ``Est.''  in the ``Ref.''\ column was estimated as explained in Section~\ref{sec:estimatingRates}.}
\label{tab:bigReactionTableLast}
\end{table*}

\label{lastpage}

\end{document}